\documentclass[aps,prd,onecolumn,groupedaddress,nofootinbib,preprintnumbers,superscriptaddress]{revtex4-2}  
\usepackage{graphicx}
\usepackage{amsmath,amssymb,latexsym,bbold}
\usepackage{amsfonts}
\usepackage[colorlinks=true,citecolor=red,linkcolor=blue,breaklinks=true]{hyperref}
\usepackage[figure, table]{hypcap}
\usepackage{xcolor}
\usepackage{subfigure}
\usepackage{setspace}
\usepackage{footnote}
\usepackage{multirow}
\usepackage[normalem]{ulem}
\usepackage[utf8]{inputenc}
\usepackage{mathrsfs}
\usepackage{slashed}
\usepackage{longtable}
\usepackage{enumitem}
\usepackage[scr=boondox]{mathalpha}
\usepackage[many]{tcolorbox}
\numberwithin{equation}{section}

\allowdisplaybreaks

\definecolor{tabred}{rgb}{0.8392156862745098, 0.15294117647058825, 0.1568627450980392}
\definecolor{tabblue}{rgb}{0.12156862745098039, 0.4666666666666667, 0.7058823529411765}
\hypersetup{colorlinks=true, linkcolor={tabred}, citecolor={tabblue}, urlcolor={tabblue}} 

\usepackage{orcidlink}

\usepackage{fontawesome5}
\definecolor{color_git}{rgb}{0.098, 0.160, 0.345}
\newcommand{\gitlink}{\href{https://github.com/mariofnavarro/COFLASY}{\textsc{g}it\textsc{h}ub {\large\color{color_git}\faGithub}}}

\newcommand{\orcid}[1]{\,\orcidlink{#1}}

\newcommand{\dd}{\mathrm{d}}
\newcommand{\fd}{f_\mathrm{fd}}

\begin{document}

\preprint{CERN-TH-2025-010}
\preprint{LA-UR-25-20235}

\vspace*{1cm}
\title{
Lepton Flavor Asymmetries: from the early Universe to BBN
}

\author{Valerie Domcke\orcid{0000-0002-7208-4464}}
\email{valerie.domcke@cern.ch}
\affiliation{Theoretical Physics Department, CERN, 1 Esplanade des Particules, CH-1211 Geneva 23, Switzerland}

\author{Miguel Escudero\orcid{0000-0002-4487-8742}}
\email{miguel.escudero@cern.ch}
\affiliation{Theoretical Physics Department, CERN, 1 Esplanade des Particules, CH-1211 Geneva 23, Switzerland}

\author{Mario Fern\'andez Navarro\orcid{0000-0002-8796-0172}}
\email{mario.fernandeznavarro@glasgow.ac.uk}
\affiliation{School of Physics \& Astronomy, University of Glasgow, Glasgow G12 8QQ, UK}

\author{Stefan Sandner\orcid{0000-0002-1802-9018}\,}
\email{stefan.sandner@lanl.gov}
\affiliation{Theoretical Division, Los Alamos National Laboratory, Los Alamos, NM 87545, USA}

\begin{abstract}
\vspace*{1cm}
\noindent
Large primordial lepton flavor asymmetries with almost vanishing total baryon-minus-lepton number can evade the usual BBN and CMB constraints if neutrino oscillations lead to perfect flavor equilibration. Solving the momentum averaged quantum kinetic equations (QKEs) describing neutrino oscillations and interactions, we perform the first systematic investigation of this scenario, uncovering a rich flavor structure in stark contradiction to the assumption of simple flavor equilibration. We find  (i) a particular direction in flavor space, $\Delta n_e \simeq -2/3 \, (-1)  \Delta n_\mu$ for normal (inverted) neutrino mass hierarchy, in which the flavor equilibration is efficient and primordial asymmetries are essentially unconstrained, (ii) a minimal washout factor, $\Delta n_e^2|_\mathrm{BBN} \leq 0.03 \, (0.016) \sum_\alpha \Delta n_\alpha^2|_{\mathrm{ini}}$ yielding a conservative estimate for the allowed primordial asymmetries in a generic flavor direction, and (iii) particularly strong or weak washout if one of the initial flavor asymmetries vanishes due to non-adiabatic muon- or electron-driven MSW transitions.  These results open up the possibility of a first-order QCD phase transition facilitated by large lepton asymmetries as well as baryogenesis from large and compensated $\Delta n_e =- \Delta n_\mu$ asymmetries.  Our systematic approach of deriving momentum averaged QKEs includes collision terms beyond the damping approximation, energy transfer between the neutrino and electron-photon plasma, and provides a fast and reliable way to investigate the impact of primordial lepton asymmetries at the time of BBN.
We publicly release the Mathematica code \texttt{COFLASY-M} on~\gitlink\;which solves the QKEs numerically.
\end{abstract}

\maketitle

\newpage
{
\hypersetup{linkcolor=black}
\tableofcontents
}

\noindent\makebox[\linewidth]{\rule{0.8 \paperwidth}{0.2pt}}
\vspace{1 mm}
\vspace{-4.5mm}

\section{Introduction}
\label{sec:intro}

\setlength{\parskip}{1pt}

The presence of a matter-antimatter asymmetry in our Universe, parametrized by the baryon-to-photon ratio $ n_b/n_\gamma = (6.12 \pm 0.04) \times 10^{-10}$~\cite{Planck:2018vyg} is evidence of $\mathrm{CP}$ violation in our cosmic history~\cite{sakharov}. The smallness of this number may be either pointing to a process which only mildly violates $\mathrm{CP}$ (such as in traditional thermal leptogenesis~\cite{Fukugita:1986hr}) or alternatively to an inefficient conversion of primordial $\mathrm{CP}$ asymmetries to baryon number (see e.g. \cite{Kuzmin:1987wn}). Since baryon-minus-lepton number ($B$$-$$L$) is conserved in the Standard Model (SM) while $B$$+$$L$ asymmetries are erased by sphaleron processes, the latter case would point to the presence of large primordial quark and lepton flavor asymmetries in the early Universe, subject to the condition of (approximately) zero total $B$$-$$L$ and charge neutrality. The possibility of large flavor asymmetries in the early Universe has intriguing consequences for baryogenesis (see e.g.~\cite{Cohen:1987vi,Cohen:1988kt,Davidson:1994gn,March-Russell:1999hpw,Yamaguchi:2002vw,Kawasaki:2002hq,Chiba:2003vp,Takahashi:2003db,Asaka:2005pn,Shaposhnikov:2008pf,Laine:2008pg,Kamada:2018tcs,Domcke:2019mnd,Domcke:2020quw,Mukaida:2021sgv}), the nature of the QCD phase transition~\cite{Gao:2021nwz,Gao:2023djs,Gao:2024fhm}, dark matter production~\cite{Shi:1998km,Shaposhnikov:2023hrx,Stuke:2011wz}, the detection prospects of the cosmic neutrino background~\cite{Stodolsky:1974aq,Duda:2001hd}, the presence of cosmic magnetic fields~\cite{Joyce:1997uy,Boyarsky:2011uy,Akamatsu:2013pjd,Hirono:2015rla,Rogachevskii:2017uyc,Domcke:2022uue}, the abundance of light elements~\cite{Froustey:2021azz,Froustey:2024mgf,Escudero:2022okz,Burns:2022hkq,Kawasaki:2022hvx,March-Russell:1999hpw,Kohri:1996ke} and motivates searches for traces for $\mathrm{CP}$ violation in the cosmic microwave~\cite{Minami:2020odp} or gravitational wave backgrounds~\cite{Seto:2006hf,Seto:2006dz,Domcke:2019zls}. As the Universe cools, taking into account in particular the sphaleron processes of the SM which erase any $B$$+$$L$ charge as well as imposing charge neutrality, large flavor asymmetries can only be stored in the neutrino sector at temperatures below the QCD phase transition.

Observational constraints on lepton flavor asymmetries are surprisingly scarce. The most robust probes of these are Big Bang Nucleosynthesis (BBN) and observations of the cosmic microwave background (CMB). The former is primarily sensitive to asymmetries in the electron neutrino flavor at temperatures around $1\,\mathrm{MeV}$, which according to the latest global analysis is constrained to $-0.025 < (\mu_e/T)^\mathrm{BBN} < 0.027$~\cite{Froustey:2024mgf} at $95$\% CL, with $\mu_e$ denoting the electron neutrino chemical potential, see also~\cite{Escudero:2022okz,Burns:2022hkq}. In addition, both BBN and the CMB probe the total radiation content of the Universe, which receives contributions from lepton flavor asymmetries at ${\cal O}(\mu_\alpha/T)^2$.

Critically, BBN and CMB observations are only sensitive to the primordial lepton asymmetries after neutrino decoupling at $T\simeq 1\,{\rm MeV}$. However, given the observed pattern of neutrino masses and mixings~\cite{Esteban:2024eli}, neutrinos start oscillating in the early Universe at temperatures $T \sim 15\,{\rm MeV}$. These flavor changing oscillations effectively lead to a washout of primordial lepton asymmetries. If this washout were perfect, BBN would not constrain primordial asymmetries with approximately vanishing total lepton number.

Pioneering work~\cite{Dolgov:2002ab,Pastor:2008ti,Mangano:2010ei,Castorina:2012md} indicated that the assumption of total flavor equilibration does not hold in this case. These works moreover revealed a strong sensitivity to the mixing angle $\theta_{13}$, which was poorly measured at the time. Subsequent recent work by Froustey \& Pitrou~\cite{Froustey:2021azz,Froustey:2024mgf} has significantly improved the methodology of solving the quantum kinetic equations (QKEs), including retaining the full momentum dependence of the distribution functions as well as using exact collision terms describing the interactions of the neutrinos among themselves and with the thermal plasma. The resulting numerical codes represent the state-of-the-art understanding of QKEs, but are numerically expensive as oscillations and interactions with vastly different time scales need to be tracked for all momentum modes. The case of interest here, featuring significant lepton flavor asymmetries with vanishing total lepton number, has proven particularly challenging~\cite{Froustey:2024mgf} and has not been analyzed in a systematic way.

\begin{figure}[!t]
\centering
\begin{tabular}{cc}
\hspace{-0.075cm}\includegraphics[width=0.5\textwidth]{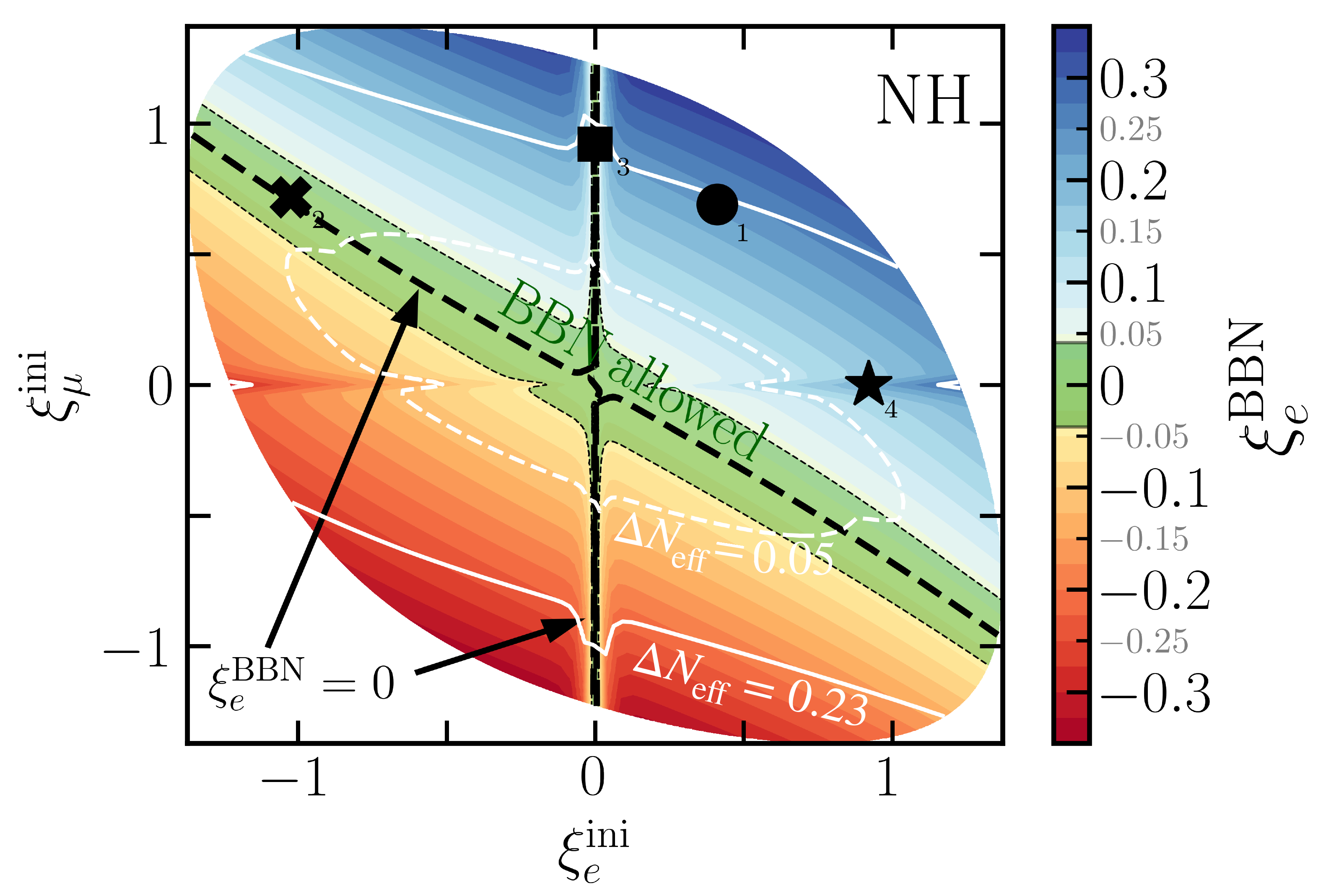}
\hspace{-0.075cm}\includegraphics[width=0.5\textwidth]{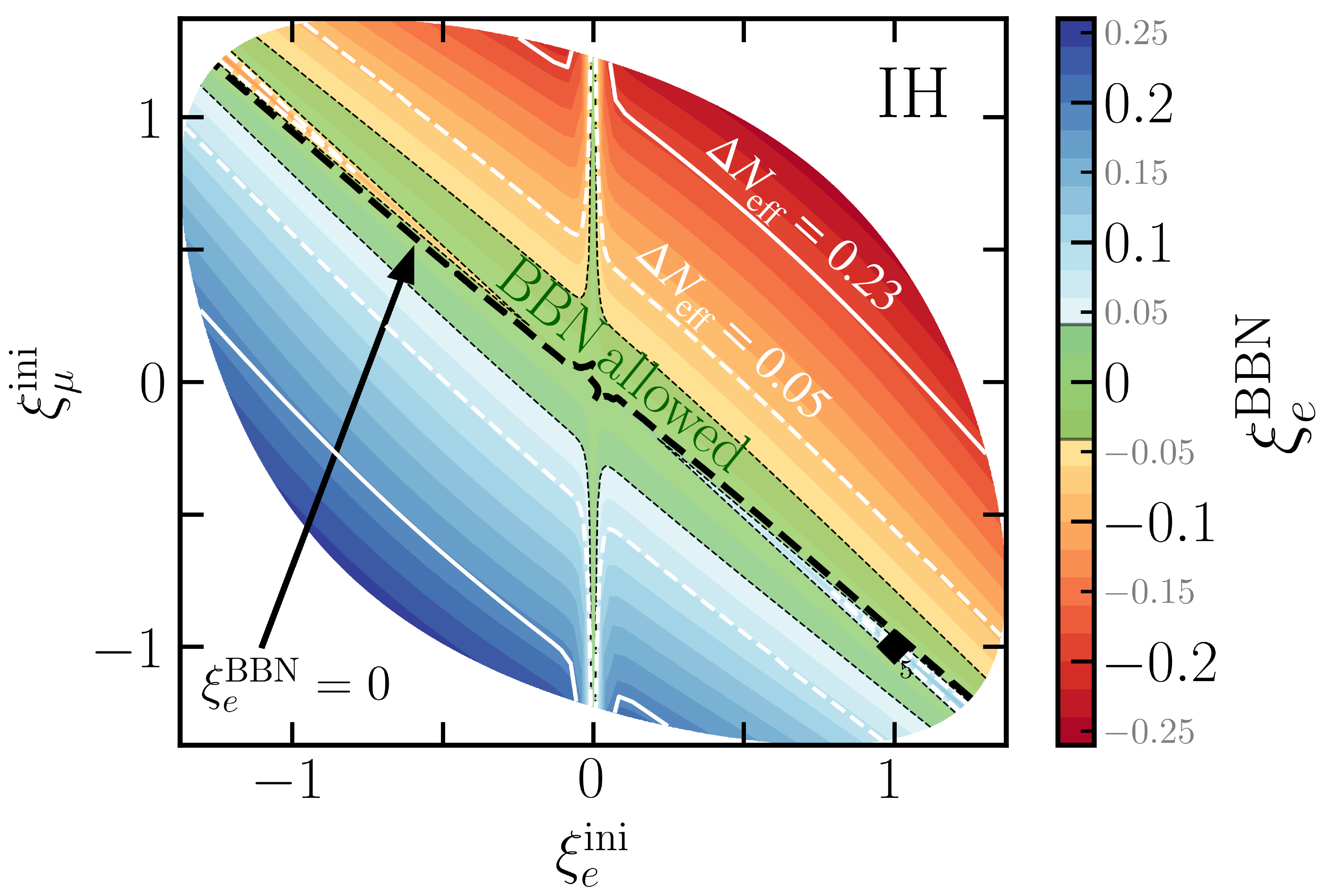}
\end{tabular}
\vspace{-0.2cm}
\caption{Asymmetry in the electron neutrino sector at the onset of BBN ($\xi_e^\mathrm{BBN} = \mu_e/T|_\mathrm{BBN}$) as a function of the primordial lepton flavor asymmetries $\xi_e^\mathrm{ini}$ and $\xi_\mu^\mathrm{ini}$ (with $\Delta n_e^\mathrm{ini} + \Delta n_\mu^\mathrm{ini} +  \Delta n_\tau^\mathrm{ini} = 0$) present before the onset of neutrino oscillations at $20\,\mathrm{MeV}$. The green region indicates the region currently allowed by BBN constraints~\cite{Froustey:2024mgf} at $95$\% CL and the black symbols indicate particular parameter points that will be discussed in detail in Sec.~\ref{sec:results}. Almost the entire parameter space shown is compatible with Planck CMB observations ($\Delta N_{\rm eff} < 0.23$). The two panels show our result for normal (left) and inverted (right) neutrino mass hierarchy.
}
\label{fig:constraints_intro}
\end{figure}

The goal of this work is to address precisely this short-coming and answer the following question: What is the upper bound on primordial lepton flavor asymmetries (subject to approximately vanishing $B$$-$$L$) which are compatible with BBN and CMB observations? A summary of our results, answering this question, is shown in Fig.~\ref{fig:constraints_intro}. 
We find that the answer depends on the flavor decomposition of the primordial asymmetries as well as on the choice of normal hierarchy (NH) or inverted hierarchy (IH) in the neutrino masses. However, for both mass hierarchies, there is a particular direction in flavor space (shown as the green band in Fig.~\ref{fig:constraints_intro}) for which flavor equilibration is efficient and the asymmetry in the electron neutrino evaluated at $T = 1$~MeV ($\xi_e^\mathrm{BBN} = (\mu_e/T)^\mathrm{BBN}$) is (nearly) completely erased. Further substructures in flavor space appear due to non-adiabatic muon- and electron-driven Mikheyev-Smirnonv-Wolfenstein (MSW) transitions.

Our results allow to confront early Universe cosmologies featuring significant primordial asymmetries with constraints from the light element abundances produced during BBN. For example, a particularly interesting scenario is the proposal of large lepton flavor asymmetries leading to a first-order QCD phase transition~\cite{Gao:2021nwz,Gao:2024fhm,Zheng:2024tib,Cline:2025bwe} and which could simultaneously yield successful baryogenesis through leptoflavorgenesis~\cite{Kuzmin:1987wn,Khlebnikov:1988sr,March-Russell:1999hpw,Laine:1999wv,Shu:2006mm,Gu:2010dg,Mukaida:2021sgv}. Notably, the tauphobic direction, i.e.\ $\xi_e^\mathrm{ini} = -\xi_\mu^\mathrm{ini}$ and $\xi_\tau^\mathrm{ini} = 0$, can achieve both with asymmetries of $|\xi_{e,\mu}| = {\cal O}(0.1)$. The new constraints resulting from our analysis are precisely at this level, and we find this scenario to be compatible with current BBN limits and testable in the near future, in particular for the case of IH.\footnote{Stringent constraints on this scenario have been set in Ref.~\cite{Domcke:2022uue}. They may however be avoided if the large lepton asymmetries are sourced just before the electroweak phase transition, at temperatures $T \simeq 10^2..10^5$~GeV.}

The method developed in this work is based on momentum averaged QKEs. We provide a self-contained, systematic derivation of these equations, including expressions for the collision terms which are exact under this ansatz and notably are not of the damping form often assumed in earlier studies. These collision terms also control the energy transfer between the neutrino sector and the electron-photon plasma, enabling us to dynamically track the respective temperatures, crucial for computing the contribution to $N_\text{eff}$. We discuss cross-checks of this procedure with the full momentum dependent code used in Ref.~\cite{Froustey:2021azz,Froustey:2024mgf}. Our procedure gives a fast and reliable estimate of the final lepton asymmetries and allows for an extremely efficient exploration of the full parameter space of early universe neutrino oscillations including lepton flavor asymmetries, both with and without a non-vanishing total lepton number.

\vspace{0.2cm}

The rest of this paper is organized as follows. We introduce the system of QKEs at the heart of this study in Sec.~\ref{sec:qke}, explaining the momentum-averaging procedure implemented in this paper and in particular the resulting collision terms and initial conditions. We further give details on the numerical strategy employed in solving these equations. Subsequently, Sec.~\ref{sec:results} presents our results: Solving the QKEs from before the onset of neutrino oscillations until BBN, we derive limits on primordial asymmetries with vanishing total lepton number as a function of the flavor composition of the initial primordial asymmetries. Interesting features are associated with the dynamics of the different MSW transitions. We conclude this section by demonstrating that our framework can also efficiently solve the QKEs in the case of non-vanishing total lepton number and demonstrate excellent agreement with the results obtained in Refs.~\cite{Froustey:2021azz,Froustey:2024mgf} which retain the full neutrino momentum dependence. Sec.~\ref{sec:implications} discusses the implications of our results for different early Universe scenarios involving large lepton flavor asymmetries, before we conclude in Sec.~\ref{sec:conclu}.
\\ \noindent
Four appendices are dedicated to important technical details of the QKEs, aiming to provide all necessary steps for deriving, implementing and interpreting our results: App.~\ref{app:conventions} lists our conventions and parameter choices, App.~\ref{app:collisionterms} provides an ab-initio derivation of the collision terms in our momentum averaged formalism, App.~\ref{sec:Cterms_TnuneqTgamma} generalizes these collision terms allowing for different neutrino and photon temperatures, App.~\ref{app:SUN-decomposition} gives the explicit $SU(N)$ decomposition of the density matrix employed in our numerical study and App.~\ref{app:slowness-factor} is dedicated to the dynamics of the MSW transitions. 
\\ \noindent
Last, but not least, we highlight that we release the Mathematica code \texttt{COFLASY-M} available on~\gitlink\;which implements and solves a simplified version of the quantum kinetic equations numerically.

\section{Kinetic equations}
\label{sec:qke}

In the Standard Model of particle physics at temperatures below the QCD phase transition, sizable particle asymmetries can only be stored in the neutrino sector due to charge neutrality of the plasma. Their evolution becomes challenging to track once neutrino oscillations set in at around $15\,\mathrm{MeV}$, which is when the frequency of vacuum oscillations becomes comparable to the matter potentials. This dynamics is best described introducing density matrices, whose diagonal components give the abundances of neutrinos (or anti-neutrinos) while the off-diagonal components account for neutrino oscillations. These density matrices are a priori time and momentum dependent and evolve following Schrödinger-type equations, referred to as QKEs~\cite{Sigl:1993ctk}. The key terms entering these equations are the vacuum Hamiltonian, containing the neutrino mass splittings and mixing angles and which is responsible for neutrino oscillations in vacuum, as well as the matter potentials and collision terms, accounting for neutrino self-energies, self-interactions as well as interactions with charged leptons (see also  Fig.~\ref{fig:rates}). These matter potentials are key to MSW transitions, while the collision terms lead to a damping of neutrino oscillations. In the following we discuss all these terms in more detail, largely following~\cite{Froustey:2021azz, Froustey:2020mcq} for the momentum dependent equations, before introducing the momentum average which is key to the framework developed here.

Our starting point are the full momentum dependent QKEs governing the evolution of the hermitian (anti-)neutrino density matrices $\rho$ ($\bar \rho$)~\cite{Sigl:1993ctk,Volpe:2013uxl,Blaschke:2016xxt,Bennett:2020zkv,Froustey:2021azz, Froustey:2020mcq, Li:2024gzf},\footnote{We follow the convention introduced in the seminal work of Sigl \& Raffelt~\cite{Sigl:1993ctk}. In particular, in the definition of the anti-neutrino density matrix we assume a reversed flavor index ordering of the creation and annihilation operator compared to the neutrino case, see Eq.~(2.2) of~\cite{Sigl:1993ctk}. On the level of the kinetic equations~\eqref{eq:QKE}, this, most strikingly, leads to $-i \mathcal{H}_0 \mapsto +i \mathcal{H}_0$ when changing from neutrino to anti-neutrino. This convention is also followed in most of the literature on the present topic~\cite{Froustey:2020mcq, Froustey:2021azz, Froustey:2024mgf, deSalas:2016ztq,Pastor:2008ti, Lesgourgues:2013sjj, Li:2024gzf}, as well as in some more recent works on leptogenesis via right-handed neutrino oscillations (ARS)~\cite{deVries:2024rfh, Klaric:2021cpi, Klaric:2020phc}. The alternative assumption of having the same flavor index ordering for neutrino and anti-neutrino is mainly adapted in ARS~\cite{Canetti:2012kh, Hernandez:2015wna, Hernandez:2016kel, Hernandez:2022ivz, Sandner:2023tcg, Abada:2018oly, Asaka:2005pn, Asaka:2011wq, Ghiglieri:2017gjz, Ghiglieri:2018wbs, Drewes:2017zyw} and results in changing $-i \mathcal{H}_0 \mapsto -i \mathcal{H}_0^*$ for neutrino to anti-neutrino. Any observable is of course unaffected by the choice and alternative forms of deriving the kinetic equations have been presented e.g. in some of the aforementioned references. }
\begin{align}
\frac{d \rho}{dt} = - i [{\cal H}, \rho] + \mathcal{I}\left[ \rho\right] \,, \quad  \frac{d \bar \rho}{dt} = +i [{\cal H}, \bar  \rho]  + \bar{\mathcal{I}}\left[ \bar{\rho}\right] \,.
\label{eq:QKE}
\end{align}
The Hamiltonian $\mathcal{H} = {\cal H}_0 + {\cal V}$ is described in the following, while the collision terms $\mathcal{I}\left[ \rho\right],\,\bar{\mathcal{I}}\left[ \bar{\rho}\right]$ are discussed in more detail in Sec.~\ref{subsec:collisions} and App.~\ref{app:collisionterms}.

In the flavor basis, the non-commuting contribution of the vacuum Hamiltonian can be expressed as
\begin{align}
\label{eq:def_H0}
	{\cal H}_0 = U \frac{M^2}{2k} U^\dagger \equiv {\cal H}_0^{\mathrm{even}} + {\cal H}_0^{\mathrm{odd}}\,,
\end{align}
with $k = |\vec{k}|$ indicating the absolute value of the neutrino momentum, $M^2 = \mathrm{diag}(0, \Delta m_{21}^2, \Delta m_{31}^2)$ accounting for neutrino mass splittings and $U$ is the PMNS matrix describing the rotation with angles $\theta_{ij}$ and Dirac CP phase $\delta_{\mathrm{CP}}$ between mass and flavor basis.\footnote{Note that possible Majorana phases contained in $U$ do not affect neutrino oscillations.} It is always possible to decompose ${\cal H}_0 $ into a part which is CP-even (symmetric under $\delta_{\mathrm{CP}} \leftrightarrow -\delta_{\mathrm{CP}}$) and real, ${\cal H}_0^{\mathrm{even}}$, and CP-odd (anti-symmetric under $\delta_{\mathrm{CP}} \leftrightarrow -\delta_{\mathrm{CP}}$) and imaginary, ${\cal H}_0^{\mathrm{odd}}$. The values for the neutrino mass splittings and mixings used in this paper are specified in App.~\ref{app:conventions} and we follow the PDG convention for the PMNS matrix~\cite{ParticleDataGroup:2024cfk}. We note that although we present the kinetic equations and the numerical strategy on how to integrate them in full generality, our results in Sec.~\ref{sec:results} are given for the $\mathrm{CP}$-conserving case of $\delta_{\mathrm{CP}} = 0$. We, however, explicitly checked that a non-zero $\delta_{\mathrm{CP}}$ has only minimal impact on cosmological observables, a result in agreement with expectations from previous studies~\cite{Gava:2010kz, Froustey:2021azz, Akhmedov:2002zj,Balantekin:2007es}.

The matter potentials, ${\cal V} = V_c + V_s$, are most easily obtained in the flavor basis and account for neutrino self-energy corrections arising from charged and neutral leptons respectively~\cite{Sigl:1993ctk,Notzold:1987ik,Froustey:2020mcq},
\begin{align}
\label{eq:Vc_def}
    V_{\mathrm{c}} &=  - 2 \sqrt{2} G_F \frac{k}{m_W^2}  (\mathbb{E} + \mathbb{P})_{\alpha \alpha} T_\gamma^4 = - 2 \sqrt{2} G_F \frac{k}{m_W^2}  4
      \int \frac{\mathrm{d}^3k'}{(2\pi)^3} \left(\sqrt{{k'}^2 + m_\alpha^2} + \frac{{k'}^2}{3 \sqrt{{k'}^2 + m_\alpha^2}} \right) f_\mathrm{fd}({k'}/T_\gamma,0)  \,, \\
\label{eq:Vs_def}
    V_{\mathrm{s}} & = \sqrt{2} G_F  \int \frac{\mathrm{d}^3k'}{(2 \pi)^3} (\rho - \bar{\rho}) \,,
\end{align}
with $\alpha = {e,\mu,\tau}$, the Fermi-Dirac equilibrium distribution
\begin{align}
    f_{\mathrm{fd}}(k/T,\mu/T) = \left[ \exp\left( \sqrt{\frac{k^2}{T^2} + \frac{m^2}{T^2}} - \frac{\mu}{T} \right) + 1 \right]^{-1}\,,
\end{align}
and the summed dimensionless energy-pressure densities of both charged leptons and anti-leptons
\begin{align}
\label{eq:EP_approx}
    (\mathbb{E} + \mathbb{P})_{ee} \simeq \frac{7}{45} \pi^2  - \frac{1}{6} m_e^2/T_\gamma^2\,,\quad \,
    (\mathbb{E} + \mathbb{P})_{\mu\mu} \simeq  \frac{2}{\pi^2}(m_\mu/T_\gamma)^3 K_3(m_\mu/T_\gamma) \,.
\end{align}
In the last step we introduced the modified Bessel function of the second kind which is a result of approximating the full integral expression of $V_c$ with Maxwell-Boltzmann statistics as relevant for muons. $T_\gamma$ denotes the temperature of the plasma containing the photons and charged leptons. For $1\,\mathrm{MeV} \leq T_\gamma\leq 20\,\mathrm{MeV}$, the region of interest in the present study, the approximations agree with the full result at a level of $\mathcal{O}(0.1\%)$ or better. Furthermore, for the $\tau$ flavor we can safely assume $ (\mathbb{E} + \mathbb{P})_{\tau\tau} = 0$, as well as $(\mathbb{E} + \mathbb{P})_{\alpha\beta} = 0$ for $\alpha \neq \beta$. Moreover, $G_F = 1.166 \cdot 10^{-5}/\mathrm{GeV}^2$ is Fermi's constant and $m_W = 80.377\,\mathrm{GeV}$~\cite{ParticleDataGroup:2024cfk}. We note that there is a further ${\cal O}(G_F^2)$ contribution to the matter potentials proportional the sum of the neutrino and anti-neutrino density matrices. It is however subdominant compared to the term in Eq.~\eqref{eq:Vs_def} since $G_F T_\gamma^2 \lesssim 10^{-8}$. Nevertheless, in the absence of (large) asymmetries this term can indeed become relevant. Moreover, we can also safely neglect a possible contribution from the charged lepton asymmetry in $V_s$ since at $T_\gamma \leq 20\,\mathrm{MeV}$, sizable lepton asymmetries can only occur in the neutrino sector. It is for this reason that in $V_c$ the Fermi-Dirac distribution is taken with $\mu=0$. These approximations are in accordance with previously performed studies, see for example~\cite{Froustey:2021azz}. For anti-neutrinos the $V_c$ potential is the same as for neutrinos, but $V_s$ changes sign under $\rho \leftrightarrow \bar{\rho}$, i.e.\ $\bar{V}_s = -V_s$~\cite{Notzold:1987ik}.

Solving the kinetic equation~\eqref{eq:QKE} is extremely challenging for three main reasons: i) it is an integro-differential equation as $\rho$ and $\bar{\rho}$ depend upon momenta and the collision terms are integrals of the density matrices themselves, ii) it is a stiff system due to the different time scales evolved, and iii) it is non-linear, since the $V_s$ potential depends upon the density matrices and the collision terms are non-linear. In consequence, solving these QKEs is computationally a very expensive task~\cite{Bennett:2020zkv,Akita:2020szl,Froustey:2021azz,Froustey:2024mgf,deSalas:2016ztq, Li:2024gzf}.\footnote{See, however, the recent Refs.~\cite{Ovchynnikov:2024rfu,Ovchynnikov:2024xyd} which propose new and fast alternative solutions based on Monte Carlo simulations.} Instead, we proceed here by considering momentum averaged equations, which drastically reduce the computational cost while, as we will demonstrate, providing a sufficiently accurate description of the system. Although in principle all the neutrino modes oscillate independently, the physical scenario at hand provides further justification for this approximation as all momentum modes undergo coherent synchronous oscillations~\cite{Pastor:2001iu}. Thus, we approximate the neutrino and anti-neutrino density matrices as a Fermi-Dirac distribution with vanishing chemical potential and with temperature $T_\nu$, multiplied by time-dependent matrices ($r(x)$ and $\bar{r}(x)$, respectively) which encode the neutrino asymmetries as well as flavor correlations,
\begin{align}
\label{eq:ansatz}
 \rho(x,y) = r(x) \, f_\mathrm{fd}(y/z_\nu,0) \,, \quad \bar \rho(x,y) = \bar r(x) \, f_\mathrm{fd}(y/z_\nu,0)\,, \qquad (\text{key ansatz})
\end{align}
where we introduced the scale factor $x$ as a measure of time or comoving temperature $T_\mathrm{cm}$, as well as the dimensionless momentum variable $y$,
\begin{align}
x = T_\mathrm{ref}/T_\mathrm{cm}\,,\;\;\; y = k/T_\mathrm{cm}\,,
\end{align}
with $T_\mathrm{ref}$ a reference temperature which can be arbitrarily chosen. Neutrino oscillations, together with neutrino-electron interactions, imply that the neutrino temperature $T_\nu$, the photon temperature $T_\gamma$ and the comoving temperature $T_\text{cm}$ may differ, which we parametrize by the time-dependent dimensionless variables
\begin{align}
 z_\gamma = T_\gamma/T_\mathrm{cm} \,, \quad z_\nu = T_\nu / T_\mathrm{cm} \,.
\end{align}
To translate the time derivative of the kinetic equation~\eqref{eq:QKE} to the new variables, we use the Friedmann equation to obtain
\begin{align}
\label{eq:dt_to_dx}
    \frac{\mathrm{d}\overset{(-)}{\rho}}{\mathrm{d}t}= \left. H x \frac{\mathrm{d}\overset{(-)}{\rho}}{\mathrm{d}x} \right\vert_{y\,\mathrm{fixed}}\,,
\end{align}
with the Hubble parameter $H$ determined by the total energy density of the Universe,
\begin{align}
    H^2 = \frac{\varepsilon_\gamma + \varepsilon_e + \varepsilon_\nu}{3 M_P^2}\,,
\end{align}
where $M_P = 2.435 \cdot 10^{18}\,\mathrm{GeV}$ is the reduced Planck mass. The energy densities in turn are given by integrating over the first moment of the respective distribution functions,
\begin{align}
 \varepsilon_\nu = \int \frac{\mathrm{d}^3 k }{(2\pi)^3}  \, k \,{\rm Tr}[ \rho+\bar{\rho}] = \frac{7 \pi^2}{240} \, \text{Tr}[r + \bar r] \, T_\nu^4
 \,, \quad \varepsilon_e = 4\int \frac{\mathrm{d}^3 k }{(2\pi)^3} \, k f^{m_e}_\mathrm{fd}(k/T_\gamma,0)
 \,, \quad \varepsilon_\gamma = 2\frac{\pi^2}{30} T_\gamma^4 \,.
\end{align}
Note that while we take into account the contributions of muons in the thermal potential $V_c$ (since this is critical for the correct modeling of (non-) adiabatic MSW transitions as explained in Sec.~\ref{subsec:MSW}) their contribution to the expansion rate of the Universe is always negligible for the temperatures of interest in this study.

The ansatz of Eq.~\eqref{eq:ansatz} allows to explicitly integrate over $y$ and we obtain a momentum averaged version of the quantum kinetic equations~\eqref{eq:QKE} in dimensionless form:
\begin{align}
\label{eq:kinetic_eq_flavor}
\begin{split}
     Hx r' + 3 H x \, r \, \frac{z_\nu'}{z_\nu}  &= -i [\langle \mathcal{H}_0 \rangle + \langle V_{\mathrm{c}} \rangle + \langle V_{\mathrm{s}} \rangle , r] + \langle \mathcal{I} \rangle\,,\\
    Hx \bar{r}' + 3 H x \, \bar{r} \, \frac{z_\nu'}{z_\nu}  &= +i [\langle \mathcal{H}_0 \rangle + \langle V_{\mathrm{c}} \rangle - \langle V_{\mathrm{s}} \rangle, \bar{r}] + \langle \bar{\mathcal{I}} \rangle\,,
\end{split}
\end{align}
where we denote $'=\partial_x$ and the momentum average corresponds to 
\begin{align}
\label{SM:eq:def_momentum_ave}
    \langle ... \rangle \equiv  \frac{\int\mathrm{d}y\, ... y^2 \fd(y/z_\nu,0)  }{\int\mathrm{d}y\, y^2 \fd(y/z_\nu,0)} = \frac{2}{3 \zeta(3) z_\nu^3} \int\mathrm{d}y\, ... y^2 \fd(y/z_\nu,0) \,,
\end{align}
with $\zeta(3) \simeq 1.20206$. 
In addition, we need to track the energy transfer between the neutrino sector and electron-photon plasma. We do this via the continuity equation which relates
\begin{align}
 \frac{d (\varepsilon_\gamma + \varepsilon_e)}{dt} + 4 H \varepsilon_\gamma + 3 H (\varepsilon_e + p_e) = - \frac{\delta \varepsilon}{\delta t} \,, \quad \frac{d \varepsilon_\nu}{dt} + 4 H \varepsilon_\nu = \frac{\delta \varepsilon}{\delta t}\,,
\end{align}
where $(\mathbb{E} + \mathbb{P})_{ee} = (\varepsilon_e + p_e)/T_\gamma^4$. The energy transfer, $\delta \varepsilon/ \delta t$, is determined by the diagonal elements of the collision terms,
\begin{align}
\label{eq:def_energy_transfer}
 \frac{\delta \varepsilon}{\delta t} =  \int \frac{\dd^3 k}{(2\pi)^3} \, k   \, \text{Tr}[ {\cal I} + \bar{\cal I}] \,.
\end{align}
In a more explicit form we obtain the coupled set of equations,
\begin{align}
\label{eq:master}
\begin{split}
    H x r' & + 3 H x r \frac{z_\nu'}{z_\nu}  =  \\
    & -i  \left[  \frac{\pi^2}{36 \zeta(3)} UM^2U^\dagger \frac{1}{z_\nu}\left(\frac{x}{T_\mathrm{ref}} \right)   -  2 \sqrt{2} \frac{7 \pi^4}{180 \zeta(3)} \frac{G_F}{m_W^2}  z_\nu z_\gamma^4 \left( \frac{T_\mathrm{ref}}{x} \right)^5 (\mathbb{E} + \mathbb{P}) -   \frac{3 \zeta(3) G_F}{2 \pi^2 \sqrt{2}} z_\nu^3  \left( \frac{T_\mathrm{ref}}{x} \right)^3 \bar{r} , r\right] +  \langle {\cal I} \rangle\,, \\
    H x \bar{r}' & +  3 H x \bar{r} \frac{z_\nu'}{z_\nu} =   \\
    & +i \left[ \frac{\pi^2}{36 \zeta(3)} UM^2U^\dagger \frac{1}{z_\nu}\left(\frac{x}{T_\mathrm{ref}} \right)  -2 \sqrt{2} \frac{7 \pi^4}{180 \zeta(3)}  \frac{G_F}{m_W^2}  z_\nu z_\gamma^4 \left( \frac{T_\mathrm{ref}}{x} \right)^5 (\mathbb{E} + \mathbb{P}) -  \frac{3 \zeta(3) G_F}{2 \pi^2 \sqrt{2}} z_\nu^3\left( \frac{T_\mathrm{ref}}{x} \right)^3 r, \bar{r} \right] +  \langle \bar{\cal I} \rangle\,.
\end{split}
\end{align}
and
\begin{align}
    z_\nu' = - \frac{z_\nu}{4} \left[ \frac{\mathrm{Tr}[r' + \bar{r}']}{\mathrm{Tr}[r+\bar{r}]} - \frac{ \frac{1}{H}\frac{\delta \varepsilon}{\delta t}}{x \varepsilon_\nu} \right]\,, \quad
    z_\gamma' = \frac{z_\gamma}{x} - \frac{4 \varepsilon_\gamma + 3 (\varepsilon_e + p_e) - \frac{1}{H} \frac{\delta \varepsilon}{\delta t}}{T_{\mathrm{ref}} \frac{\partial}{\partial T_\gamma} (\varepsilon_\gamma + \varepsilon_e)} \,.
    \label{eq:z-evolution}
\end{align}
The collision integrals $\langle \mathcal{I} \rangle$, $\langle \bar{\mathcal{I}} \rangle$ and the energy transfer rate $\delta \varepsilon / \delta t$ are discussed in the following subsection~\ref{subsec:collisions}.

Our main interest will be the asymmetries and the excess radiation in the neutrino sector. Given our ansatz  for the neutrino distribution functions in Eq.~\eqref{eq:ansatz}, the number density of neutrinos and anti-neutrinos of flavor $\alpha$ are given by
\begin{align}
    n_\alpha =  \int \frac{\mathrm{d}^3k}{(2 \pi)^3}\, \rho_{\alpha\alpha} = \frac{3 \zeta(3)}{4\pi^2} T_\nu^3\, r_{\alpha\alpha}\,, \qquad
    \bar{n}_\alpha =  \int \frac{\mathrm{d}^3k}{(2 \pi)^3} \, \bar{\rho}_{\alpha\alpha} = \frac{3 \zeta(3)}{4\pi^2} T_\nu^3 \,\bar{r}_{\alpha\alpha}\,.
    \label{eq:def_n}
\end{align}
Another key quantity is the energy density in neutrinos
\begin{align}
\label{eq:def_dNeff2}
 N_\text{eff} = \frac{8}{7} \left( \frac{11}{4} \right)^{4/3} \frac{\varepsilon_\nu}{\varepsilon_\gamma} = \left( \frac{11}{4} \right)^{4/3}  \left(\frac{z_\nu}{z_\gamma} \right)^4 \frac{\text{Tr}[r + \bar r]}{2}\,,
\end{align}
normalized to count the effective number of neutrino species. The factor $(11/4)^{4/3}$ accounts for the heating of the photon bath from electron-positron annihilation at $T_\gamma \sim m_e$, yielding the asymptotic value of $N_\text{eff}$ at $T \ll m_e$. In practice we will solve Eqs.~\eqref{eq:master} and \eqref{eq:z-evolution} until $T_{\mathrm{cm}} = 1.5\,\mathrm{MeV}$, while at lower temperatures, since $\delta \varepsilon/\delta t \simeq 0$, it suffices to track the decoupled evolution of $z_\gamma$.
The excess radiation is parametrized by
\begin{align}
\label{eq:def_dNeff}
    \Delta N_{\mathrm{eff}} = N_{\mathrm{eff}} - N_{\mathrm{eff}}^{\mathrm{SM}}\,, 
\end{align}
with $N_{\mathrm{eff}}^{\mathrm{SM}} = 3.044$ the SM value.\footnote{NLO QED corrections to this number are currently being actively discussed in the literature~\cite{Cielo:2023bqp,Jackson:2023zkl,Jackson:2024gtr,Drewes:2024wbw,Drewes:2024nbg}, with the most complete and thorough analysis~\cite{Jackson:2024gtr} highlighting that they most likely lead to $\sim 10^{-4}$ corrections to $N_{\rm eff}$ and which are substantially smaller than those first estimated in~\cite{Cielo:2023bqp}.}

\subsection{Collision terms}
\label{subsec:collisions}

Neutrino interactions are of critical importance for neutrino oscillations in the early Universe as they lead to flavor decoherence~\cite{Sigl:1993ctk}. Their effects are encoded in collision terms in the QKEs and there are three main contributions: those arising from $\nu e \leftrightarrow \nu e$ scatterings, $\nu \bar{\nu} \leftrightarrow e^+e^-$ annihilation, and those from neutrino self-interactions, $\nu\nu \leftrightarrow \nu \nu$ and $\nu\bar{\nu} \leftrightarrow \nu \bar{\nu}$. The most complicated ones are those from neutrino self-interactions and to the best of our knowledge only a handful of references have so far performed calculations using them~\cite{Froustey:2020mcq,Bennett:2020zkv,Froustey:2021azz,Froustey:2024mgf, Li:2024gzf}. The rest of the studies have either disregarded them or implemented them approximately using a damping ansatz as derived in~\cite{McKellar:1992ja}. In the context of large primordial lepton asymmetries, it has been recently discovered and emphasized in Ref.~\cite{Froustey:2021azz} that considering the exact collision terms is critical as the use of damping coefficients from Ref.~\cite{McKellar:1992ja} leads to unphysical effects such as strong decoherence in the $\mu-\tau$ subsystem. Given that $\nu_\mu$ and $\nu_\tau$ at temperatures below $20\,{\rm MeV}$ have the same interactions with the medium, this decoherence cannot be physical.

Given the importance of accurately accounting for interactions and their impact on neutrino oscillation coherence, we perform a first-principles calculation of the collision terms starting from the already simplified expressions outlined in Ref.~\cite{deSalas:2016ztq} and applying our ansatz for the density matrix in Eq.~\eqref{eq:ansatz}. The derivation of the collision terms is rather lengthy and we relegate the details to App.~\ref{app:collisionterms} and App.~\ref{sec:Cterms_TnuneqTgamma}. In the limit of Maxwell-Boltzmann statistics for the distribution functions and neglecting the electron mass we are able to cast the momentum averaged collision terms entering Eqs.~\eqref{eq:master} in a rather compact form,
\begin{align}
\label{eq:I_nuenue}
\!\!\! \langle {\cal I} \rangle^{\nu e \leftrightarrow \nu e} &=   \frac{32 G_F^2}{\pi^3} \left(\frac{T_{\mathrm{ref}}}{x}\right)^5 z_\gamma^4 z_\nu  \left(2 G^L r G^L - \left[rG^LG^L + G^L G^Lr \right] \right)\,,  \\
\begin{split}
\label{eq:I_nunubar_annepem}
\!\!\!\!\! \langle {\cal I} \rangle^{\nu \bar{\nu} \leftrightarrow e^+e^-}   &=  \frac{8 G_F^2}{\pi^3} \left(\frac{T_{\mathrm{ref}}}{x}\right)^5 \frac{1}{z_\nu^3}\left(  2 G^L  G^L z_\gamma^8  - \left[ r G^L \bar{r} G^L + G^L \bar{r}  G^L r \right] z_\nu^8 \right) \\
&+ \frac{8 G_F^2}{\pi^3} \left(\frac{T_{\mathrm{ref}}}{x}\right)^5 \frac{1}{z_\nu^3} \left( 2 G^R   G^R  z_\gamma^8 - \left[ r G^R \bar{r} G^R + G^R \bar{r}  G^R r \right]z_\nu^8 \right)\,,
\end{split}
\\
\!\!\! \langle {\cal I} \rangle^{\nu \bar{\nu} \leftrightarrow \nu \bar{\nu}}   &= \frac{1}{4} \frac{8 G_F^2}{\pi^3}  \left(\frac{T_{\mathrm{ref}}}{x}\right)^5 z_\nu^5 \left( \mathbb{1} \, {\rm Tr}[r \bar{r}+\bar{r} r]   -  \left[r \bar{r}+\bar{r} r\right]{\rm Tr}[\mathbb{1}]    \right)\,.
\label{eq:I_nu_pair}
\end{align} 
We note that $\langle {\cal I} \rangle^{\nu \bar{\nu} \leftrightarrow \nu \bar{\nu}}$ does not receive contributions from neutrino self-scatterings within our momentum averaged ansatz~\eqref{eq:ansatz}, see App.~\ref{app:subsec:collision_rmat_FD}. Here $\mathbb{1} = {\rm diag}(1,1,1)$,  $G^L = {\rm diag}(g_L, g_L -1, g_L -1)$ and $G^R = {\rm diag}(g_R, g_R, g_R)$ with $g_L = s_W^2+1/2$ and $g_R=s_W^2$. We fix $s_W^2 = 0.22305$ for the square of the sine of the weak mixing angle~\cite{ParticleDataGroup:2024cfk}. The $-1$ in the $G^L$ matrix is explained by the absence of $\mu$ and $\tau$ leptons in the thermal plasma and thus there are only charged current interactions for electron flavored leptons. We note that there are no contributions from $G^R$ to neutrino-electron scatterings. For $m_e\to 0$ chiral symmetry is exact and the collision terms for anti-neutrinos are written simply by doing the replacement $r \leftrightarrow \bar{r}$. Note that with the exception of the collision term for $\nu e\leftrightarrow \nu e$ interactions the collision terms are non-linear.

A key property of these collision terms is that they feature a $U(2)$ symmetry in the $\mu-\tau$ subspace, as emphasized in~\cite{Froustey:2021azz}. Without such a symmetry, the flavor oscillations would be significantly more damped than with the actual collision terms. In addition, in App.~\ref{app:collisionterms} we highlight how to approximate them in a damping form which is simple, generalizes the results of~\cite{McKellar:1992ja}, and leads to relatively accurate results for the final lepton asymmetries.

Following the same procedure (see App.~\ref{sec:Cterms_TnuneqTgamma}), we obtain expressions for the energy transfer rate in Eq.~\eqref{eq:z-evolution} in the Maxwell-Boltzmann and $m_e\to 0$ limit,
\begin{align}
\label{eq:energy-transfer}
\begin{split}
    \frac{\delta \varepsilon}{\delta t} &= 128 \frac{ G_F^2}{\pi^5}  \left(\frac{T_{\mathrm{ref}}}{x}\right)^9  {\rm Tr}[z_\gamma^9( G^L   G^L+G^RG^R)   - z_\nu^9 ( G^L \bar{r}  G^L r+G^R \bar{r}  G^R r)]  \\
   &+ 112 \frac{ G_F^2}{\pi^5} \left(\frac{T_{\mathrm{ref}}}{x}\right)^9   z_\gamma^4 z_\nu^4 (z_\gamma - z_\nu){\rm Tr}[(r+\bar{r}) G^L G^L + (r+\bar{r})G^RG^R)]  \,,
\end{split}
\end{align}
where here the first line corresponds to $e^+e^-\leftrightarrow\nu\bar{\nu}$ processes and the second to $\nu e\leftrightarrow\nu e$ ones.

Corrections to the momentum averaged collision terms and to the energy transfer rate from Pauli blocking and the finite electron mass arise. The former is fully taken into account in the numerical treatment of the momentum averaged collision terms $\langle{\cal I}\rangle$, whereas for the energy transfer rate we find it to be sufficient to include leading order Fermi-Dirac corrections to the Maxwell-Boltzmann approximation, for reasons outlined in App.~\ref{sec:Cterms_TnuneqTgamma}. In App.~\ref{app:collisionterms}, we perform a perturbative expansion in terms of spin statistics corrections, demonstrating that these represent only a small correction to the collision terms determining the evolution of the density matrices. The leading order corrections are implemented in \texttt{COFLASY-M}. The impact of the finite electron mass is expected to be small, since at neutrino decoupling we still have $T_{\nu\,\mathrm{dec}} \sim 2\,\mathrm{MeV} \gg m_e$~\cite{Escudero:2018mvt,EscuderoAbenza:2020cmq}. In our numerical analysis we include the electron mass in the phase space integration for the $\nu \bar \nu \leftrightarrow  e^+ e^-$ rates, but for computational efficiency reasons neglect the subleading impact on the other rates as well as the chiral symmetry breaking contributions to the interaction rates.\footnote{We explicitly checked that for the majority of the parameter space studied our results for the primordial lepton asymmetry evolution do not change when considering the effect of a finite electron mass. However, we also encountered special flavor directions for which the time evolution of the lepton flavor asymmetries are substantially altered compared to the fully relativistic limit, while the final value of $\xi_e$ and $\Delta N_{\mathrm{eff}}$ remain essentially unaffected.} The full expressions for the collision terms and the energy transfer rate taking into account these corrections are quite cumbersome. For this reason we present here, and in particular in Apps.~\ref{app:collisionterms} and \ref{sec:Cterms_TnuneqTgamma}, different levels of approximations that lead to compact expressions which capture the relevant dynamics to good accuracy.

\subsection{Initial conditions}
\label{subsec:initial_cond}

In the momentum averaged kinetic equations any prescription for setting the initial conditions based on a set of chemical potentials is necessarily an approximation to the full momentum dependent Fermi-Dirac distribution function with non-vanishing chemical potentials. A mapping which captures well all key physical aspects can be obtained using the following guiding principles. Initially, the comoving, photon and neutrino temperatures are equal,
\begin{align}
T_\mathrm{cm} = T_\gamma = T_\nu \, \quad \mathrm{(initial\,conditions)} \,,
\end{align}
and neutrino oscillations are overdamped in the sense that the mean free path between two collisions is much shorter than the oscillation length. This is true for sufficiently large initial temperatures, which motivates the choice of $T_{\mathrm{ini}} = 20\,\mathrm{MeV}$ when solving the kinetic equation, see Fig.~\ref{fig:rates}, and results in vanishing off-diagonal elements of the density matrix,
\begin{align}
\label{eq:ini_flavor_off_diag}
    r_{\alpha \beta}^\mathrm{ini} = \bar{r}_{\alpha \beta}^\mathrm{ini} = 0\, \quad \mathrm{(initial\,conditions)} \,.
\end{align}
Finally, the difference between the diagonal terms of $r$ and $\bar{r}$ is the measure for the (initial) asymmetry. Comparing to the number densities obtained for a Fermi-Dirac distribution with initial temperature $T_\text{cm}$,
\begin{align}
 \overset{(-)\;\;}{n_\alpha} = \int \frac{\mathrm{d}^3 k}{(2\pi)^3}  \, f_\text{fd}(k/T_\text{cm}, \pm \mu_\alpha/T_\text{cm}) = - \frac{T_\text{cm}^3}{\pi^2}\mathrm{Li}_{3} \left( - e^{\pm \mu_\alpha/T_\text{cm}}\right) \,,
\end{align}
where $\mathrm{Li}_{3}$ is a polilogarithm, leads to initial conditions for the diagonal elements in our momentum averaged ansatz~\eqref{eq:ansatz} (see Eq.~\eqref{eq:def_n}) given by
\begin{align}
\label{eq:ini_flavor_diag}
r_{\alpha \alpha}^\mathrm{ini} = - \frac{4 }{3 \zeta(3)} \mathrm{Li}_{3}\left( -\mathrm{e}^{\xi_\alpha}\right)
\,, \quad \bar r_{\alpha \alpha}^\mathrm{ini} = - \frac{4 }{3 \zeta(3)} \mathrm{Li}_{3}\left( -\mathrm{e}^{-\xi_\alpha}\right)
\, \quad \mathrm{(initial\,conditions)} \,,
\end{align}
with $\xi_\alpha = \mu_\alpha/T_\text{cm}$.

Given our ansatz for the distribution functions, these initial conditions cannot fully reproduce those of an actual Fermi-Dirac distribution. As such, they don't exactly satisfy the requirement of thermal and chemical equilibrium, in which $\langle {\cal I} \rangle_{\alpha \alpha}$, $\langle \bar {\cal I} \rangle_{\alpha \alpha}$ and $\delta \varepsilon/\delta t$ should all vanish. The system dynamically accounts for this by very rapidly relaxing to this equilibrium solution, amounting to tiny corrections in $r, \bar r, z_\nu$ and $z_\gamma$. Physically, this in particular amounts to a 'mismatch' in the interpretation of the neutrino temperature, which we correct for by normalizing the final energy density obtained for vanishing lepton asymmetries to the Standard Model value,
\begin{align}
    N_\text{eff}(\mu_\alpha) \rightarrow \frac{N_{\mathrm{eff}}^{\mathrm{SM}}}{N_\text{eff}(\mu_\alpha = 0)} \, N_\text{eff}(\mu_\alpha) \,.
\end{align}

The numerical stability and speed of the numerical code is improved by algebraically determining the modified initial conditions which respect thermal and chemical equilibrium. Concretely, we require $\langle {\cal I} \rangle_{\alpha \alpha} = 0 $ ($\langle \bar {\cal I} \rangle_{\alpha \alpha} = 0 $ follows automatically), $\delta \varepsilon/\delta t = 0$ while maintaining Eq.~\eqref{eq:ini_flavor_off_diag} but modifying Eq.~\eqref{eq:ini_flavor_diag} to read
\begin{align}
\label{eq:ini_flavor_diag_mod}
r_{\alpha \alpha}^\mathrm{ini} = - \frac{4 z_\nu^{-3}}{3 \zeta(3)} \mathrm{Li}_{3}\left( -\mathrm{e}^{\xi_\alpha}\right) + \epsilon^\alpha_\text{db}
\,, \quad \bar r_{\alpha \alpha}^\mathrm{ini} = - \frac{4 z_\nu^{-3} }{3 \zeta(3)} \mathrm{Li}_{3}\left( -\mathrm{e}^{-\xi_\alpha}\right) + \epsilon^\alpha_\text{db}
\, \quad \mathrm{(initial\,conditions)} \,.
\end{align}
Together with the requirement of matching the Hubble expansion as $H(z_\nu,z_\gamma) = H(1,1)$ this gives 5 algebraic relations for the 5 unknowns $\epsilon_\text{db}^\alpha$, $z_\nu$ and $z_\gamma$ and determines their deviation from the default values of $\epsilon_\text{db} = \vec 0$, $z_\nu = 1$ and $z_\gamma = 1$, respectively.
We note that the deviation from the default values of any of these parameters is always less than $2\%$ within the whole parameter space studied here. As an example of this procedure, we provide the explicit expressions which correct the density matrices in the limit of neglecting the energy transfer ($z_\nu = z_\gamma = 1$ throughout) in Eq.~\eqref{app:eq:db} of App.~\ref{app:collisionterms}.

As an aside, we note that the difference of the diagonal elements of the neutrino and anti-neutrino matrix takes a much simpler form, $\Delta r_{\alpha} = (r - \bar r)_{\alpha \alpha} = 4 \pi^2/(18 \zeta(3)) (\xi_\alpha + \xi_\alpha^3/\pi^2)$, see Eq.~\eqref{eq:def_n}. It is thus tempting to simplify the mapping between $r$ and $\xi_\alpha$ to set $r_{\alpha \alpha} = 1 + \Delta r_\alpha/2$ and $\bar r_{\alpha \alpha} = 1 - \Delta r_\alpha/2$. Despite that this at first glance seems a good approximation to Eq.~\eqref{eq:ini_flavor_diag}, this choice eliminates the synchronous oscillations, as can be seen explicitly by the analytical expressions for the frequency of these oscillations given in App.~\ref{app:slowness-factor}. This approximation thus leads to unphysical behavior in the evolution of the neutrino asymmetries and we do not employ it in the following.

\subsection{Numerical strategy}
\label{subsec:num_strategy}

Even though the momentum averaged QKE is significantly simpler than the momentum dependent one, it is still numerically rather difficult to handle due to its non-linearity and the presence of various different timescales. Following earlier studies~\cite{Bell:1998ds,ArguellesDelgado:2014rca}, we find that it is numerically convenient to exploit the hermitian structure of $r$ and $\bar{r}$, and thus of all matrices appearing on the right hand side of the kinetic equation, and to expand the kinetic equation in terms of $SU(N)$ basis matrices $\lambda_\alpha$. For a generic hermitian matrix $\mathcal{O}$ this decomposition can be written as
\begin{align}
\label{eq:SUNdecompositionDef}
    \mathcal{O}  = \frac{1}{N} \mathrm{Tr}[\mathcal{O}] \mathbb{1} + \frac{1}{2} \mathrm{Tr}[\mathcal{O} \lambda^i] \lambda_i \equiv o^0 \mathbb{1} + o^i \lambda_i \,,
\end{align}
where the coefficients $o^0,\,o^i$ are real numbers and summation over repeated indices is assumed. For the minimal case of only two neutrinos, $\lambda_\alpha$ are the Pauli matrices, while for the SM three neutrino case these are the Gell-Mann matrices. Using the commutator as well as the orthogonality relation
\begin{align}
    [\lambda_i, \lambda_j] = 2 i f^{i j k} \lambda_k\,,\;\;\; \mathrm{Tr}[\lambda_i \lambda_j] = 2 \delta_{i j}\,,
\end{align}
where $f$ are the structure constants of $SU(N)$, we can write the kinetic equation~\eqref{eq:kinetic_eq_flavor} as a differential equation for the real coefficients:
\begin{align}
\label{eq:kinetic_eq_SUN_flavor}
\begin{split}
    r^{0}{'} &= c^0 \,,\quad r^{i}{'} = 2 f^{i j k} r^k \left(h_{0}^j +  v_c^j - v_s \bar{r}^j \right) + c^i\,,\\
    \bar{r}^{0}{'} &= \bar{c}^0 \,, \quad \bar{r}^{i}{'} = -2 f^{i j k} \bar{r}^k \left( h_{0}^j + v_c^j - v_s r^j \right) + \bar{c}^i\,.
\end{split}
\end{align}
which are complemented as before by Eqs.~\eqref{eq:z-evolution} for  $z_\nu$ and $z_\gamma$. Here, $h_{0}^j , v_c^j,c^j,\bar{c}^j$ are the respective $SU(N)$ coefficients of the operators $\langle \mathcal{H}_0 \rangle/(xH),\langle V_{\mathrm{c}} \rangle/(xH), \langle \mathcal{I} \rangle/(xH), \langle \bar{\mathcal{I}} \rangle/(xH)$ which can be obtained via Eq.~\eqref{eq:SUNdecompositionDef}, $r^i$ ($\bar r^i$) are the corresponding coefficients obtained expanding the matrices $r$ ($\bar r$), and
\begin{align}
    v_s = \frac{1}{xH} \frac{3 \zeta(3)}{2 \pi^2 \sqrt{2}} G_F z_\nu^3 \left(\frac{T_{\mathrm{ref}}}{x}\right)^3\,.
    \label{eq:vs}
\end{align}
We provide explicit expressions for these coefficients in App.~\ref{app:SUN-decomposition}. We solve Eq.~\eqref{eq:kinetic_eq_SUN_flavor} from $T_{\mathrm{ini}} = 20\,\mathrm{MeV}$ down to the comoving temperature of $T_{\mathrm{ave}} = 1.5\,\mathrm{MeV}$. At $T_{\mathrm{ave}}$ all MSW transitions have terminated, see Fig.~\ref{fig:rates}, and the evolution of the system is mainly affected by the (extremely) fast oscillations arising from vacuum flavor transitions. Any observable which is integrated over some time larger than the oscillation frequency will only be sensitive to the average spectrum. Hence, we take the solution of the kinetic equation~\eqref{eq:kinetic_eq_SUN_flavor} at $T = T_{\mathrm{ave}}$ and analytically calculate the asymptotic solution in the following way: Having chosen $T_{\mathrm{ave}} = 1.5\,\mathrm{MeV}$ (sufficiently below neutrino decoupling), to good approximation the off-diagonal elements of $r,\bar{r}$ in the mass basis vanish and the asymptotic flavor asymmetry is given by
\begin{align}
\label{eq:ave}
    \Delta r_\alpha = \left( U \langle U^\dagger \Delta r_{\alpha}^{T_{\mathrm{ave}}} U \rangle_{\mathrm{ave}} U^\dagger \right)_{\alpha\alpha}\,,
\end{align}
where $\Delta r_{\alpha}^{T_{\mathrm{ave}}}$ is the asymmetry at $T = T_{\mathrm{ave}}$ obtained by solving the kinetic equation~\eqref{eq:kinetic_eq_SUN_flavor} and $\langle .. \rangle_{\mathrm{ave}}$ means that we only keep the diagonal elements.\footnote{An alternative procedure for the oscillation averaging is to change from the Schr\"odinger picture to the interaction picture~\cite{ArguellesDelgado:2014rca}, which is more suitable if $T_{\mathrm{ave}}$ is chosen such that neutrinos do not have fully decoupled yet. We also numerically employed the oscillation average in the interaction picture and find that it agrees with Eq.~\eqref{eq:ave} for $T_{\mathrm{ave}} = 1.5\,\mathrm{MeV}$.}

In order to sample the entire parameter space we have created a code in C$++$ where the numerical integration is carried out using the GNU Scientific Library~\cite{gnu}. Specifically, we use the variable-coefficient linear multistep backward differentiation formula (BDF). It uses an explicit BDF step as a predictor and an implicit BDF as a corrector. The implicit step requires to define the Jacobian of the kinetic equation~\eqref{eq:kinetic_eq_SUN_flavor}. In principle, this $20\times 20$ matrix can be found semi-analytically using our momentum average ansatz, but we rely on a simple finite difference method to numerically obtain it. We impose that the Jacobian finite difference is always at least two orders of magnitude smaller than the maximally allowed time step of the BDF solver, which we set to $\mathrm{d}x_{\mathrm{step}} \leq 10^{-4}$. The initial step size is taken to be $10^{-5}$, with absolute and relative error tolerance of $10^{-6}$. With these settings the code finishes the integration on average in less than $25\,\mathrm{seconds}$. This allows us to perform detailed scans of the parameter space as outlined in Sec.~\ref{sec:results}. In subsection~\ref{subsec:full_qke_comparison} we show that the momentum average solution is in good agreement with the full momentum dependent solution of the kinetic equations. We will release the C$++$ code in an upcoming  publication. In the meantime, we release here a Mathematica code which can reproduce all of our results for the lepton asymmetries in the $z_\nu = z_\gamma = 1$ limit.\footnote{The code \texttt{COFLASY-M} can be found here:~\gitlink. Typical execution times are $\mathcal{O}(3)\,{\rm mins}$ per parameter point.}

\section{Results}
\label{sec:results}

\begin{figure}[!t]
 \includegraphics[width=0.48 \textwidth]{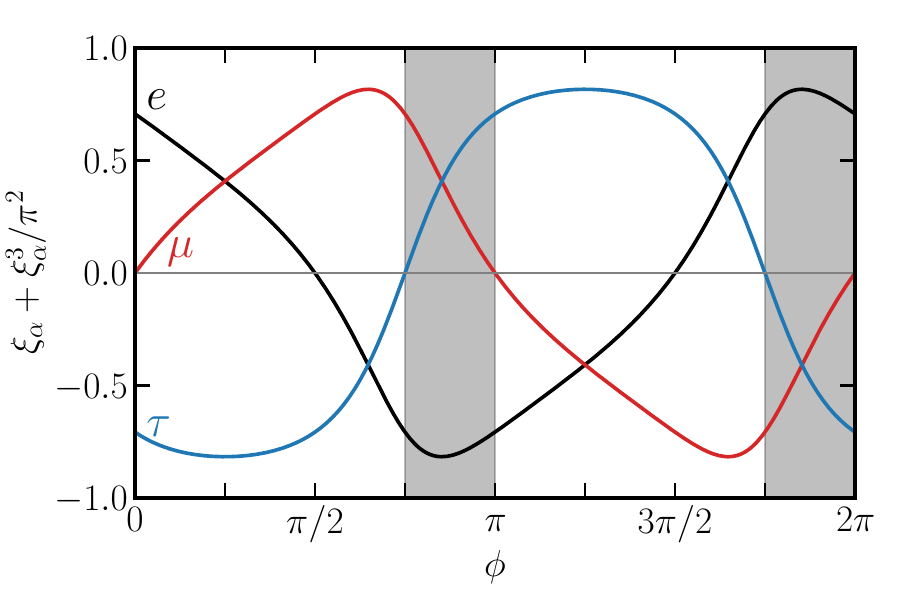} \hfill
 \caption{Flavor decomposition of the total asymmetry as a function of the direction in flavor space for vanishing total lepton number and normalized to $A=1$, see Eqs.~\eqref{eq:def_L0}-\eqref{eq:parametrization_theta}. The gray vertical bands indicate $\mathrm{sign}(\xi_e^\mathrm{ini}) = - \mathrm{sign}( \xi_\mu^\mathrm{ini})= - \mathrm{sign}( \xi_\tau^\mathrm{ini})$ and are discussed in Sec.~\ref{subsec:limits}.
 }
 \label{fig:parametrization}
\end{figure}

The asymmetry in the number densities of the neutrino flavor $\alpha$ is given by\footnote{
This expression assumes a Fermi-Dirac distribution for the neutrinos. For our initial conditions, this is of course the case. At later terms, we use Eq.~\eqref{eq:def_n} to determine the asymmetry in the number densities from our solutions for $r, \bar r$ and $z_\nu$. When quoting results for the final values of $\xi_e$, we use Eq.~\eqref{eq:expr_asymmetry} as the definition of this quantity.}
\begin{align}
\Delta n_\alpha \equiv \frac{(n - \bar n)_{\alpha \alpha}}{T_\mathrm{cm}^3} = \frac{z_\nu^3}{6} \left( \xi_\alpha + \xi_\alpha^3/\pi^2 \right)\,.
\label{eq:expr_asymmetry}
\end{align}
We see that the asymmetry is at leading order dictated linearly by $\xi_\alpha$, but can receive significant contributions from $\xi_\alpha^3$ if the chemical potential is large. This implies that the scenario of compensated chemical potentials, $\sum_\alpha \xi_\alpha = 0$, does not necessarily imply zero lepton number. Our aim is to constrain the initial $\xi_\alpha$ at $T_{\mathrm{cm}}=20\,\mathrm{MeV}$ given the BBN constraint at $T_{\mathrm{cm}}=1\,\mathrm{MeV}$ on $\xi_e$ (and CMB constraint on $\Delta N_{\mathrm{eff}}$, defined in Eq.~\eqref{eq:def_dNeff}, for $T_{\mathrm{cm}} \ll m_e$) while imposing vanishing total lepton number
\begin{align}
\label{eq:def_L0}
    \Delta n = \sum_\alpha \Delta n_\alpha = 0\,.
\end{align}
To efficiently cover the parameter space of vanishing total lepton number we introduce a parametrization in spherical coordinates for the three flavors $\alpha = e,\mu,\tau$:
\begin{align}
 \xi_e^\mathrm{ini} + (\xi_e^\mathrm{ini})^3/\pi^2  = A \cos \phi \sin \theta  \,, \quad
 \xi^\mathrm{ini}_\mu + (\xi^\mathrm{ini}_\mu)^3/\pi^2  = A \sin \phi  \sin \theta \,, \quad
 \xi_\tau^\mathrm{ini} + (\xi_\tau^\mathrm{ini})^3/\pi^2 = A \cos\theta \,,
 \label{eq:sph}
\end{align}
with the asymmetry amplitude
\begin{align}
\label{eq:def_A}
    A = \sqrt{ \sum_\alpha \left(\xi^\mathrm{ini}_\alpha + (\xi^\mathrm{ini}_\alpha)^3/\pi^2 \right)^2 }\,,
\end{align}
and the two angles ($\theta, \phi$) encoding the direction in flavor space. Imposing the zero total lepton number constraint of Eq.~\eqref{eq:def_L0}, we can eliminate one of the angles
\begin{align}
\label{eq:parametrization_theta}
    \theta = \arctan\left( -\cos \phi - \sin \phi\;,\; 1\right)\,,
\end{align}
where $\arctan(x,y)$ represents the principal value of the arc tangent of $y/x$. Fig.~\ref{fig:parametrization} shows the flavor direction as parametrized by $\phi$.
We note, however, that the framework developed in this study can be employed (and has been tested to work) equally well for non-vanishing total lepton number. This scenario will be studied in more detail in an upcoming publication.

\subsection{BBN limits on primordial asymmetries with vanishing total lepton number}
\label{subsec:limits}

\begin{figure}[!t]
\centering
\begin{tabular}{cc}
\hspace{-0.cm}\includegraphics[width=0.5\textwidth]{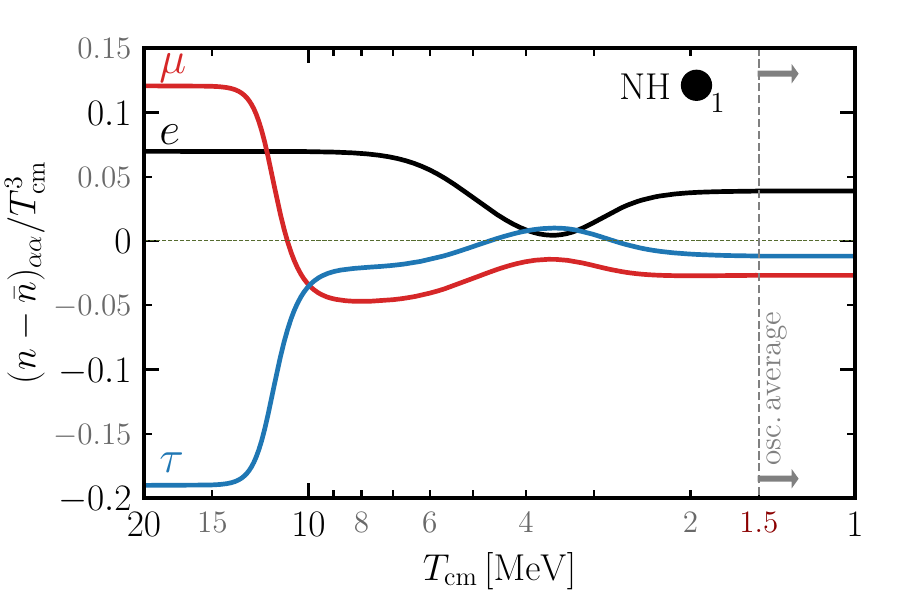}
\hspace{-0.cm}\includegraphics[width=0.5\textwidth]{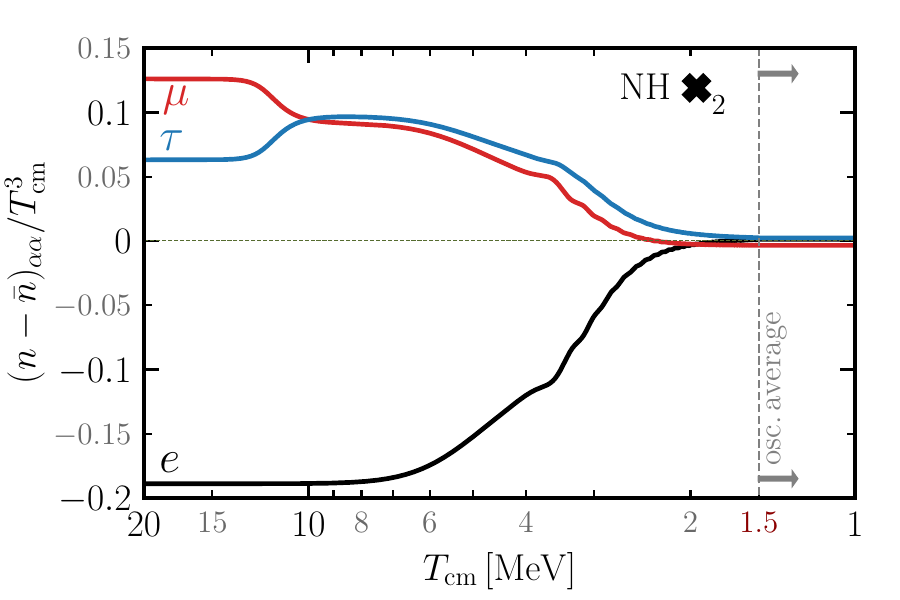}
\end{tabular}
\caption{Typical examples of the time evolution for large amplitude of initial flavor asymmetries in normal hierarchy.
 In the left (right) plot we choose $A = \sqrt{2}, \phi = \pi/3$ ($A = \sqrt{2}, \phi = \mathrm{arctan}(-3,2) \simeq 2.55$). 
 The full solution depicted here essentially overlaps with the adiabatic approximation $(V_s =0)$.
 The symbol in the top right of the plot indicates the positions of this specific initial condition within the contour plots in Figs.~\ref{fig:constraints} and \ref{fig:constraints_intro}.
 }
\label{fig:time-evolution}
\end{figure}

The evolution of neutrino asymmetries from the onset of neutrino oscillations (around $15\,\mathrm{MeV}$) until neutrino decoupling (around $2\,\mathrm{MeV}$) is governed by several key features related to the relative importance of the different terms in Eq.~\eqref{eq:master}, see Fig.~\ref{fig:rates} for a visualization. Initially, both the collision term ${\cal I}$ and the vacuum Hamiltonian ${\cal H}_0$ can be neglected, and the evolution is dominated instead by the matter potentials ${\cal V} = V_c + V_s$, which are diagonal in flavor space.  Neutrino oscillations begin once off-diagonal components of the Hamiltonian start to play a role, which happens once the mass splittings contained in the vacuum Hamiltonian become large compared to the charged current contributions to the matter potentials in $V_c$. For the muon contribution, this occurs at around $12\,\mathrm{MeV}$ (for the large mass splitting $\Delta m^2_{31}$) due to the Boltzmann suppression of the muons, and is referred to as muon-driven MSW transition. For the electron contribution, this occurs (again for the large mass splitting) at around $5\,\mathrm{MeV}$, due to the $T^5$ scaling of the matter potential and the $1/T$ scaling of vacuum Hamiltonian, and is identified as the electron-driven MSW transition (see Eq.~\eqref{eq:master}). The time scales of these transitions need to be compared with the frequency of neutrino oscillations. We give analytic expressions for this frequency in App.~\ref{app:slowness-factor}, generalizing results obtained in the two-flavor case~\cite{Froustey:2021azz}. For large neutrino asymmetries this typically results in adiabatic MSW transitions, with typical examples of the evolution of the lepton flavor asymmetries $\Delta n_\alpha$ shown in Fig.~\ref{fig:time-evolution}. In this case, the oscillations in the off-diagonal components of the density matrix are so fast that their effect averages out in the evolution of the diagonal components which enter the lepton flavor asymmetries $\Delta n_\alpha$. The evolution is well described by the adiabatic approximation described in more detail below.
\vspace{5mm}

\paragraph{Adiabatic approximation.}
For the neutrino asymmetries of interest in this paper, $|\xi| \gtrsim 10^{-3}$, the self-interaction potential $V_s$ dominates over all other terms in the Hamiltonian which leads to the effect of synchronous neutrino oscillations~\cite{Pastor:2001iu,Dolgov:2002ab,Abazajian:2002qx,Wong:2002fa,Mangano:2010ei,Froustey:2021azz}: All momentum modes become aligned and perform synchronous oscillations, which can be interpreted as a precession of the asymmetry vector around the Hamiltonian in $SU(N)$ space (see App.~\ref{app:slowness-factor}). This phenomenon has two important consequences: First, it is one of the reasons why the momentum averaged QKEs capture the dynamics of the full system to very good accuracy. Second, it ensures that for sufficiently large neutrino asymmetries, the density matrix is aligned with the Hamiltonian (i.e.\ $\rho$ and $V_s$ commute). Thus starting from initial conditions which are diagonal in flavor space, we can initially drop $V_s$ from the right-hand side of Eq.~\eqref{eq:master}. Once the density matrix develops off-diagonal elements it no longer commutes with $V_s$, however if these off-diagonal elements are rapidly oscillating it is justified to consider only their average impact, which vanishes. Thus, dropping $V_s$ \textit{throughout the evolution} yields an adiabatic approximation for the evolution of the diagonal entries of $\rho$. This approximation reproduces the time evolution shown in Fig.~\ref{fig:time-evolution} to very good accuracy and corresponds to the adiabatic transfer of averaged oscillations (more precisely the ATAO-${\cal V}$) scheme described in Ref.~\cite{Froustey:2021azz}. This procedure enables a rapid and reliable evaluation of the QKEs for adiabatic MSW transitions.

The adiabatic approximation described above becomes inappropriate once the frequency of synchronous neutrino oscillations (discussed in App.~\ref{app:slowness-factor}) becomes slow compared to the time scale of the MSW transitions. In this case, the MSW transitions act as an abrupt change to the dynamics, to which the system responds with rapid oscillations of the asymmetry vector. This notably occurs in two cases, both of which we will discuss in more detail below: (i) small neutrino asymmetries which reduce the overall impact of $V_s$ and (ii) specific flavor directions in which the leading order term to the synchronous oscillation frequency vanishes.

\vspace{5mm}
\paragraph{Energy transfer.}
The (partial) equilibration of neutrino asymmetries is a non-reversible process which leads to an increase of total comoving entropy~\cite{Froustey:2024mgf}. This results in a heating of the neutrino sector and thus an energy transfer between the neutrino and electron-photon plasma. The collision terms allow us to track this energy transfer while taking into account the process of neutrino decoupling. The net increase of energy in the neutrino sector leads to a value of $N_\text{eff}$ which is always larger than the SM value. As an aside, we note that for the purpose of computing only the asymmetries~\eqref{eq:expr_asymmetry}, tracking this energy transfer is not necessary. In this case, one can set $z_\nu = z_\gamma = 1$ throughout and drop Eq.~\eqref{eq:z-evolution}. However, for estimating the contribution to $N_\text{eff}$, evolving the neutrino and photon temperature dynamically is crucial. 

The energy transfer occurs in a small time window between the onset of neutrino oscillations ($T \simeq 15$~MeV) and neutrino decoupling ($T \simeq 5$~MeV). This leads to interesting differences between the results for NH versus IH. In IH the transitions are resonant (there is level crossing) and for specific scenarios one can visibly see that the flavor evolution of the asymmetries is significantly faster, see e.g. Fig.~\ref{fig:delayed_electron_MSW}. This implies that flavor equilibration can happen when $e^+e^-\leftrightarrow\nu\bar{\nu}$ interactions are still efficient, and consequently the system will flow towards pure thermal equilibrium with $N_{\rm eff}\simeq 3$. This property is clearly seen in the right panel of Fig.~\ref{fig:constraints_intro}. There the region with $\xi_e^{\rm BBN}\simeq 0$ (perfect flavor equilibration) also correspond to $\Delta N_{\rm eff}\simeq 0$. In NH we do not see this effect as the flavor transitions for the region where $\xi_e^{\rm BBN} \simeq 0$ occurs more slowly and at relatively low temperatures ($T\simeq 2-3\,{\rm MeV}$) and hence the flow of energy is significantly less efficient resulting in $\Delta N_{\rm eff} > 0$.

\begin{figure}[!t]
\centering
\begin{tabular}{cc}
\hspace{-0.075cm}\includegraphics[width=0.5\textwidth]{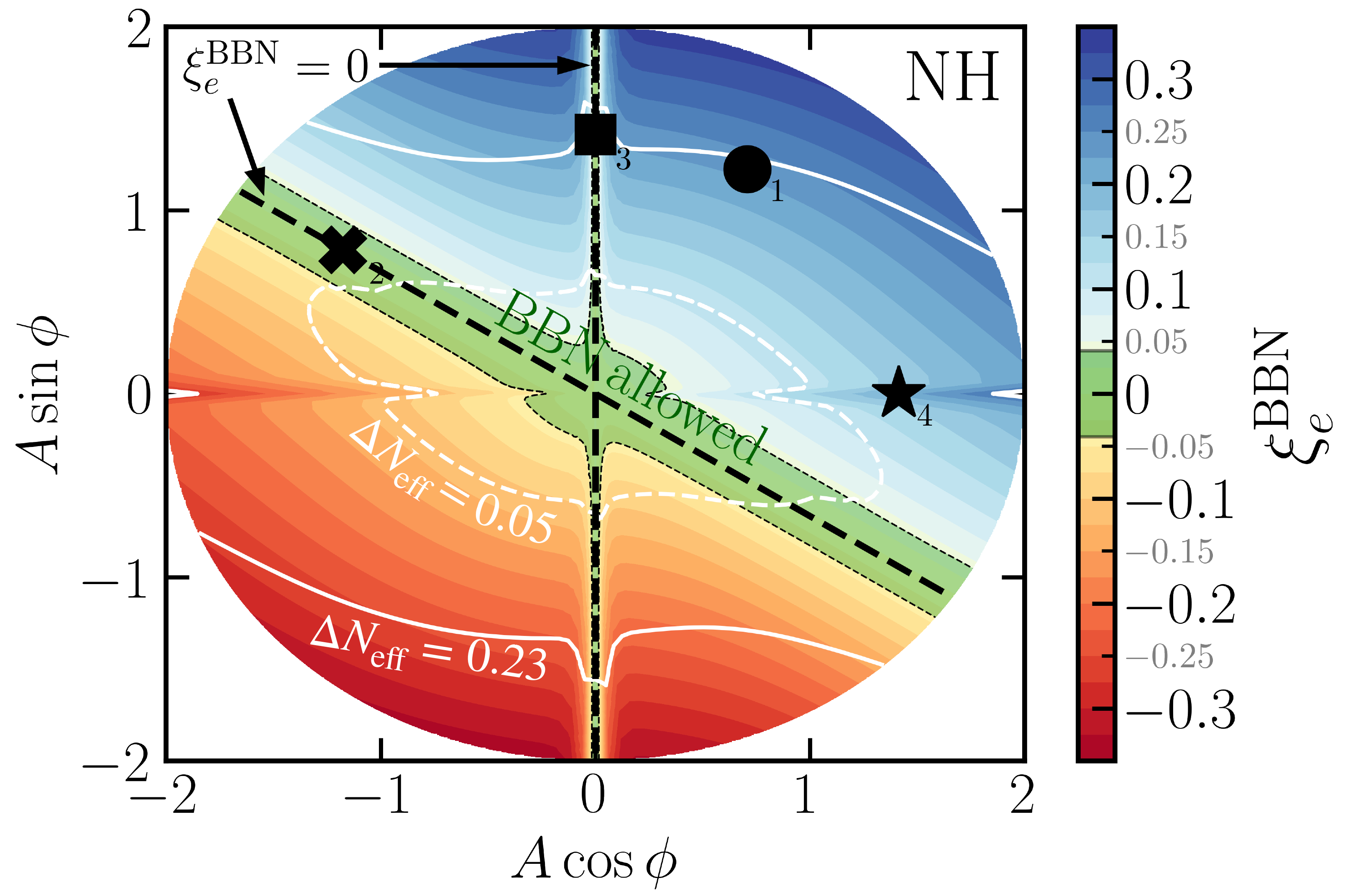}
\hspace{-0.075cm}\includegraphics[width=0.5\textwidth]{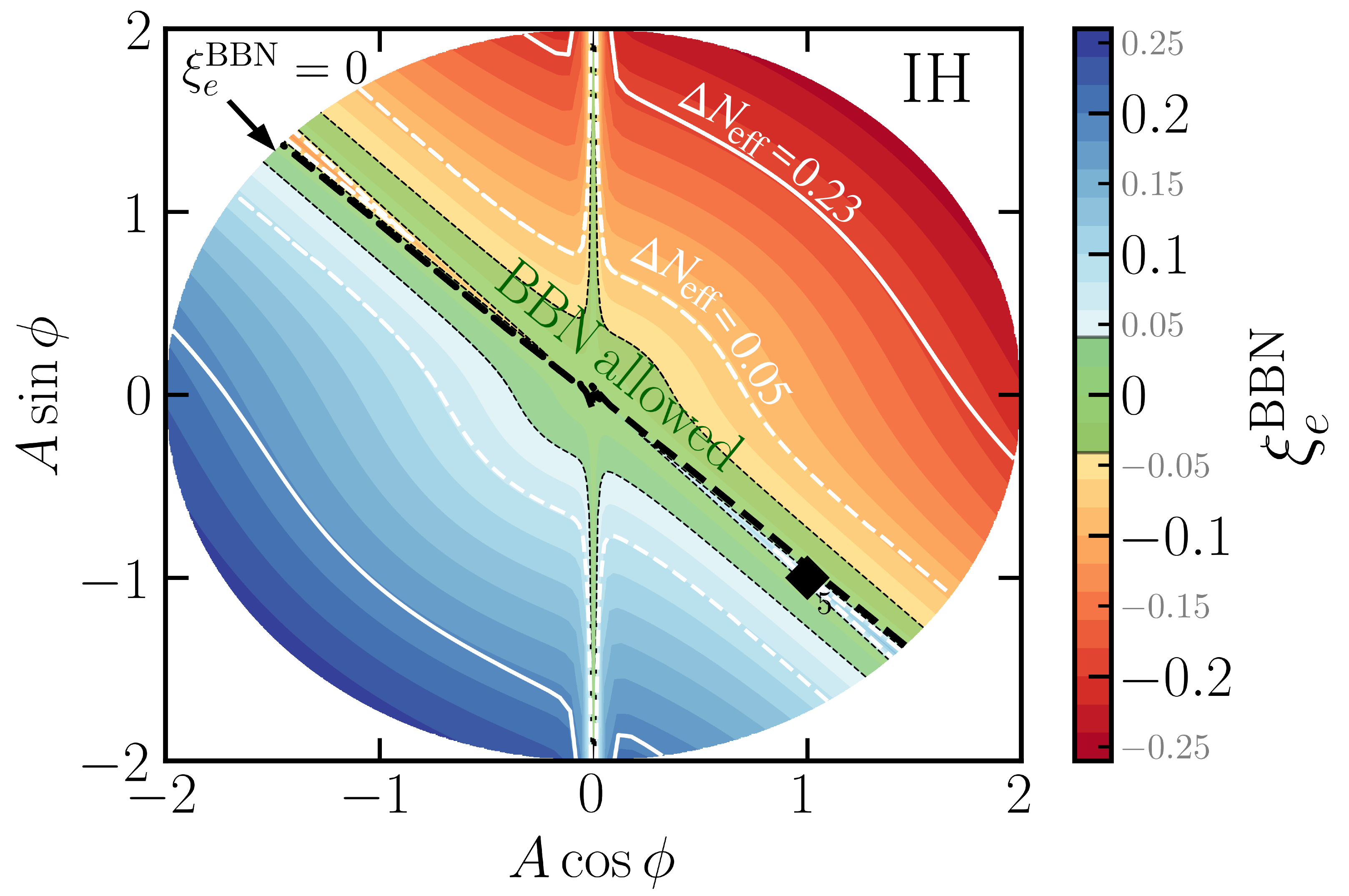} \\
\hspace{-0.075cm}\includegraphics[width=0.5\textwidth]{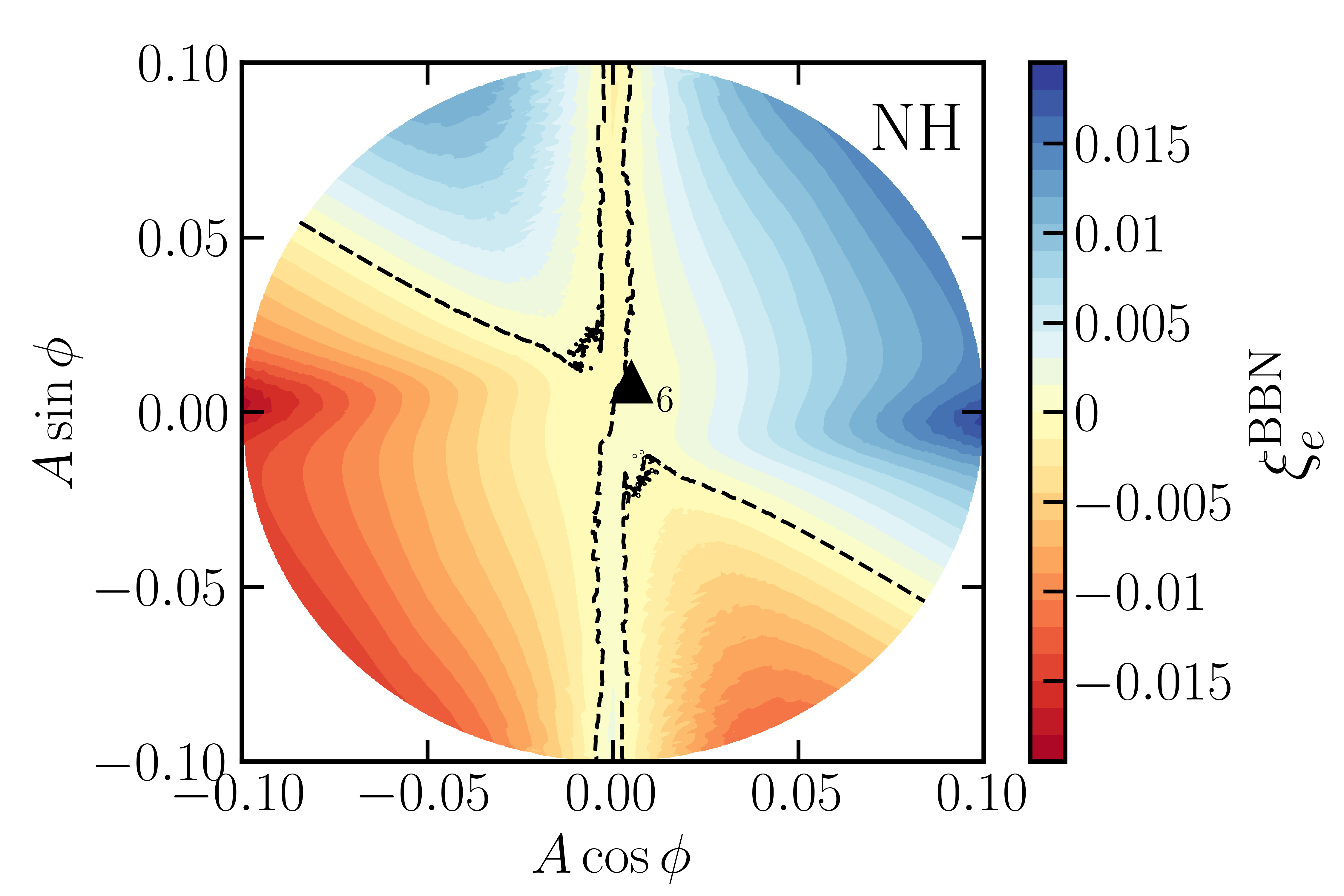}
\hspace{-0.075cm}\includegraphics[width=0.5\textwidth]{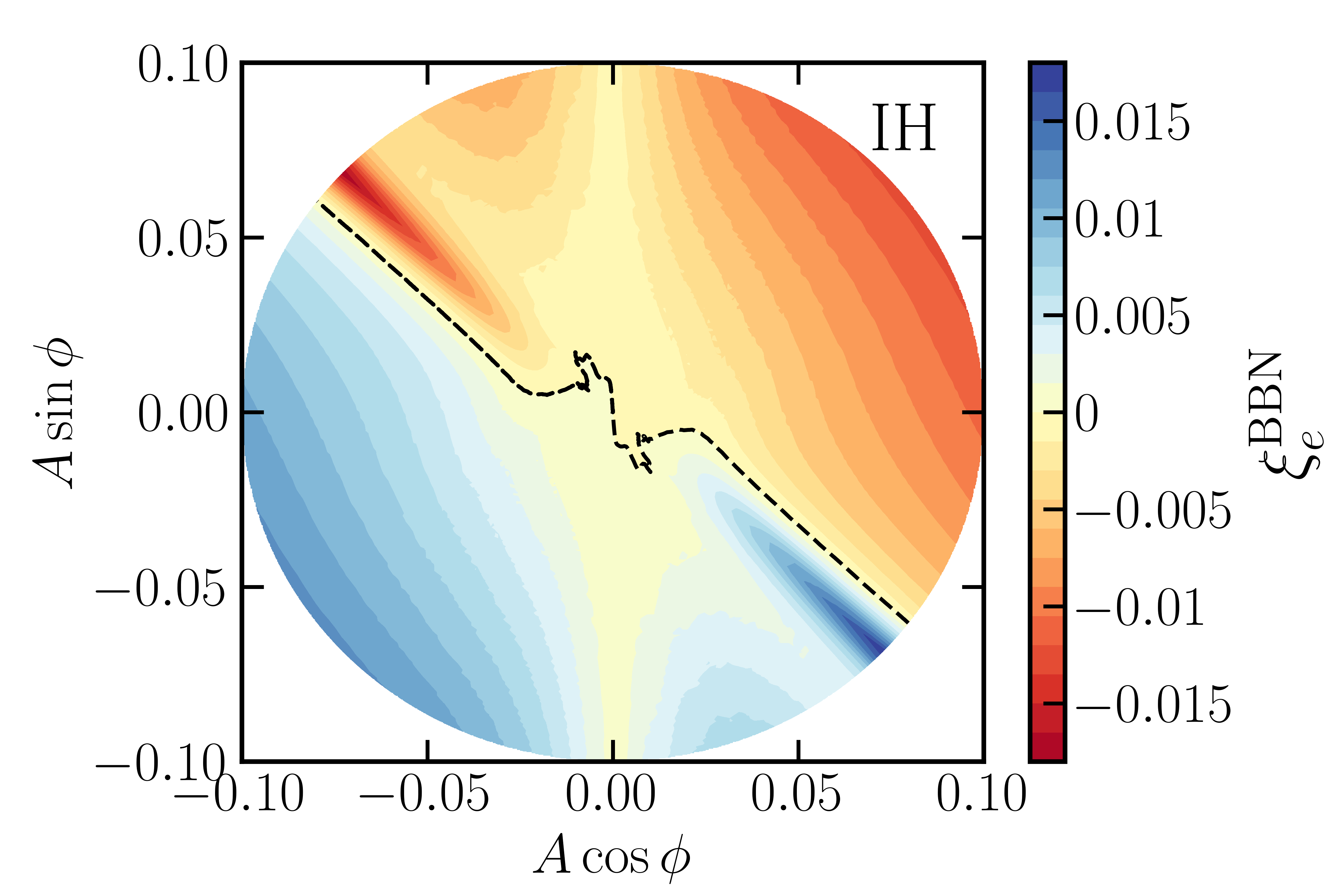}
\end{tabular}
 \caption{Asymmetry in the electron neutrino flavor at the onset of BBN as a function of primordial $e,\mu,\tau$ asymmetries featuring zero total lepton number. Left and right panels correspond to NH and IH respectively, and the lower panels are close-ups of the central regions of the upper panels. The variables $A$ and $\phi$ are related to the chemical potentials via Eq.~\eqref{eq:sph} with $\theta$ given Eq.~\eqref{eq:parametrization_theta}, see also Fig.~\ref{fig:parametrization}. The values $\phi = \{0, \pi\}$ correspond to vanishing $\xi_\mu^{\rm ini}$ and values $\phi = \{ \pi/2, 3 \pi/2 \}$ correspond to vanishing $\xi_e^{\rm ini}$. In green we highlight the region of parameter space currently allowed by BBN at $2\sigma$ CL, see Eq.~\eqref{eq:xie_BBNbound}. The dashed black lines correspond to a final $\xi_e^{\rm BBN} = 0$. The white contours represent the contribution of the asymmetries to $\Delta N_{\rm eff}$, see Eq.~\eqref{eq:def_dNeff}, with the solid (dashed) white line indicating the current CMB constraint~\eqref{eq:dNeff_limit} (the target sensitivity of the Simons Observatory~\cite{SimonsObservatory:2018koc}). The black symbols mark parameter points which will be discussed in detail below, and whose time evolution is shown in Fig.~\ref{fig:time-evolution}, Fig.~\ref{fig:delayed_muon_MSW} and Fig.~\ref{fig:delayed_electron_MSW}.
 }
 \label{fig:constraints}
\end{figure}

\vspace{5mm}
\paragraph{Constraints from BBN and CMB.}
Fig.~\ref{fig:constraints} (see also Fig.~\ref{fig:constraints_intro}) summarizes our results for final asymmetries (at $T_{\mathrm{cm}} = 1\,\mathrm{MeV}$) as a function of the initial asymmetries (at $T_{\mathrm{cm}} = 20\,\mathrm{MeV}$) for the case of vanishing total lepton number, see Eq.~\eqref{eq:def_L0}. 
At late times, the two quantities of interest are the asymmetry in the electron flavor $\xi_{e}$ which is bounded by BBN~\cite{Froustey:2024mgf},
\begin{align}\label{eq:xie_BBNbound}
    \xi_e^{\mathrm{BBN}}  = 0.001 \pm 0.013 \text{ (exp)} \pm 0.016 \text{ (th)} \quad (@ \, 1\sigma)\,,
\end{align}
and the contribution to the radiation density,
\begin{align}
\label{eq:dNeff_limit}
 \Delta N_\mathrm{eff}
 \leq 0.23 
  \quad (@ \, 2\sigma)
 \,,
\end{align}
which is constrained by both BBN and the CMB, and where here we take the bound from Planck~\cite{Planck:2018vyg}. For the asymmetry in the electron flavor we have included an estimate of our theory uncertainty (see Sec.~\ref{subsec:full_qke_comparison} for details), which is comparable to the current experimental uncertainty. For $\Delta N_\text{eff}$ we find the theory error to be negligible compared to the current experimental uncertainty. We find that in most of the parameter space the evolution is well described by the adiabatic approximation, and these two quantities scale approximately linearly and quadratically with the amplitude $A$ of the initial asymmetry~\eqref{eq:def_A}, respectively, 
\begin{align}
 \xi_e^{\mathrm{BBN}}  = 0.171 \,  (0.125) \,  A \, f_1(\phi)\,, \qquad
 \Delta N_\mathrm{eff}  = 0.14 \, (0.125) \, A^2 \, f_2(\phi) \quad   (\mathrm{for} \quad A < 1)\,,
 \label{eq:minimal_washout}
\end{align}
for NH (IH). For larger values of $A$, we find that $\Delta N_\text{eff}$ continues to grow monotonically, but less rapidly with $A$. The functions $|f_{1,2}(\phi)| \leq 1$ are shown in Fig.~\ref{fig:f12} and their implications, as well as non-adiabatic deviations, are discussed in more detail in the following.  We note in particular that $f_1$ and $f_2$ in the case of IH can vanish even for large $A$, indicating flavor directions in which the initial asymmetry is unconstrained by these observations. On the contrary, $f_2$ for the case of NH is bounded by $f_2 \geq 0.2$, indicating a flavor-universal minimal $\Delta N_\text{eff}$ bound, as indicated by the closed $\Delta N_\text{eff}$ contour in Fig.~\ref{fig:constraints}.

In Fig.~\ref{fig:constraints}, the color code refers to the values of $\xi_e$ as relevant for BBN, with the green band in the upper row corresponding to the region currently allowed by observations taking into account both experimental and theoretical uncertainties. The close-up of the central region shown in the bottom row is entirely within the BBN limits. The clearly visible symmetry structure reflects a sign flip in all asymmetries, i.e.\ $\phi \mapsto \phi + \pi$ flips all signs of $\xi_\alpha$ in Eq.~\eqref{eq:sph} while preserving $|\xi_\alpha|$. Assuming a smooth evolution of the final asymmetries as a function of the initial flavor composition, this implies that there necessarily needs to be a direction (and its opposite) in flavor space for which the final electron asymmetry vanishes, i.e.\ $f_1 = 0$ in Eq.~\eqref{eq:minimal_washout}. We find this to occur for $\tan \phi \simeq -2/3$ (NH) and $\tan \phi \simeq -1$ (IH), indicated by the black dashed lines within the broad green bands, which we refer to as the direction of maximal asymmetry washout. Looking at Fig.~\ref{fig:parametrization} we note that this falls within the gray shaded regions which indicate $\mathrm{sign}( \xi_e^\mathrm{ini}) = - \mathrm{sign }( \xi_\mu^\mathrm{ini}) = - \mathrm{sign }( \xi_\tau^\mathrm{ini})$. If the evolution is adiabatic, this sign constellation is preserved throughout the evolution and leads to particularly strong washout of the electron flavor. Although the adiabatic approximation is accurate across the majority of the parameter space, there are significant deviations from this behavior when $\xi_e^\mathrm{ini} = 0$ and $\xi_\mu^\mathrm{ini} = 0\, (\xi_\tau^\mathrm{ini} = 0)$ for NH (IH). The origin of these non-adiabatic evolutions are explained in detail in Sec.~\ref{subsec:MSW} and App.~\ref{app:slowness-factor}. They happen at $\phi = \{\pi/2,0 \}$ for NH and $\phi = \{\pi/2, 3\pi/4\}$ for IH. In particular for $\phi = \pi/2$ ($\xi_e^\mathrm{ini} = 0$) in both NH and IH the non-adiabatic transitions lead to a drastic suppression of $\xi_e^{\mathrm{BBN}}$. In fact, this leads to $\xi_e^{\mathrm{BBN}} = 0 $ for NH, while for IH $\xi_e^{\mathrm{BBN}}$ is pushed close enough to zero to fulfill the BBN bound but never actually vanishes. These features are visible in the vertical axis of Fig.~\ref{fig:constraints}. On the contrary, for $\phi = 0$ in NH or $\phi =3\pi/4$ in IH, the system undergoes a non-adiabatic electron-driven MSW transition. As can be seen from Fig.~\ref{fig:delayed_electron_MSW} this leads to a larger than usual final value of $\xi_e^{\mathrm{BBN}}$. This explains the inward flowing spikes in Fig.~\ref{fig:constraints} at these angles.

The contribution to $\Delta N_\mathrm{eff}$ is shown by the white contours. The current CMB bound~\eqref{eq:dNeff_limit} (solid) is significantly less constraining than the BBN constraint on the electron asymmetry throughout the depicted parameter space. Upcoming improvements on the accuracy of $\Delta N_\text{eff}$ will however change this situation, in particular for large asymmetries along the maximal-washout direction in NH. To illustrate this, we show also the contour of $\Delta N_\text{eff} = 0.05$, which corresponds to the target sensitivity of the ongoing Simons Observatory~\cite{SimonsObservatory:2018koc}. In the case of IH, the contours for $\Delta N_\text{eff}$ are to good approximation aligned with the contours of the final electron asymmetry. We find the constraints from $\Delta N_\text{eff}$ to be less constraining than the BBN limits on $\xi_e$ even for the target sensitivity of the Simons Observatory. This is, of course, specific to the case of vanishing total lepton number. Since the total lepton number is preserved throughout this process, a finite lepton number is an obstruction to a complete washout of the asymmetries and hence such scenarios are bounded by their contribution to $\Delta N_\mathrm{eff}$ even for moderate initial asymmetries.

\vspace{5mm}
\paragraph{Summary and comparison with previous works.}
These results should be contrasted with the case $\xi_e^\mathrm{ini} \simeq \xi_\mu^\mathrm{ini} \simeq \xi_\tau^\mathrm{ini}$ considered e.g.\ recently in \cite{Escudero:2022okz,Froustey:2024mgf}, which constrain primordial lepton asymmetries to $\xi_\alpha^\mathrm{ini} \sim 0.03$ at $95$\% CL. A key result of our study is thus to note that  the primordial asymmetries in a generic total lepton number conserving flavor direction are washed out by about a factor 6 for NH and a factor 8 for IH (see Eq.~\eqref{eq:minimal_washout} for more precise numerical values) while in specific flavor directions the washout can be even stronger up to complete erasure of the primordial asymmetries. Identifying these directions is crucial to identify viable scenarios leading to large asymmetries in the early Universe which remain compatible with BBN and CMB constraints.

The case of lepton flavor asymmetries with total vanishing lepton numbers has been previously studied in Refs.~\cite{Pastor:2008ti,Mangano:2010ei,Castorina:2012md}. At that time, the mixing angle $\theta_{13}$ was not well measured and hence these references give results for two benchmark scenarios of $\sin^2\theta_{13} = 0$ and $\sin^2\theta_{13} = 0.04$. Moreover, these references only study a particular direction in flavor space, assuming that the muon-driven MSW transition fully equilibrates the $\mu$ and $\tau$ flavors, $\Delta n_\mu = \Delta n_\tau$, while leaving the primordial value of $\Delta n_e$ unchanged. The QKEs for the asymmetries are solved from an initial temperature of $T = 10$~MeV. Resulting upper bounds on the primordial asymmetry in the electron flavor varied from $\xi_e^\mathrm{ini} \lesssim 0.12$ (for $\sin^2\theta_{13} = 0$) to essentially unconstrained (for $\sin^2\theta_{13} = 0.04$). Our results go beyond these pioneering studies by using the current best-fit $\sin^2\theta_{13} \simeq  0.02$ and tracking the asymmetries through the muon-driven MSW transition while keeping the full flavor dependence. On the methodology side, as discussed above, a key improvement is the treatment of the collision term.

Nowadays, all neutrino mixing parameters apart from the $\mathrm{CP}$-violating phase are well measured~\cite{ParticleDataGroup:2024cfk}. We have explicitly checked that our results are robust against variation of these parameters within their experimental $3\sigma$ uncertainty. In particular, we find the final asymmetries to be largely insensitive to the $\mathrm{CP}$-violating phase (confirming the results of \cite{Gava:2010kz,Froustey:2021azz}) as well as to all other oscillation parameters with the exception of $\theta_{13}$. Varying the latter within the 3$\sigma$ band of experimental uncertainty~\cite{ParticleDataGroup:2024cfk}, we observe at most a ${\cal O}(10\%)$ effect on final asymmetry in the electron neutrino flavor. The impact is most relevant in the regime where the adiabatic approximation is valid and gives a relatively large asymmetry (i.e.\ it in particular impacts the upper bound on $f_1(\phi)$) while it essentially vanishes when $f_1(\phi) \simeq 0$ and in the flavor directions leading to non-adiabatic MSW transitions.

\subsection{Non-adiabatic MSW transitions}
\label{subsec:MSW}

\paragraph{Muon-driven MSW transition.} The muon-driven MSW transition occurs when the muon mean field effects (contained in $V_c$) have decreased to become comparable with the large mass gap of the vacuum Hamiltonian ${\cal H}_0$, at $T \simeq 12$~MeV, see Fig.~\ref{fig:rates} as well as Ref.~\cite{Froustey:2021azz} for a detailed discussion. The full three flavor system can be well approximated by the $\mu-\tau$ subspace for this purpose given that $V_c^{ee}$ is much greater than any oscillation frequency. If the transition is adiabatic, i.e.\ if the change in the Hamiltonian is slow compared to the frequency of synchronous oscillations, the asymmetries adapt adiabatically without exhibiting significant oscillations. The adiabaticity factor is given by~\cite{Froustey:2021azz} (see also App.~\ref{app:slowness-factor}):
 \begin{align}
  \gamma_{\mu-\mathrm{MSW}} \simeq 100  \, |\xi_\mu + \xi_\tau| \,.
 \end{align}
Here $|\xi_\mu + \xi_\tau|$ parametrizes the `slowness factor' (to leading order in $V_c/V_s$) of the synchronous oscillations induced by the muon-driven MSW transition.  For values of $\xi^\mathrm{ini} > 0.01$, of interest for BBN, we generically have $ \gamma_\mathrm{MSW} > 1$ meaning that the transition is adiabatic.

\begin{figure}[!t]
\centering
\begin{tabular}{cc}
\hspace{-0.cm}\includegraphics[width=0.5\textwidth]{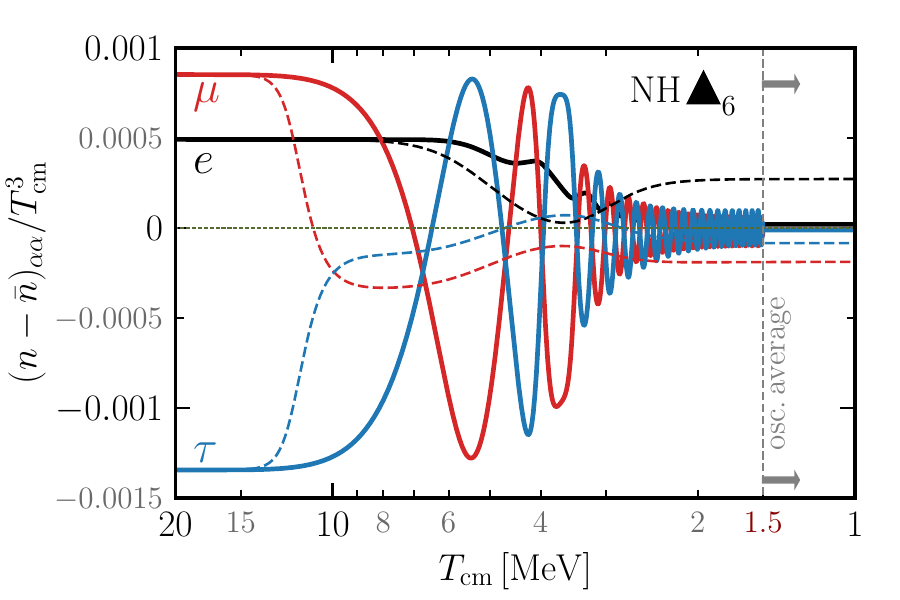}
\hspace{-0.cm}\includegraphics[width=0.5\textwidth]{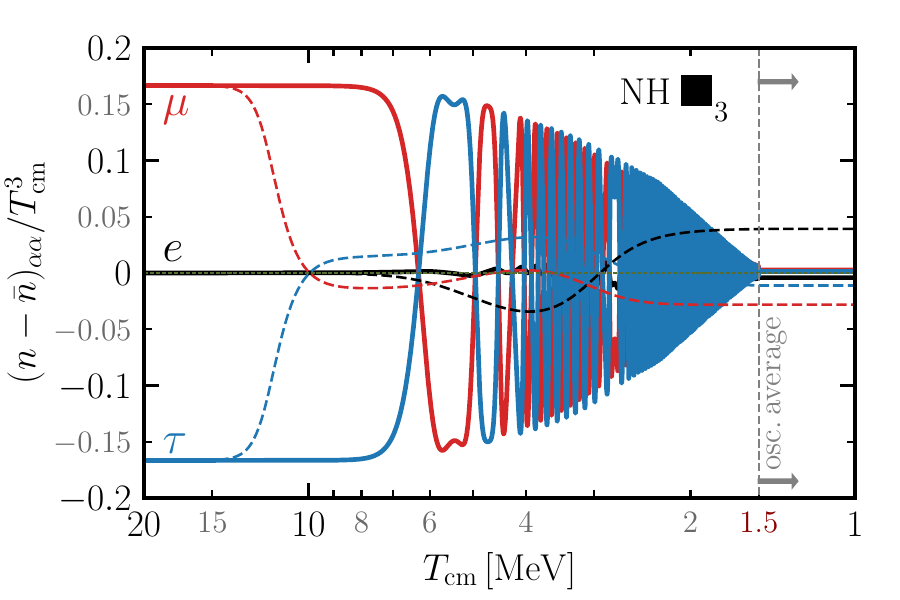}
\end{tabular}
\caption{Non-adiabatic muon-driven MSW transition due to small overall asymmetry (left: $A = 0.01, \phi = \pi/3$) and vanishing leading order slowness factor (right: $A = \sqrt{2}, \phi = \pi/2$). Solid curves indicate the full solution and dashed lines the adiabatic approximation $(V_s\to 0)$. The two clearly differ for these benchmarks. 
The symbol in the top right of the plot indicates the positions of this specific initial condition within the contour plots in Figs.~\ref{fig:constraints} and \ref{fig:constraints_intro}.
}
\label{fig:delayed_muon_MSW}
\end{figure}

We identify two cases for which the transition becomes abrupt: Small values of the overall asymmetry (shown in the left panel of Fig.~\ref{fig:delayed_muon_MSW}) or the special direction in flavor space given by $\xi_\mu^\mathrm{ini} = - \xi_\tau^\mathrm{ini}$. 
If lepton number is conserved, the latter is equivalent to $\xi_e^\mathrm{ini} = 0$. An example of this type is shown in the right panel of Fig.~\ref{fig:delayed_muon_MSW}. As expected, in these cases the evolution of the asymmetries deviates notably from the adiabatic approximation and we note a delayed response to the muon-driven MSW transitions followed by rapid oscillations between the $\mu$ and $\tau$ lepton flavor asymmetries.

The impact of non-adiabatic MSW transitions on the final lepton asymmetries can be seen in Fig.~\ref{fig:constraints}. The lower panels highlight the regime of small initial asymmetries. The case $\xi_e^{\mathrm{ini}}=0$ corresponds to the $y$-axis in Fig.~\ref{fig:constraints} and we note a significantly \textit{smaller} final electron asymmetry than obtained in the adiabatic approximation. From the right panel of Fig.~\ref{fig:delayed_muon_MSW} we can get an intuition as to why: Once the asymmetries in the $\mu$ and $\tau$ flavors start evolving, they oscillate so rapidly that the electron flavor only sees the oscillation-averaged (and hence approximately vanishing) asymmetry. The electron asymmetry thus stays very close to its initial value $\xi_e^\mathrm{ini} = 0$. In fact, in the NH scenario, this effect leads to a sign flip,  $\mathrm{sign}(\xi_e^\mathrm{ini}) = - \mathrm{sign}(\xi_e^\mathrm{BBN})$ along the y-axis, accompanied by two zero crossings $\xi_e^\mathrm{BBN} = 0$ in the immediate vicinity, as can also be nicely seen in Fig.~\ref{fig:f12}. In the bottom left panel of Fig.~\ref{fig:constraints} this corresponds to the two vertical black lines representing a vanishing final electron asymmetry (see also upper panel of Fig.~\ref{fig:constraints}). For the IH case, the asymmetry in the electron sector remains small but does not flip sign. Capturing these fine quantitative details requires solving the QKE in the non-adiabatic regime.

\vspace{5mm}
\paragraph{Electron-driven MSW transition.} The electron-driven MSW transition occurs when the electron mean field effects have decreased to become comparable with the oscillation frequency associated with the large mass gap of the vacuum Hamiltonian, which occurs at $5\,\mathrm{MeV}$. An analytic understanding of the (non-)adiabaticity of this transition is more challenging as it involves all three flavors. We provide this derivation in App.~\ref{app:eMSW} (to our knowledge for the first time) but we summarize the key points here. Assuming that the muon-driven MSW transition is adiabatic, the asymmetry matrix $\Delta r$ tracks the evolution of $\langle {\cal H}_0 \rangle + \langle V_c \rangle$ throughout the muon-driven MSW transition. Approaching the electron-driven MSW transition, we then need to determine under which conditions the right-hand side of
\begin{align}
 \frac{d {\Delta r}}{dx} \propto \left[ \langle {\cal H}_0 \rangle + \langle V_c \rangle, r + \bar{r} \right]\,,
\end{align}
vanishes (i.e.\ the frequency of synchronous oscillations vanishes). We find that this non-adiabatic transition occurs when $\Delta r_{e}^{\mathrm{ini}} + \Delta r_\tau^{\mathrm{ini}} = 0$ in NH and $\Delta r_{e}^{\mathrm{ini}} + \Delta r_\mu^{\mathrm{ini}} = 0$ in IH, while the third $\Delta r_\alpha^{\mathrm{ini}}$ can take any value. In our concrete limit of vanishing total lepton number, this implies non-adiabatic transitions for $\xi_\mu^\mathrm{ini} = 0$ in NH and $\xi_\tau^\mathrm{ini} = 0$ in IH.

\begin{figure}[!t]
\centering
\begin{tabular}{cc}
\hspace{-0.cm}\includegraphics[width=0.5\textwidth]{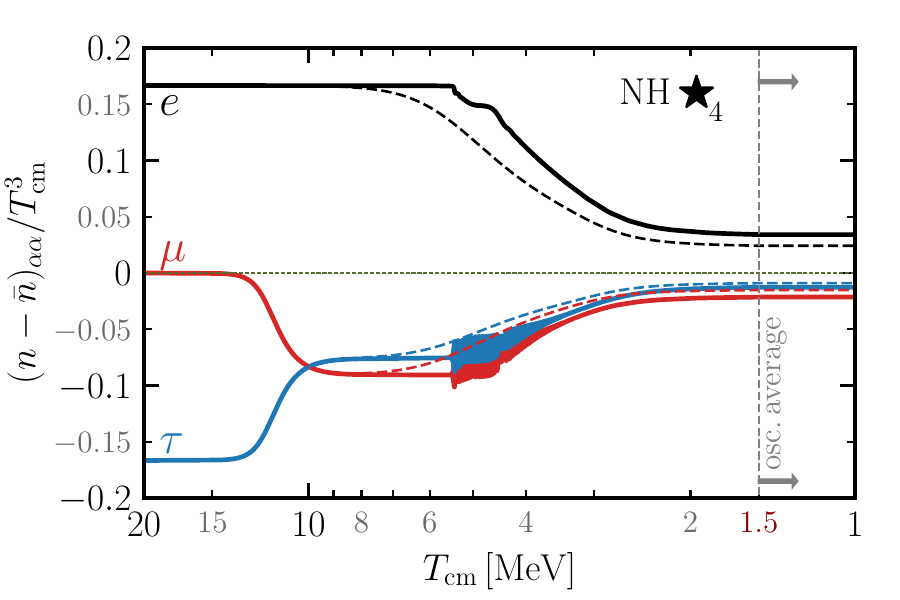}
\hspace{-0.cm}\includegraphics[width=0.5\textwidth]{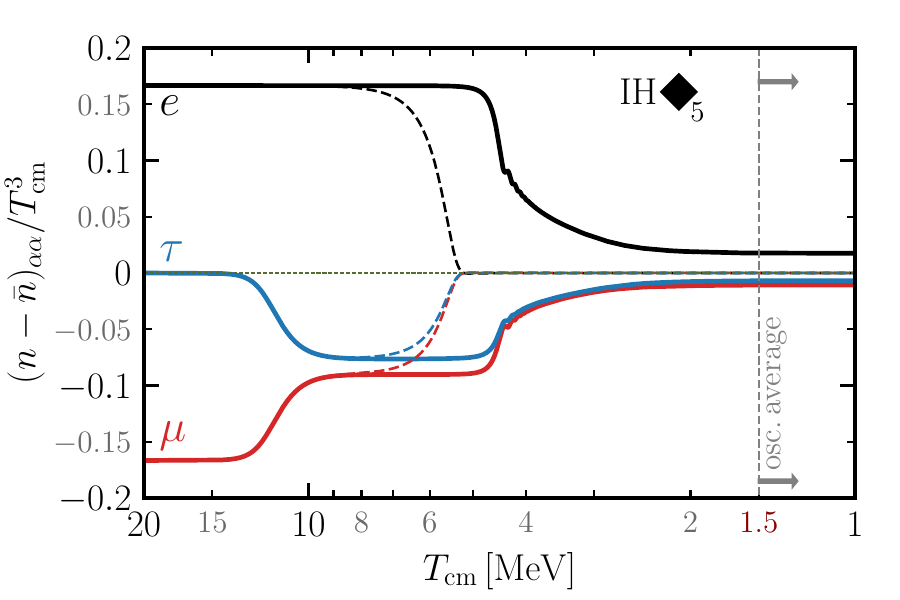}
\end{tabular}
\caption{ Non-adiabatic electron-driven MSW transition for NH (IH) in the left (right) panel with initial conditions $A = \sqrt{2}, \phi = 0$ ($A = \sqrt{2}, \phi = -\pi/4$). Solid curves indicate the full solution and dashed lines the adiabatic approximation $(V_s\to 0)$. The symbol in the top right of the plot indicates the positions of this specific initial condition within the contour plots in Figs.~\ref{fig:constraints} and \ref{fig:constraints_intro}.}
\label{fig:delayed_electron_MSW}
\end{figure}

This is illustrated in Fig.~\ref{fig:delayed_electron_MSW} for NH (left panel) and IH (right panel): We see a delayed evolution compared to the adiabatic approximation followed by rapid oscillations beginning around $5\,\mathrm{MeV}$. 
The impact on the final asymmetry is clearly visible in Fig.~\ref{fig:constraints}, where we note that along these flavor directions, the final electron asymmetry is systematically \textit{larger} compared to the adiabatic case. We can get an intuition as to why from Fig.~\ref{fig:delayed_electron_MSW}: After the muon-driven MSW we are necessarily in a situation where $\mathrm{sign}(\xi_e) = - \mathrm{sign}(\xi_\mu) = - \mathrm{sign}(\xi_\tau)$, which, as discussed above, leads to strong monotone washout of the electron asymmetry. The delay in the evolution thus leads to less washout and thus a larger asymmetry.

\vspace{5mm}
\paragraph{Consequences for BBN and CMB.} The impact of these non-adiabatic MSW transitions at large asymmetries on the constraints from BBN and CMB is twofold. First, as indicated above, these deviations from the adiabatic evolution can modify the direction of maximal washout compared to the adiabatic expectation. This occurs in particular for the electron-driven MSW transition in the IH, since the $\xi_\tau^\mathrm{ini} = 0$ direction coincides with the direction of adiabatic maximal washout. Second, the non-adiabatic muon-driven MSW transition weakens the BBN and CMB constraints for $\xi_\mu^\mathrm{ini} = - \xi_\tau^\mathrm{ini}$, such that values of $|\xi_{\mu,\tau}^\mathrm{ini}|$ of order one and larger are allowed in this particular flavor direction. This is significantly larger than the limits $ |\xi_{\mu,\tau}^\mathrm{ini}| \lesssim 0.1 (0.2)$ for NH (IH) which the adiabatic approximation would give in this direction. Conversely, the non-adiabatic electron-driven MSW transition leads to limits which are more stringent than obtained in the adiabatic approximation.

\subsection{Application to cases of non-vanishing total lepton number}

\begin{figure}[!t]
\centering
\begin{tabular}{cc}
\hspace{-0.cm}\includegraphics[width=0.5\textwidth]{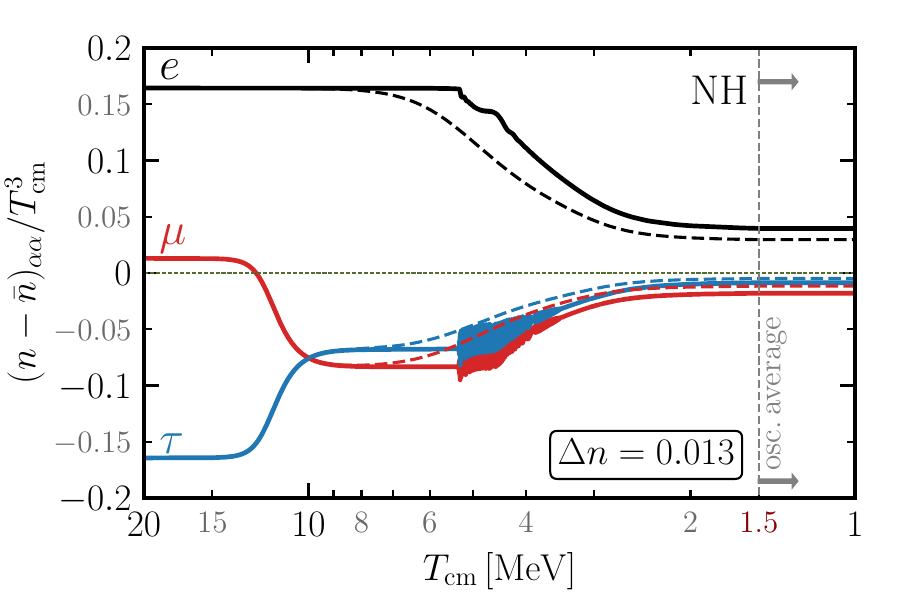}
\hspace{-0.cm}\includegraphics[width=0.5\textwidth]{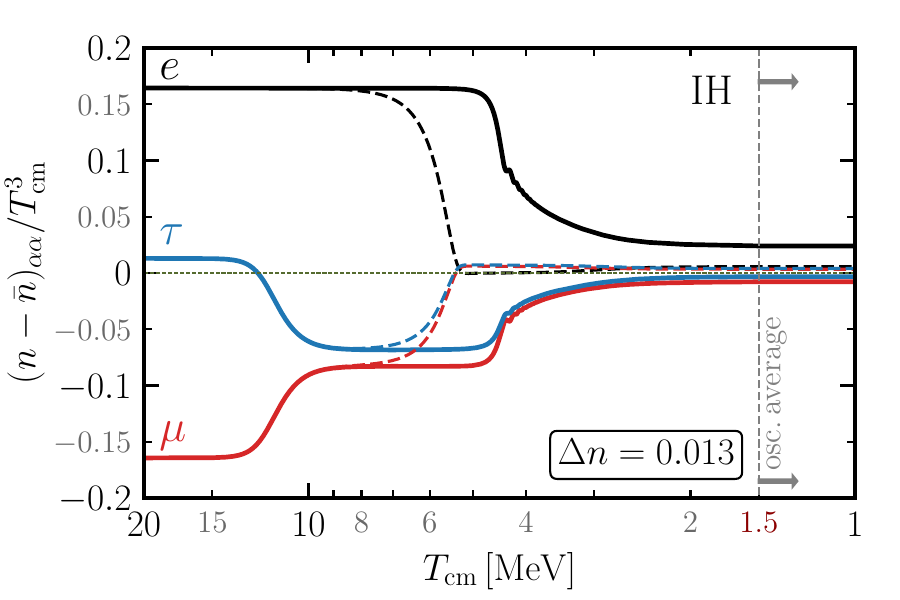}
\end{tabular}
\caption{Examples for non-adiabatic electron-driven MSW transitions for NH (left) and IH (right) for two examples with non-vanishing total lepton number $\Delta n = 0.013$ (corresponding to three times the $2\sigma$ limit on $\Delta n_e\rvert_{\mathrm{BBN}}$) and fixed amplitude $A = \sqrt{2}$ (see Eq.~\eqref{eq:def_A}). For NH this leads to $\xi_e^{\mathrm{ini}} = - \xi_\tau^{\mathrm{ini}} = 0.91$ and $\xi_\mu^{\mathrm{ini}} = 0.078$, while for IH we interchange $\mu \leftrightarrow \tau$, see App.~\ref{app:eMSW}. Results for the non-adiabatic muon-driven MSW transition for $\Delta n \neq 0$ are not shown but we explicitly checked that they generalize in the same way, see App.~\ref{app:muMSW} for the analytical derivation.}
\label{fig:nonvanishingL}
\end{figure}

The discussion in this paper largely focuses on the case of lepton flavor asymmetries subject to the condition of vanishing total lepton number. However, we stress that the numerical tools developed here work equally well for cases with a non-zero total lepton number. In fact, the vast majority of our verification points discussed in the next subsection fall in this category. As an example, Fig.~\ref{fig:nonvanishingL} illustrates non-adiabatic electron-driven MSW transitions for NH (IH) with $\sum_\alpha \Delta n_\alpha = 0.013$ and $\xi_e^\mathrm{ini} = - \xi_\tau^\mathrm{ini}$ ($\xi_e^\mathrm{ini} = - \xi_\mu^\mathrm{ini}$). 
As derived in App.~\ref{app:eMSW}, the conditions for triggering a non-adiabatic electron-driven MSW transition do not depend on the value of total lepton number, which is confirmed by the numerical results shown in Fig.~\ref{fig:nonvanishingL}. Similar conclusion also apply for the non-adiabatic muon-driven MSW transition.

\subsection{Comparison to solutions retaining the momentum dependence}
\label{subsec:full_qke_comparison}

We have verified the excellent agreement between the momentum averaged procedure developed here and the solutions obtained by solving the computationally much more demanding momentum dependent QKEs. To this end, we benchmark our results against the data set of about $8300$ points in flavor space provided as supplementary material in Ref.~\cite{Froustey:2024mgf}. Comparing lepton flavor asymmetries at BBN,
\begin{align}
 \delta n_\alpha \equiv \mathrm{max}\left[ \left|\frac{(n - \bar n)_{\alpha \alpha}^\mathrm{this \,work}}{(n - \bar n)_{\alpha \alpha}^\mathrm{Froustey+}} \right|, \left|\frac{(n - \bar n)_{\alpha \alpha}^\mathrm{Froustey+}}{(n - \bar n)_{\alpha \alpha}^\mathrm{this \, work}} \right| \right] - 1 \; \Bigg|_{\rm BBN} \,,
\end{align}
we find agreement up to a factor of two, $\delta n_\alpha \leq 1$, for $85$\% of all final lepton flavor asymmetries. Of these, about $47$\% ($11$\%) agree even up to $10$\% ($1$\%). Focusing on the remaining $15$\% which show poor agreement, we note that about $83$\% of them fall in a region which is excluded by BBN or CMB constraints (often with very large values for initial asymmetries), and hence do not impact our conclusions. Of the remaining points, about $43$\% fall in two very specific flavor directions, with $\Delta n_\tau = \Delta n_\mu \simeq - \Delta n_e/c$, with $c = 2$ or $c = 1.83$. Studying the time evolution for these flavor directions in more detail, we see no particular features, the results are well captured by the adiabatic approximations for sizable chemical potentials and the resulting final asymmetries connect smoothly to neighboring flavor directions. In fact according to the analytical derivation presented in App.~\ref{app:slowness-factor} we do not expect any non-adiabatic muon- or electron-driven MSW transition. We thus currently do not have an explanation for this discrepancy, and it would be interesting to understand why the results of the momentum dependent QKEs deviate from the adiabatic approximation in these cases. Excluding these points brings the number of points for which we do not achieve an agreement up to a factor two or better in all three flavors down to about $3.6$\%. This comparison enables us to estimate our theoretical uncertainty on the final asymmetry in the electron flavor. Focusing on the points allowed (at $95$\% CL) by BBN and CMB, we find that the deviation to the data set of Ref.~\cite{Froustey:2024mgf} is well modeled by a Gaussian distribution with zero mean and a standard deviation of 0.016. We thus consider this to be an estimate for the theoretical uncertainty of our analysis and add it in quadrature when comparing our predictions with BBN bounds, see Eq.~\eqref{eq:xie_BBNbound}.

Similarly, focusing only on the points with a flavor direction with approximately vanishing total lepton asymmetry, $\sum_\alpha \Delta n_\alpha /A < 0.01$ (about $400$ points in the data set of Ref.~\cite{Froustey:2024mgf}), we find that after excluding the peculiar flavor direction identified above, $86$\% of all points agree  up to a factor two or better in all three flavors. Overall we consider this an excellent agreement, in particular given significant differences in the implementation and the enormous gain in computing time due to the use of momentum averaged QKEs in our work.

For the contribution to $\Delta N_\text{eff}$, defining
\begin{align}
 \delta  N_\text{eff} \equiv \mathrm{max}\left[ \left|\frac{\Delta N_\text{eff}^\mathrm{this \,work}}{\Delta N_\text{eff}^\mathrm{Froustey+}} \right|, \left|\frac{\Delta N_\text{eff}^\mathrm{Froustey+}}{\Delta N_\text{eff}^\mathrm{this \, work}} \right| \right] - 1 \; \Bigg|_{T_{\mathrm{cm}} = 0.006\,\mathrm{MeV}} \,,
\end{align}
we find agreement up to a factor of two, $\delta  N_\text{eff} \leq 1$, for $99$\% of all final lepton flavor asymmetries. Of these, about $60$\% ($9$\%) agree even up to $10$\% ($1$\%). We conclude that in this case the theoretical uncertainty is subdominant compared to current experimental uncertainty.

\begin{figure}[!t]
\centering
\begin{tabular}{cc}
\hspace{-0.cm}\includegraphics[width=0.5\textwidth]{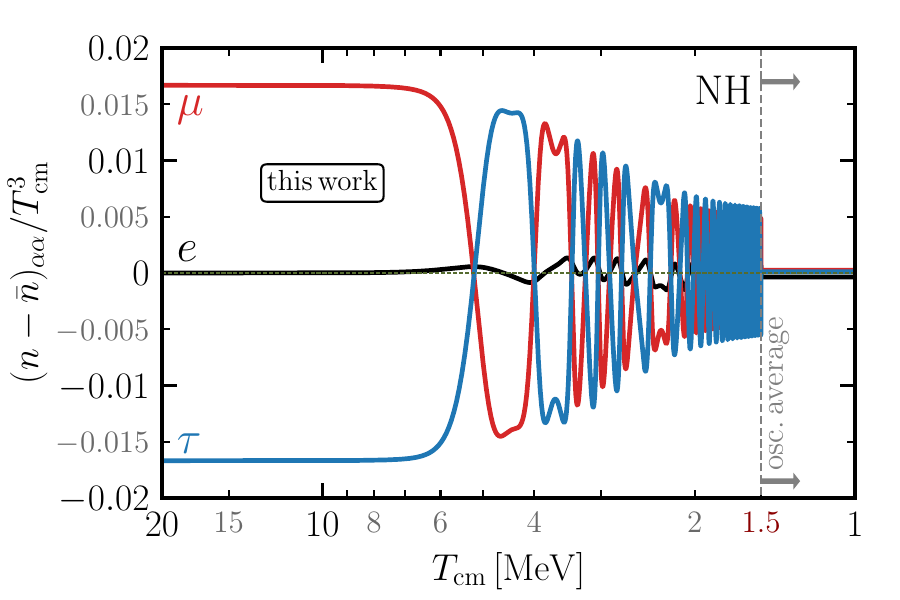}
\hspace{-0.cm}\includegraphics[width=0.5\textwidth]{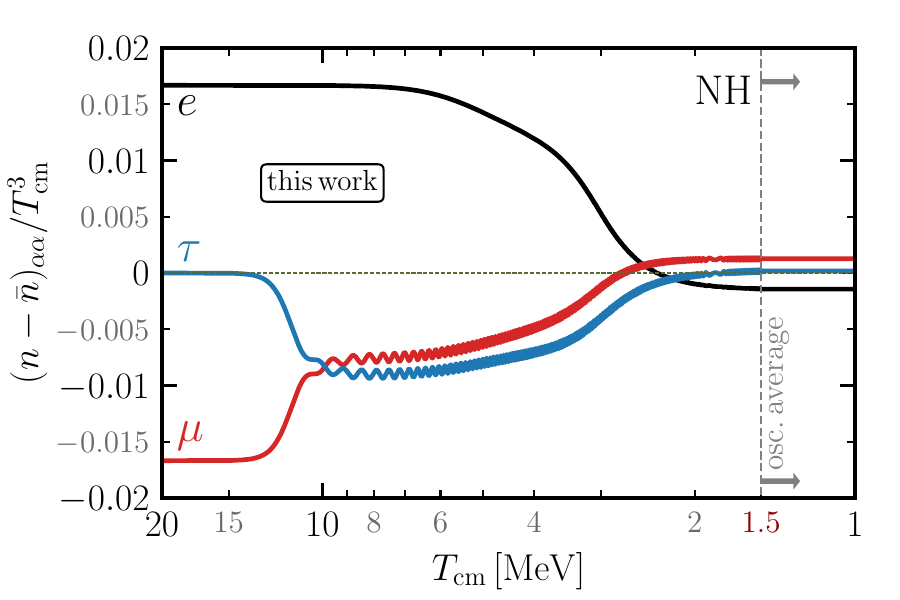} \\
\hspace{-0.cm}\includegraphics[width=0.5\textwidth]{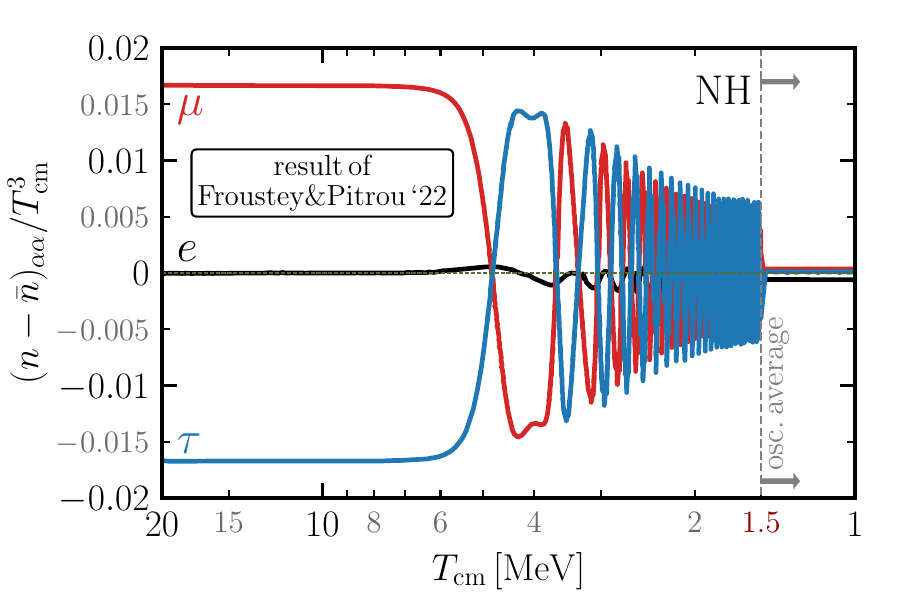}
\hspace{-0.cm}\includegraphics[width=0.5\textwidth]{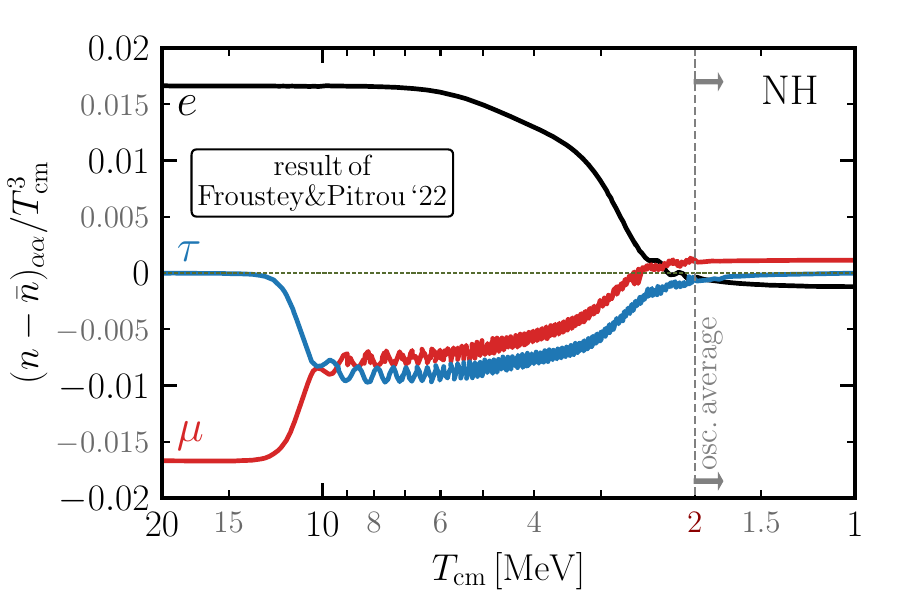}
\end{tabular}
\caption{ Time evolution of the neutrino asymmetries for two benchmarks with vanishing total lepton number as highlighted in Fig.~15 of Froustey \& Pitrou '22~\cite{Froustey:2021azz}. The lower panels show their result of solving the full momentum dependent quantum kinetic equation~\eqref{eq:QKE}, while the upper ones correspond to our results using the momentum averaged approach of Eq.~\eqref{eq:ansatz}. We can clearly appreciate the excellent agreement between our approximation and the full momentum dependent approach. As discussed in Sec.~\ref{subsec:full_qke_comparison} the good agreement extends to a vast region of parameter space for these asymmetries.}
\label{fig:comparison_full_momentum}
\end{figure}

To illustrate this global agreement we show in Fig.~\ref{fig:comparison_full_momentum} two examples of the time evolution of the lepton flavor asymmetries for vanishing total lepton number, comparing the results obtained using momentum dependent QKEs in \cite{Froustey:2021azz} to results obtained in this work. Incidentally, the left panel shows an example of an abrupt muon-driven MSW transition, and we note that even in this somewhat special case the agreement is very good. We have performed similar comparisons for all examples of time evolution shown in  \cite{Froustey:2021azz,Froustey:2024mgf} and find excellent agreement as well.

\begin{figure}[!t]
\centering
\begin{tabular}{cc}
\hspace{-0.2cm}\includegraphics[width=0.5\textwidth]{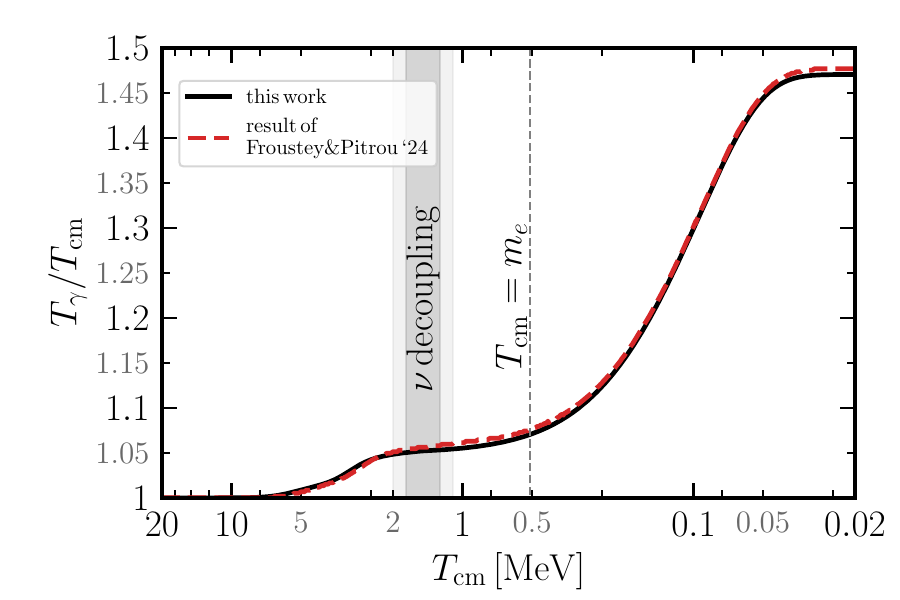}
\hspace{-0.2cm}\includegraphics[width=0.5\textwidth]{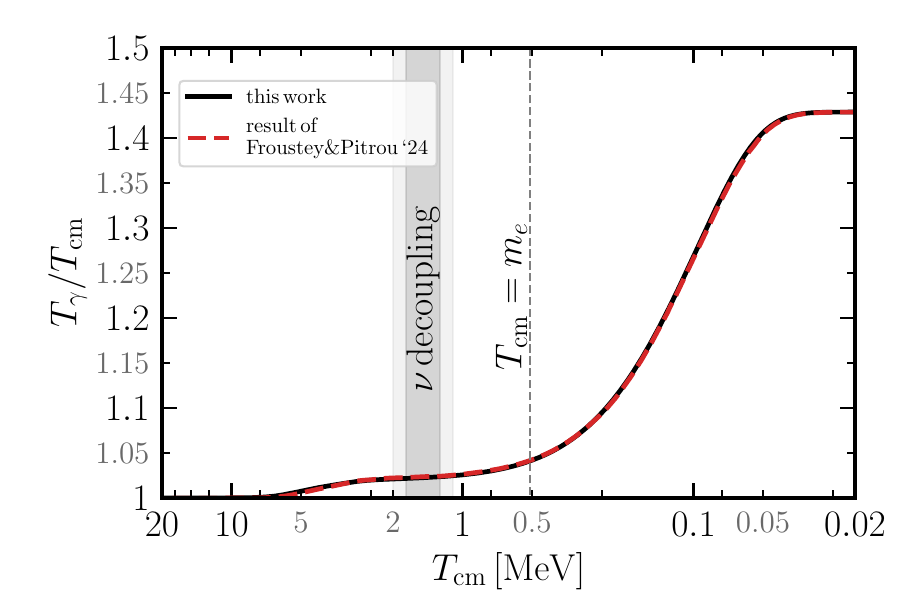}
\end{tabular}
\caption{Time evolution of the dimensionless photon temperature $T_\gamma/T_{\mathrm{cm}}$ compared to the result shown in Figs.~1 and 2 of Ref.~\cite{Froustey:2024mgf}, demonstrating the excellent agreement between our approximation and the full momentum dependent approach in describing the energy transfer between the neutrino and electron-photon sector. Initial conditions are ${\xi}_e^{\rm ini} = 0.1\,(0.67)$, ${\xi}_\mu^{\rm ini} = 1.6\,(-0.76)$ and ${\xi}_\tau^{\rm ini} = -1.5\,(-0.76)$ for the left (right) panel.}
\label{fig:comparison_full_momentum_temp}
\end{figure}

In addition Fig.~\ref{fig:comparison_full_momentum_temp} illustrates our accurate tracking of the energy transfer (and hence the photon temperature) compared to results shown in Ref.~\cite{Froustey:2024mgf} based on solving the full momentum dependent QKEs. This is crucial for accurately estimating the contribution to $\Delta N_\text{eff}$.

\section{Implications for early Universe cosmology}
\label{sec:implications}

Baryon number ($B$) and lepton number ($L$) are accidental symmetries of the SM. They are violated only at the quantum level by sphaleron processes, which preserve $B$$-$$L$ but erase $B$$+$$L$ asymmetries in the thermal plasma of the early Universe, until they freeze out at the electroweak phase transition (EWPT) at $T\sim 130\,{\rm GeV}$~\cite{DOnofrio:2014rug}. The observed baryon-to-photon ratio, $n_b/n_\gamma = (6.12 \pm 0.04) \times 10^{-10}$~\cite{Planck:2018vyg}, thus requires either the generation of a very small $B$$-$$L$ asymmetry $\xi_\mathrm{B-L} \simeq 10^{-9}$ before or at the EWPT or the generation of a $B$$+$$L$ asymmetry of similar size protected from sphaleron washout. In both cases, the total asymmetry in $B$$-$$L$, which due to charge neutrality can only be stored in neutrinos during the BBN epoch, is bounded from above by this value,  $\xi_\mathrm{B-L} \lesssim 10^{-9}$. This however does notably not preclude much larger lepton flavor asymmetries which compensate to yield an approximately vanishing lepton number, $\sum_\alpha \Delta n_\alpha \simeq 0$. This scenario is appealing theoretically as it falls naturally within the symmetry structure of the SM. There has thus been a substantial amount of interest in understanding how such large lepton flavor asymmetries might be generated in the early Universe and how they would impact early Universe cosmology. In this section we provide a brief review of the literature on this subject and contrast this with the results obtained in this work.

Large lepton asymmetries can be generated in the early Universe in multiple ways. In models of spontaneous baryogenesis~\cite{Cohen:1987vi,Cohen:1988kt} a pseudoscalar field with a time-dependent vacuum expectation value is coupled to a current and generates an asymmetry of the associated charge, as in the Affleck-Dine mechanism~\cite{Affleck:1984fy}. This initial asymmetry is then redistributed among the SM particle species according to the SM transport equations~\cite{Domcke:2020kcp}. Overproduction of baryon number can be avoided if the current carries no $B$$-$$L$ charge~\cite{Chiba:2003vp,Takahashi:2003db,Domcke:2022kfs}, by ensuring that the mechanism is efficient only after the EWPT~\cite{Yamaguchi:2002vw} or by `hiding' the $B$$+$$L$ asymmetry in Q-balls until after the sphalerons decouple~\cite{Kawasaki:2002hq}. Alternatively, large lepton flavor asymmetries with vanishing total lepton number can be generated e.g.\ through neutrino oscillations involving sterile neutrinos~\cite{Asaka:2005pn,Shaposhnikov:2008pf,Laine:2008pg}\footnote{The lepton asymmetries found in the $\nu$MSM are limited to $(n_e - \bar n_e)/s \lesssim 0.7 \times 10^{-3}$~\cite{Laine:2008pg}, and thus do not significantly affect BBN.} or the decay of heavy $B$$+$$L$ charged particles in grand unified theories~\cite{Barr:1979ye,Nanopoulos:1979gx,Yildiz:1979gx,Dreiner:1992vm,Kolb:1996jt}.

\begin{figure}[!t]
\centering
\begin{tabular}{cc}
\hspace{-0.075cm}\includegraphics[width=0.5\textwidth]{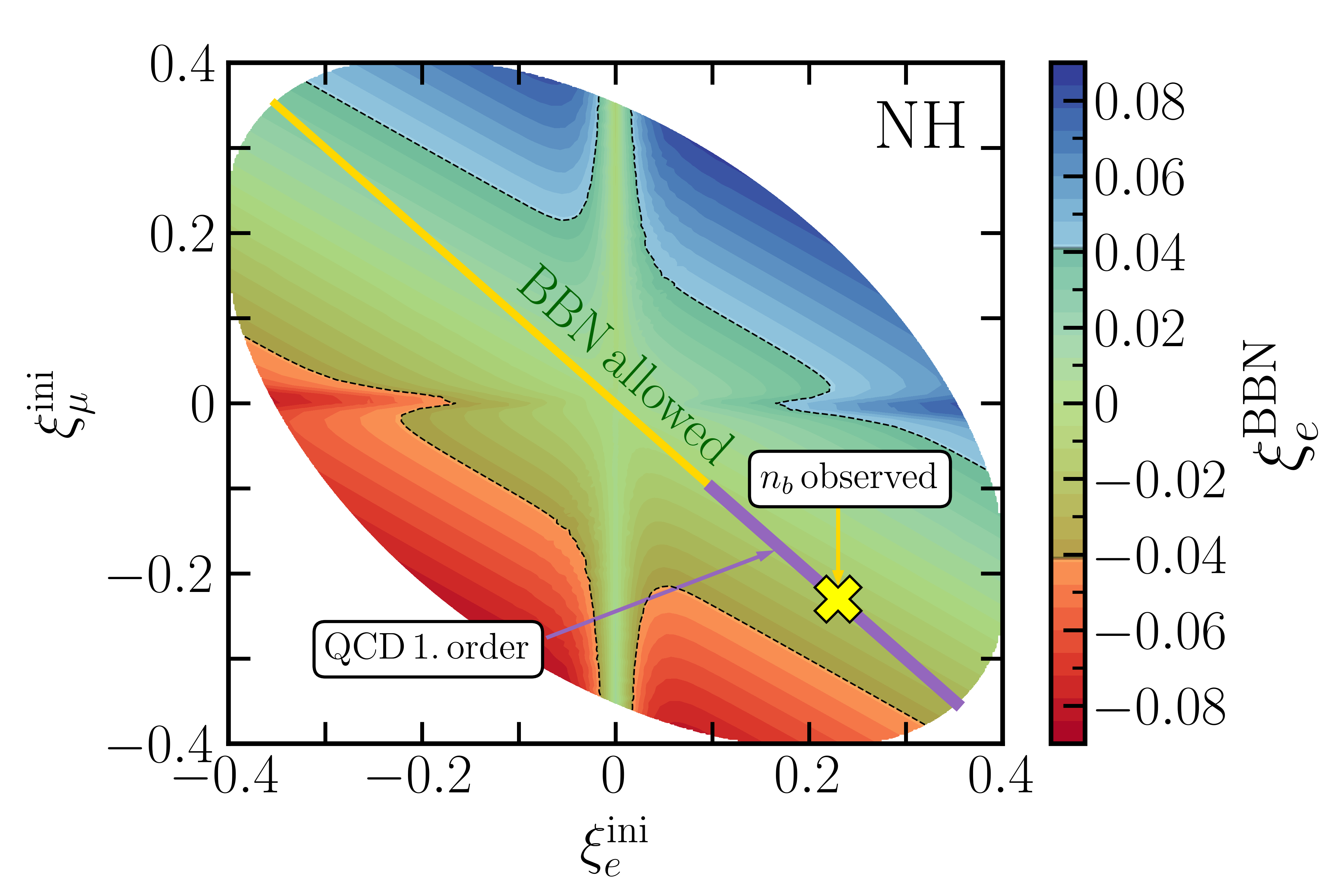}
\hspace{-0.075cm}\includegraphics[width=0.5\textwidth]{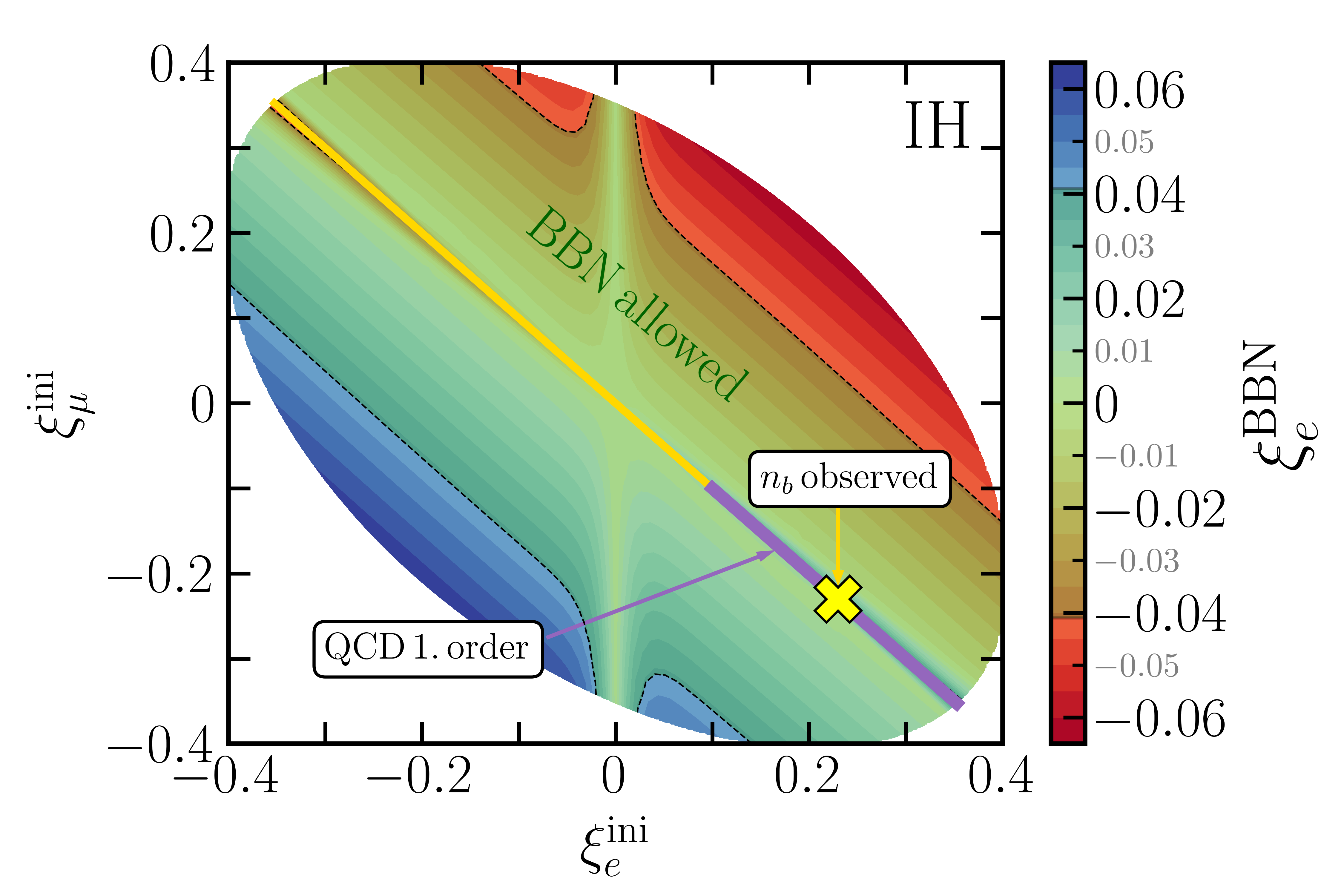}
\end{tabular}
\caption{ Zoomed in version of Fig.~\ref{fig:constraints_intro}, but showing in yellow the line of $\xi_\tau^{\mathrm{ini}} = 0$ as relevant for tauphobic leptoflavorgenesis. The yellow cross highlights the point where the observed small baryon asymmetry of the Universe can be linked to large primordial lepton asymmetries, see Eq.~\eqref{eq:xi_observed_Yb}. The purple line highlights the region for which the QCD phase transition could be first-order along the $\xi_e^{\rm ini} = -\xi_\mu^{\rm ini}$ direction~\cite{Gao:2021nwz,Gao:2024fhm}. Here the minimum value for $\xi_e^{\rm ini}$ is taken from Tab. 1 of~\cite{Gao:2021nwz} multiplied by 3 as recently recommended in~\cite{Gao:2024fhm}.}
\label{fig:constraints_cosmo}
\end{figure}

Once generated, large lepton flavor asymmetries can impact the processes in the early Universe (besides their impact on BBN) in multiple ways. Any net $B$$-$$L$ charge will contribute directly to baryogenesis. However, even for vanishing total $B$$-$$L$ number, lepton flavor asymmetries can contribute to baryogenesis. One example is wash-in leptogenesis~\cite{Domcke:2020quw}, in which an initial asymmetry with $B$$-$$L = 0$ is subject to lepton-number violating washout processes. This mechanism is typically very efficient and thus requires only relatively small asymmetries, $\xi_\alpha \sim 10^{-8}$, which are not constrained by BBN. On the contrary, leptoflavorgenesis~\cite{Kuzmin:1987wn,Khlebnikov:1988sr,March-Russell:1999hpw,Laine:1999wv,Shu:2006mm,Gu:2010dg,Mukaida:2021sgv} relies on the flavor dependence of the sphaleron processes and thus requires much larger lepton flavor asymmetries. If the initial asymmetry has a tau-component, successful baryogenesis requires $Y_\tau = (n_\tau - \bar n_\tau)/s \simeq - 3 \cdot 10^{-5}$ at the EWPT~\cite{Shu:2006mm,Gu:2010dg}, with $s$ denoting the entropy density. If it does not and hence only relies on the hierarchy between the muon and electron Yukawa coupling the initial asymmetry needs to be larger, $Y_\mu \simeq - 8 \cdot 10^{-3}$~\cite{March-Russell:1999hpw,Mukaida:2021sgv}. In the absence of lepton flavor violating processes, the yields $Y_\alpha$ are conserved until the onset of our simulations at $20\,\mathrm{MeV}$ and can be related to the respective chemical potentials as
\begin{align}
 \frac{4 \pi^2}{15} g_* Y_\alpha = \xi_\alpha + \xi_\alpha^3/\pi^2\,,
 \label{eq:yield}
\end{align}
with $g_* = 10.75$ at $20\,\mathrm{MeV}$. Successful baryogenesis hence requires
\begin{align}
\label{eq:xi_observed_Yb}
 - \xi_\mu^\mathrm{ini} = \xi_e^\mathrm{ini} = 0.23 \,, \quad \xi_\tau = 0 \quad  \mathrm{(tauphobic)} \,, \qquad - \xi_\tau^\mathrm{ini} = (\xi_e^\mathrm{ini} + \xi_\mu^\mathrm{ini}) = 9 \cdot 10^{-4} \quad \mathrm{(else)} \,.
\end{align}
While in the latter case the asymmetries are too small to impact BBN, the former lies precisely in the parameter region central to this study. For NH, our $95$\% CL BBN limits, $-0.040 < \xi_e^\mathrm{BBN} < 0.042$, translate to $-0.5 \lesssim \xi_e^\mathrm{ini} = - \xi_\mu^\mathrm{ini} \lesssim 0.5$, and thus tauphobic leptoflavorgenesis is still compatible with the current limits, but will be under scrutiny if the BBN limits improve. For IH, on the other hand, this flavor combination experiences a non-adiabatic electron-driven MSW transition leading to an increased value for the final asymmetry in the electron flavor, so that the BBN limits read $-0.3 \lesssim \xi_e^\mathrm{ini} = - \xi_\mu^\mathrm{ini} \lesssim 0.3$, approaching the parameter point corresponding to tauphobic leptogenesis in the IH. The findings are summarized in Fig.~\ref{fig:constraints_cosmo}.
We note that in the latter case, the exact value of the constraint is quite sensitive to the dynamics of the non-adiabatic electron-driven MSW transition and it would be very interesting to study this particular case with numerical code retaining the full momentum dependence of the QKEs.

Large lepton flavor asymmetries can moreover lead to a first-order QCD phase transition~\cite{Gao:2021nwz,Gao:2024fhm}. Depending on the direction in flavor space, this requires lepton flavor asymmetries of $|n_\alpha - \bar n_\alpha|/s \simeq 10^{-3..-1}$~\cite{Gao:2021nwz}, or equivalently, following Eq.~\eqref{eq:yield}, chemical potentials of $|\xi_\alpha| = 0.03..2$. A more recent analysis suggests that these values may be up to a factor 3 larger~\cite{Gao:2024fhm}. Among the specific set of flavor directions discussed in Ref.~\cite{Gao:2021nwz} the tauphobic direction with vanishing total lepton number is of particular interest, as small asymmetries of order $\xi_e = - \xi_\mu \simeq 10^{-1}$ suffice to generate a first-order QCD phase transition. This value is currently not only just within the BBN limits discussed above for this direction in flavor space, but also implies that the tauphobic leptoflavorgenesis scenario discussed above is accompanied by a first-order QCD phase transition, leading to an intriguing scenario with multiple possible observational signatures. More generally, we note the strong sensitivity to the flavor direction (for vanishing total lepton number) observed in Ref.~\cite{Gao:2021nwz}, illustrating the importance of mapping out the BBN constrains in flavor space in this case.

Large lepton flavor asymmetries have moreover been proposed as resolutions to cosmological puzzles, in particular the helium-$4$ anomaly~\cite{Matsumoto:2022tlr} (see Refs.~\cite{Kawasaki:2022hvx,Burns:2022hkq,March-Russell:1999hpw,Kohri:1996ke}) and to alleviate the Hubble tension by increasing $\Delta N_\mathrm{eff}$~\cite{Gelmini:2020ekg}. These scenarios can now be assessed directly based on our analysis above. Addressing the helium-$4$ anomaly requires a positive value of the asymmetry in electron neutrino sector at the boundary of the allowed region considered above, $\xi_e^{\mathrm{BBN}} \simeq 0.04$. On the contrary, we point out that no significant contribution to $\Delta N_\mathrm{eff}$ is possible without violating BBN constraints.

Finally, we note that very large lepton flavor asymmetries are constrained by the observed baryon number if they are present in the early Universe at temperatures before all SM interactions equilibrate at around $10^6\,\mathrm{GeV}$. In this case, the chiral plasma instability can convert the lepton flavor asymmetries to helical magnetic fields, which in turn are a source of baryon asymmetry during the EWPT. This leads to an overproduction of the baryon asymmetry for $\xi_\alpha|_{T > 10^6\,\mathrm{GeV}} \gtrsim 0.01$~\cite{Domcke:2022uue}. This constraint however leaves the loophole of generating lepton flavor asymmetries at $T \lesssim 10^6$~GeV, as well as the possibility of diluting the baryon asymmetry by an entropy injection between the EWPT and the onset of BBN. Independent constraints from BBN and CMB directly are thus crucial to our understanding of $\mathrm{CP}$ violation in the early Universe.

\section{Conclusion} 
\label{sec:conclu}

This paper presents a fast and accurate framework to study the evolution of primordial lepton flavor asymmetries from the onset of neutrino oscillations at $T\sim 15\,{\rm MeV}$ up to the time of neutrino decoupling when BBN starts, $T\sim 1\,{\rm MeV}$. We mainly focus on relatively large primordial lepton flavor asymmetries with  vanishing total lepton number, as in this case neutrino oscillations can lead to small lepton asymmetries at the time of BBN, in agreement with cosmological observations.

A key novelty of our study is the systematic derivation of momentum averaged neutrino
kinetic equations. The dynamics of neutrino oscillations is described by partial differential equations for the neutrino and anti-neutrino density matrices. As discussed in Sec.~\ref{sec:qke}, the problem can be significantly simplified by performing a thermal average of the system of equations. This reduces a system of $\mathcal{O}(100)$ stiff and non-linear integro-differential equations to a system of 20 ordinary 
differential equations. This simplification comes with a gain of several orders of magnitude in computational cost.
The system also describes the evolution of the effective neutrino and photon temperatures, which allows us to account for the increase in the energy density in the neutrino sector induced by neutrino flavor equilibration as well as the energy transfer between the neutrino and electron-photon sector. This turns out to be crucial for an accurate determination of $N_{\rm eff}$. 

Several factors play a role to justify the use of momentum averaged equations. In the presence of large primordial lepton asymmetries neutrinos undergo synchronous oscillations, during which modes with different momenta all oscillate coherently. In addition, our first-principles derivation of the collision terms respects the delicate structure of flavor decoherence via interactions with the medium, retaining all relevant symmetries of the neutrino system. This is in contrast to the so-called damping approximation employed in earlier works. As a result, and as described in Sec.~\ref{subsec:full_qke_comparison}, we find a remarkably excellent agreement with solutions to the full momentum dependent kinetic equation recently studied in~\cite{Froustey:2021azz,Froustey:2024mgf}.

The accuracy and low computational cost of our approach allows to explore the entire region of parameter space of asymmetries with zero total lepton asymmetry. Our results are summarized in Fig.~\ref{fig:constraints_intro} and described in more detail in Sec.~\ref{sec:results}. In particular we identify directions in flavor space for which the final asymmetry in the electron neutrino, $\xi_{e}^{\rm BBN}$, which is the most relevant parameter for BBN, is small or even vanishing despite large initial lepton flavor asymmetries. The regions of parameter space which evade BBN constraints are: 
\begin{subequations}
\begin{align}
    \Delta n_e^{\rm ini} &\simeq 0 \,, \,\,\,\,\,\qquad\qquad   \Delta n_\mu^{\rm ini} \simeq - \Delta n_\tau^{\rm ini} \qquad  \,\,\,\,\,{\rm [NH, IH]}  \,,\label{eq:dn_cond1} \\
    \Delta n_e^{\rm ini} &\simeq -2/3 \, \Delta n_\mu^{\rm ini}\,,\,\,\, \Delta n_\tau^{\rm ini} \simeq -1/3 \, \Delta n_\mu^{\rm ini} \,\quad [\rm{NH}] \,,\label{eq:dn_cond2a} \\
    \Delta n_e^{\rm ini} &\simeq -  \Delta n_\mu^{\rm ini}\,, \,\,\,\,\,\,\,\,\,\,\,\,\, \Delta n_\tau^{\rm ini} \simeq  0 \,\,\,\qquad\qquad\quad [\rm{IH}]\,.  \label{eq:dn_cond2b}
\end{align}
\end{subequations}
This flavor structure reflects the nature of the MSW transitions across the parameter space. In most of the parameter space, the frequency of synchronous neutrino oscillations is fast compared to the typical time scale of the MSW transitions, resulting in adiabatically evolving neutrino asymmetries. This allows for a particularly efficient method of solving the momentum averaged quantum kinetic equations (as the matter potential arising from self-interactions can be neglected) and approximately explains the latter two directions in flavor space. However, in specific flavor directions the frequency of synchronous neutrino oscillations becomes particularly slow, resulting in non-adiabatic MSW transitions. The non-adiabatic electron-driven MSW transition explains the difference between NH and IH in Eqs.~\eqref{eq:dn_cond2a} and \eqref{eq:dn_cond2b}, whereas the Eq.~\eqref{eq:dn_cond1} coincides with the condition of a non-adiabatic muon-driven MSW transition.

Implications of our results for early Universe cosmology are discussed in Sec.~\ref{sec:implications}. We highlight that the region with $\Delta n_e \simeq -\Delta n_\mu$ and $\Delta n_\tau \simeq 0$ allows for relatively large primordial asymmetries to be compatible with BBN. This in particular opens up the possibility for a first-order QCD phase transition and for scenarios where the small baryon asymmetry of the Universe originates from a large lepton asymmetry at the time of sphaleron freeze-out. In addition, such large flavor asymmetries allow for the possibility of significantly enhancing the production of sterile neutrino dark matter.

We have drawn the conclusion of our study by using the results on the electron neutrino asymmetry from the latest global BBN analyses. In this context, either improving the measurements of the cosmic helium-$4$ abundance or the knowledge of the nuclear reaction rates that control the deuterium abundance can have relevant implications for our conclusions. On the theory side, we have estimated our accuracy to predict $\xi_e^{\rm BBN}$ to be $\simeq 0.016$. Further refinements in our modeling or more careful comparisons with full solutions of the momentum dependent QKEs may significantly reduce this number and allow for more accurate cosmological inferences. In particular, as can be seen in Fig.~\ref{fig:constraints_cosmo}, the scenario where large lepton asymmetries can be connected with the observed baryon asymmetry is very close to be tested. Hence, improvements on these inferences may have profound implications for early Universe cosmology as they could either rule out or rule in this interesting scenario.

To conclude, in this study we have performed a comprehensive analysis of the allowed ranges of the primordial lepton flavor asymmetries given current BBN limits focusing on the case where total lepton number is zero. Looking forward, an important question is the generalization of these results to the case of non-zero total lepton number (see also Refs.~\cite{Froustey:2021azz,Froustey:2024mgf,Barenboim:2016shh}). We plan to return to this in an upcoming publication.

\vspace{0.5 cm}
\begin{center}
\textbf{Acknowledgments}
\end{center}

We thank Luke Johns and Kai Schmitz for useful discussions and/or clarifications. M.F.N. and S.S. would like to thank the CERN Theory group for hospitality and financial support  while part of this work was performed. We acknowledge support from the DOE Topical Collaboration ``Nuclear Theory for New Physics,'' award No.~DE-SC0023663. M.F.N. is supported by the STFC under grant ST/X000605/1. S.S. was supported by the US Department of Energy Office and by the Laboratory Directed Research and Development (LDRD) program of Los Alamos National Laboratory under project numbers 20230047DR and 20250164ER. Los Alamos National Laboratory is operated by Triad National Security, LLC, for the National Nuclear Security Administration of the U.S. Department of Energy (Contract No. 89233218CNA000001).

\newpage
\addtocontents{toc}{\vspace{0.25em}}  
\appendix

\setcounter{section}{0}
\setcounter{equation}{0}
\setcounter{table}{0}

\noindent\makebox[\linewidth]{\rule{0.8 \paperwidth}{0.2pt}}
\vspace{1 mm}
\vspace{-4.5mm}

\section{Neutrino oscillation parameters}
\label{app:conventions}

We fix the following values for the neutrino mass splittings and mixing angles independent of the neutrino hierarchy:
\begin{align}
 \Delta m^2_{21} = 7.53 \cdot 10^{-5}~\mathrm{eV}^2 \,,\quad  
 \theta_{12} = 0.587 \,, \quad \theta_{13} = 0.148 \,, \quad \delta_\mathrm{CP} = 0 \,,
\end{align}
with $\Delta m_{ij}^2 = m_{i}^2 - m_{j}^2$. Because $m_1 < m_3\,(m_1>m_3)$ in NH (IH) we have
\begin{align}
    \Delta m_{31}^{\mathrm{NH}} = 2.534 \cdot 10^{-3}\,\mathrm{eV}^2\,,\quad \Delta m_{31}^{\mathrm{IH}} = -2.46 \cdot 10^{-3}\,\mathrm{eV}^2\,,
\end{align}
and for the $2$$-$$3$ mixing angle we set
\begin{align}
    \theta_{23}^{\mathrm{NH}} = 0.831 \,,\quad \theta_{23}^{\mathrm{IH}} = 0.824\,.
\end{align}
The values of the mixing angles and mass splittings are within the $1\sigma$ preferred region of the latest global neutrino oscillation fit (excluding SK atmospheric data) performed in~\cite{Esteban:2024eli} and in particular the same as used in Ref.~\cite{Froustey:2021azz}. Setting the Dirac CP phase $\delta_{\mathrm{CP}}=0$ is a choice of simplicity, but as has been shown in~\cite{Froustey:2021azz} $\delta_{\mathrm{CP}} \neq 0$ does not appreciably impact cosmological observables. The PMNS matrix is the same as defined in the PDG~\cite{ParticleDataGroup:2024cfk}:
\begin{align}
\label{app:eq:neutrino:UPMNS}
U = 
\begin{pmatrix}
1 & 0 & 0 \\
0 & c_{23} & s_{23} \\
0 & -s_{23} & c_{23} 
\end{pmatrix}
\begin{pmatrix}
c_{13} & 0 & s_{13} e^{i \delta_{\mathrm{CP}}} \\
0 & 0 & 0 \\
-s_{13} e^{i\delta_{\mathrm{CP}}} & 0  & c_{13} 
\end{pmatrix}
\begin{pmatrix}
c_{12} & s_{12} & 0 \\
-s_{12} & c_{12} & 0 \\
0 & 0  & 0 
\end{pmatrix}
\,,
\end{align}
with $s_{ij} = \sin \theta_{ij}\,,\, c_{ij} = \cos\theta_{ij}$.

\begin{figure}[!b]
\includegraphics[width = 0.55 \textwidth]{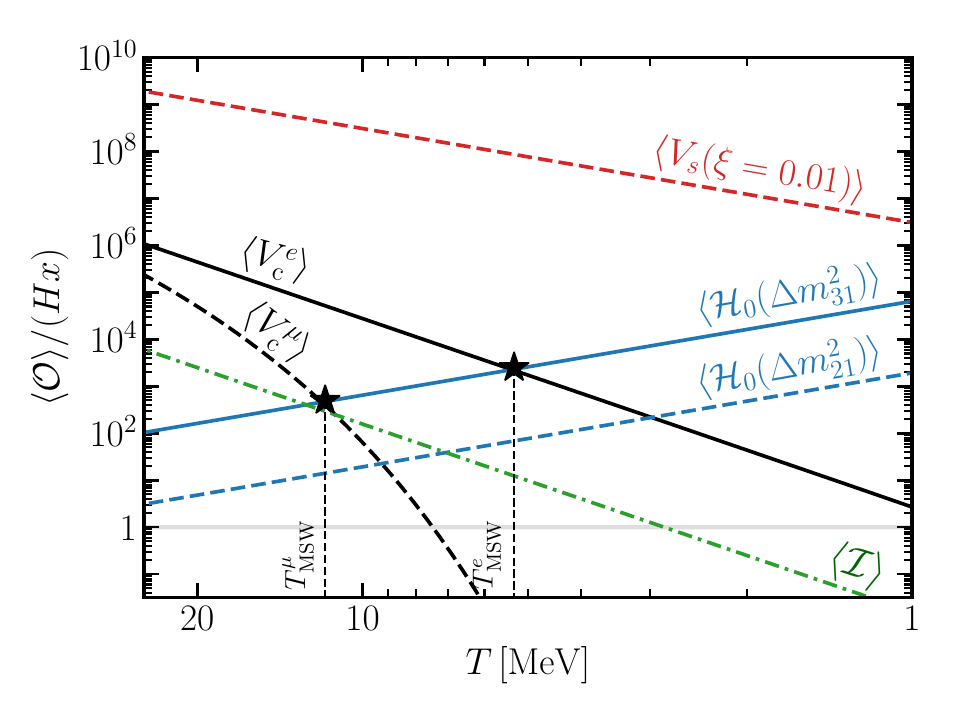}
\caption{Illustration of the different contributions to the momentum averaged Hamiltonian and interactions, see Eq.~\eqref{eq:master}. For the large chemical potentials of interest to this work, the self-interaction matter potential $V_s$ dominates the Hamiltonian throughout the evolution. The muon- and electron-driven MSW transitions occur when the vacuum oscillations, controlled by the neutrino mass splittings contained in $\mathcal{H}_0$, overcome the muon and electron contribution to the charged current matter potential $V_c$. The collision terms $\mathcal{I}$ play a crucial role in accurately capturing the flavor decoherence and are active until $T\sim 2\,{\rm MeV}$, the time of neutrino decoupling.  Adapted from Refs.~\cite{Lesgourgues:2013sjj, Froustey:2021azz}. }
\label{fig:rates}
\end{figure}

A visualization of the different contributions to the Hamiltonian in Eq.~\eqref{eq:master} for these parameters is given in Fig.~\ref{fig:rates}, based on Eqs.~\eqref{eq:def_H0}, \eqref{eq:Vc_def} and \eqref{eq:Vs_def}.
For the collision term we show the leading contribution arising from $\nu e \leftrightarrow \nu e$ interactions given in Eq.~\eqref{eq:I_nuenue}. For this figure, we normalize all rates by the Hubble expansion rate $1/(xH)$, take the effective number of degrees of freedom $g_* = 10.75$ (SM value) and assume $T_\nu = T_\gamma = T_{\mathrm{cm}}$.

\section{Collision terms: Derivation, approximations \& comparison with previous literature} \label{app:collisionterms}

The neutrino collision terms were first written by Sigl and Raffelt~\cite{Sigl:1993ctk} and have been explicitly worked out in several other references, see~\cite{Blaschke:2016xxt,deSalas:2016ztq,Gariazzo:2019gyi,Froustey:2020mcq,Bennett:2020zkv,Froustey:2021azz}. There are three main contributions: those arising from $\nu e \leftrightarrow \nu e$ scatterings, $\nu \bar{\nu} \leftrightarrow e^+e^-$ annihilation, and those from neutrino self-interactions, $\nu\nu \leftrightarrow \nu \nu$ and $\nu\bar{\nu} \leftrightarrow \nu \bar{\nu}$. In this appendix we provide explicit derivations of the various collision terms based on the ansatz~\eqref{eq:ansatz}, i.e.\ that all entries of the density matrices feature the same momentum dependence given by the Fermi-Dirac distribution with zero chemical potential. As a warm-up, we will start by first deriving the collision terms assuming that all particles follow Maxwell-Boltzmann statistics and setting the electron mass to zero. These two approximations are fairly good and allow to compute the full collision terms entirely analytically. Then we will add Fermi-Dirac corrections in a way such that the exact result can still be written analytically. We will also make contact with the work of Thomson \& McKellar~\cite{McKellar:1992ja} who were the first to compute approximate damping terms for neutrino oscillations in the early Universe and which have been used in many references in the literature. We show that only for $\nu e \leftrightarrow \nu e$ interactions can the collision term be written in a damping form and that in the relevant limit our results precisely match the ones of Ref.~\cite{McKellar:1992ja}.

In this appendix we will work in the limit where $T_\nu = T_\gamma$ for all the equations which greatly simplifies the calculations. In the next App.~\ref{sec:Cterms_TnuneqTgamma} we generalize these collision terms to include the full temperature dependence. There we will also calculate the energy transfer rate.

\subsection{Generalities and phase space integration of the collision terms}

Assuming CP-conserving interactions, the collision term for particle 1 resulting from a $1+2\leftrightarrow 3+4$ annihilation/scattering process can be written as~\cite{Dolgov:2002wy}:
\begin{align}
\label{eq:Definitionofcollterm}
{\mathcal I}_{12\leftrightarrow 34} = \frac{1}{2E_1} \int  \frac{\dd^3 \vec{k}_2}{(2\pi)^3 2 E_2}\frac{\dd^3 \vec{k}_3}{(2\pi)^3 2 E_3}\frac{\dd^3 \vec{k}_4}{(2\pi)^3 2 E_4}  (2\pi)^4 \delta^4(p_1+p_2-p_3-p_4) \times S|M|^2  \times {F}(f_1,f_2,f_3,f_4) \,,
\end{align}
where here $S$ is a symmetry factor, $M$ is the amplitude for the given process, and $f_i$ are the distribution functions of the particles that depend upon $|\vec{k}_i| = k_i$ and $p_i$ is the $4$-momentum of particle $i$. For fermions that do not oscillate, the function ${F}$ reads:
\begin{align}
\label{eq:Fdefinition}
{F}(f_1,f_2,f_3,f_4) = f_3f_4(1-f_1)(1-f_2) - f_1f_2(1-f_3)(1-f_4)\,.
\end{align}
Note that we are particularly interested in the momentum averaged collision term (see Eq.~\eqref{eq:master}) which reads:
\begin{align}
\label{eq:average_n}
\langle {\mathcal I}_{12\leftrightarrow 34} \rangle = \frac{1}{\int \frac{\dd^3 \vec{k}_1}{(2\pi)^3} f_1} \int    \frac{\dd^3 \vec{k}_1}{(2\pi)^3 2 E_1} \frac{\dd^3 \vec{k}_2}{(2\pi)^3 2 E_2}\frac{\dd^3 \vec{k}_3}{(2\pi)^3 2 E_3}\frac{\dd^3 \vec{k}_4}{(2\pi)^3 2 E_4}  (2\pi)^4 \delta^4(p_1+p_2-p_3-p_4) \times S|M|^2  \times {F}(f_1,f_2,f_3,f_4)\,.
\end{align}
We see that a priori the phase space integration is 12 dimensional. However, it has been shown that for a four-fermion interaction as relevant for neutrino-electron and neutrino-neutrino interactions it can be reduced to a three dimensional integral, see Ref.~\cite{Hannestad:1995rs} for the first full calculation, Refs.~\cite{Fradette:2018hhl,Kreisch:2019yzn} for clear descriptions of the method of integration, and Ref.~\cite{Dolgov:1997mb} for an alternative and powerful way of computing the collision term by rewriting the delta function as a Fourier integral. The matrix elements of the collision terms relevant for neutrino-electron and neutrino-neutrino interactions are outlined in Tab.~2 of~\cite{Dolgov:2002wy} and reproduced in our Tab.~\ref{tab:amplitudes-nu-e0} for convenience. In the $m_e\to 0$ limit they are all proportional to only three four-momentum scalar combinations: $(p_1 \cdot p_2) ( p_3 \cdot p_4) $, $(p_1 \cdot p_3) ( p_2 \cdot p_4)$ and $(p_1 \cdot p_4) ( p_2 \cdot p_3)$.

\begin{table}
\begin{center}
\begin{tabular}{cccc|cc}
& Process &&&& $2^{-5} G^{-2}_{F} S \left| M \right| ^{2}$
\\ \hline\hline
$\nu _{e} + \bar{\nu}_{e} $ & $\rightarrow$ & $  \nu _{e} + \bar{\nu}_{e} $&&&
$4 (p_{1} \cdot p_{4}) (p_{2} \cdot p_{3})$ \\

$\nu _{e} + \nu_{e} $ & $\rightarrow$ & $  \nu _{e} + \nu_{e}$ &&&
$2  (p_{1} \cdot p_{2}) (p_{3} \cdot p_{4})$\\

$\nu _{e}+\bar{\nu}_{e} $ & $\rightarrow$ &
$ \nu _{\mu (\tau)}+\bar{\nu}_{\mu (\tau)} $&&&
$ (p_{1} \cdot p_{4}) (p_{2} \cdot p_{3}) $\\

$\nu_{e} + \bar{\nu}_{\mu (\tau)} $ & $\rightarrow$&
$\nu _{e}+\bar{\nu}_{\mu (\tau)} $&&&
$ (p_{1} \cdot p_{4}) (p_{2} \cdot p_{3})$\\

$\nu _{e} + \nu_{\mu (\tau)} $ & $\rightarrow$ &
$ \nu _{e}+\nu_{\mu (\tau)} $&&&
$ (p_{1} \cdot p_{2}) (p_{3} \cdot p_{4})$\\

$\nu _{e} + \bar{\nu}_{e} $ & $\rightarrow$ & $  e^{+} + e^{-}$ &&&
$4[ ( g_{L}^{2} (p_{1} \cdot p_{4}) (p_{2} \cdot p_{3}) + g_{R}^{2}  (p_{1} \cdot p_{3}) (p_{2} \cdot p_{4}) + g_{L} g_{R} m _{e}^{2} (p_{1} \cdot p_{2})]$\\

$\nu _{e} +  e^{-} $ & $\rightarrow$ & $ \nu _{e} +  e^{-} $&&&
$4 [ g_{L}^{2} (p_{1} \cdot p_{2}) (p_{3} \cdot p_{4}) +g_{R}^{2}  (p_{1} \cdot p_{4}) (p_{2} \cdot p_{3}) -g_{L} g_{R} m _{e}^{2} (p_{1} \cdot p_{3}) ] $\\

$\nu _{e} +  e^{+} $ & $\rightarrow$ & $ \nu _{e} +  e^{+} $&&&

$4 [ g_{R} ^{2} (p_{1} \cdot p_{2}) (p_{3} \cdot p_{4}) +g_{L}^{2} (p_{1} \cdot p_{4}) (p_{2} \cdot p_{3}) -g_{L} g_{R} m _{e}^{2} (p_{1} \cdot p_{3}) ] $ 
\end{tabular}
\caption{Matrix elements squared for reactions with electron neutrinos as reported in Tab.~2 of~\cite{Dolgov:2002wy}; $S$ is the symmetrization factor related to identical particles in the initial or final state, $g_{L} = \frac{1}{2} + \sin^{2}\theta_{W}$ and $g_{R} = \sin^{2} \theta_{W}$. Matrix elements for muon or tau neutrino processes are obtained by the substitutions $\nu_e \rightarrow \nu_{\mu,\tau}$ and $g_{L} \rightarrow \tilde{g}_{L} =g_{L} - 1$.}
\label{tab:amplitudes-nu-e0}
\end{center}
\end{table}

\subsection{Maxwell-Boltzmann case}

A nice property of averaging the collision term is that the integral over phase space is symmetric over all the particles which allows for further simplifications. In particular, if one considers Maxwell-Boltzmann statistics the integrations in the collision terms can be performed analytically. The integration in this case proceeds analogously to the thermal averaging of WIMP annihilations, see~\cite{Gondolo:1990dk,Edsjo:1997bg}. For the case of Maxwell-Boltzmann statistics with thermal equilibrium distributions of the same temperature we have for the first term in Eq.~\eqref{eq:Fdefinition}:
\begin{align}
     {F}|_{\rm MB}^{\rm forward} = f_3 f_4 = f_1f_2 = e^{-(E_1+E_2)/T} \,.
\end{align}
This greatly simplifies the calculations as we can trivially integrate over the $34$ phase space by using the techniques outlined in~\cite{Edsjo:1997bg}. In the center of mass frame and for massless states we have:
\begin{align}
\!\!\!\! \int \frac{\dd^3 \vec{k}_3}{(2\pi)^3 2 E_3}\frac{\dd^3 \vec{k}_4}{(2\pi)^3 2 E_4}  (2\pi)^4 \delta^4(p_1+p_2-p_3-p_4) S|M|^2  =  \int \frac{\dd^3 \vec{k}_3}{(2\pi)^2 2 E_3 2E_4}\delta(\sqrt{s}-2E_3) S|M|^2 =   \int \frac{\dd\Omega}{32\pi^2}  S|M|^2\,,
\end{align}
with the Mandelstam variable $s = (p_1+p_2)^2=(p_3+p_4)^2$ (and $t = (p_1-p_3)^2=(p_2-p_4)^2$ accordingly). We can then perform the rest of the integration by using the phase space integration strategy of~\cite{Gondolo:1990dk,Edsjo:1997bg}. In particular,
\begin{align}
\langle {\mathcal I}_{12\leftrightarrow 34} \rangle|_{\rm MB}^{\rm forward} &=   \frac{1}{\int \frac{\dd^3 \vec{k}_1}{(2\pi)^3} f_1} \int    \frac{\dd^3 \vec{k}_1}{(2\pi)^3 2 E_1} \frac{\dd^3 \vec{k}_2}{(2\pi)^3 2 E_2}  \times e^{-(E_1+E_2)/T} \int \frac{\dd\Omega}{32\pi^2}  S|M|^2 \,, \\
&=  \frac{1}{\int \frac{\dd^3 \vec{k}_1}{(2\pi)^3} f_1}  \frac{1}{(2\pi)^4} \frac{1}{8}  \int dE_+dE_-ds  \times e^{-E_+/T}  \int \frac{\dd\Omega}{32\pi^2}  S|M|^2 \,,
\end{align}
where $E_+ = E_1+E_2$ and $E_- = E_1-E_2$. For the massless particles we consider we have as limits of integration $s \geq 0$, $E_+ \geq \sqrt{s}$, $|E_-| \leq \sqrt{E_+^2-s}$ (see Eq. 51 of~\cite{Edsjo:1997bg}). The integration over $E_-$ is trivial:
\begin{align}
\langle {\mathcal I}_{12\leftrightarrow 34}\rangle|_{\rm MB}^{\rm forward}  &=    \frac{1}{\int \frac{\dd^3 \vec{k}_1}{(2\pi)^3} f_1}  \frac{1}{(2\pi)^4} \frac{1}{8} \int \dd E_+ \dd s \, 2  \sqrt{E_+^2-s} \times e^{-E_+/T}  \int \frac{\dd \Omega}{32\pi^2}  S|M|^2 \,.
\end{align}
Next we perform the angular integration in the center of mass frame for the momentum combinations appearing in the various matrix elements. Taking $\dd \Omega = \dd\cos\theta \dd\phi$ and defining $\theta$ as the angle subtended between $\vec{k}_1$ and $\vec{k}_3$, one finds:
\begin{align}
     &\int \frac{\dd\Omega}{32\pi^2}  (p_{1} \cdot p_{2}) (p_{3} \cdot p_{4}) = \int \frac{\dd\Omega}{32\pi^2}  \frac{s^2}{4} = \frac{1}{32\pi} s^2  \,,\\
     &\int \frac{\dd\Omega}{32\pi^2}  (p_{1} \cdot p_{3}) (p_{2} \cdot p_{4}) = \int \frac{\dd\Omega}{32\pi^2}  \frac{s^2}{16}(1-\cos\theta)^2 = \frac{1}{3} \frac{1}{32\pi} s^2 \,,\\
     &\int \frac{\dd\Omega}{32\pi^2}  (p_{1} \cdot p_{4}) (p_{2} \cdot p_{3}) = \int \frac{\dd\Omega}{32\pi^2}  \frac{s^2}{16}(1+\cos\theta)^2 = \frac{1}{3} \frac{1}{32\pi} s^2 \,.
\end{align}
The last integral to be performed is over $E_+$ and reads:
\begin{align}
    \langle {\mathcal I}_{12\leftrightarrow 34} \rangle|_{\rm MB}^{\rm forward} &=  \frac{1}{\int \frac{\dd^3 \vec{k}_1}{(2\pi)^3} f_1}  \frac{T}{64\pi^4}  \int \dd s   \sqrt{s} K_{1}(\sqrt{s}/T) \int \frac{\dd \Omega}{32\pi^2} S|M(s)|^2\,,
\end{align}
where $K_1$ is a modified Bessel function of the second kind of order 1.

Considering as an example the case of $\nu +e \to \nu +e $ scatterings, taking the matrix element in Tab.~\ref{tab:amplitudes-nu-e0} and summing over the contributions of electrons and positrons yields:
\begin{align}
    \int \frac{\dd\Omega}{32\pi^2} S|M|^2 = 2^5 G_F^2 4 (g_L^2+g_R^2) \frac{1}{32\pi}  s^2 4/3 =G_F^2 \frac{16}{3\pi} (g_L^2+g_R^2)s^2\,.
\end{align}
The only remaining integral that remains to be performed is over $s$ and in this simple case reads,
\begin{align}
    \int \dd s s^2\sqrt{s} K_1(\sqrt{s}/T) = 768 \, T^7\,.
\end{align}
so that in summary, the forward collision term for $\nu e\to \nu e$ scatterings which reads:
\begin{align}
   \langle{\mathcal I}_{\nu e\to\nu e}\rangle|_{\rm MB}^{\rm forward} &= \frac{1}{\int \frac{\dd^3 \vec{k}_1}{(2\pi)^3} f_1} \times 2\times 4 \times (g_L^2+g_R^2)\frac{8G_F^2T^8}{\pi^5} = 2\times 4  \times (g_L^2+g_R^2) \frac{8G_F^2T^5}{\pi^3}\,,
\end{align}
where in the last step we have considered the case of Maxwell-Boltzmann statistics for the integral over the $f_1$ distribution function, $n = T^3/\pi^2$. 

Proceeding in this way we can easily obtain all the collision terms in the Maxwell-Boltzmann approximation. In particular, given that there are only three types of matrix elements one can simply use the following replacement rule:
\begin{tcolorbox}[ams align, boxrule=0.3pt, arc=0.5mm, colback=white!100!white, colbacktitle=white!100!white, colframe=black!100!black, coltitle=black]
\label{eq:Easysubstitution_1}
    S|M|^2 &= 2^5 G_F^2 (p_{1} \cdot p_{2}) (p_{3} \cdot p_{4}) \quad \longrightarrow \quad \langle {\mathcal I}_{12\leftrightarrow 34} \rangle|^{\rm forward}_{\rm MB} =3 \frac{4 G_F^2 T^5}{\pi^3}  \,,\\
\label{eq:Easysubstitution_2}
     S|M|^2 &= 2^5 G_F^2 (p_{1} \cdot p_{3}) (p_{2} \cdot p_{4})\quad \longrightarrow \quad  \langle {\mathcal I}_{12\leftrightarrow 34} \rangle|^{\rm forward}_{\rm MB} = \,\,\, \frac{4 G_F^2 T^5}{\pi^3} \,,\\
\label{eq:Easysubstitution_3}
    S|M|^2 &= 2^5 G_F^2 (p_{1} \cdot p_{4}) (p_{2} \cdot p_{3}) \quad \longrightarrow \quad  \langle {\mathcal I}_{12\leftrightarrow 34} \rangle|^{\rm forward}_{\rm MB} =\,\,\, \frac{4 G_F^2 T^5}{\pi^3} \,.
\end{tcolorbox}

To obtain the full final formulae for the collision terms the two remaining items to be dealt with are: 1) the spin-statistic corrections to these collision terms, and 2) once we allow for neutrino oscillations, the collision terms will become matrices that depend upon the entire neutrino density matrices. We address how to implement these two features in what follows.

\subsection{Including spin-statistics in the collision terms} 
\label{sec:Spinstatistics}

The compactness of the formulae in the previous section is due to the simplification that Maxwell-Boltzmann statistics brings to the collision terms, but in general there are spin-statistic corrections that need to be accounted for. For fermions (as relevant for the system of interest here) the forward collision term can be written schematically as:
\begin{align}
\label{eq:F_function}
    {F} = f_1f_2(1-f_3)(1-f_4) = f_1f_2(1-f_3-f_4 -f_3 f_4)\,.
\end{align}
In fact, the actual collision terms that we are interested in are more complicated as they feature particle correlations and these $f_i$ functions would in general become $3\times3$ matrices and not just numbers. Written in a schematic way the collision term including particle correlations will have a kernel of the form:
\begin{align}\label{eq:F_function_matrices}
    {F} = \rho_1 \rho_2(1-\rho_3)(1-\rho_4) =  \rho_1 \rho_2(1-\rho_3 - \rho_4-\rho_3\rho_4)\,,
\end{align}
where $\rho_i$ are density matrices that depend upon momentum. 

Critically, the terms that contain higher powers of the distribution functions in Eq.~\eqref{eq:F_function} or Eq.~\eqref{eq:F_function_matrices} will be numerically smaller. Following our momentum average ansatz of Eq.~\eqref{eq:ansatz}, we will assume that any density matrix can be decomposed as $\rho_i = A_i \, f^{\rm FD}$, where the massless Fermi-Dirac distribution with zero chemical potential $f^{\rm FD}$ contains all momentum dependence. For neutrinos $A_i$ can be precisely identified $A_i \mapsto r_i$, see Eq.~\eqref{eq:ansatz}, that is $A_i$ is only temperature dependent and encodes information about the lepton asymmetries and flavor correlations. For charged leptons on the other hand we have $A_i \mapsto 1$, that is we consider a plasma without muon and tau leptons. This allows us to write the collision term as:
\begin{align}
{\mathcal I} = \int A_1 A_2 f_1^{\rm FD}f_2^{\rm FD} (1- A_4 f_4^{\rm FD} -A_3 f_3^{\rm FD} - A_3 A_4 f_3^{\rm FD} f_4^{\rm FD}) \,.
\end{align}
Now, taking all particles to be massless there is no other energy scale than the temperature and this means that upon integration of the phase space we schematically obtain:
\begin{align}
\label{eq:Cterm_FDform}
\langle {\mathcal I}\rangle  = \frac{\int {\mathcal I} \dd^3\vec{k}_1}{n_{\rm FD}} =  A_1 A_2 B_{\rm FD}(T) \left(1- B_4 A_4 - B_3 A_3 - B_{34} A_3 A_4 \right)\,.
\end{align}
Importantly, while the $B_{\rm FD}(T)$ factor does depend upon temperature, $B_4$, $B_3$ and $B_{34}$ do not. This happens as a result of the fact that the only energy scale is $T$. In addition, we expect $1\gg B_{3,4} \gg B_{34}$ because they arise from terms containing higher powers of the distribution function.

The final step is to compute these numbers. In order to do that one needs to perform numerically the full phase space integration. Here we follow the original method used by Hannestad and Madsen~\cite{Hannestad:1995rs} to do as many integrals as possible analytically (see also~\cite{Yueh1976} for similar strategies in the context of supernova neutrinos and ~\cite{Fradette:2018hhl,Kreisch:2019yzn} for modern references with pedagogical descriptions of the method). By doing this we find the following results for all the four-fermion matrix elements relevant for neutrino interactions:
\begin{align}
B_{\rm FD}(T) & = 0.995 B_{\rm MB}(T) \,,\\
B_3 = B_4 &= 0.0490 \,,\\
B_{34} &= 0.00230\,.
\end{align}
Thus, quite remarkably, the leading order terms can be obtained by just multiplying the Maxwell-Boltzmann results by $F_{\rm FD} = 0.995$. An important ingredient regarding the other terms is to note that
\begin{align}
B_{34} &\simeq B_{3}^2 = B_4^2\,.
\end{align}
This means that we can effectively treat the higher powers of the distribution functions as a power series with expansion coefficient:
\begin{align}
\epsilon = 0.049 = B_3 = B_4 \simeq \sqrt{B_{34}}\,.
\end{align}
Hence, the key rule that can be derived from this reasoning is that for any additional power of $f^{\rm FD}$ in the collision term we will simply add an $\epsilon$ suppression factor in front. This, together with the analytical results for Maxwell-Boltzmann statistics from the previous section, allow to construct fully analytical formulae for the thermally averaged collision terms taking into account the correct spin statistics factors for fermions, i.e.\ including Pauli blocking. In particular, we will have collision terms featuring integrands proportional to:
\begin{align}
    {F} = A_1 f^{\rm FD}_1  A_2 f^{\rm FD}_2(1-A_3 f_3^{\rm FD})(1-A_4 f_4^{\rm FD}) \,,
\end{align}
where here $A_i$ are matrices (without any momentum dependence). The momentum integrations can be performed by following these rules to obtain:
\begin{tcolorbox}[ams align, boxrule=0.3pt, arc=0.5mm, colback=white!100!white, colbacktitle=white!100!white, colframe=black!100!black, coltitle=black]\label{eq:Cterm_FDform_simple}
\langle {\mathcal I}\rangle = {N}_{\rm norm} F_{\rm FD}  A_1 A_2  (1- \epsilon A_3)(1- \epsilon A_4) \,.
\end{tcolorbox}
The overall normalization factor, ${N}_{\rm norm}$, can be obtained from the particular matrix element in question from Eqs.~\eqref{eq:Easysubstitution_1}-\eqref{eq:Easysubstitution_3}.

\subsection{Collision terms including density matrices and spin-statistics}
\label{app:subsec:collision_rmat_FD}

By collecting the results from the previous subsections we can build a key recipe to build collision terms including spin statistics and allowing for the possibility of off-diagonal components in the density matrix. The recipe reads:\\

{\textbf{The Collision Term Recipe}}
\begin{enumerate}
    \item Compare a given collision term with our defining equation Eq.~\eqref{eq:Definitionofcollterm}. 
    \item Identify the relevant matrix elements and use Eqs.~\eqref{eq:Easysubstitution_1}-\eqref{eq:Easysubstitution_3} to obtain the prefactor of the averaged collision term that corresponds to it.
    \item Add Fermi-Dirac corrections to the collision term following Eq.~\eqref{eq:Cterm_FDform_simple}. 
\end{enumerate}
In what follows we apply these rules to build all the relevant collision terms affecting neutrino oscillations in the Standard Model.

\subsubsection{Neutrino-electron scattering}

We start considering the collision term for neutrino-electron scattering. It is contained in many references but for concreteness we start from Eq.~2.8 and 2.9 of~\cite{deSalas:2016ztq} and follow their notation. In the massless electron limit and summing over both $\nu e^- \leftrightarrow \nu e^-$ and $\nu e^+ \leftrightarrow \nu e^+$ scatterings the collision term reads:
\begin{align}
\label{eq:I_nuenuescatt}
{\mathcal I}^{\nu e \leftrightarrow \nu e}(|\vec{k}_1|) &= \frac{1}{2}\frac{2^5 G_{\rm F}^2}{2 |\vec{k}_1|} \int \frac{\dd^3 \vec{k}_2}{(2\pi)^3 2 E_2}\frac{\dd^3 \vec{k}_3}{(2\pi)^3 2 |\vec{k}_3|}\frac{\dd^3 \vec{k}_4}{(2\pi)^3 2 E_4} (2\pi)^4 \delta^{(4)} (p_1 + p_2 - p_3 - p_4) \\
& \times \left\{ 4 [(p_1\cdot p_4) (p_2 \cdot p_3) + (p_1 \cdot p_2) ( p_3 \cdot p_4) ] \left[F^{LL}_{\rm sc} (\nu^{(1)}, {e}^{(2)}, \nu^{(3)}, {e}^{(4)}) +F^{RR}_{\rm sc} (\nu^{(1)}, {e}^{(2)}, \nu^{(3)}, {e}^{(4)}) ) \right] \right\} \,, \nonumber \\ \nonumber
\end{align}
where
\begin{align}
\label{eq:Fscat}
F^{ab}_{\rm sc} (\nu^{(1)}, e^{(2)}, \nu^{(3)}, e^{(4)}) &= f_4 (1-f_2) \left( G^a \rho_3 G^b (1-\rho_1) + (1-\rho_1) G^b \rho_3 G^a \right)\nonumber\\
&- f_2 (1-f_4) \left( \rho_1 G^b (1-\rho_3) G^a + G^a (1-\rho_3) G^b \rho_1 \right)\,,
\end{align}
where $f$ denotes the distribution functions of electrons and $\rho$ are the neutrino density matrices, which all depend upon the magnitude of the three-momentum of the corresponding particle. In addition, the coupling matrices $G$ read:
\begin{align}\label{eq:Gmatricesdef}
G^L = {\rm diag}(g_L, \tilde{g}_L, \tilde{g}_L)\,,\qquad 
G^R = {\rm diag}(g_R, g_R, g_R) 
\end{align}
with $g_L = s_W^2 + 1/2$, $\tilde{g}_L = s_W^2 - 1/2$, $g_R = s_W^2$, and where we use $s_W^2 \simeq 0.23$. The difference of +1 between $g_L$ and $\tilde{g}_L$ simply corresponds to charged current interactions between $\nu_e$ neutrinos and $e^\pm$. 

The work of the previous subsections pays off precisely at this point and we follow the recipe outlined at the beginning of the subsection. By comparing Eq.~\eqref{eq:I_nuenuescatt} with Eq.~\eqref{eq:Definitionofcollterm} we can clearly see that in terms of momentum dependencies we have the very same types of integrals we solved in the previous subsections. Our ansatz of taking the momentum dependence of the neutrino density matrix to be a Fermi-Dirac distribution with vanishing chemical potential allows us to simply use the formulae we had before but retaining the matrix that comes with it. In particular, this means that we can use Eq.~\eqref{eq:Cterm_FDform_simple}. We see that for this collision term the momentum dependence is of the form $(p_1\cdot p_4) (p_2 \cdot p_3) + (p_1 \cdot p_2) ( p_3 \cdot p_4)$. From Eqs.~\eqref{eq:Easysubstitution_1}-\eqref{eq:Easysubstitution_3} this fixes the overall normalization coefficient to be
\begin{align}
    {N}_{\rm norm} = \frac{1}{2}\,4 \,4 \, (3+1) G_F^2T^5/\pi^3 = 8 \times \frac{4 G_F^2 T^5}{\pi^3}\,.
\end{align}
In this expression the first factor of $1/2$ comes from Eq.~\eqref{eq:I_nuenuescatt}, the second from the second line of Eq.~\eqref{eq:I_nuenuescatt} and the rest from summing over the relevant terms in Eqs.~\eqref{eq:Easysubstitution_1}-\eqref{eq:Easysubstitution_3}. Finally plugging in the whole result of Eqs.~\eqref{eq:Cterm_FDform_simple} we find the entire collision term to be:
\begin{align}\label{eq:Ctermnuscatt}
   \!\!\! \langle {\cal I} \rangle^{\nu e \leftrightarrow \nu e} &= 4  F_{\rm FD}  \frac{8G_F^2 T^5}{\pi^3} (1-\epsilon) \left\{[G^L r G^L (\mathbb{1}- \epsilon\, r ) + (\mathbb{1}-  \epsilon\, r) G^L r G^L]  - [ r G^L (\mathbb{1}-\epsilon\,  r) G^L + G^L (\mathbb{1}-\epsilon \, r) G^L r] \right\}  \,,
\end{align}
where in this expression we have written it in terms of the relevant $r$ and $\bar{r}$ matrices. Neglecting spin statistics corrections, $F_\text{FD} \rightarrow 1$ and $\epsilon \rightarrow 0$, this yields Eq.~\eqref{eq:I_nuenue} in the main text. Note here that only the matrix $G_L$ appears because the terms associated with $G_R$ being proportional to the unit matrix end up yielding vanishing contributions.

\subsubsection{Neutrino-electron scattering: damping form}

We can easily expand the collision term in Eq.~\eqref{eq:Ctermnuscatt} in matrix form. We explicitly find:
\begin{align}
\langle {\mathcal I} \rangle^{\nu e \leftrightarrow \nu e} = -G_{\nu e} \begin{pmatrix}
0 &  r_{12} & r_{13}\\
r_{12}^* & 0 & 0 \\
r_{13}^* & 0 & 0
\end{pmatrix} \,,
\end{align}
where we can clearly see that this is of the damping form. This is: $\langle \mathcal{I}_{\alpha \beta}\rangle = - d_{\alpha \beta} r_{\alpha \beta} $ for $\beta \neq \alpha$, i.e.\ proportional to the off-diagonal elements of the density matrices and hence leading to the damping of flavor coherence. Critically, it is easy to see that this collision term respects the $U(2)$ symmetry in the $\mu-\tau$ subspace. This is as it should be because neutrino-electron interactions cannot tell $\mu$ and $\tau$ neutrinos apart. The damping rate in this expression is given by:
\begin{align}
G_{\nu e} =  F_{\rm FD}  \frac{32}{\pi^3} G_F^2 T^5 (1-\epsilon)\,,
\end{align}
which reproduces the result found in Thomson and McKellar~\cite{McKellar:1992ja}, up to a spin statistics correction of ${\cal O}(\epsilon)$ which was not accounted for in~\cite{McKellar:1992ja}.

\subsubsection{Electron-positron annihilations into neutrinos}

The collision term for $e^+e^- \leftrightarrow \nu \bar{\nu}$ is given e.g.\ in Eqs.~(2.4) and (2.5) of~\cite{deSalas:2016ztq} and in the limit of $m_e\to 0$ reads:
\begin{align}
\label{eq:I_ann_dS}
\begin{split}
\mathcal{I}^{\nu \bar{\nu} \leftrightarrow e^- e^+} &= \frac{1}{2}\frac{2^5 G_{\rm F}^2}{2 |\vec{k}_1|} \int \frac{\dd^3 \vec{k}_2}{(2\pi)^3 2 |\vec{k}_2|}\frac{\dd^3 \vec{k}_3}{(2\pi)^3 2 E_3}\frac{\dd^3 \vec{k}_4}{(2\pi)^3 2 E_4} (2\pi)^4 \delta^{(4)} (p_1 + p_2 - p_3 - p_4) \\
& \times \left\{ 4(p_1\cdot p_4)(p_2\cdot p_3) F^{LL}_{\rm ann}(\nu^{(1)}, \bar{\nu}^{(2)},e^{(3)}, \bar{e}^{(4)})  + 4 (p_1\cdot p_3) (p_2\cdot p_4) F^{RR}_{\rm ann}(\nu^{(1)}, \bar{\nu}^{(2)},e^{(3)}, \bar{e}^{(4)}) \right\}\,,
\end{split}
\end{align}
where
\begin{align}\label{eq:Fann}
\begin{split}
F^{ab}_{\rm ann}(\nu^{(1)}, \bar{\nu}^{(2)},e^{(3)}, \bar{e}^{(4)}) &= f_3 \bar{f}_4 \left( G^a (1-\bar{\rho}_2) G^b (1-\rho_1) +(1-\rho_1) G^b (1-\bar{\rho}_2) G^a \right) \\
&-\, (1-f_3) (1-\bar{f}_4) \left( \rho_1 G^b \bar{\rho}_2 G^a + G^a \bar{\rho}_2 G^b \rho_1 \right)\,.
\end{split}
\end{align}
Here the matrices $G$ are defined as in the previous subsection.

By using the recipe outlined above we 1) compare Eq.~\eqref{eq:I_ann_dS} with Eq.~\eqref{eq:Definitionofcollterm} and 2) by using the formulae in Eqs.~\eqref{eq:Easysubstitution_1}-\eqref{eq:Easysubstitution_3} together with 3) Eq.~\eqref{eq:Cterm_FDform_simple} we can write down the full collision term analytically. The $N_{\rm norm}$ coefficient in this case is $N_{\rm norm} = 8G_F^2 T^5/\pi^3 $
%$N_{\rm norm} = \frac{1}{2} 4 \frac{4G_F^2 T^5}{\pi^3} $ 
and we find:
\begin{align}
     \! \!\!\!\!\!  \langle {\cal I} \rangle^{\nu \bar{\nu} \leftrightarrow e^+e^-}   &= F_{\rm FD} \frac{8 G_F^2 T^5}{\pi^3}  \left\{ [ G^L (\mathbb{1}-\epsilon \, \bar{r}  ) G^L  (\mathbb{1}-\epsilon\, r  ) +   (\mathbb{1}-\epsilon \, r  ) G^L (\mathbb{1}-\epsilon\, \bar{r}  ) G^L ]  - (1-\epsilon )^2 [ r G^L \bar{r} G^L + G^L \bar{r}  G^L r] \right\}   \nonumber \\
&+F_{\rm FD} \frac{8 G_F^2 T^5}{\pi^3}  \left\{ [ G^R (\mathbb{1}-\epsilon\, \bar{r}  ) G^R  (\mathbb{1}-\epsilon \, r  ) +   (\mathbb{1}-\epsilon \, r  ) G^R (\mathbb{1}-\epsilon \,\bar{r}  ) G^R ]  - (1-\epsilon )^2 [ r G^R \bar{r} G^R + G^R \bar{r}  G^R r]\right\} .
\label{eq:finalI_annepem}
\end{align}
Neglecting spin statistics corrections, $F_\text{FD} \rightarrow 1$ and $\epsilon \rightarrow 0$, this gives Eq.~\eqref{eq:I_nunubar_annepem} in the main text. We note that this collision term is non-linear and cannot be written in the damping form as in the previous case.

\subsubsection{Neutrino-neutrino annihilation and scatterings}

The final (and most complicated) collision term arises from neutrino-self interactions. The full form was first written in Eq.~(5.11) of~\cite{Sigl:1993ctk} and then reproduced in many others references, see e.g.\ Ref.~\cite{Bennett:2020zkv} (see Eqs. (A8)-(A11)) or  Ref.~\cite{Froustey:2020mcq} (see Eqs.~(C.1)-(C.2)). In these references the collision terms are expressed for the case of $\rho = \bar{\rho}$ but we can easily generalize it to the case where $\rho\neq \bar{\rho}$, see also~\cite{Li:2024gzf}. In particular, generalizing the results of Eqs.~(A8)-(A11) of~\cite{Bennett:2020zkv} we can write:
\begin{align}
\mathcal{I}^{\nu \bar{\nu} \leftrightarrow \nu \bar{\nu}} &= \frac{1}{2}\frac{2^5 G_{\rm F}^2}{2 |\vec{k}_1|} \int \frac{\dd^3 \vec{k}_2}{(2\pi)^3 2 |\vec{k}_2|}\frac{\dd^3 \vec{k}_3}{(2\pi)^3 2 E_3}\frac{\dd^3 \vec{k}_4}{(2\pi)^3 2 E_4} (2\pi)^4 \delta^{(4)} (p_1 + p_2 - p_3 - p_4) \nonumber\\
& \times \left\{ (p_1\cdot p_4)(p_2\cdot p_3) F_{\rm pair}^{\nu\nu} \left(\rho^{(1)}, \bar{\rho}^{(2)}, \rho^{(3)}, \bar{\rho}^{(4)}\right) \right\}, \label{eq:I_nuann_dS}
\end{align} 
where the phase space matrices are
\begin{equation}
\begin{aligned}
F^{\nu\nu}_{\rm pair}
\equiv& \, (\mathbb{1}-\rho^{(1)}) (\mathbb{1}-
{\bar{\rho}}^{(2)}) \! \left [{\bar{\rho}}^{(4)} \rho^{(3)} + {\rm Tr}(\cdots) \mathbb{1} \right]
\! -\! \rho^{(1)} {\bar{\rho}}^{(2)} \left [(\mathbb{1}- {\bar{\rho}}^{(4)} )(\mathbb{1}- \rho^{(3)} )+ {\rm Tr}(\cdots) \mathbb{1} \right] \\
+&
(\mathbb{1}-\rho^{(1)}) \rho^{(3)} \! \left [ {\bar{\rho}}^{(4)}(\mathbb{1}-{\bar{\rho}}^{(2)}) + {\rm Tr}(\cdots) \mathbb{1} \right]
\! -\! \rho^{(1)} (\mathbb{1}- \rho^{(3)}) \! \left [(\mathbb{1}- {\bar{\rho}}^{(4)} ) \bar{{\rho}}^{(2)} + {\rm Tr}(\cdots) \mathbb{1}  \right] + {\rm h.c.},
\label{eq:phasespace}
\end{aligned}
\end{equation}
and ${\rm Tr}(\cdots)$ denotes the trace of the preceding term and $\mathrm{h.c.}$ is shorthand for the anti-commutator, following the notation of~\cite{Bennett:2020zkv}. We note that within our momentum average approximation, the collision term associated with neutrino-neutrino scatterings vanishes:
\begin{align}
F_{\rm sc}^{\nu\nu} \equiv & \, (\mathbb{1}-\rho^{(1)}) \rho^{(3)} \! \left [(\mathbb{1}- \bar{\rho}^{(2)}) \bar{\rho}^{(4)} + {\rm Tr}(\cdots) \mathbb{1}  \right]
\!-\! \rho^{(1)} (\mathbb{1}- \rho^{(3)})\! \left [\bar{\rho}^{(2)} (\mathbb{1}- \bar{\rho}^{(4)} )+ {\rm Tr}(\cdots) \mathbb{1}  \right] + {\rm h.c.} \; \to \; 0\,.
\end{align}
Physically this is because the t-channel of neutrino self-scatterings only contributes to a change of momenta, which after averaging is zero. We note, however, that this is due to our momentum average ansatz and that in a full momentum dependent calculation $F_{\rm sc}^{\nu\nu} \neq 0$ in general.

Concentrating then in the case of $\nu\bar{\nu} $ annihilation, again, by following the recipe and 1) comparing Eq.~\eqref{eq:I_nuann_dS} with Eq.~\eqref{eq:Definitionofcollterm} and 2) by using the formulae in Eqs.~\eqref{eq:Easysubstitution_1}-\eqref{eq:Easysubstitution_3} together with 3) Eq.~\eqref{eq:Cterm_FDform_simple} allows us to write down the full collision term analytically. The $N_{\rm norm}$ coefficient in this case is $N_{\rm norm} = 2 G_F^2 T^5/\pi^3$ and we then find:
\begin{align}
\begin{split}
    \langle \mathcal{I}\rangle^{\nu \bar{\nu} \leftrightarrow \nu \bar{\nu}} &=  \frac{1}{4} F_{\rm FD} \frac{8G_F^2 T^5}{\pi^3}\left[ (\mathbb{1}-\epsilon \, r) (\mathbb{1}-
\epsilon \, \bar{r}) \! \left [\bar{r} r + {\rm Tr}\{\bar{r} r\} \right]
\! -\! r \bar{r} \left [(\mathbb{1}-\epsilon \, \bar{r} )(\mathbb{1}-\epsilon \, {r} )+ {\rm Tr}\{(\mathbb{1}-\epsilon \, \bar{r} )(\mathbb{1}-\epsilon \, {r} )\} \right] \right. \\
&+\left. (\mathbb{1}-\epsilon \, r ) r \! \left [ \bar{r}(\mathbb{1}-\epsilon \,\bar{r}) + {\rm Tr}\{\bar{r}(\mathbb{1}-\epsilon \,\bar{r}) \} \right]
\! -\! r (\mathbb{1}- \epsilon \, r) \! \left [(\mathbb{1}- \epsilon \, \bar{r} ) \bar{r} + {\rm Tr}\{(\mathbb{1}- \epsilon \, \bar{r} ) \bar{r}\} \right]  \right]+ {\rm h.c.}\,.
\end{split}
\end{align}
We readily see that the second line vanishes and the first one can be simplified to yield:
\begin{align}
    \langle \mathcal{I}\rangle^{\nu \bar{\nu} \leftrightarrow \nu \bar{\nu}} =  \frac{1}{4} F_{\rm FD} \frac{8G_F^2 T^5}{\pi^3}\left[([{\rm Tr}[r\bar{r}]-r\bar{r}{\rm Tr}[\mathbb{1}]) + \epsilon (r\bar{r}{\rm Tr}[r+\bar{r}]-(r+\bar{r}){\rm Tr}[r\bar{r}]) +\epsilon (r \bar{r}\bar{r}-\bar{r}\bar{r}r) \right] + {\rm h.c.}\,,
\end{align}
Upon explicitly writing the hermitian conjugate the last set of terms with three density matrices vanishes and we end up with the final result:
\begin{align}
\label{eq:Cterm_nunubar}
    \!\!\! \langle {\cal I} \rangle^{\nu \bar{\nu} \leftrightarrow \nu \bar{\nu}}   = \frac{1}{4} F_{\rm FD}  \frac{8 G_F^2 T^5}{\pi^3}  \left\{2{\rm Tr}[r \bar{r}] [\mathbb{1}-\epsilon\, (r + \bar{r})] -  (r \bar{r}+\bar{r} r) ({\rm Tr}[\mathbb{1}] - \epsilon \, {\rm Tr}[r+\bar{r}])   \right\}\,.
\end{align}
Neglecting spin statistics corrections, $F_\text{FD} \rightarrow 1$ and $\epsilon \rightarrow 0$, this yields Eq.~\eqref{eq:I_nu_pair} in the main text. Again, we see that this collision term is non-linear.

\subsection{Accuracy of the Collision terms: damping form, detailed balance and full comparisons}

\subsubsection{An approximate damping-like collision term}

As described in the previous sections the collision terms are non-linear and they cannot be expressed as simple temperature dependent damping coefficients multiplying the respective density matrix elements, $\langle {\cal I} \rangle_{\alpha \beta} \simeq -d_{\alpha \beta} \rho_{\alpha \beta} $. However, collision terms of the damping form are significantly simpler to implement when solving the QKEs and it is thus useful to derive approximate results of this form.
In this context, we provide a damping term approximation taking into account all the collision terms based on linearizing the full results above. To this end we consider the density matrices $r$ and $\bar{r}$ to have $1$'s in the diagonal and then include the leading order off-diagonal components from either $r$ or $\bar{r}$. This procedure incorrectly leads to a breaking of the $U(2)$ symmetry in the $\mu-\tau$ subspace. To amend it, we simply ignore the damping term for this entry and thus we can write:
\begin{align}
\label{app:eq:damping_approx}
\langle {\mathcal I} \rangle^{\rm damping \, approx} = -D \begin{pmatrix}
0 &  r_{12} & r_{13}\\
r_{12}^* & 0 & 0 \\
r_{13}^* & 0 & 0
\end{pmatrix} \,, \qquad {\rm with} \qquad D =  F_{\rm FD}  \frac{16}{\pi^3} G_F^2 T_\gamma^5 (1-\epsilon)(3+2s_W^2)\,. 
\end{align}
This results in a damping rate which is a factor 1.553 larger than the one from $\nu e\leftrightarrow \nu e$ scatterings only.

\begin{figure}[!t]
\centering
\begin{tabular}{cc}
\hspace{-0.cm}\includegraphics[width=0.5\textwidth]{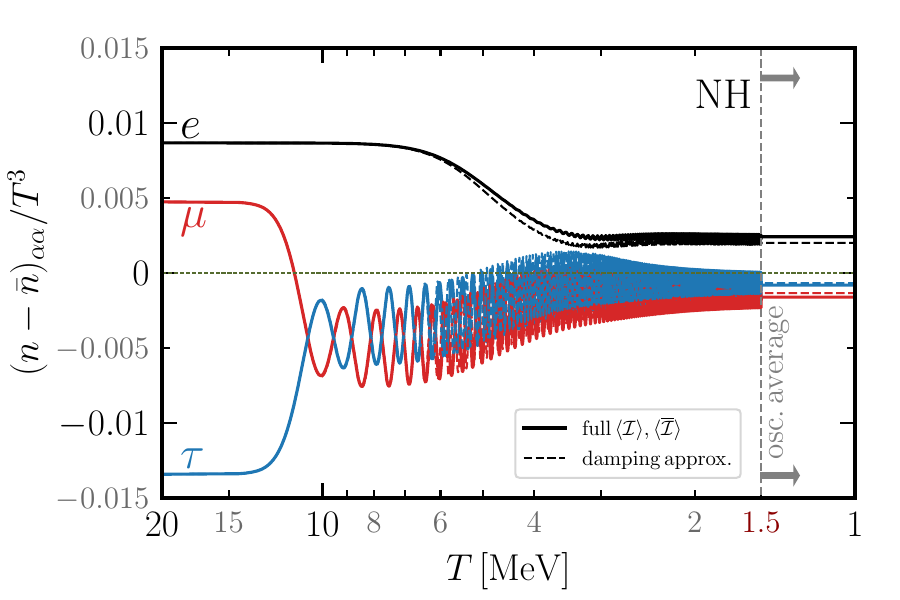}
\hspace{-0.cm}\includegraphics[width=0.5\textwidth]{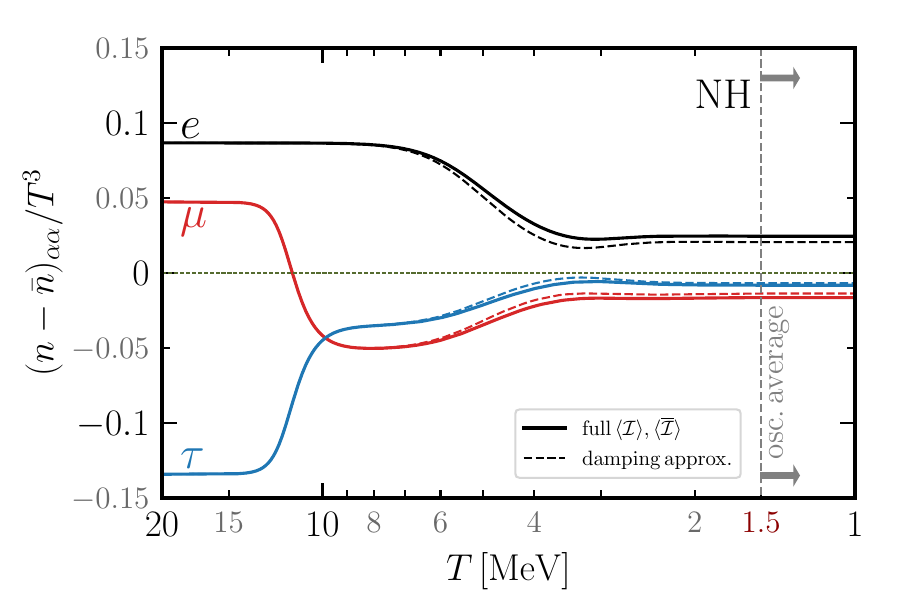}
\end{tabular}
\caption{Comparison of the time evolution of neutrino flavor asymmetries obtained by using the full collision term (solid), see Eq.~\eqref{eq:Cterm_FDform_simple} and below, and the damping approximation (dashed), see Eq.~\eqref{app:eq:damping_approx}. We specify $\Delta n = 0$, $\phi= 0.5$ and $A = 0.1\,(1)$ in left (right) plot. According to the parametrization of Eqs.~\eqref{eq:sph}-\eqref{eq:parametrization_theta} this translates to $\xi_e = 0.0520,\,\xi_\mu = 0.0284,\,\xi_\tau = -0.0805$ in the left panel and $\xi_e = 0.5075,\,\xi_\mu = 0.282,\,\xi_\tau = -0.760$ in the right panel.}
\label{app:fig:damping_time_evol}
\end{figure}

\begin{figure}[!t]
\centering
\begin{tabular}{cc}
\hspace{-0.cm}\includegraphics[width=0.5\textwidth]{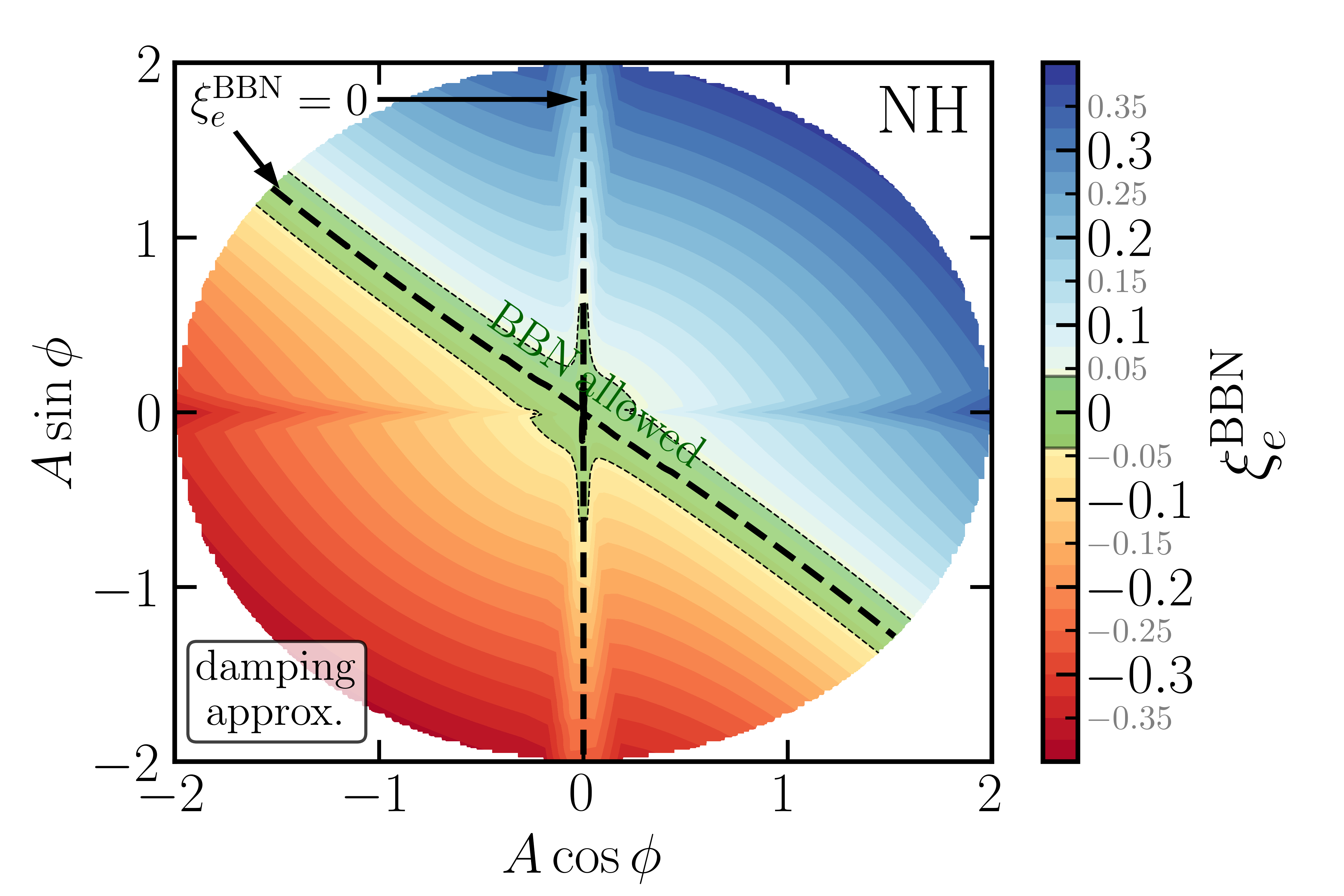}
\hspace{-0.cm}\includegraphics[width=0.5\textwidth]{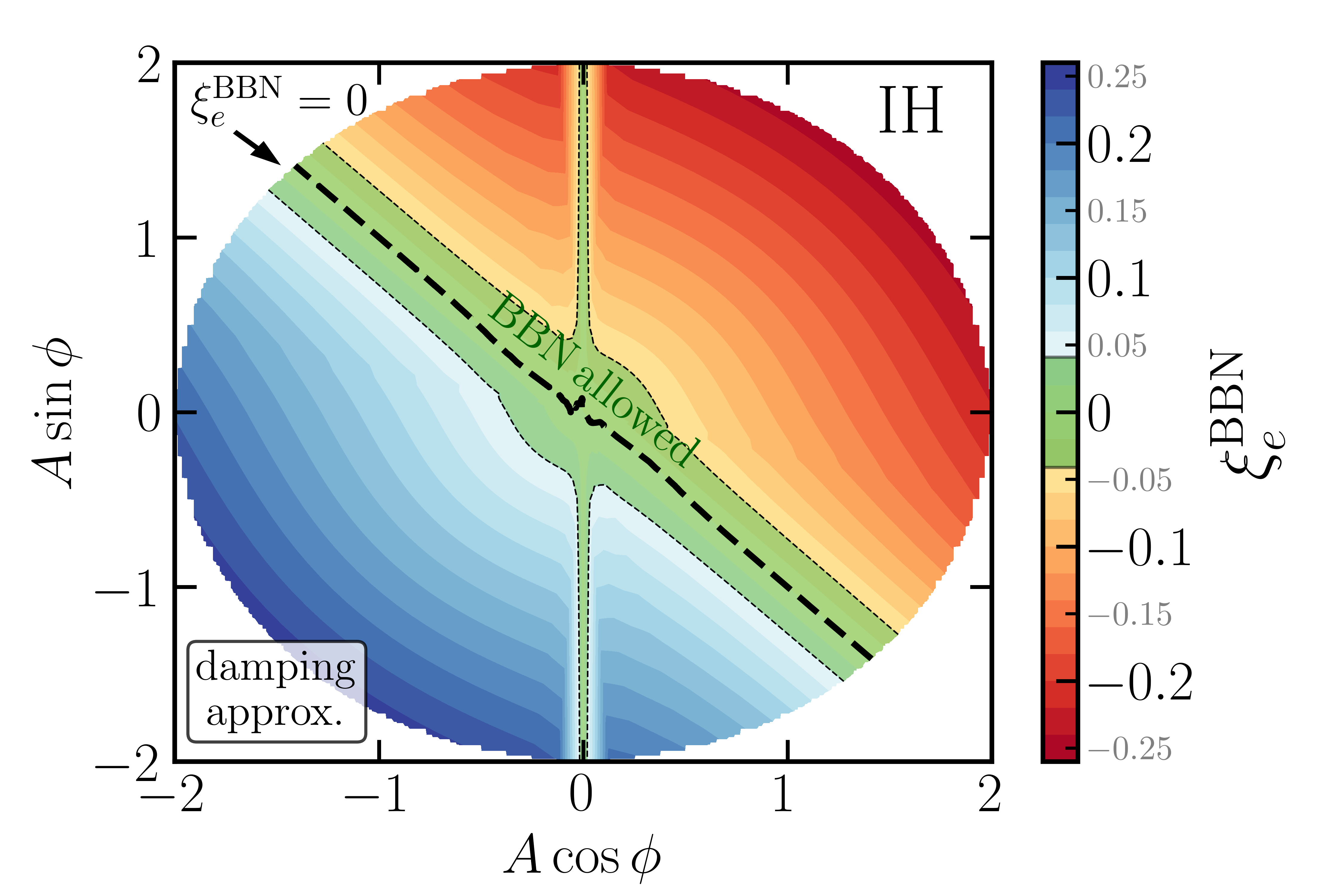}
\end{tabular}
\caption{ Resulting asymmetry in the electron neutrino flavor as relevant for BBN from initial $e,\mu,\tau$ asymmetries with zero total lepton number. Same as Fig.~\ref{fig:constraints} but using the damping approximation of Eq.~\eqref{app:eq:damping_approx} for the collision term.}
\label{app:fig:damping_contour}
\end{figure}

In Fig.~\ref{app:fig:damping_time_evol} we compare the evolution of two benchmark point using the full collision term versus the one using the damping form~\eqref{app:eq:damping_approx}. 
We see that they are in rather good agreement. The impact of this approximation across the entire parameter space is shown in Fig.~\ref{app:fig:damping_contour} where one sees that the results from using the damping approximation roughly resemble the full results and the main effect is a slight tilt in the parameter space region that leads to small $\xi_e$ at BBN times.
Note Eq.~\eqref{app:eq:damping_approx} was derived neglecting the energy transfer between the neutrino and the photon/electron sector. As such, while it gives reasonably accurate results for the final neutrino asymmetries, it should not be employed to estimate $\Delta N_\text{eff}$, as it would result in a significant overestimation of this quantity.

\subsubsection{Detailed balance}
\label{app:subsubsec:db}

Our ansatz that the entire density matrix is proportional to the Fermi-Dirac distribution function with vanishing chemical potential has proven very useful to derive compact, simple and yet accurate collision terms. However, we know that by doing this the diagonal components of the collision terms for annihilation will a priori contain non-zero entries. This is simply because $f^{\rm FD}(E, T,\mu)$ cannot be written as a product of a function that depends upon $E/T$ and one that depends upon $\mu$ (unless $\mu = 0$). This feature of the collision terms in not ideal, because it indicates that initial conditions given by diagonal matrices (since neutrino oscillations are overdamped at $T \gtrsim 15\,\mathrm{MeV}$, i.e.\  the mean free path between two collisions is much shorter than the oscillation length) and obtained by performing the momentum overage over the Fermi-Dirac distribution following our ansatz~\eqref{eq:ansatz},
\begin{align}
\label{app:eq:rfd_ini}
    r_{\alpha \alpha }^{\rm fd} = -\frac{4}{3\zeta(3)} \mathrm{Li}_3\left(- \mathrm{e}^{\xi_\alpha}\right)\,, \quad \text{and  }  \; \xi_\alpha \mapsto - \xi_\alpha\;  \text{   for   } \; \bar{r}_{\alpha\alpha}^{\rm fd},
\end{align}
do not satisfy the detailed balance condition, i.e.\ they do not represent an equilibrium state. Thus, in order to fulfill detailed balance in this regime where the neutrino density matrix is diagonal, we analyze the diagonal entries of the collision term:
\begin{align}
    \langle {\cal I} \rangle^{\nu \bar{\nu} \leftrightarrow e^+e^-} |_{\alpha \alpha} \propto[1-r_{\alpha \alpha} \bar{r}_{\alpha \alpha}+ \epsilon (2 r_{\alpha \alpha} \bar{r}_{\alpha \alpha}  -r_{\alpha \alpha} - \bar{r}_{\alpha \alpha})]\,.
\end{align}
The first thing to note is that in the limit in which the distribution functions are of Maxwell-Boltzmann type $\epsilon \to 0$, $r_{\alpha \alpha} = e^{\xi_\alpha}$ and $\bar{r}_{\alpha \alpha} = e^{-\xi_\alpha}$ and they do indeed fulfill detailed balance. For $\epsilon \neq 0$ this is not the case anymore reflecting that the Fermi-Dirac distribution with a non-zero chemical potential cannot be factorized. Hence the collision terms will dynamically enforce an appropriate new detailed balance condition. Importantly, since the collision terms cannot possibly alter the neutrino asymmetries (all SM interactions are lepton number conserving) this will only impact the symmetric components $r_{\alpha \alpha} + \bar{r}_{\alpha \alpha}$. In particular, by writing $r_{\alpha \alpha }^{\rm ini} = r_{\alpha \alpha }^{\rm fd} + \epsilon_{\mathrm{db},\alpha}$ and $\bar{r}_{\alpha \alpha }^{\rm ini} = \bar{r}_{\alpha \alpha }^{\rm fd} + \epsilon_{\mathrm{db},\alpha}$, requiring the collision term to vanish yields
\small
\begin{align}
\label{app:eq:db}
\epsilon_{\mathrm{db},\alpha} = \frac{-\sqrt{4 \epsilon^2 \left((\bar{r}_{\alpha\alpha}^{\rm fd}-r_{\alpha\alpha}^{\rm fd})^2+1\right)-4 \epsilon \left((\bar{r}_{\alpha\alpha}^{\rm fd}-r_{\alpha\alpha}^{\rm fd})^2+2\right)+(\bar{r}_{\alpha\alpha}^{\rm fd}-r_{\alpha\alpha}^{\rm fd})^2+4}-2 \epsilon (\bar{r}_{\alpha\alpha}^{\rm fd}+r_{\alpha\alpha}^{\rm fd}-1)+\bar{r}_{\alpha\alpha}^{\rm fd}+r_{\alpha\alpha}^{\rm fd}}{4 \epsilon-2}\,,
\end{align}
\normalsize
with $r_{\alpha \alpha }^{\rm fd}$ given in Eq.~\eqref{app:eq:rfd_ini}.

In summary, we have shown that at high temperatures when neutrino density matrices are diagonal a first-principle calculation of the collision terms leads to a small violation of detailed balance within the momentum average ansatz of Eq.~\eqref{eq:ansatz} (unless $r = \bar{r}$). Although we find that the quantum kinetic equations~\eqref{eq:master} dynamically converge to a state which fulfills the condition of detailed balance, we derived an analytical expression to correct for the initial mismatch. By imposing the initial conditions of Eq.~\eqref{app:eq:db}-\eqref{app:eq:rfd_ini}, we ensure that detailed balance is fulfilled at all times. As sketched in Sec.~\ref{subsec:initial_cond}, this strategy can be extended to allow for different temperatures in the neutrino and electron-photon bath. This results in an additional constraint of vanishing net energy transfer between the two baths in equilibrium.

\section{Collision term with $T_\nu \neq T_\gamma$}
\label{sec:Cterms_TnuneqTgamma}

In the previous appendix we have worked out systematically the momentum averaged collision terms for the case of equal temperature for neutrinos and the thermal electron-photon plasma. In this appendix we relax this assumption and also compute the energy transfer rate resulting from the fact that $T_\nu \neq T_\gamma$. We recall that the origin of this temperature difference is the non-reversible process of flavor equilibration through neutrino oscillations in the neutrino sector. This entropy injection effectively heats up the neutrino bath. Since this process occurs close to neutrino decoupling, this energy increase is only partially shared with the tightly coupled system of photons and charged leptons. For the case of Maxwell-Boltzmann statistics the relevant collision integrals can be computed analytically. If, for simplicity, one assumes that the Pauli blocking effects are temperature independent (i.e. $T_\gamma \simeq T_\nu$) then both the Fermi-Dirac corrections and the leading temperature dependence can be also obtained analytically.

\subsection{Momentum averaged collision term}

The integrals needed to obtain the right temperature scaling can be obtained analytically using the phase space integration method described in Sec.~\ref{sec:Spinstatistics} and read:
\begin{align}
   \frac{\int \frac{\dd^3\vec{k}}{(2\pi)^3} {\mathcal I}^{\nu e \leftrightarrow \nu e}(|\vec{k}|)_{{\rm Forward-MB},{f_1(T_\nu),f_2(T_\gamma)}}}{\int \frac{\dd^3\vec{k}}{(2\pi)^3} {\mathcal I}^{\nu e \leftrightarrow \nu e}(|\vec{k}|)_{{\rm Forward-MB},{f_1(T),f_2(T)}}}&= \left[\frac{T_\gamma}{T}\right]^4\left[\frac{T_\nu}{T}\right]^4 \,,\\
   \frac{\int \frac{\dd^3\vec{k}}{(2\pi)^3} {\mathcal I}^{\nu e \leftrightarrow \nu e}(|\vec{k}|)_{{\rm Backward-MB},{f_3(T_\nu),f_4(T_\gamma)}}}{\int \frac{\dd^3\vec{k}}{(2\pi)^3} {\mathcal I}^{\nu e \leftrightarrow \nu e}(|\vec{k}|)_{{\rm Backward-MB},{f_3(T),f_4(T)}}}&= \left[\frac{T_\gamma}{T}\right]^4\left[\frac{T_\nu}{T}\right]^4\,, \\
      \frac{\int \frac{\dd^3\vec{k}}{(2\pi)^3} {\mathcal I}^{\nu \bar{\nu} \leftrightarrow e^+ e^-}(|\vec{k}|)_{{\rm Forward-MB},{f_1(T_\nu),f_2(T_\nu)}}}{\int \frac{\dd^3\vec{k}}{(2\pi)^3} {\mathcal I}^{\nu \bar{\nu} \leftrightarrow e^+ e^-}(|\vec{k}|)_{{\rm Forward-MB},{f_1(T),f_2(T)}}}&= \left[\frac{T_\nu}{T}\right]^8\,,\\
   \frac{\int \frac{\dd^3\vec{k}}{(2\pi)^3} {\mathcal I}^{\nu \bar{\nu} \leftrightarrow e^+ e^-}(|\vec{k}|)_{{\rm Backward-MB},{f_3(T_\gamma),f_4(T_\gamma)}}}{\int \frac{\dd^3\vec{k}}{(2\pi)^3} {\mathcal I}^{\nu \bar{\nu} \leftrightarrow e^+ e^-}(|\vec{k}|)_{{\rm Backward-MB},{f_3(T),f_4(T)}}}&= \left[\frac{T_\gamma}{T}\right]^8 \,,
\end{align} 
where we have normalized the expressions to the equal temperature limit. Given these results, and assuming that the Pauli blocking factors are temperature independent (which is a good approximation if $(T_\gamma-T_\nu)/T_\gamma \ll 1$), then we can easily obtain the momentum averaged collision terms from the previous expressions. In particular, collecting the results from~\eqref{eq:Ctermnuscatt}, ~\eqref{eq:finalI_annepem}, and~\eqref{eq:Cterm_nunubar} the averaged collision terms read:
\small
\begin{align}
\label{eq:I_nuenue_T}
\!\!\! \langle {\cal I} \rangle^{\nu e \leftrightarrow \nu e} &= 4  F_{\rm FD}  \frac{8G_F^2 T^5}{\pi^3} z_\gamma^4 z_\nu (1-\epsilon) \left\{[G^L r G^L (\mathbb{1}- \epsilon\, r ) + (\mathbb{1}-  \epsilon\, r) G^L r G^L]  - [ r G^L (\mathbb{1}-\epsilon\,  r) G^L + G^L (\mathbb{1}-\epsilon \, r) G^L r] \right\}  \\
\begin{split}
\label{eq:I_nunubar_annepem_T}
\!\!\!\!\! \langle {\cal I} \rangle^{\nu \bar{\nu} \leftrightarrow e^+e^-}   &= F_{\rm FD} \frac{8 G_F^2 T^5}{\pi^3} \frac{1}{z_\nu^3} \left\{ z_\gamma^8[ G^L (\mathbb{1}-\epsilon \, \bar{r}  ) G^L  (\mathbb{1}-\epsilon\, r  ) +   (\mathbb{1}-\epsilon \, r  ) G^L (\mathbb{1}-\epsilon\, \bar{r}  ) G^L ]  - z_\nu^8(1-\epsilon )^2 [ r G^L \bar{r} G^L + G^L \bar{r}  G^L r] \right\} \\
&+F_{\rm FD} \frac{8 G_F^2 T^5}{\pi^3} \frac{1}{z_\nu^3} \left\{ z_\gamma^8[ G^R (\mathbb{1}-\epsilon\, \bar{r}  ) G^R  (\mathbb{1}-\epsilon \, r  ) +   (\mathbb{1}-\epsilon \, r  ) G^R (\mathbb{1}-\epsilon \,\bar{r}  ) G^R ]  - z_\nu^8(1-\epsilon )^2 [ r G^R \bar{r} G^R + G^R \bar{r}  G^R r]\right\}    
\end{split}
\\
\!\!\! \langle {\cal I} \rangle^{\nu \bar{\nu} \leftrightarrow \nu \bar{\nu}}   &= \frac{1}{4} F_{\rm FD}  \frac{8 G_F^2 T^5}{\pi^3} z_\nu^5 \left\{2{\rm Tr}[r \bar{r}] [\mathbb{1}-\epsilon\, (r + \bar{r})] -  (r \bar{r}+\bar{r} r) ({\rm Tr}[\mathbb{1}] - \epsilon \, {\rm Tr}[r+\bar{r}])   \right\}\,,
\label{eq:I_nu_pair_T}
\end{align}
\normalsize
where as in the main text $T = T_{\rm cm}$,  $z_\gamma \equiv T_\gamma/T_{\rm cm}$ and $z_\nu \equiv T_\nu/T_{\rm cm}$. In the limit $F_{\rm FD}\to 1$ and $\epsilon\to 0$, we recover the MB limit outlined in the main text in Eqs.~\eqref{eq:I_nuenue},~\eqref{eq:I_nunubar_annepem}, and \eqref{eq:I_nu_pair}, respectively.

\subsection{Energy transfer rate}

Having calculated in the previous subsection the collision terms when $T_\nu \neq T_\gamma$ in this section we calculate the energy transfer rate $\delta \varepsilon/\delta t$ needed in order to track the joint evolution of $T_\gamma$ and $T_\nu$. As required by the continuity equation and the Liouville equation, and as outlined in the main text, the energy transfer rate is given by Eq.~\eqref{eq:def_energy_transfer} and reads:
\begin{align}
 \frac{\delta \varepsilon}{\delta t} =  \int \frac{\dd^3 k}{(2\pi)^3} \, k   \, \text{Tr}[ {\cal I} + \bar{\cal I}] \,,
 \nonumber
\end{align}
where here ${\cal I}$ and $\bar{\cal I}$ are the total un-averaged collision terms. In the Maxwell-Boltzmann limit it is particularly easy to obtain analytically this rate of energy transfer from the expressions we had for the collision terms. In particular, the integrals needed are:
\begin{align}
   \frac{\int \frac{\dd^3\vec{k}}{(2\pi)^3} |\vec{k}|{\mathcal I}^{\nu e \leftrightarrow \nu e}(|\vec{k}|)_{{\rm Forward-MB},{f_1(T_\nu),f_2(T_\gamma)}}}{\int \frac{\dd^3\vec{k}}{(2\pi)^3} {\mathcal I}^{\nu e \leftrightarrow \nu e}(|\vec{k}|)_{{\rm Forward-MB},{f_1(T),f_2(T)}}}&= \frac{T_\nu^4 T_\gamma^4}{4T^8} (16 T_\nu) \,,\\
   \frac{\int \frac{\dd^3\vec{k}}{(2\pi)^3}  |\vec{k}|{\mathcal I}^{\nu e \leftrightarrow \nu e}(|\vec{k}|)_{{\rm Backward-MB},{f_3(T_\nu),f_4(T_\gamma)}}}{\int \frac{\dd^3\vec{k}}{(2\pi)^3} {\mathcal I}^{\nu e \leftrightarrow \nu e}(|\vec{k}|)_{{\rm Backward-MB},{f_3(T),f_4(T)}}}&= \frac{T_\nu^4 T_\gamma^4 }{4 T^8} (9 T_\nu+7 T_\gamma) \,,  \\
      \frac{\int \frac{\dd^3\vec{k}}{(2\pi)^3} |\vec{k}| {\mathcal I}^{\nu \bar{\nu} \leftrightarrow e^+ e^-}(|\vec{k}|)_{{\rm Forward-MB},{f_1(T_\nu),f_2(T_\nu)}}}{\int \frac{\dd^3\vec{k}}{(2\pi)^3} {\mathcal I}^{\nu \bar{\nu} \leftrightarrow e^+ e^-}(|\vec{k}|)_{{\rm Forward-MB},{f_1(T),f_2(T)}}}&= 4 \frac{T_\nu^9}{T^8}\,,\\
      \frac{\int \frac{\dd^3\vec{k}}{(2\pi)^3} |\vec{k}| {\mathcal I}^{\nu \bar{\nu} \leftrightarrow e^+ e^-}(|\vec{k}|)_{{\rm Backward-MB},{f_3(T_\gamma),f_4(T_\gamma)}}}{\int \frac{\dd^3\vec{k}}{(2\pi)^3} {\mathcal I}^{\nu \bar{\nu} \leftrightarrow e^+ e^-}(|\vec{k}|)_{{\rm Backward-MB},{f_3(T),f_4(T)}}}&= 4 \frac{T_\gamma^9}{T^8} \,,
\end{align} 
where again, these integrals are obtained analytically using the phase space integration method described in Sec.~\ref{sec:Spinstatistics}.

With this, it is straightforward to obtain the energy transfer rate from the collision terms we had derived in the previous App.~\ref{app:collisionterms} as one simply needs to update the various pre-factors using these previous equations and multiply by $n_{\nu}^{\rm FD} = T_\nu^3/\pi^2$ to compensate for the denominator used in the averaging procedure. Summing over neutrinos and anti-neutrinos, we then find in the Maxwell-Boltzmann approximation the following expression for the energy density transfer rate:
\begin{align}
\begin{split}
    \frac{\delta \varepsilon}{\delta t} &= 128 \frac{ G_F^2 T^9}{\pi^5}    {\rm Tr}[z_\gamma^9( G^L   G^L+G^RG^R)   - z_\nu^9 ( G^L \bar{r}  G^L r+G^R \bar{r}  G^R r)]  \\
   &+\, 112 \frac{ G_F^2 T^9}{\pi^5}   z_\gamma^4 z_\nu^4 (z_\gamma - z_\nu){\rm Tr}[(r+\bar{r}) G^L G^L + (r+\bar{r})G^RG^R)] \,,
\end{split}
\end{align}
where here the first line corresponds to annihilation and the second one to scatterings (the $\nu \bar{\nu} \leftrightarrow \nu \bar{\nu}$ term does not contribute). This expression matches Eq.~\eqref{eq:energy-transfer} in the main text.

We note that this energy transfer rate does contain the $G^R$ coupling for scattering interactions (which did not lead to decoherence for scatterings). Importantly, in the limit in which $r = \bar{r} = \mathbb{1}$ we recover precisely the Maxwell-Boltzmann energy transfer rates obtained in~\cite{Escudero:2018mvt,EscuderoAbenza:2020cmq}, see Eqs (3.4)-(3.6) of~\cite{EscuderoAbenza:2020cmq}. We have also checked that this rate agrees with the direct numerical computation using the collision terms from~\cite{deSalas:2016ztq}.

We note that while in principle Fermi-Dirac corrections to these terms can be included, due to our factorization ansatz of the distribution function it can lead to situations where e.g. for $z_\nu=z_\gamma=1$ in thermal equilibrium $\delta \varepsilon/\delta t \neq 0$, indicating a mismatch in the interpretation of $T_{\gamma,\nu}$ as physical photon and neutrino temperatures, respectively. This mismatch is only very minor when including the leading order Fermi-Dirac corrections, but becomes numerically significant at next-to-leading order (and large chemical potentials). Since this mismatch is unphysical and purely arises due to our ansatz~\eqref{eq:ansatz}, we omit to include the next-to-leading order Fermi-Dirac corrections. In any case, it can be explicitly checked that for the non-averaged energy transfer rate these corrections are indeed small and thus omitting them does not impact any of our results and conclusions.

\section{SU(3) decomposition of the kinetic equation}
\label{app:SUN-decomposition}
In this appendix we provide some more detailed guidance on how to implement the SU($3$) decomposition of the momentum averaged kinetic equations in flavor space. Our starting point is Eq.~\eqref{eq:master}, which given the decomposition~\eqref{eq:SUNdecompositionDef} can be recast as Eq.~\eqref{eq:kinetic_eq_SUN_flavor}. This implies in particular that we express the matrices $r$ and  $\bar r$ as
\begin{align}
 r(x) = r^0(x) \mathbb{1} + r^i(x) \lambda_i  \,,  \quad  \bar r(x) = \bar r^0(x) \mathbb{1} + \bar r^i(x) \lambda_i \,,
\end{align}
with $\lambda_{i}$, $i = \{1..8\}$ denoting the Gell-Mann matrices and $r^0,r^i$ being real coefficients. The kinetic equation for these coefficients is given in Eq.~\eqref{eq:kinetic_eq_SUN_flavor}. In the following we will give the coefficients appearing in Eq.~\eqref{eq:kinetic_eq_SUN_flavor}, which are obtained via Eq.~\eqref{eq:SUNdecompositionDef}.

The flavor structure of the matter potential of the neutrino self-interactions is contained in the non-linear contribution of the density matrix itself, leading to a universal coefficient $v_s$ given in Eq.~\eqref{eq:vs} which multiplies the SU($3$) coefficients of the matrix $\bar r$ or $r$, respectively.  The matter potential generated by the charged currents is diagonal in flavor space, and dropping the term proportional to the unit matrix as it does not contribute to the commutator on the right-hand side of Eq.~\eqref{eq:master}, we can write 
\begin{align}
 \frac{\langle V_c\rangle}{xH} = - \frac{1}{xH}  2 \sqrt{2} \frac{7 \pi^4}{180 \zeta(3)}  \frac{G_F}{m_W^2} z_\nu z_\gamma^4  \left( \frac{T_{\mathrm{ref}}}{x} \right)^5  (\mathbb{E} + \mathbb{P}) \rightarrow v_c^3 \lambda_3 + v_c^8 \lambda_8
\end{align}
with
\begin{align}
\begin{split}
 v_c^3 &= - \frac{1}{2} \frac{1}{xH}  2 \sqrt{2} \frac{7 \pi^4}{180 \zeta(3)}  \frac{G_F}{m_W^2}  z_\nu z_\gamma^4 \left( \frac{T_{\mathrm{ref}}}{x} \right)^5  [(\mathbb{E} + \mathbb{P})_e - (\mathbb{E} + \mathbb{P})_\mu] \,, \\
  \sqrt{3} \,  v_c^8 &= - \frac{1}{2} \frac{1}{xH}  2 \sqrt{2} \frac{7 \pi^4}{180 \zeta(3)}  \frac{G_F}{m_W^2} z_\nu z_\gamma^4 \left( \frac{T_{\mathrm{ref}}}{x} \right)^5  [(\mathbb{E} + \mathbb{P})_e + (\mathbb{E} + \mathbb{P})_\mu] \,.
\end{split}
\end{align}
The vacuum Hamiltonian will involve both diagonal and symmetric off-diagonal components,
\begin{align}
 \frac{\langle {\cal H}_0 \rangle}{xH} = \frac{1}{xH} \frac{1}{z_\nu} \left(\frac{x}{T_\mathrm{ref}} \right) \frac{\pi^2}{36 \zeta(3)} UM^2U^\dagger = \frac{\langle {\cal H}_0^{\mathrm{even}} \rangle + \langle {\cal H}_0^{\mathrm{odd}} \rangle}{xH} \rightarrow (h^i_{0,\mathrm{even}} + h^i_{0,\mathrm{odd}}) \lambda_i\,,
\end{align}
where we explicitly decompose the vacuum Hamiltonian into CP-even and CP-odd parts, see Eq.~\eqref{eq:def_H0}. Introducing the dimensionless effective mass splitting
\begin{align}
    \Delta_{ij} \equiv \frac{1}{xH} \frac{1}{z_\nu}\left(\frac{x}{T_\mathrm{ref}} \right) \frac{\pi^2}{36 \zeta(3)} \Delta m_{ij}^2\,,
\end{align}
the CP-even components are:
\begin{align}
2 h^1_{0,\mathrm{even}} &= \Delta_{31} c_\delta s_{2 \theta_{13}} s_{\theta_{23}} + \Delta_{21} s_{2\theta_{12}} c_{\theta_{23}} (c_{\theta_{12}} - c_\delta s_{2\theta_{13}} t_{\theta_{12}} t_{\theta_{23}}/2)\,,\\
8 h^3_{0,\mathrm{even}} &= \Delta_{31} (2 c_{\theta_{23}}^2 + c_{2\theta_{12}} [c_{2\theta_{23}} - 3]) + \Delta_{21} \left( c_{\theta_{13}} s_{\theta_{12}}^2(3-c_{\theta_{23}}) - c_{\theta_{23}}^2 \left[ 1 + c_{\theta_{12}} (3-8 c_{\delta} s_{\theta_{13}} s_{\theta_{12}} t_{\theta_{23}} ) \right]
\right) \,,\\
2 h^4_{0,\mathrm{even}} &= \Delta_{31} c_\delta s_{2\theta_{13}} c_{\theta_{23}} - \Delta_{21} c_{\theta_{13}} c_{\theta_{23}} s_{2\theta_{12}} (c_\delta s_{\theta_{13}} t_{\theta_{12}} + t_{\theta_{23}}) \,,\\
2 h^6_{0,\mathrm{even}} &= \Delta_{31} c_{\theta_{13}}^2 s_{2 \theta_{23}} + \Delta_{21} \left( s_{2 \theta_{23}} [s_{\theta_{13}}^2 s_{\theta_{12}}^2 - c_{\theta_{13}}^2] - c_\delta c_{2\theta_{23}} s_{2\theta_{12}} s_{\theta_{13}} \right) \,,\\
\begin{split}
4 \sqrt{3} h^8_{0,\mathrm{even}} &= \Delta_{31} (2 s_{\theta_{13}}^2 - c_{\theta_{13}}^2 [1+3c_{2\theta_{13}}]) + \Delta_{21} \left( c_{\theta_{12}}^2 (3c_{2\theta_{23}} - 1) + \right. \\
&\left. \;\;\; s_{\theta_{12}}^2 [2 c_{\theta_{13}}^2 - s_{\theta_{13}}^2 \{3 c_{2\theta_{23}} + 1 \}] - 3 c_\delta s_{2\theta_{12}} s_{2\theta_{23}} s_{\theta_{13}} \right) \,,
\end{split}
\\
 h^2_{0,\mathrm{even}} &=  h^5_{0,\mathrm{even}} =  h^7_{0,\mathrm{even}} = 0 \,.
\end{align}
While the CP-odd components are:
\begin{align}
    2 h^2_{0,\mathrm{odd}} &= s_\delta s_{2 \theta_{13}} s_{\theta_{23}} (\Delta_{31} - \Delta_{21} s_{\theta_{12}}^2) \,,\\
    2h^5_{0,\mathrm{odd}} &= s_\delta s_{2 \theta_{13}} c_{\theta_{23}} (\Delta_{31} - \Delta_{21} s_{\theta_{12}}^2) \,,\\
    2h^7_{0,\mathrm{odd}} &= - s_\delta s_{\theta_{13}} s_{2\theta_{12}} \Delta_{21} \,,\\
h^1_{0,\mathrm{odd}} &=  h^3_{0,\mathrm{odd}} =  h^4_{0,\mathrm{odd}} = h^6_{0,\mathrm{odd}} =  h^8_{0,\mathrm{odd}} = 0 \,.
\end{align}

The expressions for the collision terms are in general lengthy, but can be obtained following the same procedure as outlined in Eq.~\eqref{eq:SUNdecompositionDef}. Note, however, that contrary to the Hamiltonian terms, the full collision matrix has a non-vanishing trace and thus there will also be also a non-zero coefficient $c^0\,(\bar{c}^0)$. When using the damping approximation the collision matrix is assumed to be fully non-diagonal in flavor space and hence the trace vanishes. That is one of the non-trivial differences between using the full collision term and the damping approximation.

\section{Frequency of synchronous oscillations and (non-)adiabatic MSW transitions}
\label{app:slowness-factor}

In this appendix we extend the discussion of the frequency of synchronous oscillations, key to distinguishing adiabatic from abrupt MSW transitions, from the two flavor case discussed in Ref.~\cite{Froustey:2021azz} to three (or more) neutrino flavors. For this purpose we will neglect the collision term and the difference in the neutrino and photon temperatures. Defining the asymmetry matrix $\Delta r$ and the total (anti-) neutrino density matrix $r_+$, respectively
\begin{align}
 \Delta r \equiv r - \bar{r}\,, \quad r_+ \equiv r + \bar{r}\,,
\end{align}
we use the momentum averaged kinetic equation~\eqref{eq:kinetic_eq_flavor} to obtain the evolution equation of the asymmetry
\begin{align}
\label{app:eq:Deltar_equations}
 \frac{ \mathrm{d} \Delta r}{\mathrm{d}x} = \frac{-i}{x H} \left[ \langle {\cal H}_0 \rangle + \langle V_c \rangle, r_+ \right] \,, \quad \frac{ \mathrm{d} r_+}{\mathrm{d} x} = -i \frac{1}{xH} [ \langle \mathcal{H}_0 \rangle + \langle V_c\rangle, \Delta r] - i v_s [\Delta r, r_+]\,,
\end{align}
with $v_s$ an $x$-dependent scalar defined in Eq.~\eqref{eq:vs} parametrizing the neutrino self-interaction potential. This represents an alternative way of expressing the kinetic equation~\eqref{eq:master} of the main text. To identify (non-)adiabatic transitions in the evolution equations for $\Delta r$, we would like to express the right-hand side in terms of its value at time $x$, i.e.\
\begin{align}
\label{app:eq:fslow_ansatz}
    \frac{\mathrm{d} \Delta r}{ \mathrm{d} x} =  \frac{-i}{xH} \mathcal{F}_{\mathrm{slow}} \left[ \langle {\cal H}_0 \rangle + \langle V_c \rangle, \Delta r \right] \,,
\end{align}
Comparing Eqs.~\eqref{app:eq:Deltar_equations} and \eqref{app:eq:fslow_ansatz} and using the $SU(N)$ decomposition described in Sec.~\ref{subsec:num_strategy} as well as App.~\ref{app:SUN-decomposition}, we can write
\begin{align}
 {\cal F}_\mathrm{slow} = \left[ \langle {\cal H}_0 \rangle + \langle V_c \rangle, r_+ \right] \left(\left[ \langle {\cal H}_0 \rangle + \langle V_c \rangle, \Delta r \right]\right)^{-1}
 = \left(f_{ijk} (h_0^j + v_c^j) r_+^k \lambda^i \right) \left(f_{ijk} (h_0^j + v_c^j) \Delta r^k \lambda^i \right)^{-1} \,.
 \label{app:eq:Fslow}
\end{align}
As we will discuss below, in the situations of interest in the SM this general expression can be greatly simplified. Due to the flavor patterns in ${\cal H}_0$, $V_c$, $\Delta r$ and $r_+$, together with the hierarchies in the entries of ${\cal H}_0$ and $V_c$, we can identify which component of the $SU(N)$ decomposition (i.e.\ which Gell-Mann matrix $\lambda_i$) dominates the expressions in the angular brackets in Eq.~\eqref{app:eq:Fslow}. If this is the same Gell-Mann matrix in both angular brackets, the `slowness factor' $\cal{F}_\mathrm{slow}$ is simply a scalar (multiplied by a unit matrix), accounting for the frequency of the synchronous neutrino oscillations relative the naive frequency set by $\langle {\cal H}_0 \rangle + \langle V_c \rangle$.

Due to the temperature dependence of the eigenvalues of ${\cal H}_0$ and $V_c$ (leading to MSW transitions), as well as the evolving flavor structure in $r$ and $\bar r$, different subspaces, and thus different slowness factors, are relevant at different times. A non-adiabatic MSW transition is encountered when there is a level crossing in the eigenvalues of ${\cal H}_0$ and $V_c$ (MSW transitions, see Fig.~\ref{fig:rates}) while simultaneously ${\cal F}_{\mathrm{slow}} \to 0$, indicating a particularly slow frequency for the synchronous neutrino oscillations. In the following we will apply this framework to identify the conditions which lead to non-adiabatic muon- and electron-driven MSW transitions.

\subsection{Non-adiabatic muon-driven MSW transition}
\label{app:muMSW}

The first level crossing of the Hamiltonian eigenvalues, see Fig.~\ref{fig:rates}, happens at around $12\,\mathrm{MeV}$ and is referred to as the muon-driven MSW transition.
The relevant timescales are thus set by $\Delta m_{31}^2$ and the muon component of the charged loop currents $V_c$.
The initial density matrices $r,\bar{r}$ are diagonal in flavor space, see Eq.~\eqref{app:eq:Deltar_equations}, and so is $V_c$.
This means that their commutator vanishes in the kinetic equation~\eqref{subsec:num_strategy} and neutrino oscillations are triggered only by the off-diagonal components of the vacuum Hamiltonian. The vacuum Hamiltonian ${\cal H}_0$ is a hermitian matrix whose off-diagonal components can be expressed in terms of the three off-diagonal symmetric Gell-Mann matrices in the three $\alpha\beta$-subspaces. Inserting this into Eq.~\eqref{app:eq:Fslow}, we obtain for the slowness factors in respective $\alpha\beta$ subspaces
\begin{align}
\label{eq:fslow_mu}
    {\cal F}_{\mathrm{slow},\alpha\beta} =  \frac{ r^{\mathrm{ini}}_{\alpha \alpha} + \bar r^{\mathrm{ini}}_{\alpha \alpha} - (r^{\mathrm{ini}}_{\beta \beta} + \bar r^{\mathrm{ini}}_{\beta \beta})}{r^{\mathrm{ini}}_{\alpha \alpha} - \bar r^{\mathrm{ini}}_{\alpha \alpha} - (r^{\mathrm{ini}}_{\beta \beta} - \bar r^{\mathrm{ini}}_{\beta \beta})} \,,
\end{align}
which, upon expansion in $\xi_\alpha^\mathrm{ini}$ reads
\begin{align}
  {\cal F}_{\mathrm{slow},\alpha\beta} \simeq 6 \log 2 \frac{\xi_\alpha^\mathrm{ini} + \xi_\beta^\mathrm{ini}}{\pi^2 + (\xi_\alpha^\mathrm{ini})^2 +(\xi_\beta^\mathrm{ini})^2 + \xi_\alpha^\mathrm{ini} \xi_\beta^\mathrm{ini}}\,.
\end{align}
very similar to the result derived in~\cite{Froustey:2021azz} for the case of two neutrino flavors, with the difference in the numerical prefactor only arising due to the ansatz of momentum-averaging. Notably, the slowness factor becomes very small for small values of $\xi^\mathrm{ini}$ or for $\xi_\alpha^\mathrm{ini} = - \xi_\beta^\mathrm{ini}$, resulting in non-adiabatic MSW transitions where the frequency of the synchronous oscillations are no longer sufficient to adiabatically track the changes in the Hamiltonian. In the SM, the mass hierarchy and mixing angles (see App.~\ref{app:conventions}) dictate that the $\mu-\tau$ oscillations are the first to set in, and hence we expect a non-adiabatic muon MSW transition for
\begin{align}
    \xi_\mu^{\mathrm{ini}} = -\xi_\tau^{\mathrm{ini}} \quad (\mathrm{non}\mathrm{-}\mathrm{adiabatic}\,\mu\,\mathrm{MSW})\,,
\end{align}
independent of the total lepton number. This is nicely seen in Fig.~\ref{fig:delayed_muon_MSW}.

\subsection{Non-adiabatic electron-driven MSW transition}
\label{app:eMSW}

The muon-driven MSW transition discussed above is only the first level crossing of the eigenvalues of the Hamiltonian, see Fig.~\ref{fig:rates}. Provided that the muon-driven MSW transition in fact proceeded adiabatically, we would now like to understand under which circumstances the electron-driven MSW transition at $T\sim 5\,\mathrm{MeV}$ will be non-adiabatic. The general formula~\eqref{app:eq:Fslow} to estimate non-adiabatic transitions still holds. But to make sense of it we need to know the concrete form of $r,\bar{r}$ after the adiabatic muon MSW transition, which will be our task in the following. For simplicity the calculations we will present here assumes $\Delta m_{21}^2 = 0$ (as this mass splitting is subdominant), $\theta_{12} = 0$ and $\theta_{23} = \pi/4$ (which is experimentally allowed).  A generalization however is straightforward.

We start by considering temperatures above the electron-driven MSW transition, which implies $\langle V_{c}^e \rangle \gg \langle \mathcal{H}_0 \rangle, \langle V_c^\mu \rangle$. Given the SM neutrino oscillation parameters, $\theta_{13} \ll \theta_{23} (= \pi/4)$, we can effectively consider only $\mu-\tau$ oscillations in the beginning and the sub-system is described by
\begin{align}\label{eq:thermalpotential2x2}
    \mathcal{H}_0 = \frac{\Delta m_{31}^2}{2k} 
    \begin{pmatrix}
    1 &1\\
    1   &  1
    \end{pmatrix} \,,\quad   
    V_c =   
    \begin{pmatrix}
    V_\mu  & 0\\
    0   &  0
    \end{pmatrix}\,.
\end{align}
Initially, in the $\mu-\tau$ sector $\Delta r$ and $r_+$ are diagonal and are thus both proportional to the  $SU(2)$ basis matrix $\sigma_3$, where $\sigma_i$ represent the Pauli matrices (plus a component proportional to the unit matrix which drops out in the commutator of the evolution equation). The same is true, at leading order, for $\mathcal{H}_0 + V_c$ before the $\mu$ MSW transition, because $\mathcal{H}_0 \ll V_c$. However, after an adiabatic muon-driven MSW transition, $\mathcal{H}_0 \gg V_c$ and the leading order term in the $SU(2)$ decomposition is given by $\sigma_1$. Due to the adiabatic transition, it is expected that $\Delta r$ and $r_+$ will stay aligned with $\mathcal{H}_0 + V_c$ after the $\mu$ MSW transition. This implies that the density matrices undergo a rotation from $\sigma_3 \mapsto \pm \sigma_1$, which is simply a $\mp 45^\circ$ rotation, and the initially diagonal density matrices~\eqref{eq:ini_flavor_diag} become
\begin{align}
\label{eq:r_rbar_muMSW}
    r &= \begin{pmatrix} 
   r^{\mathrm{ini}}_e & 0 & 0 \\ 
    0 &   \frac{1}{2}(r^{\mathrm{ini}}_\mu + r^{\mathrm{ini}}_\tau)  & \pm \frac{1}{2}(r^{\mathrm{ini}}_\tau - r^{\mathrm{ini}}_\mu)  \\
    0 &   \pm \frac{1}{2}(r^{\mathrm{ini}}_\tau - r^{\mathrm{ini}}_\mu)  &  \frac{1}{2}(r^{\mathrm{ini}}_\mu + r^{\mathrm{ini}}_\tau)
    \end{pmatrix}\,,  \quad
    \bar{r} = \begin{pmatrix} 
   \bar{r}^{\mathrm{ini}}_e & 0 & 0 \\ 
    0 &  \frac{1}{2}(\bar{r}^{\mathrm{ini}}_\mu +\bar{r}^{\mathrm{ini}}_\tau) & \pm \frac{1}{2}(\bar{r}^{\mathrm{ini}}_\tau -\bar{r}^{\mathrm{ini}}_\mu) \\
     0 &  \pm \frac{1}{2}(\bar{r}^{\mathrm{ini}}_\tau -\bar{r}^{\mathrm{ini}}_\mu) &\frac{1}{2}(\bar{r}^{\mathrm{ini}}_\mu +\bar{r}^{\mathrm{ini}}_\tau)
    \end{pmatrix}  \,\quad \mathrm{at} \,\, 10\,{\rm MeV} \gtrsim T \gtrsim 5 \,{\rm MeV}\,,
\end{align}
where $+(-)$ applies for NH (IH) due to the different sign of $\Delta m_{31}^2$ in both hierarchies. These are the density matrices after the muon-driven MSW transition and are the initial conditions to calculate the slowness factor for the subsequent electron-driven MSW transition. We do so by using these matrices to calculate the slowness factor via Eq.~\eqref{app:eq:Fslow}, which in particular amounts to evaluating the commutator $   [ \langle \mathcal{H}_0 \rangle + \langle V_{\mathrm{c}} \rangle,r+\bar{r}]$. Noting that at this point the $V_\mu$ potential is vanishingly small and given our approximations of the mixing angles and mass splittings, the vacuum Hamiltonian and the potentials are
\begin{align}
 \mathcal{H}_0 =   \frac{\Delta m_{31}^2}{2k}\left(
\begin{array}{ccc}
  s_{\theta_{13}}^2 &  \frac{1}{2} s_{\theta_{13}} c_{\theta_{13}} & \frac{1}{2} s_{\theta_{13}} c_{\theta_{13}} \\
 \frac{1}{2} s_{\theta_{13}} c_{\theta_{13}} & \frac{1}{2} c_{\theta_{13}}^2 & \frac{1}{2} c_{\theta_{13}} c_{\theta_{13}}^2 \\
 \frac{1}{2} s_{\theta_{13}} c_{\theta_{13}} & \frac{1}{2} c_{\theta_{13}}^2 & \frac{1}{2} c_{\theta_{13}}^2 \\
\end{array}
\right)\,, \quad V_c = {\rm diag}(V_e,0,0) \,,
\end{align}
and calculation of the commutator yields,
\begin{align}\label{eq:commutatorcase}
    [ \langle \mathcal{H}_0 \rangle + \langle V_{\mathrm{c}} \rangle,r+\bar{r}] & \propto \Delta m_{31}^2 s_{2\theta_{13}} (r^{\mathrm{ini}}_e+\bar{r}^{\mathrm{ini}}_e-r^{\mathrm{ini}}_\tau-\bar{r}^{\mathrm{ini}}_\tau)    \left(
\begin{array}{ccc}
 0 & -1 & -1\\
 1& 0 &  0\\
 1& 0 &  0\\
\end{array}
\right)  \qquad \mathrm{(NH)} \,, \\
    [ \langle \mathcal{H}_0 \rangle + \langle V_{\mathrm{c}} \rangle,r+\bar{r}] & \propto \Delta m_{31}^2 s_{2\theta_{13}} (r^{\mathrm{ini}}_e+\bar{r}^{\mathrm{ini}}_e-r^{\mathrm{ini}}_\mu-r^{\mathrm{ini}}_\mu)    \left(
\begin{array}{ccc}
 0 &  -1 & -1\\
 1& 0 &  0\\
 1& 0 &  0\\
\end{array}
\right) \qquad \mathrm{(IH)}  \,.
\end{align}
Using Eq.~\eqref{eq:ini_flavor_diag} for the initial condition for $r,\bar{r}$, the commutator vanishes for $\xi^{\mathrm{ini}}_e = \pm \xi^{\mathrm{ini}}_\tau$ for NH and $\xi^{\mathrm{ini}}_e = \pm \xi^{\mathrm{ini}}_\mu$ for IH.
Determining the slowness factor via Eq.~\eqref{app:eq:Fslow}, we obtain that the $+$ solution remains finite, while a vanishing slowness factor, and thus a non-adiabatic electron driven MSW transition, only occurs if 
\begin{align}
\label{app:eq:fslow_e}
\begin{split}
    \xi_e^{\mathrm{ini}} &= -\xi_\tau^{\mathrm{ini}} \quad (\mathrm{NH}\;\mathrm{non}\mathrm{-}\mathrm{adiabatic}\,e\,\mathrm{MSW})\,, \\
    \xi_e^{\mathrm{ini}} &= -\xi_\mu^{\mathrm{ini}} \quad (\mathrm{IH}\;\mathrm{non}\mathrm{-}\mathrm{adiabatic}\,e\,\mathrm{MSW})\,.
\end{split}
\end{align}
These findings have been numerically confirmed in e.g. Fig.~\ref{fig:delayed_electron_MSW}. We further note that Eqs.~\eqref{eq:fslow_mu},\eqref{app:eq:fslow_e} also hold for non-vanishing total lepton number.

\subsection{Phenomenology of non-adiabatic MSW transitions}

\begin{figure}[!h]
\centering
\begin{tabular}{cc}
\hspace{-0.cm}\includegraphics[width=0.5\textwidth]{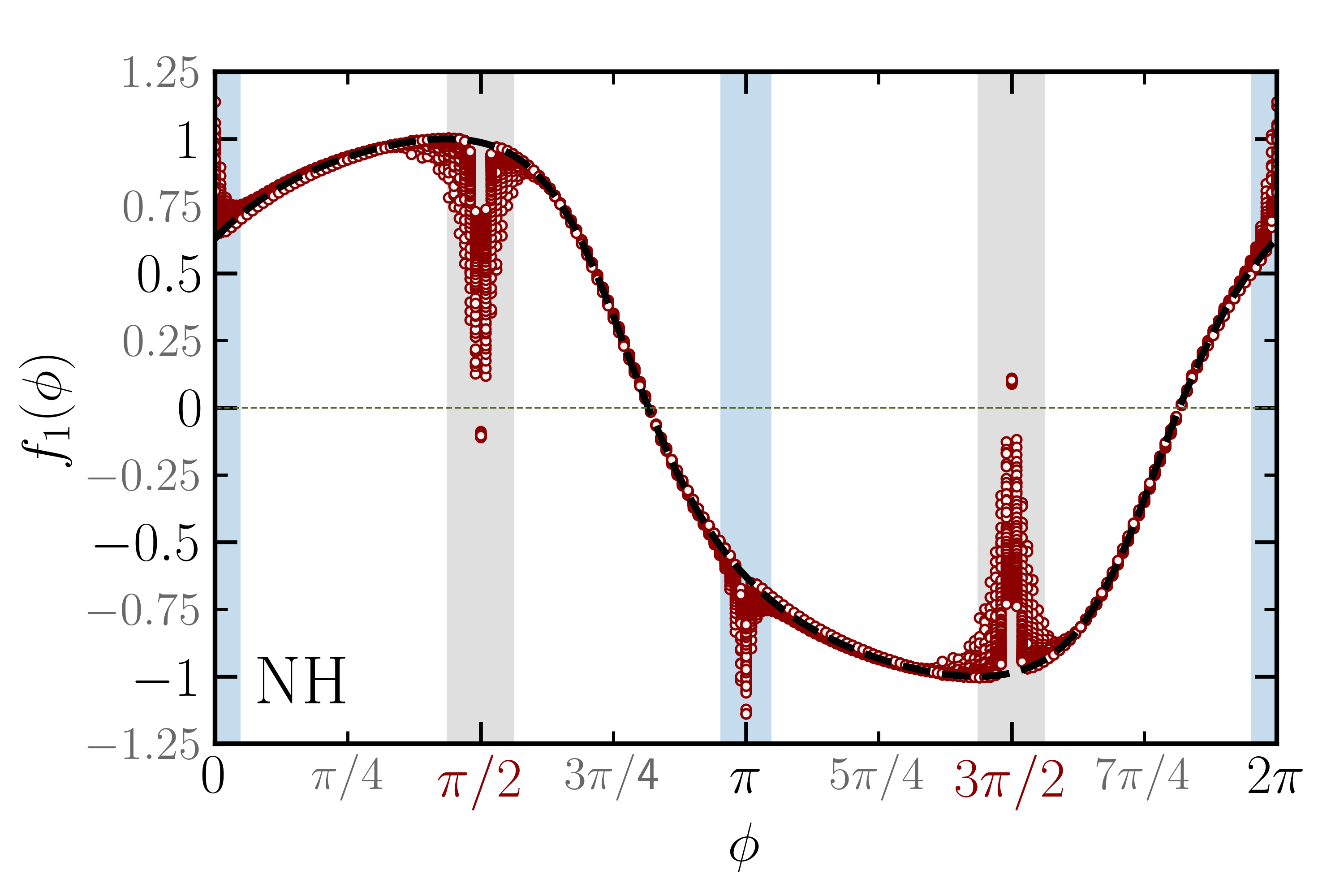}
\hspace{-0.cm}\includegraphics[width=0.5\textwidth]{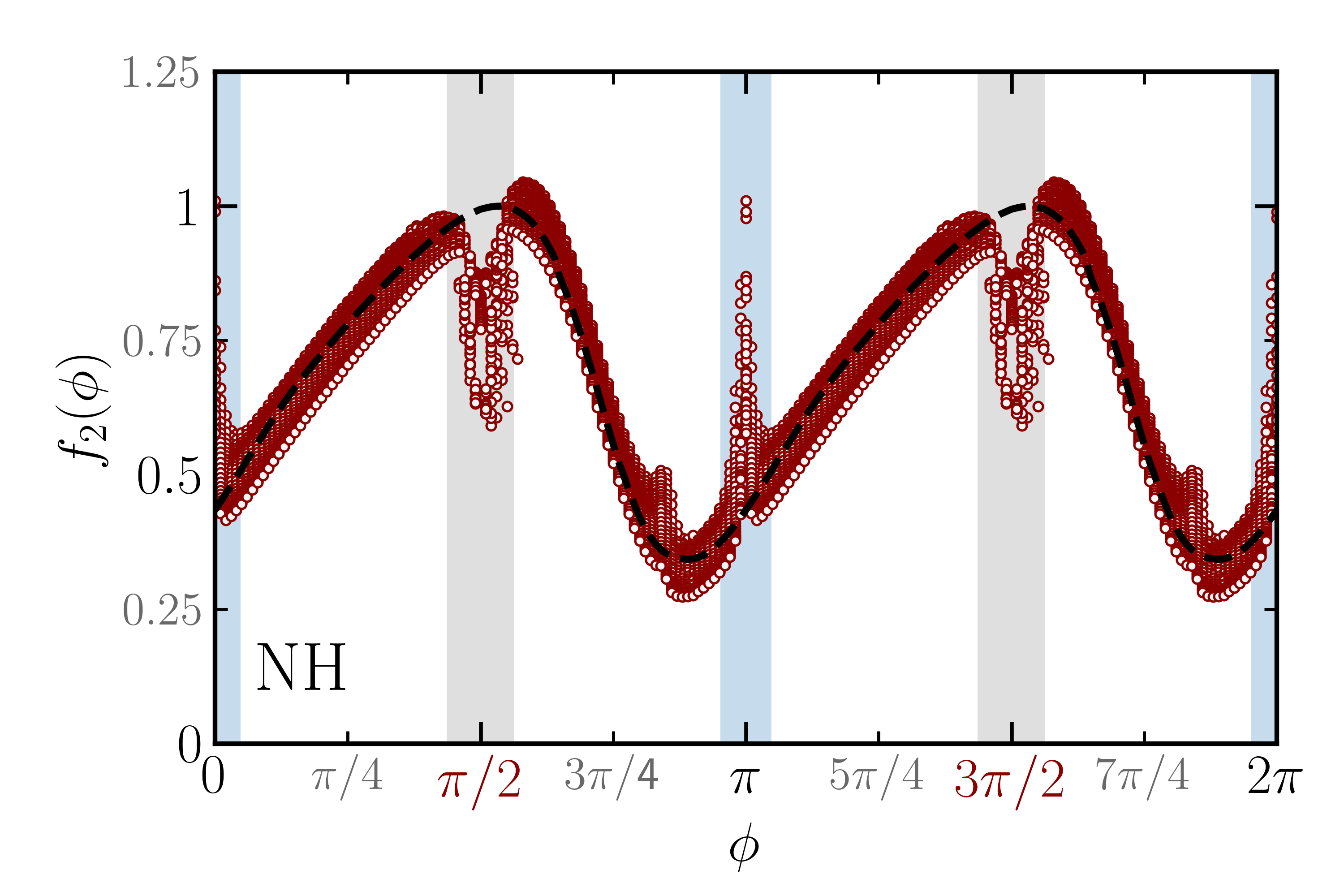} \\
\hspace{-0.cm}\includegraphics[width=0.5\textwidth]{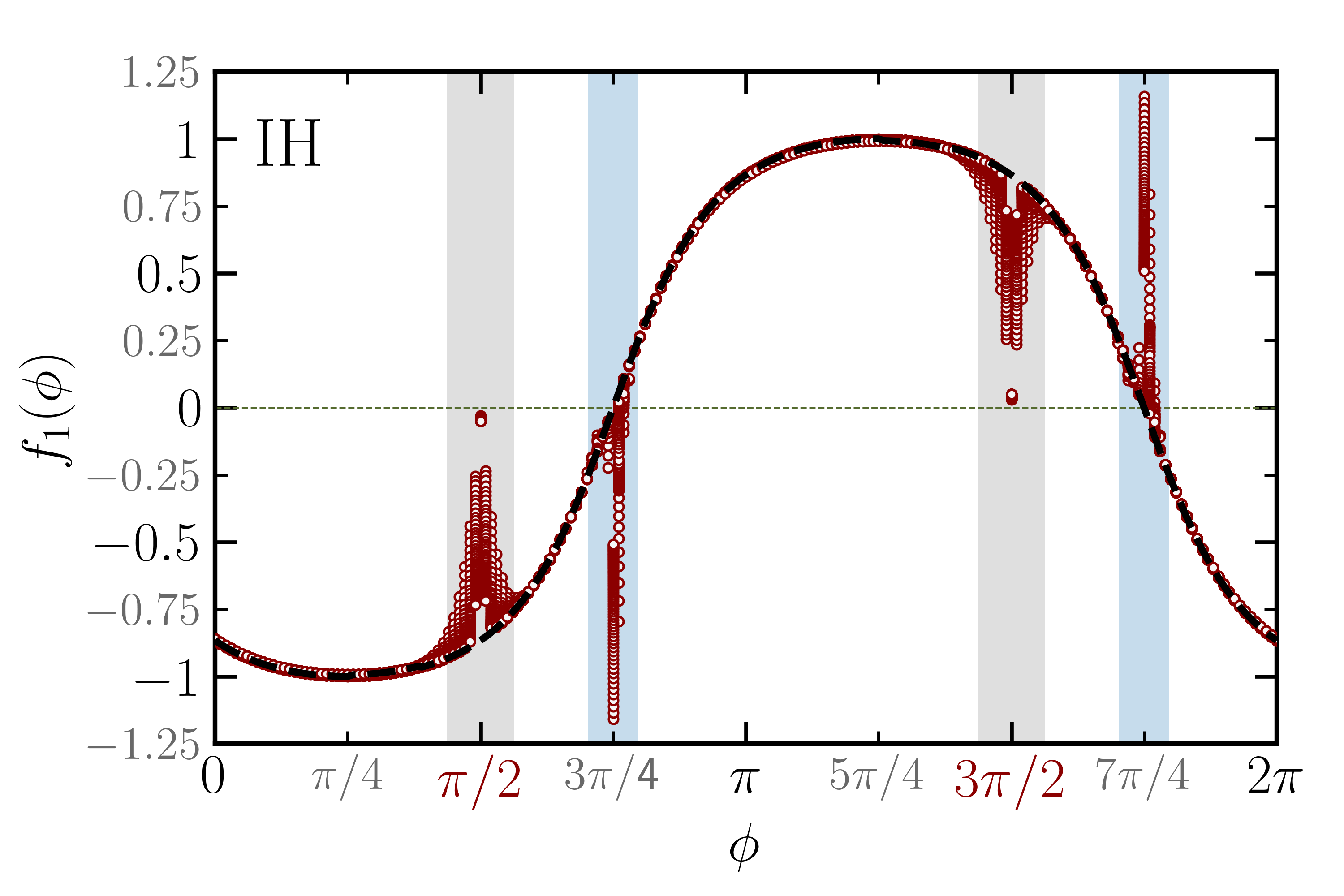}
\hspace{-0.cm}\includegraphics[width=0.5\textwidth]{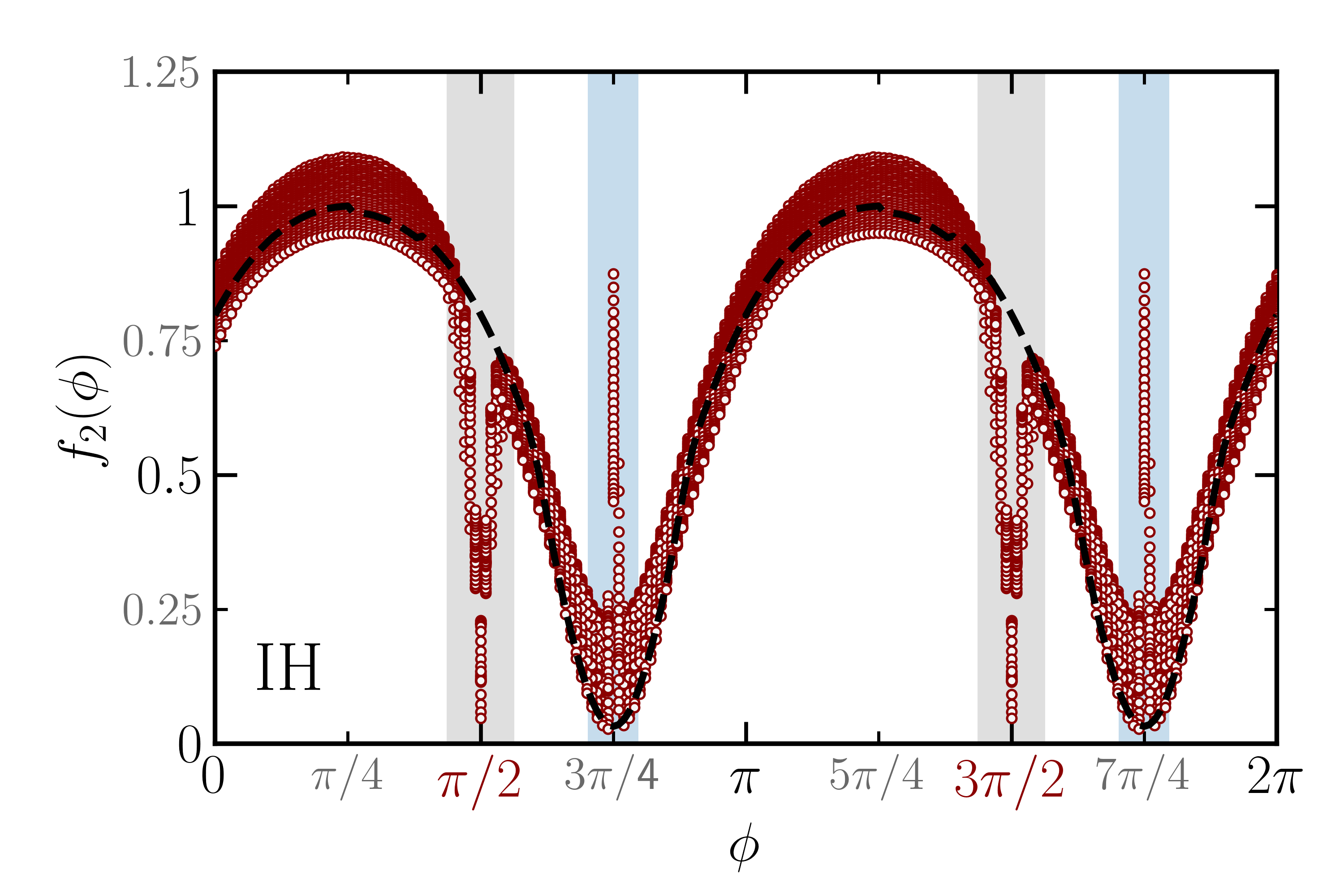} \\
\end{tabular}
\caption{
Functions $f_{1,2}(\phi)$ in NH and IH defined in Eq.~\eqref{eq:minimal_washout}, parametrizing flavor dependence of the asymmetry washout for vanishing total lepton number. The dashed black line corresponds to the adiabatic approximation, the red points to the result of our numerical evaluation of the QKEs for $A > 0.25$ (and $A<1$ for the right column). The vertical gray bands indicate the region of non-adiabatic muon-driven MSW transitions at $\xi_e^\mathrm{ini} = 0$ ($\phi = \{\pi/2, 3 \pi/2 \}$, the vertical blue bands indicate the region of non-adiabatic electron-driven MSW transitions at $\xi_\mu^\mathrm{ini} = 0$ ($\phi = \{0, \pi \}$) for NH and $\xi_\tau^\mathrm{ini} = 0$ ($ \phi = \{3 \pi/4 , 7 \pi/4 \} $) for IH.
 }
\label{fig:f12}
\end{figure}

As discussed in the main text, the non-adiabatic MSW transitions significantly impact the evolution of the lepton flavor asymmetries, leading to final asymmetries that can be either smaller or larger than the results obtained assuming an adiabatic evolution. This is visualized in Fig.~\ref{fig:f12} which shows the adiabatic as well as non-adiabatic solution for the functions $f_{1,2}(\phi)$ in NH and IH defined in Eq.~\eqref{eq:minimal_washout}.

\newpage
\bibliography{biblio}

\begin{thebibliography}{119}%
\makeatletter
\providecommand \@ifxundefined [1]{%
 \@ifx{#1\undefined}
}%
\providecommand \@ifnum [1]{%
 \ifnum #1\expandafter \@firstoftwo
 \else \expandafter \@secondoftwo
 \fi
}%
\providecommand \@ifx [1]{%
 \ifx #1\expandafter \@firstoftwo
 \else \expandafter \@secondoftwo
 \fi
}%
\providecommand \natexlab [1]{#1}%
\providecommand \enquote  [1]{``#1''}%
\providecommand \bibnamefont  [1]{#1}%
\providecommand \bibfnamefont [1]{#1}%
\providecommand \citenamefont [1]{#1}%
\providecommand \href@noop [0]{\@secondoftwo}%
\providecommand \href [0]{\begingroup \@sanitize@url \@href}%
\providecommand \@href[1]{\@@startlink{#1}\@@href}%
\providecommand \@@href[1]{\endgroup#1\@@endlink}%
\providecommand \@sanitize@url [0]{\catcode `\\12\catcode `\$12\catcode
  `\&12\catcode `\#12\catcode `\^12\catcode `\_12\catcode `\%12\relax}%
\providecommand \@@startlink[1]{}%
\providecommand \@@endlink[0]{}%
\providecommand \url  [0]{\begingroup\@sanitize@url \@url }%
\providecommand \@url [1]{\endgroup\@href {#1}{\urlprefix }}%
\providecommand \urlprefix  [0]{URL }%
\providecommand \Eprint [0]{\href }%
\providecommand \doibase [0]{https://doi.org/}%
\providecommand \selectlanguage [0]{\@gobble}%
\providecommand \bibinfo  [0]{\@secondoftwo}%
\providecommand \bibfield  [0]{\@secondoftwo}%
\providecommand \translation [1]{[#1]}%
\providecommand \BibitemOpen [0]{}%
\providecommand \bibitemStop [0]{}%
\providecommand \bibitemNoStop [0]{.\EOS\space}%
\providecommand \EOS [0]{\spacefactor3000\relax}%
\providecommand \BibitemShut  [1]{\csname bibitem#1\endcsname}%
\let\auto@bib@innerbib\@empty
%</preamble>
\bibitem [{\citenamefont {Aghanim}\ \emph {et~al.}(2020)\citenamefont {Aghanim}
  \emph {et~al.}}]{Planck:2018vyg}%
  \BibitemOpen
  \bibfield  {author} {\bibinfo {author} {\bibfnamefont {N.}~\bibnamefont
  {Aghanim}} \emph {et~al.} (\bibinfo {collaboration} {Planck}),\ }\bibfield
  {title} {\bibinfo {title} {{Planck 2018 results. VI. Cosmological
  parameters}},\ }\href {https://doi.org/10.1051/0004-6361/201833910}
  {\bibfield  {journal} {\bibinfo  {journal} {Astron. Astrophys.}\ }\textbf
  {\bibinfo {volume} {641}},\ \bibinfo {pages} {A6} (\bibinfo {year} {2020})},\
  \bibinfo {note} {[Erratum: Astron.Astrophys. 652, C4 (2021)]},\ \Eprint
  {https://arxiv.org/abs/1807.06209} {arXiv:1807.06209 [astro-ph.CO]}
  \BibitemShut {NoStop}%
\bibitem [{\citenamefont {Sakharov}(1967)}]{sakharov}%
  \BibitemOpen
  \bibfield  {author} {\bibinfo {author} {\bibfnamefont {A.~D.}\ \bibnamefont
  {Sakharov}},\ }\bibfield  {title} {\bibinfo {title} {{Violation of CP
  Invariance, c Asymmetry, and Baryon Asymmetry of the Universe}},\ }\href
  {https://doi.org/10.1070/PU1991v034n05ABEH002497} {\bibfield  {journal}
  {\bibinfo  {journal} {Pisma Zh. Eksp. Teor. Fiz.}\ }\textbf {\bibinfo
  {volume} {5}},\ \bibinfo {pages} {32} (\bibinfo {year} {1967})},\ \bibinfo
  {note} {[Usp. Fiz. Nauk161,61(1991)]}\BibitemShut {NoStop}%
%%CITATION = ZFPRA,5,32;%%
\bibitem [{\citenamefont {Fukugita}\ and\ \citenamefont
  {Yanagida}(1986)}]{Fukugita:1986hr}%
  \BibitemOpen
  \bibfield  {author} {\bibinfo {author} {\bibfnamefont {M.}~\bibnamefont
  {Fukugita}}\ and\ \bibinfo {author} {\bibfnamefont {T.}~\bibnamefont
  {Yanagida}},\ }\bibfield  {title} {\bibinfo {title} {{Baryogenesis Without
  Grand Unification}},\ }\href {https://doi.org/10.1016/0370-2693(86)91126-3}
  {\bibfield  {journal} {\bibinfo  {journal} {Phys. Lett. B}\ }\textbf
  {\bibinfo {volume} {174}},\ \bibinfo {pages} {45} (\bibinfo {year}
  {1986})}\BibitemShut {NoStop}%
\bibitem [{\citenamefont {Kuzmin}\ \emph {et~al.}(1987)\citenamefont {Kuzmin},
  \citenamefont {Rubakov},\ and\ \citenamefont {Shaposhnikov}}]{Kuzmin:1987wn}%
  \BibitemOpen
  \bibfield  {author} {\bibinfo {author} {\bibfnamefont {V.~A.}\ \bibnamefont
  {Kuzmin}}, \bibinfo {author} {\bibfnamefont {V.~A.}\ \bibnamefont
  {Rubakov}},\ and\ \bibinfo {author} {\bibfnamefont {M.~E.}\ \bibnamefont
  {Shaposhnikov}},\ }\bibfield  {title} {\bibinfo {title} {{Anomalous
  Electroweak Baryon Number Nonconservation and GUT Mechanism for
  Baryogenesis}},\ }\href {https://doi.org/10.1016/0370-2693(87)91340-2}
  {\bibfield  {journal} {\bibinfo  {journal} {Phys. Lett. B}\ }\textbf
  {\bibinfo {volume} {191}},\ \bibinfo {pages} {171} (\bibinfo {year}
  {1987})}\BibitemShut {NoStop}%
\bibitem [{\citenamefont {Cohen}\ and\ \citenamefont
  {Kaplan}(1987)}]{Cohen:1987vi}%
  \BibitemOpen
  \bibfield  {author} {\bibinfo {author} {\bibfnamefont {A.~G.}\ \bibnamefont
  {Cohen}}\ and\ \bibinfo {author} {\bibfnamefont {D.~B.}\ \bibnamefont
  {Kaplan}},\ }\bibfield  {title} {\bibinfo {title} {{Thermodynamic Generation
  of the Baryon Asymmetry}},\ }\href
  {https://doi.org/10.1016/0370-2693(87)91369-4} {\bibfield  {journal}
  {\bibinfo  {journal} {Phys. Lett. B}\ }\textbf {\bibinfo {volume} {199}},\
  \bibinfo {pages} {251} (\bibinfo {year} {1987})}\BibitemShut {NoStop}%
\bibitem [{\citenamefont {Cohen}\ and\ \citenamefont
  {Kaplan}(1988)}]{Cohen:1988kt}%
  \BibitemOpen
  \bibfield  {author} {\bibinfo {author} {\bibfnamefont {A.~G.}\ \bibnamefont
  {Cohen}}\ and\ \bibinfo {author} {\bibfnamefont {D.~B.}\ \bibnamefont
  {Kaplan}},\ }\bibfield  {title} {\bibinfo {title} {{SPONTANEOUS
  BARYOGENESIS}},\ }\href {https://doi.org/10.1016/0550-3213(88)90134-4}
  {\bibfield  {journal} {\bibinfo  {journal} {Nucl. Phys. B}\ }\textbf
  {\bibinfo {volume} {308}},\ \bibinfo {pages} {913} (\bibinfo {year}
  {1988})}\BibitemShut {NoStop}%
\bibitem [{\citenamefont {Davidson}\ \emph {et~al.}(1994)\citenamefont
  {Davidson}, \citenamefont {Kainulainen},\ and\ \citenamefont
  {Olive}}]{Davidson:1994gn}%
  \BibitemOpen
  \bibfield  {author} {\bibinfo {author} {\bibfnamefont {S.}~\bibnamefont
  {Davidson}}, \bibinfo {author} {\bibfnamefont {K.}~\bibnamefont
  {Kainulainen}},\ and\ \bibinfo {author} {\bibfnamefont {K.~A.}\ \bibnamefont
  {Olive}},\ }\bibfield  {title} {\bibinfo {title} {{Protecting the Baryon
  asymmetry with thermal masses}},\ }\href
  {https://doi.org/10.1016/0370-2693(94)90361-1} {\bibfield  {journal}
  {\bibinfo  {journal} {Phys. Lett. B}\ }\textbf {\bibinfo {volume} {335}},\
  \bibinfo {pages} {339} (\bibinfo {year} {1994})},\ \Eprint
  {https://arxiv.org/abs/hep-ph/9405215} {arXiv:hep-ph/9405215} \BibitemShut
  {NoStop}%
\bibitem [{\citenamefont {March-Russell}\ \emph {et~al.}(1999)\citenamefont
  {March-Russell}, \citenamefont {Murayama},\ and\ \citenamefont
  {Riotto}}]{March-Russell:1999hpw}%
  \BibitemOpen
  \bibfield  {author} {\bibinfo {author} {\bibfnamefont {J.}~\bibnamefont
  {March-Russell}}, \bibinfo {author} {\bibfnamefont {H.}~\bibnamefont
  {Murayama}},\ and\ \bibinfo {author} {\bibfnamefont {A.}~\bibnamefont
  {Riotto}},\ }\bibfield  {title} {\bibinfo {title} {{The Small observed baryon
  asymmetry from a large lepton asymmetry}},\ }\href
  {https://doi.org/10.1088/1126-6708/1999/11/015} {\bibfield  {journal}
  {\bibinfo  {journal} {JHEP}\ }\textbf {\bibinfo {volume} {11}},\ \bibinfo
  {pages} {015}},\ \Eprint {https://arxiv.org/abs/hep-ph/9908396}
  {arXiv:hep-ph/9908396} \BibitemShut {NoStop}%
\bibitem [{\citenamefont {Yamaguchi}(2003)}]{Yamaguchi:2002vw}%
  \BibitemOpen
  \bibfield  {author} {\bibinfo {author} {\bibfnamefont {M.}~\bibnamefont
  {Yamaguchi}},\ }\bibfield  {title} {\bibinfo {title} {{Generation of
  cosmological large lepton asymmetry from a rolling scalar field}},\ }\href
  {https://doi.org/10.1103/PhysRevD.68.063507} {\bibfield  {journal} {\bibinfo
  {journal} {Phys. Rev. D}\ }\textbf {\bibinfo {volume} {68}},\ \bibinfo
  {pages} {063507} (\bibinfo {year} {2003})},\ \Eprint
  {https://arxiv.org/abs/hep-ph/0211163} {arXiv:hep-ph/0211163} \BibitemShut
  {NoStop}%
\bibitem [{\citenamefont {Kawasaki}\ \emph {et~al.}(2002)\citenamefont
  {Kawasaki}, \citenamefont {Takahashi},\ and\ \citenamefont
  {Yamaguchi}}]{Kawasaki:2002hq}%
  \BibitemOpen
  \bibfield  {author} {\bibinfo {author} {\bibfnamefont {M.}~\bibnamefont
  {Kawasaki}}, \bibinfo {author} {\bibfnamefont {F.}~\bibnamefont
  {Takahashi}},\ and\ \bibinfo {author} {\bibfnamefont {M.}~\bibnamefont
  {Yamaguchi}},\ }\bibfield  {title} {\bibinfo {title} {{Large lepton asymmetry
  from Q balls}},\ }\href {https://doi.org/10.1103/PhysRevD.66.043516}
  {\bibfield  {journal} {\bibinfo  {journal} {Phys. Rev. D}\ }\textbf {\bibinfo
  {volume} {66}},\ \bibinfo {pages} {043516} (\bibinfo {year} {2002})},\
  \Eprint {https://arxiv.org/abs/hep-ph/0205101} {arXiv:hep-ph/0205101}
  \BibitemShut {NoStop}%
\bibitem [{\citenamefont {Chiba}\ \emph {et~al.}(2004)\citenamefont {Chiba},
  \citenamefont {Takahashi},\ and\ \citenamefont {Yamaguchi}}]{Chiba:2003vp}%
  \BibitemOpen
  \bibfield  {author} {\bibinfo {author} {\bibfnamefont {T.}~\bibnamefont
  {Chiba}}, \bibinfo {author} {\bibfnamefont {F.}~\bibnamefont {Takahashi}},\
  and\ \bibinfo {author} {\bibfnamefont {M.}~\bibnamefont {Yamaguchi}},\
  }\bibfield  {title} {\bibinfo {title} {{Baryogenesis in a flat direction with
  neither baryon nor lepton charge}},\ }\href
  {https://doi.org/10.1103/PhysRevLett.92.011301} {\bibfield  {journal}
  {\bibinfo  {journal} {Phys. Rev. Lett.}\ }\textbf {\bibinfo {volume} {92}},\
  \bibinfo {pages} {011301} (\bibinfo {year} {2004})},\ \bibinfo {note}
  {[Erratum: Phys.Rev.Lett. 114, 209901 (2015)]},\ \Eprint
  {https://arxiv.org/abs/hep-ph/0304102} {arXiv:hep-ph/0304102} \BibitemShut
  {NoStop}%
\bibitem [{\citenamefont {Takahashi}\ and\ \citenamefont
  {Yamaguchi}(2004)}]{Takahashi:2003db}%
  \BibitemOpen
  \bibfield  {author} {\bibinfo {author} {\bibfnamefont {F.}~\bibnamefont
  {Takahashi}}\ and\ \bibinfo {author} {\bibfnamefont {M.}~\bibnamefont
  {Yamaguchi}},\ }\bibfield  {title} {\bibinfo {title} {{Spontaneous
  baryogenesis in flat directions}},\ }\href
  {https://doi.org/10.1103/PhysRevD.69.083506} {\bibfield  {journal} {\bibinfo
  {journal} {Phys. Rev. D}\ }\textbf {\bibinfo {volume} {69}},\ \bibinfo
  {pages} {083506} (\bibinfo {year} {2004})},\ \Eprint
  {https://arxiv.org/abs/hep-ph/0308173} {arXiv:hep-ph/0308173} \BibitemShut
  {NoStop}%
\bibitem [{\citenamefont {Asaka}\ and\ \citenamefont
  {Shaposhnikov}(2005)}]{Asaka:2005pn}%
  \BibitemOpen
  \bibfield  {author} {\bibinfo {author} {\bibfnamefont {T.}~\bibnamefont
  {Asaka}}\ and\ \bibinfo {author} {\bibfnamefont {M.}~\bibnamefont
  {Shaposhnikov}},\ }\bibfield  {title} {\bibinfo {title} {{The $\nu$MSM, dark
  matter and baryon asymmetry of the universe}},\ }\href
  {https://doi.org/10.1016/j.physletb.2005.06.020} {\bibfield  {journal}
  {\bibinfo  {journal} {Phys. Lett. B}\ }\textbf {\bibinfo {volume} {620}},\
  \bibinfo {pages} {17} (\bibinfo {year} {2005})},\ \Eprint
  {https://arxiv.org/abs/hep-ph/0505013} {arXiv:hep-ph/0505013} \BibitemShut
  {NoStop}%
\bibitem [{\citenamefont {Shaposhnikov}(2008)}]{Shaposhnikov:2008pf}%
  \BibitemOpen
  \bibfield  {author} {\bibinfo {author} {\bibfnamefont {M.}~\bibnamefont
  {Shaposhnikov}},\ }\bibfield  {title} {\bibinfo {title} {{The nuMSM, leptonic
  asymmetries, and properties of singlet fermions}},\ }\href
  {https://doi.org/10.1088/1126-6708/2008/08/008} {\bibfield  {journal}
  {\bibinfo  {journal} {JHEP}\ }\textbf {\bibinfo {volume} {08}},\ \bibinfo
  {pages} {008}},\ \Eprint {https://arxiv.org/abs/0804.4542} {arXiv:0804.4542
  [hep-ph]} \BibitemShut {NoStop}%
\bibitem [{\citenamefont {Laine}\ and\ \citenamefont
  {Shaposhnikov}(2008)}]{Laine:2008pg}%
  \BibitemOpen
  \bibfield  {author} {\bibinfo {author} {\bibfnamefont {M.}~\bibnamefont
  {Laine}}\ and\ \bibinfo {author} {\bibfnamefont {M.}~\bibnamefont
  {Shaposhnikov}},\ }\bibfield  {title} {\bibinfo {title} {{Sterile neutrino
  dark matter as a consequence of nuMSM-induced lepton asymmetry}},\ }\href
  {https://doi.org/10.1088/1475-7516/2008/06/031} {\bibfield  {journal}
  {\bibinfo  {journal} {JCAP}\ }\textbf {\bibinfo {volume} {06}},\ \bibinfo
  {pages} {031}},\ \Eprint {https://arxiv.org/abs/0804.4543} {arXiv:0804.4543
  [hep-ph]} \BibitemShut {NoStop}%
\bibitem [{\citenamefont {Kamada}(2018)}]{Kamada:2018tcs}%
  \BibitemOpen
  \bibfield  {author} {\bibinfo {author} {\bibfnamefont {K.}~\bibnamefont
  {Kamada}},\ }\bibfield  {title} {\bibinfo {title} {{Return of grand unified
  theory baryogenesis: Source of helical hypermagnetic fields for the baryon
  asymmetry of the universe}},\ }\href
  {https://doi.org/10.1103/PhysRevD.97.103506} {\bibfield  {journal} {\bibinfo
  {journal} {Phys. Rev. D}\ }\textbf {\bibinfo {volume} {97}},\ \bibinfo
  {pages} {103506} (\bibinfo {year} {2018})},\ \Eprint
  {https://arxiv.org/abs/1802.03055} {arXiv:1802.03055 [hep-ph]} \BibitemShut
  {NoStop}%
\bibitem [{\citenamefont {Domcke}\ \emph {et~al.}(2019)\citenamefont {Domcke},
  \citenamefont {von Harling}, \citenamefont {Morgante},\ and\ \citenamefont
  {Mukaida}}]{Domcke:2019mnd}%
  \BibitemOpen
  \bibfield  {author} {\bibinfo {author} {\bibfnamefont {V.}~\bibnamefont
  {Domcke}}, \bibinfo {author} {\bibfnamefont {B.}~\bibnamefont {von Harling}},
  \bibinfo {author} {\bibfnamefont {E.}~\bibnamefont {Morgante}},\ and\
  \bibinfo {author} {\bibfnamefont {K.}~\bibnamefont {Mukaida}},\ }\bibfield
  {title} {\bibinfo {title} {{Baryogenesis from axion inflation}},\ }\href
  {https://doi.org/10.1088/1475-7516/2019/10/032} {\bibfield  {journal}
  {\bibinfo  {journal} {JCAP}\ }\textbf {\bibinfo {volume} {10}},\ \bibinfo
  {pages} {032}},\ \Eprint {https://arxiv.org/abs/1905.13318} {arXiv:1905.13318
  [hep-ph]} \BibitemShut {NoStop}%
\bibitem [{\citenamefont {Domcke}\ \emph {et~al.}(2021)\citenamefont {Domcke},
  \citenamefont {Kamada}, \citenamefont {Mukaida}, \citenamefont {Schmitz},\
  and\ \citenamefont {Yamada}}]{Domcke:2020quw}%
  \BibitemOpen
  \bibfield  {author} {\bibinfo {author} {\bibfnamefont {V.}~\bibnamefont
  {Domcke}}, \bibinfo {author} {\bibfnamefont {K.}~\bibnamefont {Kamada}},
  \bibinfo {author} {\bibfnamefont {K.}~\bibnamefont {Mukaida}}, \bibinfo
  {author} {\bibfnamefont {K.}~\bibnamefont {Schmitz}},\ and\ \bibinfo {author}
  {\bibfnamefont {M.}~\bibnamefont {Yamada}},\ }\bibfield  {title} {\bibinfo
  {title} {{Wash-In Leptogenesis}},\ }\href
  {https://doi.org/10.1103/PhysRevLett.126.201802} {\bibfield  {journal}
  {\bibinfo  {journal} {Phys. Rev. Lett.}\ }\textbf {\bibinfo {volume} {126}},\
  \bibinfo {pages} {201802} (\bibinfo {year} {2021})},\ \Eprint
  {https://arxiv.org/abs/2011.09347} {arXiv:2011.09347 [hep-ph]} \BibitemShut
  {NoStop}%
\bibitem [{\citenamefont {Mukaida}\ \emph {et~al.}(2022)\citenamefont
  {Mukaida}, \citenamefont {Schmitz},\ and\ \citenamefont
  {Yamada}}]{Mukaida:2021sgv}%
  \BibitemOpen
  \bibfield  {author} {\bibinfo {author} {\bibfnamefont {K.}~\bibnamefont
  {Mukaida}}, \bibinfo {author} {\bibfnamefont {K.}~\bibnamefont {Schmitz}},\
  and\ \bibinfo {author} {\bibfnamefont {M.}~\bibnamefont {Yamada}},\
  }\bibfield  {title} {\bibinfo {title} {{Baryon Asymmetry of the Universe from
  Lepton Flavor Violation}},\ }\href
  {https://doi.org/10.1103/PhysRevLett.129.011803} {\bibfield  {journal}
  {\bibinfo  {journal} {Phys. Rev. Lett.}\ }\textbf {\bibinfo {volume} {129}},\
  \bibinfo {pages} {011803} (\bibinfo {year} {2022})},\ \Eprint
  {https://arxiv.org/abs/2111.03082} {arXiv:2111.03082 [hep-ph]} \BibitemShut
  {NoStop}%
\bibitem [{\citenamefont {Gao}\ and\ \citenamefont
  {Oldengott}(2022)}]{Gao:2021nwz}%
  \BibitemOpen
  \bibfield  {author} {\bibinfo {author} {\bibfnamefont {F.}~\bibnamefont
  {Gao}}\ and\ \bibinfo {author} {\bibfnamefont {I.~M.}\ \bibnamefont
  {Oldengott}},\ }\bibfield  {title} {\bibinfo {title} {{Cosmology Meets
  Functional QCD: First-Order Cosmic QCD Transition Induced by Large Lepton
  Asymmetries}},\ }\href {https://doi.org/10.1103/PhysRevLett.128.131301}
  {\bibfield  {journal} {\bibinfo  {journal} {Phys. Rev. Lett.}\ }\textbf
  {\bibinfo {volume} {128}},\ \bibinfo {pages} {131301} (\bibinfo {year}
  {2022})},\ \Eprint {https://arxiv.org/abs/2106.11991} {arXiv:2106.11991
  [hep-ph]} \BibitemShut {NoStop}%
\bibitem [{\citenamefont {Gao}\ \emph {et~al.}(2023)\citenamefont {Gao},
  \citenamefont {Harz}, \citenamefont {Hati}, \citenamefont {Lu}, \citenamefont
  {Oldengott},\ and\ \citenamefont {White}}]{Gao:2023djs}%
  \BibitemOpen
  \bibfield  {author} {\bibinfo {author} {\bibfnamefont {F.}~\bibnamefont
  {Gao}}, \bibinfo {author} {\bibfnamefont {J.}~\bibnamefont {Harz}}, \bibinfo
  {author} {\bibfnamefont {C.}~\bibnamefont {Hati}}, \bibinfo {author}
  {\bibfnamefont {Y.}~\bibnamefont {Lu}}, \bibinfo {author} {\bibfnamefont
  {I.~M.}\ \bibnamefont {Oldengott}},\ and\ \bibinfo {author} {\bibfnamefont
  {G.}~\bibnamefont {White}},\ }\bibfield  {title} {\bibinfo {title}
  {{Sphaleron freeze-in baryogenesis with gravitational waves from the QCD
  transition}},\ }\href@noop {} {\  (\bibinfo {year} {2023})},\ \Eprint
  {https://arxiv.org/abs/2309.00672} {arXiv:2309.00672 [hep-ph]} \BibitemShut
  {NoStop}%
\bibitem [{\citenamefont {Gao}\ \emph {et~al.}(2024)\citenamefont {Gao},
  \citenamefont {Harz}, \citenamefont {Hati}, \citenamefont {Lu}, \citenamefont
  {Oldengott},\ and\ \citenamefont {White}}]{Gao:2024fhm}%
  \BibitemOpen
  \bibfield  {author} {\bibinfo {author} {\bibfnamefont {F.}~\bibnamefont
  {Gao}}, \bibinfo {author} {\bibfnamefont {J.}~\bibnamefont {Harz}}, \bibinfo
  {author} {\bibfnamefont {C.}~\bibnamefont {Hati}}, \bibinfo {author}
  {\bibfnamefont {Y.}~\bibnamefont {Lu}}, \bibinfo {author} {\bibfnamefont
  {I.~M.}\ \bibnamefont {Oldengott}},\ and\ \bibinfo {author} {\bibfnamefont
  {G.}~\bibnamefont {White}},\ }\bibfield  {title} {\bibinfo {title}
  {{Baryogenesis and first-order QCD transition with gravitational waves from a
  large lepton asymmetry}},\ }\href@noop {} {\  (\bibinfo {year} {2024})},\
  \Eprint {https://arxiv.org/abs/2407.17549} {arXiv:2407.17549 [hep-ph]}
  \BibitemShut {NoStop}%
\bibitem [{\citenamefont {Shi}\ and\ \citenamefont
  {Fuller}(1999)}]{Shi:1998km}%
  \BibitemOpen
  \bibfield  {author} {\bibinfo {author} {\bibfnamefont {X.-D.}\ \bibnamefont
  {Shi}}\ and\ \bibinfo {author} {\bibfnamefont {G.~M.}\ \bibnamefont
  {Fuller}},\ }\bibfield  {title} {\bibinfo {title} {{A New dark matter
  candidate: Nonthermal sterile neutrinos}},\ }\href
  {https://doi.org/10.1103/PhysRevLett.82.2832} {\bibfield  {journal} {\bibinfo
   {journal} {Phys. Rev. Lett.}\ }\textbf {\bibinfo {volume} {82}},\ \bibinfo
  {pages} {2832} (\bibinfo {year} {1999})},\ \Eprint
  {https://arxiv.org/abs/astro-ph/9810076} {arXiv:astro-ph/9810076}
  \BibitemShut {NoStop}%
\bibitem [{\citenamefont {Shaposhnikov}\ and\ \citenamefont
  {Smirnov}(2024)}]{Shaposhnikov:2023hrx}%
  \BibitemOpen
  \bibfield  {author} {\bibinfo {author} {\bibfnamefont {M.}~\bibnamefont
  {Shaposhnikov}}\ and\ \bibinfo {author} {\bibfnamefont {A.~Y.}\ \bibnamefont
  {Smirnov}},\ }\bibfield  {title} {\bibinfo {title} {{Sterile neutrino dark
  matter, matter-antimatter separation, and the QCD phase transition}},\ }\href
  {https://doi.org/10.1103/PhysRevD.110.063520} {\bibfield  {journal} {\bibinfo
   {journal} {Phys. Rev. D}\ }\textbf {\bibinfo {volume} {110}},\ \bibinfo
  {pages} {063520} (\bibinfo {year} {2024})},\ \Eprint
  {https://arxiv.org/abs/2309.13376} {arXiv:2309.13376 [hep-ph]} \BibitemShut
  {NoStop}%
\bibitem [{\citenamefont {Stuke}\ \emph {et~al.}(2012)\citenamefont {Stuke},
  \citenamefont {Schwarz},\ and\ \citenamefont {Starkman}}]{Stuke:2011wz}%
  \BibitemOpen
  \bibfield  {author} {\bibinfo {author} {\bibfnamefont {M.}~\bibnamefont
  {Stuke}}, \bibinfo {author} {\bibfnamefont {D.~J.}\ \bibnamefont {Schwarz}},\
  and\ \bibinfo {author} {\bibfnamefont {G.}~\bibnamefont {Starkman}},\
  }\bibfield  {title} {\bibinfo {title} {{WIMP abundance and lepton (flavour)
  asymmetry}},\ }\href {https://doi.org/10.1088/1475-7516/2012/03/040}
  {\bibfield  {journal} {\bibinfo  {journal} {JCAP}\ }\textbf {\bibinfo
  {volume} {03}},\ \bibinfo {pages} {040}},\ \Eprint
  {https://arxiv.org/abs/1111.3954} {arXiv:1111.3954 [astro-ph.CO]}
  \BibitemShut {NoStop}%
\bibitem [{\citenamefont {Stodolsky}(1975)}]{Stodolsky:1974aq}%
  \BibitemOpen
  \bibfield  {author} {\bibinfo {author} {\bibfnamefont {L.}~\bibnamefont
  {Stodolsky}},\ }\bibfield  {title} {\bibinfo {title} {{Speculations on
  Detection of the Neutrino Sea}},\ }\href
  {https://doi.org/10.1103/PhysRevLett.34.110} {\bibfield  {journal} {\bibinfo
  {journal} {Phys. Rev. Lett.}\ }\textbf {\bibinfo {volume} {34}},\ \bibinfo
  {pages} {110} (\bibinfo {year} {1975})},\ \bibinfo {note} {[Erratum:
  Phys.Rev.Lett. 34, 508 (1975)]}\BibitemShut {NoStop}%
\bibitem [{\citenamefont {Duda}\ \emph {et~al.}(2001)\citenamefont {Duda},
  \citenamefont {Gelmini},\ and\ \citenamefont {Nussinov}}]{Duda:2001hd}%
  \BibitemOpen
  \bibfield  {author} {\bibinfo {author} {\bibfnamefont {G.}~\bibnamefont
  {Duda}}, \bibinfo {author} {\bibfnamefont {G.}~\bibnamefont {Gelmini}},\ and\
  \bibinfo {author} {\bibfnamefont {S.}~\bibnamefont {Nussinov}},\ }\bibfield
  {title} {\bibinfo {title} {{Expected signals in relic neutrino detectors}},\
  }\href {https://doi.org/10.1103/PhysRevD.64.122001} {\bibfield  {journal}
  {\bibinfo  {journal} {Phys. Rev. D}\ }\textbf {\bibinfo {volume} {64}},\
  \bibinfo {pages} {122001} (\bibinfo {year} {2001})},\ \Eprint
  {https://arxiv.org/abs/hep-ph/0107027} {arXiv:hep-ph/0107027} \BibitemShut
  {NoStop}%
\bibitem [{\citenamefont {Joyce}\ and\ \citenamefont
  {Shaposhnikov}(1997)}]{Joyce:1997uy}%
  \BibitemOpen
  \bibfield  {author} {\bibinfo {author} {\bibfnamefont {M.}~\bibnamefont
  {Joyce}}\ and\ \bibinfo {author} {\bibfnamefont {M.~E.}\ \bibnamefont
  {Shaposhnikov}},\ }\bibfield  {title} {\bibinfo {title} {{Primordial magnetic
  fields, right-handed electrons, and the Abelian anomaly}},\ }\href
  {https://doi.org/10.1103/PhysRevLett.79.1193} {\bibfield  {journal} {\bibinfo
   {journal} {Phys. Rev. Lett.}\ }\textbf {\bibinfo {volume} {79}},\ \bibinfo
  {pages} {1193} (\bibinfo {year} {1997})},\ \Eprint
  {https://arxiv.org/abs/astro-ph/9703005} {arXiv:astro-ph/9703005}
  \BibitemShut {NoStop}%
\bibitem [{\citenamefont {Boyarsky}\ \emph {et~al.}(2012)\citenamefont
  {Boyarsky}, \citenamefont {Frohlich},\ and\ \citenamefont
  {Ruchayskiy}}]{Boyarsky:2011uy}%
  \BibitemOpen
  \bibfield  {author} {\bibinfo {author} {\bibfnamefont {A.}~\bibnamefont
  {Boyarsky}}, \bibinfo {author} {\bibfnamefont {J.}~\bibnamefont {Frohlich}},\
  and\ \bibinfo {author} {\bibfnamefont {O.}~\bibnamefont {Ruchayskiy}},\
  }\bibfield  {title} {\bibinfo {title} {{Self-consistent evolution of magnetic
  fields and chiral asymmetry in the early Universe}},\ }\href
  {https://doi.org/10.1103/PhysRevLett.108.031301} {\bibfield  {journal}
  {\bibinfo  {journal} {Phys. Rev. Lett.}\ }\textbf {\bibinfo {volume} {108}},\
  \bibinfo {pages} {031301} (\bibinfo {year} {2012})},\ \Eprint
  {https://arxiv.org/abs/1109.3350} {arXiv:1109.3350 [astro-ph.CO]}
  \BibitemShut {NoStop}%
\bibitem [{\citenamefont {Akamatsu}\ and\ \citenamefont
  {Yamamoto}(2013)}]{Akamatsu:2013pjd}%
  \BibitemOpen
  \bibfield  {author} {\bibinfo {author} {\bibfnamefont {Y.}~\bibnamefont
  {Akamatsu}}\ and\ \bibinfo {author} {\bibfnamefont {N.}~\bibnamefont
  {Yamamoto}},\ }\bibfield  {title} {\bibinfo {title} {{Chiral Plasma
  Instabilities}},\ }\href {https://doi.org/10.1103/PhysRevLett.111.052002}
  {\bibfield  {journal} {\bibinfo  {journal} {Phys. Rev. Lett.}\ }\textbf
  {\bibinfo {volume} {111}},\ \bibinfo {pages} {052002} (\bibinfo {year}
  {2013})},\ \Eprint {https://arxiv.org/abs/1302.2125} {arXiv:1302.2125
  [nucl-th]} \BibitemShut {NoStop}%
\bibitem [{\citenamefont {Hirono}\ \emph {et~al.}(2015)\citenamefont {Hirono},
  \citenamefont {Kharzeev},\ and\ \citenamefont {Yin}}]{Hirono:2015rla}%
  \BibitemOpen
  \bibfield  {author} {\bibinfo {author} {\bibfnamefont {Y.}~\bibnamefont
  {Hirono}}, \bibinfo {author} {\bibfnamefont {D.}~\bibnamefont {Kharzeev}},\
  and\ \bibinfo {author} {\bibfnamefont {Y.}~\bibnamefont {Yin}},\ }\bibfield
  {title} {\bibinfo {title} {{Self-similar inverse cascade of magnetic helicity
  driven by the chiral anomaly}},\ }\href
  {https://doi.org/10.1103/PhysRevD.92.125031} {\bibfield  {journal} {\bibinfo
  {journal} {Phys. Rev. D}\ }\textbf {\bibinfo {volume} {92}},\ \bibinfo
  {pages} {125031} (\bibinfo {year} {2015})},\ \Eprint
  {https://arxiv.org/abs/1509.07790} {arXiv:1509.07790 [hep-th]} \BibitemShut
  {NoStop}%
\bibitem [{\citenamefont {Rogachevskii}\ \emph {et~al.}(2017)\citenamefont
  {Rogachevskii}, \citenamefont {Ruchayskiy}, \citenamefont {Boyarsky},
  \citenamefont {Fr\"ohlich}, \citenamefont {Kleeorin}, \citenamefont
  {Brandenburg},\ and\ \citenamefont {Schober}}]{Rogachevskii:2017uyc}%
  \BibitemOpen
  \bibfield  {author} {\bibinfo {author} {\bibfnamefont {I.}~\bibnamefont
  {Rogachevskii}}, \bibinfo {author} {\bibfnamefont {O.}~\bibnamefont
  {Ruchayskiy}}, \bibinfo {author} {\bibfnamefont {A.}~\bibnamefont
  {Boyarsky}}, \bibinfo {author} {\bibfnamefont {J.}~\bibnamefont
  {Fr\"ohlich}}, \bibinfo {author} {\bibfnamefont {N.}~\bibnamefont
  {Kleeorin}}, \bibinfo {author} {\bibfnamefont {A.}~\bibnamefont
  {Brandenburg}},\ and\ \bibinfo {author} {\bibfnamefont {J.}~\bibnamefont
  {Schober}},\ }\bibfield  {title} {\bibinfo {title} {{Laminar and turbulent
  dynamos in chiral magnetohydrodynamics-I: Theory}},\ }\href
  {https://doi.org/10.3847/1538-4357/aa886b} {\bibfield  {journal} {\bibinfo
  {journal} {Astrophys. J.}\ }\textbf {\bibinfo {volume} {846}},\ \bibinfo
  {pages} {153} (\bibinfo {year} {2017})},\ \Eprint
  {https://arxiv.org/abs/1705.00378} {arXiv:1705.00378 [physics.plasm-ph]}
  \BibitemShut {NoStop}%
\bibitem [{\citenamefont {Domcke}\ \emph
  {et~al.}(2023{\natexlab{a}})\citenamefont {Domcke}, \citenamefont {Kamada},
  \citenamefont {Mukaida}, \citenamefont {Schmitz},\ and\ \citenamefont
  {Yamada}}]{Domcke:2022uue}%
  \BibitemOpen
  \bibfield  {author} {\bibinfo {author} {\bibfnamefont {V.}~\bibnamefont
  {Domcke}}, \bibinfo {author} {\bibfnamefont {K.}~\bibnamefont {Kamada}},
  \bibinfo {author} {\bibfnamefont {K.}~\bibnamefont {Mukaida}}, \bibinfo
  {author} {\bibfnamefont {K.}~\bibnamefont {Schmitz}},\ and\ \bibinfo {author}
  {\bibfnamefont {M.}~\bibnamefont {Yamada}},\ }\bibfield  {title} {\bibinfo
  {title} {{New Constraint on Primordial Lepton Flavor Asymmetries}},\ }\href
  {https://doi.org/10.1103/PhysRevLett.130.261803} {\bibfield  {journal}
  {\bibinfo  {journal} {Phys. Rev. Lett.}\ }\textbf {\bibinfo {volume} {130}},\
  \bibinfo {pages} {261803} (\bibinfo {year} {2023}{\natexlab{a}})},\ \Eprint
  {https://arxiv.org/abs/2208.03237} {arXiv:2208.03237 [hep-ph]} \BibitemShut
  {NoStop}%
\bibitem [{\citenamefont {Froustey}\ and\ \citenamefont
  {Pitrou}(2022)}]{Froustey:2021azz}%
  \BibitemOpen
  \bibfield  {author} {\bibinfo {author} {\bibfnamefont {J.}~\bibnamefont
  {Froustey}}\ and\ \bibinfo {author} {\bibfnamefont {C.}~\bibnamefont
  {Pitrou}},\ }\bibfield  {title} {\bibinfo {title} {{Primordial neutrino
  asymmetry evolution with full mean-field effects and collisions}},\ }\href
  {https://doi.org/10.1088/1475-7516/2022/03/065} {\bibfield  {journal}
  {\bibinfo  {journal} {JCAP}\ }\textbf {\bibinfo {volume} {03}}\bibfield
  {number} {\bibinfo  {number} { (03)},\ \bibinfo {pages} {065}},\ }\Eprint
  {https://arxiv.org/abs/2110.11889} {arXiv:2110.11889 [hep-ph]} \BibitemShut
  {NoStop}%
\bibitem [{\citenamefont {Froustey}\ and\ \citenamefont
  {Pitrou}(2024)}]{Froustey:2024mgf}%
  \BibitemOpen
  \bibfield  {author} {\bibinfo {author} {\bibfnamefont {J.}~\bibnamefont
  {Froustey}}\ and\ \bibinfo {author} {\bibfnamefont {C.}~\bibnamefont
  {Pitrou}},\ }\bibfield  {title} {\bibinfo {title} {{Constraints on primordial
  lepton asymmetries with full neutrino transport}},\ }\href
  {https://doi.org/10.1103/PhysRevD.110.103551} {\bibfield  {journal} {\bibinfo
   {journal} {Phys. Rev. D}\ }\textbf {\bibinfo {volume} {110}},\ \bibinfo
  {pages} {103551} (\bibinfo {year} {2024})},\ \Eprint
  {https://arxiv.org/abs/2405.06509} {arXiv:2405.06509 [hep-ph]} \BibitemShut
  {NoStop}%
\bibitem [{\citenamefont {Escudero}\ \emph {et~al.}(2023)\citenamefont
  {Escudero}, \citenamefont {Ibarra},\ and\ \citenamefont
  {Maura}}]{Escudero:2022okz}%
  \BibitemOpen
  \bibfield  {author} {\bibinfo {author} {\bibfnamefont {M.}~\bibnamefont
  {Escudero}}, \bibinfo {author} {\bibfnamefont {A.}~\bibnamefont {Ibarra}},\
  and\ \bibinfo {author} {\bibfnamefont {V.}~\bibnamefont {Maura}},\ }\bibfield
   {title} {\bibinfo {title} {{Primordial lepton asymmetries in the precision
  cosmology era: Current status and future sensitivities from BBN and the
  CMB}},\ }\href {https://doi.org/10.1103/PhysRevD.107.035024} {\bibfield
  {journal} {\bibinfo  {journal} {Phys. Rev. D}\ }\textbf {\bibinfo {volume}
  {107}},\ \bibinfo {pages} {035024} (\bibinfo {year} {2023})},\ \Eprint
  {https://arxiv.org/abs/2208.03201} {arXiv:2208.03201 [hep-ph]} \BibitemShut
  {NoStop}%
\bibitem [{\citenamefont {Burns}\ \emph {et~al.}(2023)\citenamefont {Burns},
  \citenamefont {Tait},\ and\ \citenamefont {Valli}}]{Burns:2022hkq}%
  \BibitemOpen
  \bibfield  {author} {\bibinfo {author} {\bibfnamefont {A.-K.}\ \bibnamefont
  {Burns}}, \bibinfo {author} {\bibfnamefont {T.~M.~P.}\ \bibnamefont {Tait}},\
  and\ \bibinfo {author} {\bibfnamefont {M.}~\bibnamefont {Valli}},\ }\bibfield
   {title} {\bibinfo {title} {{Indications for a Nonzero Lepton Asymmetry from
  Extremely Metal-Poor Galaxies}},\ }\href
  {https://doi.org/10.1103/PhysRevLett.130.131001} {\bibfield  {journal}
  {\bibinfo  {journal} {Phys. Rev. Lett.}\ }\textbf {\bibinfo {volume} {130}},\
  \bibinfo {pages} {131001} (\bibinfo {year} {2023})},\ \Eprint
  {https://arxiv.org/abs/2206.00693} {arXiv:2206.00693 [hep-ph]} \BibitemShut
  {NoStop}%
\bibitem [{\citenamefont {Kawasaki}\ and\ \citenamefont
  {Murai}(2022)}]{Kawasaki:2022hvx}%
  \BibitemOpen
  \bibfield  {author} {\bibinfo {author} {\bibfnamefont {M.}~\bibnamefont
  {Kawasaki}}\ and\ \bibinfo {author} {\bibfnamefont {K.}~\bibnamefont
  {Murai}},\ }\bibfield  {title} {\bibinfo {title} {{Lepton asymmetric
  universe}},\ }\href {https://doi.org/10.1088/1475-7516/2022/08/041}
  {\bibfield  {journal} {\bibinfo  {journal} {JCAP}\ }\textbf {\bibinfo
  {volume} {08}}\bibfield  {number} {\bibinfo  {number} { (08)},\ \bibinfo
  {pages} {041}},\ }\Eprint {https://arxiv.org/abs/2203.09713}
  {arXiv:2203.09713 [hep-ph]} \BibitemShut {NoStop}%
\bibitem [{\citenamefont {Kohri}\ \emph {et~al.}(1997)\citenamefont {Kohri},
  \citenamefont {Kawasaki},\ and\ \citenamefont {Sato}}]{Kohri:1996ke}%
  \BibitemOpen
  \bibfield  {author} {\bibinfo {author} {\bibfnamefont {K.}~\bibnamefont
  {Kohri}}, \bibinfo {author} {\bibfnamefont {M.}~\bibnamefont {Kawasaki}},\
  and\ \bibinfo {author} {\bibfnamefont {K.}~\bibnamefont {Sato}},\ }\bibfield
  {title} {\bibinfo {title} {{Big bang nucleosynthesis and lepton number
  asymmetry in the universe}},\ }\href {https://doi.org/10.1086/512793}
  {\bibfield  {journal} {\bibinfo  {journal} {Astrophys. J.}\ }\textbf
  {\bibinfo {volume} {490}},\ \bibinfo {pages} {72} (\bibinfo {year} {1997})},\
  \Eprint {https://arxiv.org/abs/astro-ph/9612237} {arXiv:astro-ph/9612237}
  \BibitemShut {NoStop}%
\bibitem [{\citenamefont {Minami}\ and\ \citenamefont
  {Komatsu}(2020)}]{Minami:2020odp}%
  \BibitemOpen
  \bibfield  {author} {\bibinfo {author} {\bibfnamefont {Y.}~\bibnamefont
  {Minami}}\ and\ \bibinfo {author} {\bibfnamefont {E.}~\bibnamefont
  {Komatsu}},\ }\bibfield  {title} {\bibinfo {title} {{New Extraction of the
  Cosmic Birefringence from the Planck 2018 Polarization Data}},\ }\href
  {https://doi.org/10.1103/PhysRevLett.125.221301} {\bibfield  {journal}
  {\bibinfo  {journal} {Phys. Rev. Lett.}\ }\textbf {\bibinfo {volume} {125}},\
  \bibinfo {pages} {221301} (\bibinfo {year} {2020})},\ \Eprint
  {https://arxiv.org/abs/2011.11254} {arXiv:2011.11254 [astro-ph.CO]}
  \BibitemShut {NoStop}%
\bibitem [{\citenamefont {Seto}(2006)}]{Seto:2006hf}%
  \BibitemOpen
  \bibfield  {author} {\bibinfo {author} {\bibfnamefont {N.}~\bibnamefont
  {Seto}},\ }\bibfield  {title} {\bibinfo {title} {{Prospects for direct
  detection of circular polarization of gravitational-wave background}},\
  }\href {https://doi.org/10.1103/PhysRevLett.97.151101} {\bibfield  {journal}
  {\bibinfo  {journal} {Phys. Rev. Lett.}\ }\textbf {\bibinfo {volume} {97}},\
  \bibinfo {pages} {151101} (\bibinfo {year} {2006})},\ \Eprint
  {https://arxiv.org/abs/astro-ph/0609504} {arXiv:astro-ph/0609504}
  \BibitemShut {NoStop}%
\bibitem [{\citenamefont {Seto}(2007)}]{Seto:2006dz}%
  \BibitemOpen
  \bibfield  {author} {\bibinfo {author} {\bibfnamefont {N.}~\bibnamefont
  {Seto}},\ }\bibfield  {title} {\bibinfo {title} {{Quest for circular
  polarization of gravitational wave background and orbits of laser
  interferometers in space}},\ }\href
  {https://doi.org/10.1103/PhysRevD.75.061302} {\bibfield  {journal} {\bibinfo
  {journal} {Phys. Rev. D}\ }\textbf {\bibinfo {volume} {75}},\ \bibinfo
  {pages} {061302} (\bibinfo {year} {2007})},\ \Eprint
  {https://arxiv.org/abs/astro-ph/0609633} {arXiv:astro-ph/0609633}
  \BibitemShut {NoStop}%
\bibitem [{\citenamefont {Domcke}\ \emph
  {et~al.}(2020{\natexlab{a}})\citenamefont {Domcke}, \citenamefont
  {Garcia-Bellido}, \citenamefont {Peloso}, \citenamefont {Pieroni},
  \citenamefont {Ricciardone}, \citenamefont {Sorbo},\ and\ \citenamefont
  {Tasinato}}]{Domcke:2019zls}%
  \BibitemOpen
  \bibfield  {author} {\bibinfo {author} {\bibfnamefont {V.}~\bibnamefont
  {Domcke}}, \bibinfo {author} {\bibfnamefont {J.}~\bibnamefont
  {Garcia-Bellido}}, \bibinfo {author} {\bibfnamefont {M.}~\bibnamefont
  {Peloso}}, \bibinfo {author} {\bibfnamefont {M.}~\bibnamefont {Pieroni}},
  \bibinfo {author} {\bibfnamefont {A.}~\bibnamefont {Ricciardone}}, \bibinfo
  {author} {\bibfnamefont {L.}~\bibnamefont {Sorbo}},\ and\ \bibinfo {author}
  {\bibfnamefont {G.}~\bibnamefont {Tasinato}},\ }\bibfield  {title} {\bibinfo
  {title} {{Measuring the net circular polarization of the stochastic
  gravitational wave background with interferometers}},\ }\href
  {https://doi.org/10.1088/1475-7516/2020/05/028} {\bibfield  {journal}
  {\bibinfo  {journal} {JCAP}\ }\textbf {\bibinfo {volume} {05}},\ \bibinfo
  {pages} {028}},\ \Eprint {https://arxiv.org/abs/1910.08052} {arXiv:1910.08052
  [astro-ph.CO]} \BibitemShut {NoStop}%
\bibitem [{\citenamefont {Esteban}\ \emph {et~al.}(2024)\citenamefont
  {Esteban}, \citenamefont {Gonzalez-Garcia}, \citenamefont {Maltoni},
  \citenamefont {Martinez-Soler}, \citenamefont {Pinheiro},\ and\ \citenamefont
  {Schwetz}}]{Esteban:2024eli}%
  \BibitemOpen
  \bibfield  {author} {\bibinfo {author} {\bibfnamefont {I.}~\bibnamefont
  {Esteban}}, \bibinfo {author} {\bibfnamefont {M.~C.}\ \bibnamefont
  {Gonzalez-Garcia}}, \bibinfo {author} {\bibfnamefont {M.}~\bibnamefont
  {Maltoni}}, \bibinfo {author} {\bibfnamefont {I.}~\bibnamefont
  {Martinez-Soler}}, \bibinfo {author} {\bibfnamefont {J.~a.~P.}\ \bibnamefont
  {Pinheiro}},\ and\ \bibinfo {author} {\bibfnamefont {T.}~\bibnamefont
  {Schwetz}},\ }\bibfield  {title} {\bibinfo {title} {{NuFit-6.0: updated
  global analysis of three-flavor neutrino oscillations}},\ }\href
  {https://doi.org/10.1007/JHEP12(2024)216} {\bibfield  {journal} {\bibinfo
  {journal} {JHEP}\ }\textbf {\bibinfo {volume} {12}},\ \bibinfo {pages}
  {216}},\ \Eprint {https://arxiv.org/abs/2410.05380} {arXiv:2410.05380
  [hep-ph]} \BibitemShut {NoStop}%
\bibitem [{\citenamefont {Dolgov}\ \emph {et~al.}(2002)\citenamefont {Dolgov},
  \citenamefont {Hansen}, \citenamefont {Pastor}, \citenamefont {Petcov},
  \citenamefont {Raffelt},\ and\ \citenamefont {Semikoz}}]{Dolgov:2002ab}%
  \BibitemOpen
  \bibfield  {author} {\bibinfo {author} {\bibfnamefont {A.~D.}\ \bibnamefont
  {Dolgov}}, \bibinfo {author} {\bibfnamefont {S.~H.}\ \bibnamefont {Hansen}},
  \bibinfo {author} {\bibfnamefont {S.}~\bibnamefont {Pastor}}, \bibinfo
  {author} {\bibfnamefont {S.~T.}\ \bibnamefont {Petcov}}, \bibinfo {author}
  {\bibfnamefont {G.~G.}\ \bibnamefont {Raffelt}},\ and\ \bibinfo {author}
  {\bibfnamefont {D.~V.}\ \bibnamefont {Semikoz}},\ }\bibfield  {title}
  {\bibinfo {title} {{Cosmological bounds on neutrino degeneracy improved by
  flavor oscillations}},\ }\href
  {https://doi.org/10.1016/S0550-3213(02)00274-2} {\bibfield  {journal}
  {\bibinfo  {journal} {Nucl. Phys. B}\ }\textbf {\bibinfo {volume} {632}},\
  \bibinfo {pages} {363} (\bibinfo {year} {2002})},\ \Eprint
  {https://arxiv.org/abs/hep-ph/0201287} {arXiv:hep-ph/0201287} \BibitemShut
  {NoStop}%
\bibitem [{\citenamefont {Pastor}\ \emph {et~al.}(2009)\citenamefont {Pastor},
  \citenamefont {Pinto},\ and\ \citenamefont {Raffelt}}]{Pastor:2008ti}%
  \BibitemOpen
  \bibfield  {author} {\bibinfo {author} {\bibfnamefont {S.}~\bibnamefont
  {Pastor}}, \bibinfo {author} {\bibfnamefont {T.}~\bibnamefont {Pinto}},\ and\
  \bibinfo {author} {\bibfnamefont {G.~G.}\ \bibnamefont {Raffelt}},\
  }\bibfield  {title} {\bibinfo {title} {{Relic density of neutrinos with
  primordial asymmetries}},\ }\href
  {https://doi.org/10.1103/PhysRevLett.102.241302} {\bibfield  {journal}
  {\bibinfo  {journal} {Phys. Rev. Lett.}\ }\textbf {\bibinfo {volume} {102}},\
  \bibinfo {pages} {241302} (\bibinfo {year} {2009})},\ \Eprint
  {https://arxiv.org/abs/0808.3137} {arXiv:0808.3137 [astro-ph]} \BibitemShut
  {NoStop}%
\bibitem [{\citenamefont {Mangano}\ \emph {et~al.}(2011)\citenamefont
  {Mangano}, \citenamefont {Miele}, \citenamefont {Pastor}, \citenamefont
  {Pisanti},\ and\ \citenamefont {Sarikas}}]{Mangano:2010ei}%
  \BibitemOpen
  \bibfield  {author} {\bibinfo {author} {\bibfnamefont {G.}~\bibnamefont
  {Mangano}}, \bibinfo {author} {\bibfnamefont {G.}~\bibnamefont {Miele}},
  \bibinfo {author} {\bibfnamefont {S.}~\bibnamefont {Pastor}}, \bibinfo
  {author} {\bibfnamefont {O.}~\bibnamefont {Pisanti}},\ and\ \bibinfo {author}
  {\bibfnamefont {S.}~\bibnamefont {Sarikas}},\ }\bibfield  {title} {\bibinfo
  {title} {{Constraining the cosmic radiation density due to lepton number with
  Big Bang Nucleosynthesis}},\ }\href
  {https://doi.org/10.1088/1475-7516/2011/03/035} {\bibfield  {journal}
  {\bibinfo  {journal} {JCAP}\ }\textbf {\bibinfo {volume} {03}},\ \bibinfo
  {pages} {035}},\ \Eprint {https://arxiv.org/abs/1011.0916} {arXiv:1011.0916
  [astro-ph.CO]} \BibitemShut {NoStop}%
\bibitem [{\citenamefont {Castorina}\ \emph {et~al.}(2012)\citenamefont
  {Castorina}, \citenamefont {Franca}, \citenamefont {Lattanzi}, \citenamefont
  {Lesgourgues}, \citenamefont {Mangano}, \citenamefont {Melchiorri},\ and\
  \citenamefont {Pastor}}]{Castorina:2012md}%
  \BibitemOpen
  \bibfield  {author} {\bibinfo {author} {\bibfnamefont {E.}~\bibnamefont
  {Castorina}}, \bibinfo {author} {\bibfnamefont {U.}~\bibnamefont {Franca}},
  \bibinfo {author} {\bibfnamefont {M.}~\bibnamefont {Lattanzi}}, \bibinfo
  {author} {\bibfnamefont {J.}~\bibnamefont {Lesgourgues}}, \bibinfo {author}
  {\bibfnamefont {G.}~\bibnamefont {Mangano}}, \bibinfo {author} {\bibfnamefont
  {A.}~\bibnamefont {Melchiorri}},\ and\ \bibinfo {author} {\bibfnamefont
  {S.}~\bibnamefont {Pastor}},\ }\bibfield  {title} {\bibinfo {title}
  {{Cosmological lepton asymmetry with a nonzero mixing angle $\theta_{13}$}},\
  }\href {https://doi.org/10.1103/PhysRevD.86.023517} {\bibfield  {journal}
  {\bibinfo  {journal} {Phys. Rev. D}\ }\textbf {\bibinfo {volume} {86}},\
  \bibinfo {pages} {023517} (\bibinfo {year} {2012})},\ \Eprint
  {https://arxiv.org/abs/1204.2510} {arXiv:1204.2510 [astro-ph.CO]}
  \BibitemShut {NoStop}%
\bibitem [{\citenamefont {Zheng}\ \emph {et~al.}(2025)\citenamefont {Zheng},
  \citenamefont {Gao}, \citenamefont {Bian}, \citenamefont {Qin},\ and\
  \citenamefont {Liu}}]{Zheng:2024tib}%
  \BibitemOpen
  \bibfield  {author} {\bibinfo {author} {\bibfnamefont {H.-w.}\ \bibnamefont
  {Zheng}}, \bibinfo {author} {\bibfnamefont {F.}~\bibnamefont {Gao}}, \bibinfo
  {author} {\bibfnamefont {L.}~\bibnamefont {Bian}}, \bibinfo {author}
  {\bibfnamefont {S.-x.}\ \bibnamefont {Qin}},\ and\ \bibinfo {author}
  {\bibfnamefont {Y.-x.}\ \bibnamefont {Liu}},\ }\bibfield  {title} {\bibinfo
  {title} {{Quantitative analysis of the gravitational wave spectrum sourced
  from a first-order chiral phase transition of QCD}},\ }\href
  {https://doi.org/10.1103/PhysRevD.111.L021303} {\bibfield  {journal}
  {\bibinfo  {journal} {Phys. Rev. D}\ }\textbf {\bibinfo {volume} {111}},\
  \bibinfo {pages} {L021303} (\bibinfo {year} {2025})},\ \Eprint
  {https://arxiv.org/abs/2407.03795} {arXiv:2407.03795 [hep-ph]} \BibitemShut
  {NoStop}%
\bibitem [{\citenamefont {Cline}\ and\ \citenamefont
  {Laurent}(2025)}]{Cline:2025bwe}%
  \BibitemOpen
  \bibfield  {author} {\bibinfo {author} {\bibfnamefont {J.~M.}\ \bibnamefont
  {Cline}}\ and\ \bibinfo {author} {\bibfnamefont {B.}~\bibnamefont
  {Laurent}},\ }\bibfield  {title} {\bibinfo {title} {{Bubble wall velocity for
  first-order QCD phase transition}},\ }\href@noop {} {\  (\bibinfo {year}
  {2025})},\ \Eprint {https://arxiv.org/abs/2502.12321} {arXiv:2502.12321
  [hep-ph]} \BibitemShut {NoStop}%
\bibitem [{\citenamefont {Khlebnikov}\ and\ \citenamefont
  {Shaposhnikov}(1988)}]{Khlebnikov:1988sr}%
  \BibitemOpen
  \bibfield  {author} {\bibinfo {author} {\bibfnamefont {S.~Y.}\ \bibnamefont
  {Khlebnikov}}\ and\ \bibinfo {author} {\bibfnamefont {M.~E.}\ \bibnamefont
  {Shaposhnikov}},\ }\bibfield  {title} {\bibinfo {title} {{The Statistical
  Theory of Anomalous Fermion Number Nonconservation}},\ }\href
  {https://doi.org/10.1016/0550-3213(88)90133-2} {\bibfield  {journal}
  {\bibinfo  {journal} {Nucl. Phys. B}\ }\textbf {\bibinfo {volume} {308}},\
  \bibinfo {pages} {885} (\bibinfo {year} {1988})}\BibitemShut {NoStop}%
\bibitem [{\citenamefont {Laine}\ and\ \citenamefont
  {Shaposhnikov}(2000)}]{Laine:1999wv}%
  \BibitemOpen
  \bibfield  {author} {\bibinfo {author} {\bibfnamefont {M.}~\bibnamefont
  {Laine}}\ and\ \bibinfo {author} {\bibfnamefont {M.~E.}\ \bibnamefont
  {Shaposhnikov}},\ }\bibfield  {title} {\bibinfo {title} {{A Remark on
  sphaleron erasure of baryon asymmetry}},\ }\href
  {https://doi.org/10.1103/PhysRevD.61.117302} {\bibfield  {journal} {\bibinfo
  {journal} {Phys. Rev. D}\ }\textbf {\bibinfo {volume} {61}},\ \bibinfo
  {pages} {117302} (\bibinfo {year} {2000})},\ \Eprint
  {https://arxiv.org/abs/hep-ph/9911473} {arXiv:hep-ph/9911473} \BibitemShut
  {NoStop}%
\bibitem [{\citenamefont {Shu}\ \emph {et~al.}(2007)\citenamefont {Shu},
  \citenamefont {Tait},\ and\ \citenamefont {Wagner}}]{Shu:2006mm}%
  \BibitemOpen
  \bibfield  {author} {\bibinfo {author} {\bibfnamefont {J.}~\bibnamefont
  {Shu}}, \bibinfo {author} {\bibfnamefont {T.~M.~P.}\ \bibnamefont {Tait}},\
  and\ \bibinfo {author} {\bibfnamefont {C.~E.~M.}\ \bibnamefont {Wagner}},\
  }\bibfield  {title} {\bibinfo {title} {{Baryogenesis from an Earlier Phase
  Transition}},\ }\href {https://doi.org/10.1103/PhysRevD.75.063510} {\bibfield
   {journal} {\bibinfo  {journal} {Phys. Rev. D}\ }\textbf {\bibinfo {volume}
  {75}},\ \bibinfo {pages} {063510} (\bibinfo {year} {2007})},\ \Eprint
  {https://arxiv.org/abs/hep-ph/0610375} {arXiv:hep-ph/0610375} \BibitemShut
  {NoStop}%
\bibitem [{\citenamefont {Gu}(2010)}]{Gu:2010dg}%
  \BibitemOpen
  \bibfield  {author} {\bibinfo {author} {\bibfnamefont {P.-H.}\ \bibnamefont
  {Gu}},\ }\bibfield  {title} {\bibinfo {title} {{Large Lepton Asymmetry for
  Small Baryon Asymmetry and Warm Dark Matter}},\ }\href
  {https://doi.org/10.1103/PhysRevD.82.093009} {\bibfield  {journal} {\bibinfo
  {journal} {Phys. Rev. D}\ }\textbf {\bibinfo {volume} {82}},\ \bibinfo
  {pages} {093009} (\bibinfo {year} {2010})},\ \Eprint
  {https://arxiv.org/abs/1005.1632} {arXiv:1005.1632 [hep-ph]} \BibitemShut
  {NoStop}%
\bibitem [{\citenamefont {Sigl}\ and\ \citenamefont
  {Raffelt}(1993)}]{Sigl:1993ctk}%
  \BibitemOpen
  \bibfield  {author} {\bibinfo {author} {\bibfnamefont {G.}~\bibnamefont
  {Sigl}}\ and\ \bibinfo {author} {\bibfnamefont {G.}~\bibnamefont {Raffelt}},\
  }\bibfield  {title} {\bibinfo {title} {{General kinetic description of
  relativistic mixed neutrinos}},\ }\href
  {https://doi.org/10.1016/0550-3213(93)90175-O} {\bibfield  {journal}
  {\bibinfo  {journal} {Nucl. Phys. B}\ }\textbf {\bibinfo {volume} {406}},\
  \bibinfo {pages} {423} (\bibinfo {year} {1993})}\BibitemShut {NoStop}%
\bibitem [{\citenamefont {Froustey}\ \emph {et~al.}(2020)\citenamefont
  {Froustey}, \citenamefont {Pitrou},\ and\ \citenamefont
  {Volpe}}]{Froustey:2020mcq}%
  \BibitemOpen
  \bibfield  {author} {\bibinfo {author} {\bibfnamefont {J.}~\bibnamefont
  {Froustey}}, \bibinfo {author} {\bibfnamefont {C.}~\bibnamefont {Pitrou}},\
  and\ \bibinfo {author} {\bibfnamefont {M.~C.}\ \bibnamefont {Volpe}},\
  }\bibfield  {title} {\bibinfo {title} {{Neutrino decoupling including flavour
  oscillations and primordial nucleosynthesis}},\ }\href
  {https://doi.org/10.1088/1475-7516/2020/12/015} {\bibfield  {journal}
  {\bibinfo  {journal} {JCAP}\ }\textbf {\bibinfo {volume} {12}},\ \bibinfo
  {pages} {015}},\ \Eprint {https://arxiv.org/abs/2008.01074} {arXiv:2008.01074
  [hep-ph]} \BibitemShut {NoStop}%
\bibitem [{\citenamefont {Volpe}\ \emph {et~al.}(2013)\citenamefont {Volpe},
  \citenamefont {V\"a\"an\"anen},\ and\ \citenamefont
  {Espinoza}}]{Volpe:2013uxl}%
  \BibitemOpen
  \bibfield  {author} {\bibinfo {author} {\bibfnamefont {C.}~\bibnamefont
  {Volpe}}, \bibinfo {author} {\bibfnamefont {D.}~\bibnamefont
  {V\"a\"an\"anen}},\ and\ \bibinfo {author} {\bibfnamefont {C.}~\bibnamefont
  {Espinoza}},\ }\bibfield  {title} {\bibinfo {title} {{Extended evolution
  equations for neutrino propagation in astrophysical and cosmological
  environments}},\ }\href {https://doi.org/10.1103/PhysRevD.87.113010}
  {\bibfield  {journal} {\bibinfo  {journal} {Phys. Rev. D}\ }\textbf {\bibinfo
  {volume} {87}},\ \bibinfo {pages} {113010} (\bibinfo {year} {2013})},\
  \Eprint {https://arxiv.org/abs/1302.2374} {arXiv:1302.2374 [hep-ph]}
  \BibitemShut {NoStop}%
\bibitem [{\citenamefont {Blaschke}\ and\ \citenamefont
  {Cirigliano}(2016)}]{Blaschke:2016xxt}%
  \BibitemOpen
  \bibfield  {author} {\bibinfo {author} {\bibfnamefont {D.~N.}\ \bibnamefont
  {Blaschke}}\ and\ \bibinfo {author} {\bibfnamefont {V.}~\bibnamefont
  {Cirigliano}},\ }\bibfield  {title} {\bibinfo {title} {{Neutrino Quantum
  Kinetic Equations: The Collision Term}},\ }\href
  {https://doi.org/10.1103/PhysRevD.94.033009} {\bibfield  {journal} {\bibinfo
  {journal} {Phys. Rev. D}\ }\textbf {\bibinfo {volume} {94}},\ \bibinfo
  {pages} {033009} (\bibinfo {year} {2016})},\ \Eprint
  {https://arxiv.org/abs/1605.09383} {arXiv:1605.09383 [hep-ph]} \BibitemShut
  {NoStop}%
\bibitem [{\citenamefont {Bennett}\ \emph {et~al.}(2021)\citenamefont
  {Bennett}, \citenamefont {Buldgen}, \citenamefont {De~Salas}, \citenamefont
  {Drewes}, \citenamefont {Gariazzo}, \citenamefont {Pastor},\ and\
  \citenamefont {Wong}}]{Bennett:2020zkv}%
  \BibitemOpen
  \bibfield  {author} {\bibinfo {author} {\bibfnamefont {J.~J.}\ \bibnamefont
  {Bennett}}, \bibinfo {author} {\bibfnamefont {G.}~\bibnamefont {Buldgen}},
  \bibinfo {author} {\bibfnamefont {P.~F.}\ \bibnamefont {De~Salas}}, \bibinfo
  {author} {\bibfnamefont {M.}~\bibnamefont {Drewes}}, \bibinfo {author}
  {\bibfnamefont {S.}~\bibnamefont {Gariazzo}}, \bibinfo {author}
  {\bibfnamefont {S.}~\bibnamefont {Pastor}},\ and\ \bibinfo {author}
  {\bibfnamefont {Y.~Y.~Y.}\ \bibnamefont {Wong}},\ }\bibfield  {title}
  {\bibinfo {title} {{Towards a precision calculation of $N_{\rm eff}$ in the
  Standard Model II: Neutrino decoupling in the presence of flavour
  oscillations and finite-temperature QED}},\ }\href
  {https://doi.org/10.1088/1475-7516/2021/04/073} {\bibfield  {journal}
  {\bibinfo  {journal} {JCAP}\ }\textbf {\bibinfo {volume} {04}},\ \bibinfo
  {pages} {073}},\ \Eprint {https://arxiv.org/abs/2012.02726} {arXiv:2012.02726
  [hep-ph]} \BibitemShut {NoStop}%
\bibitem [{\citenamefont {Li}\ and\ \citenamefont {Yu}(2024)}]{Li:2024gzf}%
  \BibitemOpen
  \bibfield  {author} {\bibinfo {author} {\bibfnamefont {Y.-Z.}\ \bibnamefont
  {Li}}\ and\ \bibinfo {author} {\bibfnamefont {J.-H.}\ \bibnamefont {Yu}},\
  }\bibfield  {title} {\bibinfo {title} {{Revisiting primordial neutrino
  asymmetries, spectral distortions and cosmological constraints with full
  neutrino transport}},\ }\href@noop {} {\  (\bibinfo {year} {2024})},\ \Eprint
  {https://arxiv.org/abs/2409.08280} {arXiv:2409.08280 [hep-ph]} \BibitemShut
  {NoStop}%
\bibitem [{\citenamefont {de~Salas}\ and\ \citenamefont
  {Pastor}(2016)}]{deSalas:2016ztq}%
  \BibitemOpen
  \bibfield  {author} {\bibinfo {author} {\bibfnamefont {P.~F.}\ \bibnamefont
  {de~Salas}}\ and\ \bibinfo {author} {\bibfnamefont {S.}~\bibnamefont
  {Pastor}},\ }\bibfield  {title} {\bibinfo {title} {{Relic neutrino decoupling
  with flavour oscillations revisited}},\ }\href
  {https://doi.org/10.1088/1475-7516/2016/07/051} {\bibfield  {journal}
  {\bibinfo  {journal} {JCAP}\ }\textbf {\bibinfo {volume} {07}},\ \bibinfo
  {pages} {051}},\ \Eprint {https://arxiv.org/abs/1606.06986} {arXiv:1606.06986
  [hep-ph]} \BibitemShut {NoStop}%
\bibitem [{\citenamefont {Lesgourgues}\ \emph {et~al.}(2013)\citenamefont
  {Lesgourgues}, \citenamefont {Mangano}, \citenamefont {Miele},\ and\
  \citenamefont {Pastor}}]{Lesgourgues:2013sjj}%
  \BibitemOpen
  \bibfield  {author} {\bibinfo {author} {\bibfnamefont {J.}~\bibnamefont
  {Lesgourgues}}, \bibinfo {author} {\bibfnamefont {G.}~\bibnamefont
  {Mangano}}, \bibinfo {author} {\bibfnamefont {G.}~\bibnamefont {Miele}},\
  and\ \bibinfo {author} {\bibfnamefont {S.}~\bibnamefont {Pastor}},\
  }\href@noop {} {\emph {\bibinfo {title} {{Neutrino Cosmology}}}}\ (\bibinfo
  {publisher} {Cambridge University Press},\ \bibinfo {year}
  {2013})\BibitemShut {NoStop}%
\bibitem [{\citenamefont {de~Vries}\ \emph {et~al.}(2024)\citenamefont
  {de~Vries}, \citenamefont {Drewes}, \citenamefont {Georis}, \citenamefont
  {Klari\'c},\ and\ \citenamefont {Plakkot}}]{deVries:2024rfh}%
  \BibitemOpen
  \bibfield  {author} {\bibinfo {author} {\bibfnamefont {J.}~\bibnamefont
  {de~Vries}}, \bibinfo {author} {\bibfnamefont {M.}~\bibnamefont {Drewes}},
  \bibinfo {author} {\bibfnamefont {Y.}~\bibnamefont {Georis}}, \bibinfo
  {author} {\bibfnamefont {J.}~\bibnamefont {Klari\'c}},\ and\ \bibinfo
  {author} {\bibfnamefont {V.}~\bibnamefont {Plakkot}},\ }\bibfield  {title}
  {\bibinfo {title} {{Confronting the low-scale seesaw and leptogenesis with
  neutrinoless double beta decay}},\ }\href@noop {} {\  (\bibinfo {year}
  {2024})},\ \Eprint {https://arxiv.org/abs/2407.10560} {arXiv:2407.10560
  [hep-ph]} \BibitemShut {NoStop}%
\bibitem [{\citenamefont {Klari\'c}\ \emph
  {et~al.}(2021{\natexlab{a}})\citenamefont {Klari\'c}, \citenamefont
  {Shaposhnikov},\ and\ \citenamefont {Timiryasov}}]{Klaric:2021cpi}%
  \BibitemOpen
  \bibfield  {author} {\bibinfo {author} {\bibfnamefont {J.}~\bibnamefont
  {Klari\'c}}, \bibinfo {author} {\bibfnamefont {M.}~\bibnamefont
  {Shaposhnikov}},\ and\ \bibinfo {author} {\bibfnamefont {I.}~\bibnamefont
  {Timiryasov}},\ }\bibfield  {title} {\bibinfo {title} {{Reconciling resonant
  leptogenesis and baryogenesis via neutrino oscillations}},\ }\href
  {https://doi.org/10.1103/PhysRevD.104.055010} {\bibfield  {journal} {\bibinfo
   {journal} {Phys. Rev. D}\ }\textbf {\bibinfo {volume} {104}},\ \bibinfo
  {pages} {055010} (\bibinfo {year} {2021}{\natexlab{a}})},\ \Eprint
  {https://arxiv.org/abs/2103.16545} {arXiv:2103.16545 [hep-ph]} \BibitemShut
  {NoStop}%
\bibitem [{\citenamefont {Klari\'c}\ \emph
  {et~al.}(2021{\natexlab{b}})\citenamefont {Klari\'c}, \citenamefont
  {Shaposhnikov},\ and\ \citenamefont {Timiryasov}}]{Klaric:2020phc}%
  \BibitemOpen
  \bibfield  {author} {\bibinfo {author} {\bibfnamefont {J.}~\bibnamefont
  {Klari\'c}}, \bibinfo {author} {\bibfnamefont {M.}~\bibnamefont
  {Shaposhnikov}},\ and\ \bibinfo {author} {\bibfnamefont {I.}~\bibnamefont
  {Timiryasov}},\ }\bibfield  {title} {\bibinfo {title} {{Uniting Low-Scale
  Leptogenesis Mechanisms}},\ }\href
  {https://doi.org/10.1103/PhysRevLett.127.111802} {\bibfield  {journal}
  {\bibinfo  {journal} {Phys. Rev. Lett.}\ }\textbf {\bibinfo {volume} {127}},\
  \bibinfo {pages} {111802} (\bibinfo {year} {2021}{\natexlab{b}})},\ \Eprint
  {https://arxiv.org/abs/2008.13771} {arXiv:2008.13771 [hep-ph]} \BibitemShut
  {NoStop}%
\bibitem [{\citenamefont {Canetti}\ \emph {et~al.}(2013)\citenamefont
  {Canetti}, \citenamefont {Drewes}, \citenamefont {Frossard},\ and\
  \citenamefont {Shaposhnikov}}]{Canetti:2012kh}%
  \BibitemOpen
  \bibfield  {author} {\bibinfo {author} {\bibfnamefont {L.}~\bibnamefont
  {Canetti}}, \bibinfo {author} {\bibfnamefont {M.}~\bibnamefont {Drewes}},
  \bibinfo {author} {\bibfnamefont {T.}~\bibnamefont {Frossard}},\ and\
  \bibinfo {author} {\bibfnamefont {M.}~\bibnamefont {Shaposhnikov}},\
  }\bibfield  {title} {\bibinfo {title} {{Dark Matter, Baryogenesis and
  Neutrino Oscillations from Right Handed Neutrinos}},\ }\href
  {https://doi.org/10.1103/PhysRevD.87.093006} {\bibfield  {journal} {\bibinfo
  {journal} {Phys. Rev. D}\ }\textbf {\bibinfo {volume} {87}},\ \bibinfo
  {pages} {093006} (\bibinfo {year} {2013})},\ \Eprint
  {https://arxiv.org/abs/1208.4607} {arXiv:1208.4607 [hep-ph]} \BibitemShut
  {NoStop}%
\bibitem [{\citenamefont {Hern\'andez}\ \emph {et~al.}(2015)\citenamefont
  {Hern\'andez}, \citenamefont {Kekic}, \citenamefont {L\'opez-Pav\'on},
  \citenamefont {Racker},\ and\ \citenamefont {Rius}}]{Hernandez:2015wna}%
  \BibitemOpen
  \bibfield  {author} {\bibinfo {author} {\bibfnamefont {P.}~\bibnamefont
  {Hern\'andez}}, \bibinfo {author} {\bibfnamefont {M.}~\bibnamefont {Kekic}},
  \bibinfo {author} {\bibfnamefont {J.}~\bibnamefont {L\'opez-Pav\'on}},
  \bibinfo {author} {\bibfnamefont {J.}~\bibnamefont {Racker}},\ and\ \bibinfo
  {author} {\bibfnamefont {N.}~\bibnamefont {Rius}},\ }\bibfield  {title}
  {\bibinfo {title} {{Leptogenesis in GeV scale seesaw models}},\ }\href
  {https://doi.org/10.1007/JHEP10(2015)067} {\bibfield  {journal} {\bibinfo
  {journal} {JHEP}\ }\textbf {\bibinfo {volume} {10}},\ \bibinfo {pages}
  {067}},\ \Eprint {https://arxiv.org/abs/1508.03676} {arXiv:1508.03676
  [hep-ph]} \BibitemShut {NoStop}%
\bibitem [{\citenamefont {Hern\'andez}\ \emph {et~al.}(2016)\citenamefont
  {Hern\'andez}, \citenamefont {Kekic}, \citenamefont {L\'opez-Pav\'on},
  \citenamefont {Racker},\ and\ \citenamefont {Salvado}}]{Hernandez:2016kel}%
  \BibitemOpen
  \bibfield  {author} {\bibinfo {author} {\bibfnamefont {P.}~\bibnamefont
  {Hern\'andez}}, \bibinfo {author} {\bibfnamefont {M.}~\bibnamefont {Kekic}},
  \bibinfo {author} {\bibfnamefont {J.}~\bibnamefont {L\'opez-Pav\'on}},
  \bibinfo {author} {\bibfnamefont {J.}~\bibnamefont {Racker}},\ and\ \bibinfo
  {author} {\bibfnamefont {J.}~\bibnamefont {Salvado}},\ }\bibfield  {title}
  {\bibinfo {title} {{Testable Baryogenesis in Seesaw Models}},\ }\href
  {https://doi.org/10.1007/JHEP08(2016)157} {\bibfield  {journal} {\bibinfo
  {journal} {JHEP}\ }\textbf {\bibinfo {volume} {08}},\ \bibinfo {pages}
  {157}},\ \Eprint {https://arxiv.org/abs/1606.06719} {arXiv:1606.06719
  [hep-ph]} \BibitemShut {NoStop}%
\bibitem [{\citenamefont {Hernandez}\ \emph {et~al.}(2022)\citenamefont
  {Hernandez}, \citenamefont {Lopez-Pavon}, \citenamefont {Rius},\ and\
  \citenamefont {Sandner}}]{Hernandez:2022ivz}%
  \BibitemOpen
  \bibfield  {author} {\bibinfo {author} {\bibfnamefont {P.}~\bibnamefont
  {Hernandez}}, \bibinfo {author} {\bibfnamefont {J.}~\bibnamefont
  {Lopez-Pavon}}, \bibinfo {author} {\bibfnamefont {N.}~\bibnamefont {Rius}},\
  and\ \bibinfo {author} {\bibfnamefont {S.}~\bibnamefont {Sandner}},\
  }\bibfield  {title} {\bibinfo {title} {{Bounds on right-handed neutrino
  parameters from observable leptogenesis}},\ }\href
  {https://doi.org/10.1007/JHEP12(2022)012} {\bibfield  {journal} {\bibinfo
  {journal} {JHEP}\ }\textbf {\bibinfo {volume} {12}},\ \bibinfo {pages}
  {012}},\ \Eprint {https://arxiv.org/abs/2207.01651} {arXiv:2207.01651
  [hep-ph]} \BibitemShut {NoStop}%
\bibitem [{\citenamefont {Sandner}\ \emph {et~al.}(2023)\citenamefont
  {Sandner}, \citenamefont {Hernandez}, \citenamefont {Lopez-Pavon},\ and\
  \citenamefont {Rius}}]{Sandner:2023tcg}%
  \BibitemOpen
  \bibfield  {author} {\bibinfo {author} {\bibfnamefont {S.}~\bibnamefont
  {Sandner}}, \bibinfo {author} {\bibfnamefont {P.}~\bibnamefont {Hernandez}},
  \bibinfo {author} {\bibfnamefont {J.}~\bibnamefont {Lopez-Pavon}},\ and\
  \bibinfo {author} {\bibfnamefont {N.}~\bibnamefont {Rius}},\ }\bibfield
  {title} {\bibinfo {title} {{Predicting the baryon asymmetry with degenerate
  right-handed neutrinos}},\ }\href {https://doi.org/10.1007/JHEP11(2023)153}
  {\bibfield  {journal} {\bibinfo  {journal} {JHEP}\ }\textbf {\bibinfo
  {volume} {11}},\ \bibinfo {pages} {153}},\ \Eprint
  {https://arxiv.org/abs/2305.14427} {arXiv:2305.14427 [hep-ph]} \BibitemShut
  {NoStop}%
\bibitem [{\citenamefont {Abada}\ \emph {et~al.}(2019)\citenamefont {Abada},
  \citenamefont {Arcadi}, \citenamefont {Domcke}, \citenamefont {Drewes},
  \citenamefont {Klaric},\ and\ \citenamefont {Lucente}}]{Abada:2018oly}%
  \BibitemOpen
  \bibfield  {author} {\bibinfo {author} {\bibfnamefont {A.}~\bibnamefont
  {Abada}}, \bibinfo {author} {\bibfnamefont {G.}~\bibnamefont {Arcadi}},
  \bibinfo {author} {\bibfnamefont {V.}~\bibnamefont {Domcke}}, \bibinfo
  {author} {\bibfnamefont {M.}~\bibnamefont {Drewes}}, \bibinfo {author}
  {\bibfnamefont {J.}~\bibnamefont {Klaric}},\ and\ \bibinfo {author}
  {\bibfnamefont {M.}~\bibnamefont {Lucente}},\ }\bibfield  {title} {\bibinfo
  {title} {{Low-scale leptogenesis with three heavy neutrinos}},\ }\href
  {https://doi.org/10.1007/JHEP01(2019)164} {\bibfield  {journal} {\bibinfo
  {journal} {JHEP}\ }\textbf {\bibinfo {volume} {01}},\ \bibinfo {pages}
  {164}},\ \Eprint {https://arxiv.org/abs/1810.12463} {arXiv:1810.12463
  [hep-ph]} \BibitemShut {NoStop}%
\bibitem [{\citenamefont {Asaka}\ \emph {et~al.}(2012)\citenamefont {Asaka},
  \citenamefont {Eijima},\ and\ \citenamefont {Ishida}}]{Asaka:2011wq}%
  \BibitemOpen
  \bibfield  {author} {\bibinfo {author} {\bibfnamefont {T.}~\bibnamefont
  {Asaka}}, \bibinfo {author} {\bibfnamefont {S.}~\bibnamefont {Eijima}},\ and\
  \bibinfo {author} {\bibfnamefont {H.}~\bibnamefont {Ishida}},\ }\bibfield
  {title} {\bibinfo {title} {{Kinetic Equations for Baryogenesis via Sterile
  Neutrino Oscillation}},\ }\href
  {https://doi.org/10.1088/1475-7516/2012/02/021} {\bibfield  {journal}
  {\bibinfo  {journal} {JCAP}\ }\textbf {\bibinfo {volume} {02}},\ \bibinfo
  {pages} {021}},\ \Eprint {https://arxiv.org/abs/1112.5565} {arXiv:1112.5565
  [hep-ph]} \BibitemShut {NoStop}%
\bibitem [{\citenamefont {Ghiglieri}\ and\ \citenamefont
  {Laine}(2017)}]{Ghiglieri:2017gjz}%
  \BibitemOpen
  \bibfield  {author} {\bibinfo {author} {\bibfnamefont {J.}~\bibnamefont
  {Ghiglieri}}\ and\ \bibinfo {author} {\bibfnamefont {M.}~\bibnamefont
  {Laine}},\ }\bibfield  {title} {\bibinfo {title} {{GeV-scale hot sterile
  neutrino oscillations: a derivation of evolution equations}},\ }\href
  {https://doi.org/10.1007/JHEP05(2017)132} {\bibfield  {journal} {\bibinfo
  {journal} {JHEP}\ }\textbf {\bibinfo {volume} {05}},\ \bibinfo {pages}
  {132}},\ \Eprint {https://arxiv.org/abs/1703.06087} {arXiv:1703.06087
  [hep-ph]} \BibitemShut {NoStop}%
\bibitem [{\citenamefont {Ghiglieri}\ and\ \citenamefont
  {Laine}(2019)}]{Ghiglieri:2018wbs}%
  \BibitemOpen
  \bibfield  {author} {\bibinfo {author} {\bibfnamefont {J.}~\bibnamefont
  {Ghiglieri}}\ and\ \bibinfo {author} {\bibfnamefont {M.}~\bibnamefont
  {Laine}},\ }\bibfield  {title} {\bibinfo {title} {{Precision study of
  GeV-scale resonant leptogenesis}},\ }\href
  {https://doi.org/10.1007/JHEP02(2019)014} {\bibfield  {journal} {\bibinfo
  {journal} {JHEP}\ }\textbf {\bibinfo {volume} {02}},\ \bibinfo {pages}
  {014}},\ \Eprint {https://arxiv.org/abs/1811.01971} {arXiv:1811.01971
  [hep-ph]} \BibitemShut {NoStop}%
\bibitem [{\citenamefont {Drewes}\ \emph {et~al.}(2018)\citenamefont {Drewes},
  \citenamefont {Garbrecht}, \citenamefont {Hernandez}, \citenamefont {Kekic},
  \citenamefont {Lopez-Pavon}, \citenamefont {Racker}, \citenamefont {Rius},
  \citenamefont {Salvado},\ and\ \citenamefont {Teresi}}]{Drewes:2017zyw}%
  \BibitemOpen
  \bibfield  {author} {\bibinfo {author} {\bibfnamefont {M.}~\bibnamefont
  {Drewes}}, \bibinfo {author} {\bibfnamefont {B.}~\bibnamefont {Garbrecht}},
  \bibinfo {author} {\bibfnamefont {P.}~\bibnamefont {Hernandez}}, \bibinfo
  {author} {\bibfnamefont {M.}~\bibnamefont {Kekic}}, \bibinfo {author}
  {\bibfnamefont {J.}~\bibnamefont {Lopez-Pavon}}, \bibinfo {author}
  {\bibfnamefont {J.}~\bibnamefont {Racker}}, \bibinfo {author} {\bibfnamefont
  {N.}~\bibnamefont {Rius}}, \bibinfo {author} {\bibfnamefont {J.}~\bibnamefont
  {Salvado}},\ and\ \bibinfo {author} {\bibfnamefont {D.}~\bibnamefont
  {Teresi}},\ }\bibfield  {title} {\bibinfo {title} {{ARS Leptogenesis}},\
  }\href {https://doi.org/10.1142/S0217751X18420022} {\bibfield  {journal}
  {\bibinfo  {journal} {Int. J. Mod. Phys. A}\ }\textbf {\bibinfo {volume}
  {33}},\ \bibinfo {pages} {1842002} (\bibinfo {year} {2018})},\ \Eprint
  {https://arxiv.org/abs/1711.02862} {arXiv:1711.02862 [hep-ph]} \BibitemShut
  {NoStop}%
\bibitem [{\citenamefont {Navas}\ \emph {et~al.}(2024)\citenamefont {Navas}
  \emph {et~al.}}]{ParticleDataGroup:2024cfk}%
  \BibitemOpen
  \bibfield  {author} {\bibinfo {author} {\bibfnamefont {S.}~\bibnamefont
  {Navas}} \emph {et~al.} (\bibinfo {collaboration} {Particle Data Group}),\
  }\bibfield  {title} {\bibinfo {title} {{Review of particle physics}},\ }\href
  {https://doi.org/10.1103/PhysRevD.110.030001} {\bibfield  {journal} {\bibinfo
   {journal} {Phys. Rev. D}\ }\textbf {\bibinfo {volume} {110}},\ \bibinfo
  {pages} {030001} (\bibinfo {year} {2024})}\BibitemShut {NoStop}%
\bibitem [{\citenamefont {Gava}\ and\ \citenamefont
  {Volpe}(2010)}]{Gava:2010kz}%
  \BibitemOpen
  \bibfield  {author} {\bibinfo {author} {\bibfnamefont {J.}~\bibnamefont
  {Gava}}\ and\ \bibinfo {author} {\bibfnamefont {C.}~\bibnamefont {Volpe}},\
  }\bibfield  {title} {\bibinfo {title} {{CP violation effects on the neutrino
  degeneracy parameters in the Early Universe}},\ }\href
  {https://doi.org/10.1016/j.nuclphysb.2010.04.024} {\bibfield  {journal}
  {\bibinfo  {journal} {Nucl. Phys. B}\ }\textbf {\bibinfo {volume} {837}},\
  \bibinfo {pages} {50} (\bibinfo {year} {2010})},\ \bibinfo {note} {[Erratum:
  Nucl.Phys.B 957, 115035 (2020)]},\ \Eprint {https://arxiv.org/abs/1002.0981}
  {arXiv:1002.0981 [hep-ph]} \BibitemShut {NoStop}%
\bibitem [{\citenamefont {Akhmedov}\ \emph {et~al.}(2002)\citenamefont
  {Akhmedov}, \citenamefont {Lunardini},\ and\ \citenamefont
  {Smirnov}}]{Akhmedov:2002zj}%
  \BibitemOpen
  \bibfield  {author} {\bibinfo {author} {\bibfnamefont {E.~K.}\ \bibnamefont
  {Akhmedov}}, \bibinfo {author} {\bibfnamefont {C.}~\bibnamefont
  {Lunardini}},\ and\ \bibinfo {author} {\bibfnamefont {A.~Y.}\ \bibnamefont
  {Smirnov}},\ }\bibfield  {title} {\bibinfo {title} {{Supernova neutrinos:
  Difference of muon-neutrino - tau-neutrino fluxes and conversion effects}},\
  }\href {https://doi.org/10.1016/S0550-3213(02)00692-2} {\bibfield  {journal}
  {\bibinfo  {journal} {Nucl. Phys. B}\ }\textbf {\bibinfo {volume} {643}},\
  \bibinfo {pages} {339} (\bibinfo {year} {2002})},\ \Eprint
  {https://arxiv.org/abs/hep-ph/0204091} {arXiv:hep-ph/0204091} \BibitemShut
  {NoStop}%
\bibitem [{\citenamefont {Balantekin}\ \emph {et~al.}(2008)\citenamefont
  {Balantekin}, \citenamefont {Gava},\ and\ \citenamefont
  {Volpe}}]{Balantekin:2007es}%
  \BibitemOpen
  \bibfield  {author} {\bibinfo {author} {\bibfnamefont {A.~B.}\ \bibnamefont
  {Balantekin}}, \bibinfo {author} {\bibfnamefont {J.}~\bibnamefont {Gava}},\
  and\ \bibinfo {author} {\bibfnamefont {C.}~\bibnamefont {Volpe}},\ }\bibfield
   {title} {\bibinfo {title} {{Possible CP-Violation effects in core-collapse
  Supernovae}},\ }\href {https://doi.org/10.1016/j.physletb.2008.03.038}
  {\bibfield  {journal} {\bibinfo  {journal} {Phys. Lett. B}\ }\textbf
  {\bibinfo {volume} {662}},\ \bibinfo {pages} {396} (\bibinfo {year}
  {2008})},\ \Eprint {https://arxiv.org/abs/0710.3112} {arXiv:0710.3112
  [astro-ph]} \BibitemShut {NoStop}%
\bibitem [{\citenamefont {N\"otzold}\ and\ \citenamefont
  {Raffelt}(1988)}]{Notzold:1987ik}%
  \BibitemOpen
  \bibfield  {author} {\bibinfo {author} {\bibfnamefont {D.}~\bibnamefont
  {N\"otzold}}\ and\ \bibinfo {author} {\bibfnamefont {G.}~\bibnamefont
  {Raffelt}},\ }\bibfield  {title} {\bibinfo {title} {{Neutrino dispersion at
  finite temperature and density}},\ }\href
  {https://doi.org/10.1016/0550-3213(88)90113-7} {\bibfield  {journal}
  {\bibinfo  {journal} {Nucl. Phys. B}\ }\textbf {\bibinfo {volume} {307}},\
  \bibinfo {pages} {924} (\bibinfo {year} {1988})}\BibitemShut {NoStop}%
\bibitem [{\citenamefont {Akita}\ and\ \citenamefont
  {Yamaguchi}(2020)}]{Akita:2020szl}%
  \BibitemOpen
  \bibfield  {author} {\bibinfo {author} {\bibfnamefont {K.}~\bibnamefont
  {Akita}}\ and\ \bibinfo {author} {\bibfnamefont {M.}~\bibnamefont
  {Yamaguchi}},\ }\bibfield  {title} {\bibinfo {title} {{A precision
  calculation of relic neutrino decoupling}},\ }\href
  {https://doi.org/10.1088/1475-7516/2020/08/012} {\bibfield  {journal}
  {\bibinfo  {journal} {JCAP}\ }\textbf {\bibinfo {volume} {08}},\ \bibinfo
  {pages} {012}},\ \Eprint {https://arxiv.org/abs/2005.07047} {arXiv:2005.07047
  [hep-ph]} \BibitemShut {NoStop}%
\bibitem [{\citenamefont {Ovchynnikov}\ and\ \citenamefont
  {Syvolap}(2024{\natexlab{a}})}]{Ovchynnikov:2024rfu}%
  \BibitemOpen
  \bibfield  {author} {\bibinfo {author} {\bibfnamefont {M.}~\bibnamefont
  {Ovchynnikov}}\ and\ \bibinfo {author} {\bibfnamefont {V.}~\bibnamefont
  {Syvolap}},\ }\bibfield  {title} {\bibinfo {title} {{How new physics affects
  primordial neutrinos decoupling: Direct Simulation Monte Carlo approach}},\
  }\href@noop {} {\  (\bibinfo {year} {2024}{\natexlab{a}})},\ \Eprint
  {https://arxiv.org/abs/2409.07378} {arXiv:2409.07378 [astro-ph.CO]}
  \BibitemShut {NoStop}%
\bibitem [{\citenamefont {Ovchynnikov}\ and\ \citenamefont
  {Syvolap}(2024{\natexlab{b}})}]{Ovchynnikov:2024xyd}%
  \BibitemOpen
  \bibfield  {author} {\bibinfo {author} {\bibfnamefont {M.}~\bibnamefont
  {Ovchynnikov}}\ and\ \bibinfo {author} {\bibfnamefont {V.}~\bibnamefont
  {Syvolap}},\ }\bibfield  {title} {\bibinfo {title} {{Primordial neutrinos and
  new physics: novel approach to solving neutrino Boltzmann equation}},\
  }\href@noop {} {\  (\bibinfo {year} {2024}{\natexlab{b}})},\ \Eprint
  {https://arxiv.org/abs/2409.15129} {arXiv:2409.15129 [hep-ph]} \BibitemShut
  {NoStop}%
\bibitem [{\citenamefont {Pastor}\ \emph {et~al.}(2002)\citenamefont {Pastor},
  \citenamefont {Raffelt},\ and\ \citenamefont {Semikoz}}]{Pastor:2001iu}%
  \BibitemOpen
  \bibfield  {author} {\bibinfo {author} {\bibfnamefont {S.}~\bibnamefont
  {Pastor}}, \bibinfo {author} {\bibfnamefont {G.~G.}\ \bibnamefont
  {Raffelt}},\ and\ \bibinfo {author} {\bibfnamefont {D.~V.}\ \bibnamefont
  {Semikoz}},\ }\bibfield  {title} {\bibinfo {title} {{Physics of synchronized
  neutrino oscillations caused by selfinteractions}},\ }\href
  {https://doi.org/10.1103/PhysRevD.65.053011} {\bibfield  {journal} {\bibinfo
  {journal} {Phys. Rev. D}\ }\textbf {\bibinfo {volume} {65}},\ \bibinfo
  {pages} {053011} (\bibinfo {year} {2002})},\ \Eprint
  {https://arxiv.org/abs/hep-ph/0109035} {arXiv:hep-ph/0109035} \BibitemShut
  {NoStop}%
\bibitem [{\citenamefont {Cielo}\ \emph {et~al.}(2023)\citenamefont {Cielo},
  \citenamefont {Escudero}, \citenamefont {Mangano},\ and\ \citenamefont
  {Pisanti}}]{Cielo:2023bqp}%
  \BibitemOpen
  \bibfield  {author} {\bibinfo {author} {\bibfnamefont {M.}~\bibnamefont
  {Cielo}}, \bibinfo {author} {\bibfnamefont {M.}~\bibnamefont {Escudero}},
  \bibinfo {author} {\bibfnamefont {G.}~\bibnamefont {Mangano}},\ and\ \bibinfo
  {author} {\bibfnamefont {O.}~\bibnamefont {Pisanti}},\ }\bibfield  {title}
  {\bibinfo {title} {{Neff in the Standard Model at NLO is 3.043}},\ }\href
  {https://doi.org/10.1103/PhysRevD.108.L121301} {\bibfield  {journal}
  {\bibinfo  {journal} {Phys. Rev. D}\ }\textbf {\bibinfo {volume} {108}},\
  \bibinfo {pages} {L121301} (\bibinfo {year} {2023})},\ \Eprint
  {https://arxiv.org/abs/2306.05460} {arXiv:2306.05460 [hep-ph]} \BibitemShut
  {NoStop}%
\bibitem [{\citenamefont {Jackson}\ and\ \citenamefont
  {Laine}(2024{\natexlab{a}})}]{Jackson:2023zkl}%
  \BibitemOpen
  \bibfield  {author} {\bibinfo {author} {\bibfnamefont {G.}~\bibnamefont
  {Jackson}}\ and\ \bibinfo {author} {\bibfnamefont {M.}~\bibnamefont
  {Laine}},\ }\bibfield  {title} {\bibinfo {title} {{QED corrections to the
  thermal neutrino interaction rate}},\ }\href
  {https://doi.org/10.1007/JHEP05(2024)089} {\bibfield  {journal} {\bibinfo
  {journal} {JHEP}\ }\textbf {\bibinfo {volume} {05}},\ \bibinfo {pages}
  {089}},\ \Eprint {https://arxiv.org/abs/2312.07015} {arXiv:2312.07015
  [hep-ph]} \BibitemShut {NoStop}%
\bibitem [{\citenamefont {Jackson}\ and\ \citenamefont
  {Laine}(2024{\natexlab{b}})}]{Jackson:2024gtr}%
  \BibitemOpen
  \bibfield  {author} {\bibinfo {author} {\bibfnamefont {G.}~\bibnamefont
  {Jackson}}\ and\ \bibinfo {author} {\bibfnamefont {M.}~\bibnamefont
  {Laine}},\ }\bibfield  {title} {\bibinfo {title} {{Neutrino-antineutrino
  production, annihilation, and scattering at MeV temperatures and NLO
  accuracy}},\ }\href@noop {} {\  (\bibinfo {year} {2024}{\natexlab{b}})},\
  \Eprint {https://arxiv.org/abs/2412.03958} {arXiv:2412.03958 [hep-ph]}
  \BibitemShut {NoStop}%
\bibitem [{\citenamefont {Drewes}\ \emph
  {et~al.}(2024{\natexlab{a}})\citenamefont {Drewes}, \citenamefont {Georis},
  \citenamefont {Klasen}, \citenamefont {Wiggering},\ and\ \citenamefont
  {Wong}}]{Drewes:2024wbw}%
  \BibitemOpen
  \bibfield  {author} {\bibinfo {author} {\bibfnamefont {M.}~\bibnamefont
  {Drewes}}, \bibinfo {author} {\bibfnamefont {Y.}~\bibnamefont {Georis}},
  \bibinfo {author} {\bibfnamefont {M.}~\bibnamefont {Klasen}}, \bibinfo
  {author} {\bibfnamefont {L.~P.}\ \bibnamefont {Wiggering}},\ and\ \bibinfo
  {author} {\bibfnamefont {Y.~Y.~Y.}\ \bibnamefont {Wong}},\ }\bibfield
  {title} {\bibinfo {title} {{Towards a precision calculation of N $_{eff}$ in
  the Standard Model. Part III. Improved estimate of NLO contributions to the
  collision integral}},\ }\href {https://doi.org/10.1088/1475-7516/2024/06/032}
  {\bibfield  {journal} {\bibinfo  {journal} {JCAP}\ }\textbf {\bibinfo
  {volume} {06}},\ \bibinfo {pages} {032}},\ \Eprint
  {https://arxiv.org/abs/2402.18481} {arXiv:2402.18481 [hep-ph]} \BibitemShut
  {NoStop}%
\bibitem [{\citenamefont {Drewes}\ \emph
  {et~al.}(2024{\natexlab{b}})\citenamefont {Drewes}, \citenamefont {Georis},
  \citenamefont {Klasen}, \citenamefont {Pierobon},\ and\ \citenamefont
  {Wong}}]{Drewes:2024nbg}%
  \BibitemOpen
  \bibfield  {author} {\bibinfo {author} {\bibfnamefont {M.}~\bibnamefont
  {Drewes}}, \bibinfo {author} {\bibfnamefont {Y.}~\bibnamefont {Georis}},
  \bibinfo {author} {\bibfnamefont {M.}~\bibnamefont {Klasen}}, \bibinfo
  {author} {\bibfnamefont {G.}~\bibnamefont {Pierobon}},\ and\ \bibinfo
  {author} {\bibfnamefont {Y.~Y.~Y.}\ \bibnamefont {Wong}},\ }\bibfield
  {title} {\bibinfo {title} {{Towards a precision calculation of $N_{\rm eff}$
  in the Standard Model IV: Impact of positronium formation}},\ }\href@noop {}
  {\  (\bibinfo {year} {2024}{\natexlab{b}})},\ \Eprint
  {https://arxiv.org/abs/2411.14091} {arXiv:2411.14091 [hep-ph]} \BibitemShut
  {NoStop}%
\bibitem [{\citenamefont {McKellar}\ and\ \citenamefont
  {Thomson}(1994)}]{McKellar:1992ja}%
  \BibitemOpen
  \bibfield  {author} {\bibinfo {author} {\bibfnamefont {B.~H.~J.}\
  \bibnamefont {McKellar}}\ and\ \bibinfo {author} {\bibfnamefont {M.~J.}\
  \bibnamefont {Thomson}},\ }\bibfield  {title} {\bibinfo {title} {{Oscillating
  doublet neutrinos in the early universe}},\ }\href
  {https://doi.org/10.1103/PhysRevD.49.2710} {\bibfield  {journal} {\bibinfo
  {journal} {Phys. Rev. D}\ }\textbf {\bibinfo {volume} {49}},\ \bibinfo
  {pages} {2710} (\bibinfo {year} {1994})}\BibitemShut {NoStop}%
\bibitem [{\citenamefont {Escudero}(2019)}]{Escudero:2018mvt}%
  \BibitemOpen
  \bibfield  {author} {\bibinfo {author} {\bibfnamefont {M.}~\bibnamefont
  {Escudero}},\ }\bibfield  {title} {\bibinfo {title} {{Neutrino decoupling
  beyond the Standard Model: CMB constraints on the Dark Matter mass with a
  fast and precise $N_{\rm eff}$ evaluation}},\ }\href
  {https://doi.org/10.1088/1475-7516/2019/02/007} {\bibfield  {journal}
  {\bibinfo  {journal} {JCAP}\ }\textbf {\bibinfo {volume} {02}},\ \bibinfo
  {pages} {007}},\ \Eprint {https://arxiv.org/abs/1812.05605} {arXiv:1812.05605
  [hep-ph]} \BibitemShut {NoStop}%
\bibitem [{\citenamefont {Escudero~Abenza}(2020)}]{EscuderoAbenza:2020cmq}%
  \BibitemOpen
  \bibfield  {author} {\bibinfo {author} {\bibfnamefont {M.}~\bibnamefont
  {Escudero~Abenza}},\ }\bibfield  {title} {\bibinfo {title} {{Precision early
  universe thermodynamics made simple: $N_{\rm eff}$ and neutrino decoupling in
  the Standard Model and beyond}},\ }\href
  {https://doi.org/10.1088/1475-7516/2020/05/048} {\bibfield  {journal}
  {\bibinfo  {journal} {JCAP}\ }\textbf {\bibinfo {volume} {05}},\ \bibinfo
  {pages} {048}},\ \Eprint {https://arxiv.org/abs/2001.04466} {arXiv:2001.04466
  [hep-ph]} \BibitemShut {NoStop}%
\bibitem [{\citenamefont {Bell}\ \emph {et~al.}(1999)\citenamefont {Bell},
  \citenamefont {Volkas},\ and\ \citenamefont {Wong}}]{Bell:1998ds}%
  \BibitemOpen
  \bibfield  {author} {\bibinfo {author} {\bibfnamefont {N.~F.}\ \bibnamefont
  {Bell}}, \bibinfo {author} {\bibfnamefont {R.~R.}\ \bibnamefont {Volkas}},\
  and\ \bibinfo {author} {\bibfnamefont {Y.~Y.~Y.}\ \bibnamefont {Wong}},\
  }\bibfield  {title} {\bibinfo {title} {{Relic neutrino asymmetry evolution
  from first principles}},\ }\href {https://doi.org/10.1103/PhysRevD.59.113001}
  {\bibfield  {journal} {\bibinfo  {journal} {Phys. Rev. D}\ }\textbf {\bibinfo
  {volume} {59}},\ \bibinfo {pages} {113001} (\bibinfo {year} {1999})},\
  \Eprint {https://arxiv.org/abs/hep-ph/9809363} {arXiv:hep-ph/9809363}
  \BibitemShut {NoStop}%
\bibitem [{\citenamefont {Arg\"uelles~Delgado}\ \emph
  {et~al.}(2015)\citenamefont {Arg\"uelles~Delgado}, \citenamefont {Salvado},\
  and\ \citenamefont {Weaver}}]{ArguellesDelgado:2014rca}%
  \BibitemOpen
  \bibfield  {author} {\bibinfo {author} {\bibfnamefont {C.~A.}\ \bibnamefont
  {Arg\"uelles~Delgado}}, \bibinfo {author} {\bibfnamefont {J.}~\bibnamefont
  {Salvado}},\ and\ \bibinfo {author} {\bibfnamefont {C.~N.}\ \bibnamefont
  {Weaver}},\ }\bibfield  {title} {\bibinfo {title} {{A Simple Quantum
  Integro-Differential Solver (SQuIDS)}},\ }\href
  {https://doi.org/10.1016/j.cpc.2015.06.022} {\bibfield  {journal} {\bibinfo
  {journal} {Comput. Phys. Commun.}\ }\textbf {\bibinfo {volume} {196}},\
  \bibinfo {pages} {569} (\bibinfo {year} {2015})},\ \Eprint
  {https://arxiv.org/abs/1412.3832} {arXiv:1412.3832 [hep-ph]} \BibitemShut
  {NoStop}%
\bibitem [{\citenamefont {Gough}(2009)}]{gnu}%
  \BibitemOpen
  \bibfield  {author} {\bibinfo {author} {\bibfnamefont {B.}~\bibnamefont
  {Gough}},\ }\href@noop {} {\emph {\bibinfo {title} {GNU scientific library
  reference manual}}}\ (\bibinfo  {publisher} {Network Theory Ltd.},\ \bibinfo
  {year} {2009})\BibitemShut {NoStop}%
\bibitem [{\citenamefont {Abazajian}\ \emph {et~al.}(2002)\citenamefont
  {Abazajian}, \citenamefont {Beacom},\ and\ \citenamefont
  {Bell}}]{Abazajian:2002qx}%
  \BibitemOpen
  \bibfield  {author} {\bibinfo {author} {\bibfnamefont {K.~N.}\ \bibnamefont
  {Abazajian}}, \bibinfo {author} {\bibfnamefont {J.~F.}\ \bibnamefont
  {Beacom}},\ and\ \bibinfo {author} {\bibfnamefont {N.~F.}\ \bibnamefont
  {Bell}},\ }\bibfield  {title} {\bibinfo {title} {{Stringent Constraints on
  Cosmological Neutrino Antineutrino Asymmetries from Synchronized Flavor
  Transformation}},\ }\href {https://doi.org/10.1103/PhysRevD.66.013008}
  {\bibfield  {journal} {\bibinfo  {journal} {Phys. Rev. D}\ }\textbf {\bibinfo
  {volume} {66}},\ \bibinfo {pages} {013008} (\bibinfo {year} {2002})},\
  \Eprint {https://arxiv.org/abs/astro-ph/0203442} {arXiv:astro-ph/0203442}
  \BibitemShut {NoStop}%
\bibitem [{\citenamefont {Wong}(2002)}]{Wong:2002fa}%
  \BibitemOpen
  \bibfield  {author} {\bibinfo {author} {\bibfnamefont {Y.~Y.~Y.}\
  \bibnamefont {Wong}},\ }\bibfield  {title} {\bibinfo {title} {{Analytical
  treatment of neutrino asymmetry equilibration from flavor oscillations in the
  early universe}},\ }\href {https://doi.org/10.1103/PhysRevD.66.025015}
  {\bibfield  {journal} {\bibinfo  {journal} {Phys. Rev. D}\ }\textbf {\bibinfo
  {volume} {66}},\ \bibinfo {pages} {025015} (\bibinfo {year} {2002})},\
  \Eprint {https://arxiv.org/abs/hep-ph/0203180} {arXiv:hep-ph/0203180}
  \BibitemShut {NoStop}%
\bibitem [{\citenamefont {Ade}\ \emph {et~al.}(2019)\citenamefont {Ade} \emph
  {et~al.}}]{SimonsObservatory:2018koc}%
  \BibitemOpen
  \bibfield  {author} {\bibinfo {author} {\bibfnamefont {P.}~\bibnamefont
  {Ade}} \emph {et~al.} (\bibinfo {collaboration} {Simons Observatory}),\
  }\bibfield  {title} {\bibinfo {title} {{The Simons Observatory: Science goals
  and forecasts}},\ }\href {https://doi.org/10.1088/1475-7516/2019/02/056}
  {\bibfield  {journal} {\bibinfo  {journal} {JCAP}\ }\textbf {\bibinfo
  {volume} {02}},\ \bibinfo {pages} {056}},\ \Eprint
  {https://arxiv.org/abs/1808.07445} {arXiv:1808.07445 [astro-ph.CO]}
  \BibitemShut {NoStop}%
\bibitem [{\citenamefont {D'Onofrio}\ \emph {et~al.}(2014)\citenamefont
  {D'Onofrio}, \citenamefont {Rummukainen},\ and\ \citenamefont
  {Tranberg}}]{DOnofrio:2014rug}%
  \BibitemOpen
  \bibfield  {author} {\bibinfo {author} {\bibfnamefont {M.}~\bibnamefont
  {D'Onofrio}}, \bibinfo {author} {\bibfnamefont {K.}~\bibnamefont
  {Rummukainen}},\ and\ \bibinfo {author} {\bibfnamefont {A.}~\bibnamefont
  {Tranberg}},\ }\bibfield  {title} {\bibinfo {title} {{Sphaleron Rate in the
  Minimal Standard Model}},\ }\href
  {https://doi.org/10.1103/PhysRevLett.113.141602} {\bibfield  {journal}
  {\bibinfo  {journal} {Phys. Rev. Lett.}\ }\textbf {\bibinfo {volume} {113}},\
  \bibinfo {pages} {141602} (\bibinfo {year} {2014})},\ \Eprint
  {https://arxiv.org/abs/1404.3565} {arXiv:1404.3565 [hep-ph]} \BibitemShut
  {NoStop}%
\bibitem [{\citenamefont {Affleck}\ and\ \citenamefont
  {Dine}(1985)}]{Affleck:1984fy}%
  \BibitemOpen
  \bibfield  {author} {\bibinfo {author} {\bibfnamefont {I.}~\bibnamefont
  {Affleck}}\ and\ \bibinfo {author} {\bibfnamefont {M.}~\bibnamefont {Dine}},\
  }\bibfield  {title} {\bibinfo {title} {{A New Mechanism for Baryogenesis}},\
  }\href {https://doi.org/10.1016/0550-3213(85)90021-5} {\bibfield  {journal}
  {\bibinfo  {journal} {Nucl. Phys. B}\ }\textbf {\bibinfo {volume} {249}},\
  \bibinfo {pages} {361} (\bibinfo {year} {1985})}\BibitemShut {NoStop}%
\bibitem [{\citenamefont {Domcke}\ \emph
  {et~al.}(2020{\natexlab{b}})\citenamefont {Domcke}, \citenamefont {Ema},
  \citenamefont {Mukaida},\ and\ \citenamefont {Yamada}}]{Domcke:2020kcp}%
  \BibitemOpen
  \bibfield  {author} {\bibinfo {author} {\bibfnamefont {V.}~\bibnamefont
  {Domcke}}, \bibinfo {author} {\bibfnamefont {Y.}~\bibnamefont {Ema}},
  \bibinfo {author} {\bibfnamefont {K.}~\bibnamefont {Mukaida}},\ and\ \bibinfo
  {author} {\bibfnamefont {M.}~\bibnamefont {Yamada}},\ }\bibfield  {title}
  {\bibinfo {title} {{Spontaneous Baryogenesis from Axions with Generic
  Couplings}},\ }\href {https://doi.org/10.1007/JHEP08(2020)096} {\bibfield
  {journal} {\bibinfo  {journal} {JHEP}\ }\textbf {\bibinfo {volume} {08}},\
  \bibinfo {pages} {096}},\ \Eprint {https://arxiv.org/abs/2006.03148}
  {arXiv:2006.03148 [hep-ph]} \BibitemShut {NoStop}%
\bibitem [{\citenamefont {Domcke}\ \emph
  {et~al.}(2023{\natexlab{b}})\citenamefont {Domcke}, \citenamefont {Kamada},
  \citenamefont {Mukaida}, \citenamefont {Schmitz},\ and\ \citenamefont
  {Yamada}}]{Domcke:2022kfs}%
  \BibitemOpen
  \bibfield  {author} {\bibinfo {author} {\bibfnamefont {V.}~\bibnamefont
  {Domcke}}, \bibinfo {author} {\bibfnamefont {K.}~\bibnamefont {Kamada}},
  \bibinfo {author} {\bibfnamefont {K.}~\bibnamefont {Mukaida}}, \bibinfo
  {author} {\bibfnamefont {K.}~\bibnamefont {Schmitz}},\ and\ \bibinfo {author}
  {\bibfnamefont {M.}~\bibnamefont {Yamada}},\ }\bibfield  {title} {\bibinfo
  {title} {{Wash-in leptogenesis after axion inflation}},\ }\href
  {https://doi.org/10.1007/JHEP01(2023)053} {\bibfield  {journal} {\bibinfo
  {journal} {JHEP}\ }\textbf {\bibinfo {volume} {01}},\ \bibinfo {pages}
  {053}},\ \Eprint {https://arxiv.org/abs/2210.06412} {arXiv:2210.06412
  [hep-ph]} \BibitemShut {NoStop}%
\bibitem [{\citenamefont {Barr}\ \emph {et~al.}(1979)\citenamefont {Barr},
  \citenamefont {Segre},\ and\ \citenamefont {Weldon}}]{Barr:1979ye}%
  \BibitemOpen
  \bibfield  {author} {\bibinfo {author} {\bibfnamefont {S.~M.}\ \bibnamefont
  {Barr}}, \bibinfo {author} {\bibfnamefont {G.}~\bibnamefont {Segre}},\ and\
  \bibinfo {author} {\bibfnamefont {H.~A.}\ \bibnamefont {Weldon}},\ }\bibfield
   {title} {\bibinfo {title} {{The Magnitude of the Cosmological Baryon
  Asymmetry}},\ }\href {https://doi.org/10.1103/PhysRevD.20.2494} {\bibfield
  {journal} {\bibinfo  {journal} {Phys. Rev. D}\ }\textbf {\bibinfo {volume}
  {20}},\ \bibinfo {pages} {2494} (\bibinfo {year} {1979})}\BibitemShut
  {NoStop}%
\bibitem [{\citenamefont {Nanopoulos}\ and\ \citenamefont
  {Weinberg}(1979)}]{Nanopoulos:1979gx}%
  \BibitemOpen
  \bibfield  {author} {\bibinfo {author} {\bibfnamefont {D.~V.}\ \bibnamefont
  {Nanopoulos}}\ and\ \bibinfo {author} {\bibfnamefont {S.}~\bibnamefont
  {Weinberg}},\ }\bibfield  {title} {\bibinfo {title} {{Mechanisms for
  Cosmological Baryon Production}},\ }\href
  {https://doi.org/10.1103/PhysRevD.20.2484} {\bibfield  {journal} {\bibinfo
  {journal} {Phys. Rev. D}\ }\textbf {\bibinfo {volume} {20}},\ \bibinfo
  {pages} {2484} (\bibinfo {year} {1979})}\BibitemShut {NoStop}%
\bibitem [{\citenamefont {Yildiz}\ and\ \citenamefont
  {Cox}(1980)}]{Yildiz:1979gx}%
  \BibitemOpen
  \bibfield  {author} {\bibinfo {author} {\bibfnamefont {A.}~\bibnamefont
  {Yildiz}}\ and\ \bibinfo {author} {\bibfnamefont {P.~H.}\ \bibnamefont
  {Cox}},\ }\bibfield  {title} {\bibinfo {title} {{Net Baryon Number, {CP}
  Violation With Unified Fields}},\ }\href
  {https://doi.org/10.1103/PhysRevD.21.906} {\bibfield  {journal} {\bibinfo
  {journal} {Phys. Rev. D}\ }\textbf {\bibinfo {volume} {21}},\ \bibinfo
  {pages} {906} (\bibinfo {year} {1980})}\BibitemShut {NoStop}%
\bibitem [{\citenamefont {Dreiner}\ and\ \citenamefont
  {Ross}(1993)}]{Dreiner:1992vm}%
  \BibitemOpen
  \bibfield  {author} {\bibinfo {author} {\bibfnamefont {H.~K.}\ \bibnamefont
  {Dreiner}}\ and\ \bibinfo {author} {\bibfnamefont {G.~G.}\ \bibnamefont
  {Ross}},\ }\bibfield  {title} {\bibinfo {title} {{Sphaleron erasure of
  primordial baryogenesis}},\ }\href
  {https://doi.org/10.1016/0550-3213(93)90579-E} {\bibfield  {journal}
  {\bibinfo  {journal} {Nucl. Phys. B}\ }\textbf {\bibinfo {volume} {410}},\
  \bibinfo {pages} {188} (\bibinfo {year} {1993})},\ \Eprint
  {https://arxiv.org/abs/hep-ph/9207221} {arXiv:hep-ph/9207221} \BibitemShut
  {NoStop}%
\bibitem [{\citenamefont {Kolb}\ \emph {et~al.}(1996)\citenamefont {Kolb},
  \citenamefont {Linde},\ and\ \citenamefont {Riotto}}]{Kolb:1996jt}%
  \BibitemOpen
  \bibfield  {author} {\bibinfo {author} {\bibfnamefont {E.~W.}\ \bibnamefont
  {Kolb}}, \bibinfo {author} {\bibfnamefont {A.~D.}\ \bibnamefont {Linde}},\
  and\ \bibinfo {author} {\bibfnamefont {A.}~\bibnamefont {Riotto}},\
  }\bibfield  {title} {\bibinfo {title} {{GUT baryogenesis after preheating}},\
  }\href {https://doi.org/10.1103/PhysRevLett.77.4290} {\bibfield  {journal}
  {\bibinfo  {journal} {Phys. Rev. Lett.}\ }\textbf {\bibinfo {volume} {77}},\
  \bibinfo {pages} {4290} (\bibinfo {year} {1996})},\ \Eprint
  {https://arxiv.org/abs/hep-ph/9606260} {arXiv:hep-ph/9606260} \BibitemShut
  {NoStop}%
\bibitem [{\citenamefont {Matsumoto}\ \emph {et~al.}(2022)\citenamefont
  {Matsumoto} \emph {et~al.}}]{Matsumoto:2022tlr}%
  \BibitemOpen
  \bibfield  {author} {\bibinfo {author} {\bibfnamefont {A.}~\bibnamefont
  {Matsumoto}} \emph {et~al.},\ }\bibfield  {title} {\bibinfo {title}
  {{EMPRESS. VIII. A New Determination of Primordial He Abundance with
  Extremely Metal-poor Galaxies: A Suggestion of the Lepton Asymmetry and
  Implications for the Hubble Tension}},\ }\href
  {https://doi.org/10.3847/1538-4357/ac9ea1} {\bibfield  {journal} {\bibinfo
  {journal} {Astrophys. J.}\ }\textbf {\bibinfo {volume} {941}},\ \bibinfo
  {pages} {167} (\bibinfo {year} {2022})},\ \Eprint
  {https://arxiv.org/abs/2203.09617} {arXiv:2203.09617 [astro-ph.CO]}
  \BibitemShut {NoStop}%
\bibitem [{\citenamefont {Gelmini}\ \emph {et~al.}(2020)\citenamefont
  {Gelmini}, \citenamefont {Kawasaki}, \citenamefont {Kusenko}, \citenamefont
  {Murai},\ and\ \citenamefont {Takhistov}}]{Gelmini:2020ekg}%
  \BibitemOpen
  \bibfield  {author} {\bibinfo {author} {\bibfnamefont {G.~B.}\ \bibnamefont
  {Gelmini}}, \bibinfo {author} {\bibfnamefont {M.}~\bibnamefont {Kawasaki}},
  \bibinfo {author} {\bibfnamefont {A.}~\bibnamefont {Kusenko}}, \bibinfo
  {author} {\bibfnamefont {K.}~\bibnamefont {Murai}},\ and\ \bibinfo {author}
  {\bibfnamefont {V.}~\bibnamefont {Takhistov}},\ }\bibfield  {title} {\bibinfo
  {title} {{Big Bang Nucleosynthesis constraints on sterile neutrino and lepton
  asymmetry of the Universe}},\ }\href
  {https://doi.org/10.1088/1475-7516/2020/09/051} {\bibfield  {journal}
  {\bibinfo  {journal} {JCAP}\ }\textbf {\bibinfo {volume} {09}},\ \bibinfo
  {pages} {051}},\ \Eprint {https://arxiv.org/abs/2005.06721} {arXiv:2005.06721
  [hep-ph]} \BibitemShut {NoStop}%
\bibitem [{\citenamefont {Barenboim}\ \emph {et~al.}(2017)\citenamefont
  {Barenboim}, \citenamefont {Kinney},\ and\ \citenamefont
  {Park}}]{Barenboim:2016shh}%
  \BibitemOpen
  \bibfield  {author} {\bibinfo {author} {\bibfnamefont {G.}~\bibnamefont
  {Barenboim}}, \bibinfo {author} {\bibfnamefont {W.~H.}\ \bibnamefont
  {Kinney}},\ and\ \bibinfo {author} {\bibfnamefont {W.-I.}\ \bibnamefont
  {Park}},\ }\bibfield  {title} {\bibinfo {title} {{Resurrection of large
  lepton number asymmetries from neutrino flavor oscillations}},\ }\href
  {https://doi.org/10.1103/PhysRevD.95.043506} {\bibfield  {journal} {\bibinfo
  {journal} {Phys. Rev. D}\ }\textbf {\bibinfo {volume} {95}},\ \bibinfo
  {pages} {043506} (\bibinfo {year} {2017})},\ \Eprint
  {https://arxiv.org/abs/1609.01584} {arXiv:1609.01584 [hep-ph]} \BibitemShut
  {NoStop}%
\bibitem [{\citenamefont {Gariazzo}\ \emph {et~al.}(2019)\citenamefont
  {Gariazzo}, \citenamefont {de~Salas},\ and\ \citenamefont
  {Pastor}}]{Gariazzo:2019gyi}%
  \BibitemOpen
  \bibfield  {author} {\bibinfo {author} {\bibfnamefont {S.}~\bibnamefont
  {Gariazzo}}, \bibinfo {author} {\bibfnamefont {P.~F.}\ \bibnamefont
  {de~Salas}},\ and\ \bibinfo {author} {\bibfnamefont {S.}~\bibnamefont
  {Pastor}},\ }\bibfield  {title} {\bibinfo {title} {{Thermalisation of sterile
  neutrinos in the early Universe in the 3+1 scheme with full mixing matrix}},\
  }\href {https://doi.org/10.1088/1475-7516/2019/07/014} {\bibfield  {journal}
  {\bibinfo  {journal} {JCAP}\ }\textbf {\bibinfo {volume} {07}},\ \bibinfo
  {pages} {014}},\ \Eprint {https://arxiv.org/abs/1905.11290} {arXiv:1905.11290
  [astro-ph.CO]} \BibitemShut {NoStop}%
\bibitem [{\citenamefont {Dolgov}(2002)}]{Dolgov:2002wy}%
  \BibitemOpen
  \bibfield  {author} {\bibinfo {author} {\bibfnamefont {A.~D.}\ \bibnamefont
  {Dolgov}},\ }\bibfield  {title} {\bibinfo {title} {{Neutrinos in
  cosmology}},\ }\href {https://doi.org/10.1016/S0370-1573(02)00139-4}
  {\bibfield  {journal} {\bibinfo  {journal} {Phys. Rept.}\ }\textbf {\bibinfo
  {volume} {370}},\ \bibinfo {pages} {333} (\bibinfo {year} {2002})},\ \Eprint
  {https://arxiv.org/abs/hep-ph/0202122} {arXiv:hep-ph/0202122} \BibitemShut
  {NoStop}%
\bibitem [{\citenamefont {Hannestad}\ and\ \citenamefont
  {Madsen}(1995)}]{Hannestad:1995rs}%
  \BibitemOpen
  \bibfield  {author} {\bibinfo {author} {\bibfnamefont {S.}~\bibnamefont
  {Hannestad}}\ and\ \bibinfo {author} {\bibfnamefont {J.}~\bibnamefont
  {Madsen}},\ }\bibfield  {title} {\bibinfo {title} {{Neutrino decoupling in
  the early universe}},\ }\href {https://doi.org/10.1103/PhysRevD.52.1764}
  {\bibfield  {journal} {\bibinfo  {journal} {Phys. Rev. D}\ }\textbf {\bibinfo
  {volume} {52}},\ \bibinfo {pages} {1764} (\bibinfo {year} {1995})},\ \Eprint
  {https://arxiv.org/abs/astro-ph/9506015} {arXiv:astro-ph/9506015}
  \BibitemShut {NoStop}%
\bibitem [{\citenamefont {Fradette}\ \emph {et~al.}(2019)\citenamefont
  {Fradette}, \citenamefont {Pospelov}, \citenamefont {Pradler},\ and\
  \citenamefont {Ritz}}]{Fradette:2018hhl}%
  \BibitemOpen
  \bibfield  {author} {\bibinfo {author} {\bibfnamefont {A.}~\bibnamefont
  {Fradette}}, \bibinfo {author} {\bibfnamefont {M.}~\bibnamefont {Pospelov}},
  \bibinfo {author} {\bibfnamefont {J.}~\bibnamefont {Pradler}},\ and\ \bibinfo
  {author} {\bibfnamefont {A.}~\bibnamefont {Ritz}},\ }\bibfield  {title}
  {\bibinfo {title} {{Cosmological beam dump: constraints on dark scalars mixed
  with the Higgs boson}},\ }\href {https://doi.org/10.1103/PhysRevD.99.075004}
  {\bibfield  {journal} {\bibinfo  {journal} {Phys. Rev. D}\ }\textbf {\bibinfo
  {volume} {99}},\ \bibinfo {pages} {075004} (\bibinfo {year} {2019})},\
  \Eprint {https://arxiv.org/abs/1812.07585} {arXiv:1812.07585 [hep-ph]}
  \BibitemShut {NoStop}%
\bibitem [{\citenamefont {Kreisch}\ \emph {et~al.}(2020)\citenamefont
  {Kreisch}, \citenamefont {Cyr-Racine},\ and\ \citenamefont
  {Dor\'e}}]{Kreisch:2019yzn}%
  \BibitemOpen
  \bibfield  {author} {\bibinfo {author} {\bibfnamefont {C.~D.}\ \bibnamefont
  {Kreisch}}, \bibinfo {author} {\bibfnamefont {F.-Y.}\ \bibnamefont
  {Cyr-Racine}},\ and\ \bibinfo {author} {\bibfnamefont {O.}~\bibnamefont
  {Dor\'e}},\ }\bibfield  {title} {\bibinfo {title} {{Neutrino puzzle:
  Anomalies, interactions, and cosmological tensions}},\ }\href
  {https://doi.org/10.1103/PhysRevD.101.123505} {\bibfield  {journal} {\bibinfo
   {journal} {Phys. Rev. D}\ }\textbf {\bibinfo {volume} {101}},\ \bibinfo
  {pages} {123505} (\bibinfo {year} {2020})},\ \Eprint
  {https://arxiv.org/abs/1902.00534} {arXiv:1902.00534 [astro-ph.CO]}
  \BibitemShut {NoStop}%
\bibitem [{\citenamefont {Dolgov}\ \emph {et~al.}(1997)\citenamefont {Dolgov},
  \citenamefont {Hansen},\ and\ \citenamefont {Semikoz}}]{Dolgov:1997mb}%
  \BibitemOpen
  \bibfield  {author} {\bibinfo {author} {\bibfnamefont {A.~D.}\ \bibnamefont
  {Dolgov}}, \bibinfo {author} {\bibfnamefont {S.~H.}\ \bibnamefont {Hansen}},\
  and\ \bibinfo {author} {\bibfnamefont {D.~V.}\ \bibnamefont {Semikoz}},\
  }\bibfield  {title} {\bibinfo {title} {{Nonequilibrium corrections to the
  spectra of massless neutrinos in the early universe}},\ }\href
  {https://doi.org/10.1016/S0550-3213(97)00479-3} {\bibfield  {journal}
  {\bibinfo  {journal} {Nucl. Phys. B}\ }\textbf {\bibinfo {volume} {503}},\
  \bibinfo {pages} {426} (\bibinfo {year} {1997})},\ \Eprint
  {https://arxiv.org/abs/hep-ph/9703315} {arXiv:hep-ph/9703315} \BibitemShut
  {NoStop}%
\bibitem [{\citenamefont {Gondolo}\ and\ \citenamefont
  {Gelmini}(1991)}]{Gondolo:1990dk}%
  \BibitemOpen
  \bibfield  {author} {\bibinfo {author} {\bibfnamefont {P.}~\bibnamefont
  {Gondolo}}\ and\ \bibinfo {author} {\bibfnamefont {G.}~\bibnamefont
  {Gelmini}},\ }\bibfield  {title} {\bibinfo {title} {{Cosmic abundances of
  stable particles: Improved analysis}},\ }\href
  {https://doi.org/10.1016/0550-3213(91)90438-4} {\bibfield  {journal}
  {\bibinfo  {journal} {Nucl. Phys. B}\ }\textbf {\bibinfo {volume} {360}},\
  \bibinfo {pages} {145} (\bibinfo {year} {1991})}\BibitemShut {NoStop}%
\bibitem [{\citenamefont {Edsjo}\ and\ \citenamefont
  {Gondolo}(1997)}]{Edsjo:1997bg}%
  \BibitemOpen
  \bibfield  {author} {\bibinfo {author} {\bibfnamefont {J.}~\bibnamefont
  {Edsjo}}\ and\ \bibinfo {author} {\bibfnamefont {P.}~\bibnamefont
  {Gondolo}},\ }\bibfield  {title} {\bibinfo {title} {{Neutralino relic density
  including coannihilations}},\ }\href
  {https://doi.org/10.1103/PhysRevD.56.1879} {\bibfield  {journal} {\bibinfo
  {journal} {Phys. Rev. D}\ }\textbf {\bibinfo {volume} {56}},\ \bibinfo
  {pages} {1879} (\bibinfo {year} {1997})},\ \Eprint
  {https://arxiv.org/abs/hep-ph/9704361} {arXiv:hep-ph/9704361} \BibitemShut
  {NoStop}%
\bibitem [{\citenamefont {Yueh}\ and\ \citenamefont
  {Buchler}(1976)}]{Yueh1976}%
  \BibitemOpen
  \bibfield  {author} {\bibinfo {author} {\bibfnamefont {W.~R.}\ \bibnamefont
  {Yueh}}\ and\ \bibinfo {author} {\bibfnamefont {J.~R.}\ \bibnamefont
  {Buchler}},\ }\bibfield  {title} {\bibinfo {title} {Scattering functions for
  neutrino transport},\ }\href {https://doi.org/10.1007/BF00648341} {\bibfield
  {journal} {\bibinfo  {journal} {Astrophysics and Space Science}\ }\textbf
  {\bibinfo {volume} {39}},\ \bibinfo {pages} {429} (\bibinfo {year}
  {1976})}\BibitemShut {NoStop}%
\end{thebibliography}%

\end{document}